\documentclass[a4paper,12pt,oneside]{book}

\pdfoutput=1

\usepackage{amsmath,amssymb,amsfonts,epsfig,cite,setspace,url,color,tikz}
\usepackage{hyperref,caption,setspace}
\usepackage{pdfpages,lipsum}
\usepackage{cancel}

\usepackage[english]{babel} 

\renewenvironment{thebibliography}[1]{%
\begin{oldthebibliography}{#1}%
\setlength{\parskip}{0ex}%
\setlength{\itemsep}{0ex}%
}%
{%
\end{oldthebibliography}%
}

\newcommand{\drawsquare}[2]{\hbox{%
\rule{#2pt}{#1pt}\hskip-#2pt
\rule{#1pt}{#2pt}\hskip-#1pt
\rule[#1pt]{#1pt}{#2pt}}\rule[#1pt]{#2pt}{#2pt}\hskip-#2pt
\rule{#2pt}{#1pt}}
\newcommand{\fund}{\raisebox{-.5pt}{\drawsquare{6.5}{0.4}}}
\newcommand{\antifund}{\overline{\fund}}
\newcommand{\gen}[1]{ \langle #1 \rangle}

\definecolor{purple}{cmyk}{0,0.8,0,0.4}
\definecolor{db}{cmyk}{1,0.1,0.2,0.6}
\definecolor{dg}{cmyk}{1,0,1,0.7}
\definecolor{bl}{cmyk}{1,0.7,0.5,0.2}
\definecolor{yl}{rgb}{0.2,0.7,0.2}
\definecolor{bl2}{cmyk}{0.7,0.4,0,0.5}
\definecolor{red2}{cmyk}{0,1,1,0.8}
\definecolor{red3}{cmyk}{0,0.7,1,0.7}
\definecolor{gr2}{cmyk}{1,0.2,0.7,0.6}
\definecolor{nb}{rgb}{0.1,0.1,0.5}
\definecolor{ng}{rgb}{0,0.8,0}
\definecolor{brown}{rgb}{0.6,0.3,0.2}
\definecolor{newred}{cmyk}{0,1,1,1}

\newcommand{\red}{\color{red}}
\newcommand{\blue}{\color{blue}}

\newcommand\spec{{\rm Spec}}

\makeatletter
\renewcommand{\@chapapp}{}
\newenvironment{chapquote}[2][2em]
  {\setlength{\@tempdima}{#1}%
   \def\chapquote@author{#2}%
   \parshape 1 \@tempdima \dimexpr\textwidth-2\@tempdima\relax%
   \itshape}
  {\par\normalfont\hfill--\ \chapquote@author\hspace*{\@tempdima}\par\bigskip}
\makeatother

\newcommand{\nn}{\nonumber}
\newcommand{\tr}{{\rm Tr}}
\newcommand{\rk}{{\rm rk}}
\newcommand{\ch}{{\rm Ch}}
\newcommand{\td}{{\rm Td}}

\newcommand{\comment}[1]{}

\newcommand{\acts}{\curvearrowright }

\newcommand{\cM}{{\cal M}}
\newcommand{\cW}{{\cal W}}
\newcommand{\cN}{{\cal N}}

\newcommand{\cJ}{{\cal J}}
\newcommand{\cK}{{\cal K}}

\newcommand{\cO}{{\cal O}}

\newcommand{\cB}{{\cal B}}
\newcommand{\cC}{{\cal C}}
\newcommand{\cD}{{\cal D}}
\newcommand{\cE}{{\cal E}}

\newcommand{\cG}{{\cal G}}
\newcommand{\cI}{{\cal I}}
\newcommand{\cQ}{{\cal Q}}

\newcommand{\cV}{{\cal C}}
\newcommand{\cT}{{\cal T}}

\newcommand{\IP}{\mathbb{P}}
\newcommand{\IQ}{\mathbb{Q}}

\newcommand{\IR}{\mathbb{R}}
\newcommand{\IC}{\mathbb{C}}
\newcommand{\IF}{\mathbb{F}}

\newcommand{\II}{\mathbb{I}}
\newcommand{\IZ}{\mathbb{Z}}

\newcommand{\im}{{\rm~Im}}

\newcommand{\diff}[2]{\frac{\partial #1}{\partial #2}}

\newcommand{\mat}[1]{\left(\begin{matrix} #1 \end{matrix}\right)}

\newtheorem{definition}{\bf DEFINITION}
\newtheorem{theorem}{\bf THEOREM}
\newtheorem{conjecture}{\bf CONJECTURE}

\newtheorem{proposition}{\bf PROPOSITION}
\newtheorem{corollary}{\bf COROLLARY}
\newtheorem{observation}{\bf OBSERVATION}
\newtheorem{question}{\bf QUESTION}

\newenvironment{dedication}
{
   \cleardoublepage
   \thispagestyle{empty}
   \vspace*{\stretch{1}}
   \hfill\begin{minipage}[t]{0.66\textwidth}
   \raggedright
}%
{
   \end{minipage}
   \vspace*{\stretch{3}}
   \clearpage
}

\begin{document}

\title{{\bf {\Huge  The Calabi-Yau Landscape: }\\
	 {\LARGE from Geometry, to Physics, to Machine-Learning}}}
\author{Yang-Hui He}
\frontmatter
\date{}
\maketitle

\renewcommand{\baselinestretch}{0.75}\normalsize
\tableofcontents
\renewcommand{\baselinestretch}{1.2}\normalsize

\begin{dedication}
{\it
Catherin\ae, Iacobo, \& Elizabeth\ae,\\
Parentibusque suis,\\
pro amore Catharinae, Sanctae Alexandriae,\\
lacrimarum Mariae semper Virginis, \\
et ad Maiorem Dei Gloriam \\
hoc opusculum dedicat auctor \ldots
}
\end{dedication}

\mainmatter

\chapter*{Preface}

\begin{chapquote}{Jean d'Alembert
\footnote{``Geometry, which should only obey Physics, when united with it sometimes commands it.'' Quote from d'Alembert's {\it Essay on a New Theory of Resistance of Fluids} ({\it Essai d'une nouvelle th\'eorie de la r\'esistance des fluides}).}
}
``La G\'eom\'etrie, qui ne doit qu'ob\'eir \`a la Physique quand elle se r\'eunit avec elle, lui commande quelquefois.''
\end{chapquote}

Algebraic and differential geometry have been the {\it point d'appui} between pure mathematics and fundamental physics in the twentieth century. The two pillars of modern physics: (1) general relativity of the theory of gravitation and (2) the gauge theory of the Standard Model of elementary particles, are the physical realization of, respectively, (1) metrics on Riemannian manifolds and (2) connections of bundles with Lie structure. This fruitful dialogue persisted through the parallel development of these two strands of natural philosophy and was made evermore pertinent within the context of string theory, conceived to unify gravity with quantum field theory, and thereby providing an answer to Einstein's famous dream of a Theory of Everything (ToE).

Whether string theory - with its cornerstones of supersymmetry and extra space-time dimensions yet to be verified by some ingenious experiment - is truly ``the'' theory of Nature remains to be seen. What is indisputable is that the theory has been a Muse to pure mathematics: from enumerative geometry to Moonshine, from quantum invariants to mock modular forms, etc., the discipline has provided novel, and rather unique, methods of attack on tackling a myriad of mathematical problems.

Central to string theory is the study of Calabi-Yau manifolds, serving as a beacon to such important investigations as compactification, mirror symmetry, moduli space and duality. Since their incipience in a conjecture of E.~Calabi in 1954-7 in an esteemed tradition in  geometry of relating curvature to topology, to the celebrated proof by S.-T.~Yau in 1977-8, to the prescription thereof as a large class of vacua in string theory by Candelas-Horowitz-Strominger-Witten in 1986, Calabi-Yau manifolds have been an inspiring protagonist in modern mathematics and theoretical physics.

There have, of course, been quite a few books already on the subject.
These range from the excellent account of the master himself in 
{\it The Shape of Inner Space: String Theory and the Geometry of the Universe's Hidden Dimensions}, by S.-T.~Yau and S.Nadis \cite{yaubook}, 
to the now classic 
{\it The Elegant Universe}, by B.~Green \cite{greenebook}, both aimed at the general audience; 
from the vintage compendium {\it Calabi-Yau Manifolds, a Bestiary for Physicists}, by T.~H\"ubsch \cite{hubschbook}, to the modern lecture notes 
{\it A survey of Calabi--Yau manifolds} by S.-T.~Yau \cite{yaubook2},
{\it Calabi-Yau Manifolds and Related Geometries} by M.~Gross, D.~Huybrechts, and D.~Joyce \cite{grossbook}, as well as 
{\it Calabi-Yau Varieties: Arithmetic, Geometry and Physics}, by R.~Laza, M.~Sch\"utt, and N.~Yui (Eds. \cite{yuibook}), aimed at the specialist.

\paragraph{Why a New Book: }
Why, then, the astute reader asks, another book on Calabi-Yau manifolds?

The primary reason is that there has been a revolution over the last score of years in Science, perhaps not a paradigm shift, but certainly a transformation in style and methodology: the twenty-first century now firmly resides within the age of big data and artificial intelligence. It is therefore inevitable that mathematical physics and pure mathematics should too profit from this burgeoning enterprise. Whilst it is well-established that string theory and Calabi-Yau manifolds have been at the interface between theoretical physics and pure mathematics from the 1980s, it is less known that since around the turn of the millennium, they have also served as a bench-mark to various problems in computational mathematics as well as a passionate advocate for data-mining. 

Specifically, the author, as well as many other groups in string theory, enjoyed long collaborations with experts in large-scale computational projects, e.g., {\sf Macaulay2}, {\sf Singular}, {\sf Bertini} in algebraic geometry, {\sf GAP} in group theory, {\sf MAGMA} or {\sf PARI/GP} in number theory, and indeed the umbrella scheme of {\sf SageMath}. Furthermore, even before the establishment of such online databases such as
\begin{enumerate} 
\item ``the database of L-functions, modular forms and related objects''\\
	\url{http://www.lmfdb.org/} 
\item ``the Graded Ring Database of algebraic varieties''\\
	\url{http://www.grdb.co.uk/}
\item ``the Knot atlas'' \url{http://katlas.org/}, etc.,
\end{enumerate}
which have become standard in the mathematics literature over the last two decades or so, the now-famous 473 million hypersurfaces in toric varieties from reflexive 4-polytopes and the resulting Calabi-Yau database was born from Kreuzer-Skarke in the mid-1990’s. 
This impressive resource still remains one of the largest datasets in mathematics:\\
\url{http://hep.itp.tuwien.ac.at/~kreuzer/CY/} 

Contemporaneous to these, there have emerged an increasing number of online Calabi-Yau databases, of varying specialization and often of augmented sophistication in user interface, such as the ``Toric Calabi-Yau Database'' \cite{rossaltman}, the ``Heterotic Compactification Database'' \cite{andrelukas}, and the ``Calabi-Yau Manifold Explorer'' \cite{benjurke}, set up by various international collaboration of which the author has been a part, 
as well as ``The Calabi-Yau Operator Database'' \cite{CY-PF}, elliptically fibred Calabi-Yau manifolds \cite{CY-ell},
etc.
On a more general note, we point the reader to the ambitious and inspiring resource\\
\url{https://mathdb.mathhub.info/}\\
which attempts to link to all the various databases of pure mathematics the various communities have been compiling. 
In addition, the reader is encouraged to read the enlightening transcripts of the ICM 2018 panel discussion on machine-aided proofs and databases in mathematics \cite{icm2018}.

\paragraph{Why This Book: }
It is therefore evident that a new book on Calabi-Yau manifolds, and more widely, on algebraic geometry, is very much in demand. 
Such a book should emphasize that the subject is not only at the intersection between mathematics and physics, but resides, in fact, at the cusp of pure and applied mathematics, theoretical physics, computer science and data science. 
This book should briefly introduce the mathematical background and the physical motivation, brief because the material can be found in many textbooks. It should then delve into the heart of the matter, addressing 
\begin{enumerate}
\item how does one explicitly construct a Calabi-Yau manifold; 
\item how do these constructions lead to problems in classification, ranging from the combinatorics of lattice polytopes to the representation of quivers and finite graphs; 
\item what databases have thus far been established and 
\item how to use a combination of analytic techniques and available software to compute requisite quantities, such as Hodge numbers and bundle cohomology
\end{enumerate}
Throughout, algorithms and computerization will serve as the skein which threads the material, with an appreciation of their complexity and the indefatigability with which the community has utilized them to compile databases, in order to extract new mathematics and new physics.
In this sense, the book will focus not on the {\it theory} but the {\it practice} of Calabi-Yau manifolds.

\paragraph{The Calabi-Yau Landscape and Beyond: }
There have been many practitioners in mapping the landscape of Calabi-Yau manifolds throughout the years, exploring diverse features of interest to physicists and mathematicians alike,  such as the founders  S.-T.~Yau in mathematics and P.~Candelas in physics, as well as M.~Stillman and D.~Grayson in computational geometry.
To the fruitful collaboration with and sagacious guidance by these experts over the years the author is most grateful.

More recently, in 2017-8, a somewhat daring paradigm was proposed in \cite{He:2017aed,He:2017set}. 
A natural question was posed: confronted with the increasingly available data on various geometric properties of Calabi-Yau varieties which took many tour de force efforts to calculate and compile, can machine-learning be applied to Calabi-Yau data, and to data in pure mathematics in general?  Indeed, 
\begin{quote}
{\it
Can artificial intelligence ``learn'' algebraic geometry?
}
\end{quote}
That is, can machine-learning arrive at correct answers such as cohomology or topological invariant, {\it without} understanding the nuances of the key techniques of the trade, such as Groebner bases, resolutions or long exact sequences?
What about representation theory of groups or arithmetic-geometric problems in number theory? 
This was shown to be mysteriously possible in many cases to astounding accuracy. 
There had been some activity in the field since 2017 when \cite{He:2017aed,He:2017set,Krefl:2017yox,Ruehle:2017mzq,Carifio:2017bov} introduced modern data-science techniques to string theory, in applying big-data techniques to various problems in string and related mathematics by a host of independent groups.
The Zeistgeist of data science has thus breathed its nurturing spirit into the field of mathematical physics.

Indeed, the question can be extended to \cite{He:2017aed,mytalks} 
\begin{quote}
{\it Can AI machine-learn mathematical structures?}
\end{quote}
Such an inquiry should be taken in tandem with, but not commensurate to, the dreams of Voevodsky \cite{voevodsky} and the very active field of automated theorem proving (ATP) \cite{ATP}.
The latter has a long history in the foundation of mathematics since at least Russell and Whitehead and attempts to formulate the axiomata of mathematics in a systematic way, before the programme took a blow from G\"odel. Yet, for all practical purposes, the ATP community is thriving, to the point that Prof.~Kevin Buzzard thinks it is impossible not to have computers check proofs and generate new ones \cite{buzzard}. One could think of this direction as a ``bottom-up'' approach toward the AI mathematician. What we are proposing here can be thought of as a complementary  ``top-down'' approach, where, given enough mathematically correct statements from the literature, ML extracts enough information to find new patterns and methodology.

\paragraph{Target Audience: }
This book has grown out of a timely series of invited colloquia, seminars and lectures delivered by the author in the 2017-8 academic year, at
the National Centre for Science and Technology, Taiwan; Northeastern University, Boston; Ludwig-Maximilians-Universit\"at, Munich; Imperial College, London; L'Institut Confucius, Geneva; University of Warwick, UK; Nankai and Tianjin Universities, China; University of Plymouth, UK; Universities of Bruxelles and Leuven, Belgium; University of Nottingham, UK; Albert Einstein Institute, Potsdam; Niels Bohr Institute, Copenhagen; University of the Witwatersrand, Johannesburg; Trinity College, Dublin; Brown University, Providence;  Harvard University, Boston  and University of Pennsylvania, Philadelphia.
To the kind hospitality of the various hosts he is most grateful. 

One long version of the lecture slides, of which this book is a pedagogical expansion, is the lecture series given at L'Institut Confucius, Geneva, the PDFs of which are available at \cite{longSlides}.
A short version, on the other hand, is available from the abridged lectures notes from the 2019 PIMS Summer School
on Algebraic Geometry in High-Energy Physics at the University of Saskatchewan \cite{Bao:2020sqg} and the encyclopaedia entry \cite{He:2020bfv}.
The audience is intended to be primarily beginning graduate students in mathematics, as well as those in mathematical and theoretical physics, all with a propensity towards computers. We will take a data-driven perspective, sprinkled with history, anecdotes, and actual code. 
The reader is also referred to a recent attempt to gather the experts' voices in a collected volume on topology and physics.
\cite{Yang:2019kup}.

In a sense, the list of available books mentioned earlier is aimed either at a general audience or to a more senior graduate student.
The middle-ground of beginning PhD students, master students, or even advanced undergraduate students is left wanting.
Students with some familiarity of Reid's undergraduate algebraic geometry \cite{reid} but not yet fully prepared for Hartshorne \cite{hartshorne} and Griffiths-Harris \cite{GH}, or with some acquaintance of Artin's introductory algebra \cite{artin} but not yet fully feathered  for Fulton-Harris \cite{fulton}, may find it difficult to approach the subject matter of Calabi-Yau manifolds.
For them this writing is designed.

\paragraph{Disclaimer: }
In a way, the purpose of this book is to use Calabi-Yau databases as a playground, and explore aspects of computational algebraic geometry and data science, in the spirit of the excellent texts of Cox-Little-O'Shea \cite{CLO} and Schenck \cite{m2book}.
This book is {\it not} intended to be an introductory text in string theory, nor one in algebraic geometry.
It should service as a supplement to the aforementioned host of physics and mathematics textbooks, and to entertain the beginning researcher with a glimpse of the interactions between some modern mathematics and physics, as well as an invitation to machine-learning. Some undergraduate familiarity with manifolds and topology at the level of \cite{nakahara,reid}, as well as string theory at the level of \cite{zwiebach,kiritsis}, all excellent introductions, will more than suffice as background. 

Perhaps, due to the author's constant enjoyment as an outsider, the book is intended to be some mathematics for physicists (what are some of the techniques?), some physics for mathematicians (why are physicists interested in complex algebraic geometry?), and some data science for both (what are some of the fun and modern methods in machine-learning?).
Throughout our promenade in the landscape, there will be a continued preference of intuition over rigour, computation over formalism, and experimentation over sophistry.
It is the author's hope that we shall embark on a memorable adventure in the land of mathematics, physics and data-science, with Calabi-Yau manifolds leading as our Beatrice, and Interdisciplinarity guiding as our Virgil, and we, like Dante, diligently remarking upon our journey.

\begin{flushright}
Yang-Hui He\\
Merton College, Oxford
\end{flushright}

\newpage
\chapter*{Acknowledgements}
To the invaluable guidance and help of my mentors, official and unofficial, I owe eternal gratitude.
Each has been a guiding light at various stages of my career: Professor Amihay Hanany, Professor Burt Ovrut, Professor Philip Candelas, Professor Dame Jessica Rawson, Professor Andre Lukas, Professor Alex Scott, Professor Shing-Tung Yau, Professor Mo-Lin Ge, and Professor Minhyong Kim.
To them I raise my cup of thanksgiving.

To the host of collaborators whose friendship I have enjoyed over the years, I owe a million happy thoughts.
Their camaraderie has always offered a beacon of hope  even within the darkest moments of confusion.
To Dr.~Pierre-Philippe D\'echant, Dr.~Richard Eager, Dr.~R\'emi Lodh, Prof.~Tristan H\"ubsch, Dr.~Stephen Pietromonaco, Prof.~Hal Schenck, as well as the anonymous referees of Springer,  I am grateful for their careful reading and many valuable comments to the first draft.
To them I raise my glass of brotherly love.

To my wife, Dr.~Elizabeth Hunter He, {\it rosa Hibernia dilectissima mea}, without whose patience and support this book would not have come into being.
To my children, Katherine Kai-Lin and James Kai-Hong, without whose distractions this book would also not have come into being.
To my parents, without whose love and nurturing nothing much of mine would have ever come into being.

\chapter{Prologus Terr\ae\ Sanct\ae}

\begin{chapquote}{Burchardus de Monte Sion \footnote{
``Thus, as promised, let us now turn, to the description of the Holy Land.''
Burchard of Mount Sion, from the Prologue to his {\it Description of the Holy Land}
({\it Prologus \& Descriptio Terrae Sanctae}, 1285)
}
}
``C{\ae}terum ad veridicam, sicut iam polliciti sumus, Terr{\ae} Sanct{\ae} descriptionem, stylum vertamus.''
\end{chapquote}

That Zeitgeist should exist in the realm of scientific inquiry is perhaps more surprising than its purported influence in the growth of civilizations.
Whilst the latter could be conceived of as the emergence of vague concoction of cross-cultural ideas, the former requires precise concepts and objects to almost simultaneously appear, oftentimes independently and even mysteriously, within different disciplines of research.
In mathematics, what initially surface as ``Correspondences'' in seemingly disparate sub-fields necessitates a timely arrival of hitherto unthinkable technique for the final construction of a proof.
An archetypal example of this is {\em Moonshine}, a list of daring conjectures  by McKay-Thompson-Conway-Norton in the late 1980s relating finite groups to modular forms, which until then lived in utterly different worlds.
It so happened that Borcherds was learning quantum field theory (QFT) with conformal symmetry - yet another world - around this time, when he noted that the structure of its vertex operator algebra furnished the requisite representation and achieved his Fields medal winning proof of 1992.
The right set of ideas was indeed ``in the air''.

The involvement of physics was perhaps as surprising as the initial correspondence.
But herein lies the crux of our tale.
The past century has witnessed the cross-fertilization between mathematics and theoretical physics in a manner worthy of that illustrious tradition since Newton:  Riemannian geometry and general relativity, operator theory and quantum mechanics, fiber bundles and gauge theory of elementary particles, etc.
An impressive list of theories, such as the likes of mirror symmetry, supersymmetric indices, and Moonshine, have been the brain-children of this fruitful union.
It indeed seems that QFTs with additional symmetry, specifically, conformal symmetry and supersymmetry, is the right set of ideas for physics.

The ultimate manifestation of QFTs with such rich symmetries is, without doubt, string theory.
Marrying the conformal symmetries of the world-sheet with the supersymmetry of incorporated fermions, string theory materialized in the 1970-80s as an attempt to unify gravity as described by general relativity with quantum theory as described by the gauge theory of the Standard Model (SM) into a Theory of Everything, the holy grail of fundamental physics.
The theory still remains the best candidate for a hope of reconciling the macroscopic world of stars and galaxies with the microscopic world of elementary particles of the SM, which would in turn demystify singularities such as black-holes or the Big Bang.

To date, there is no experimental evidence for string theory, none at all for its cornerstones of supersymmetry and extra space-time dimensions, nor is it obvious that definitive signatures of either would be discovered in the near future: we must patiently await some ingenious experiment in some marvelous indirect observation.
Why, then, string theory?
While we refer the reader to the excellent response in J.~Conlon's eponymous book \cite{conlonbook}, as mathematicians  and mathematical physicists our affinity to the theory is already more than justified.
As mentioned above, the theory has not only made revolutionary contributions to theoretical physics in the guise of holography, QFT dualities and quantum entanglement, but furthermore, in an unexpected manner, it has been serving as a {\it constant Muse to pure mathematics}.

Time and again, its general framework has provided unexpected insights and results to geometry, representation theory and even number theory.
Hence, regardless of whether the theory will eventually become the unified theory of everything, its pursuit, now, in the past decades, and in foreseeable future, is just and fitting.
String theory is thus the rightful heir of that noble lineage of the interaction between geometry, algebra and physics which took a firm footing in the twentieth century, it is certainly not a fin-de-si\`ecle fancy, but has rather transitioned us, as a perspicacious mentor, into the mathematics and physics of the twenty-first century.
Central to this development is the study of {\bf Calabi-Yau varieties},  whose curvature flatness endows them with beautiful mathematical structure and physical utility, rendering them the perfect illustration of this interaction.

As we now ground ourselves into the incipience of the twenty-first century, it behoves us to contemplate as to whether a particular spirit permeates our Time.
The answer is immediate: this is the Age of Data.

From the sequencing of the Human genome, to the sifting of particle jets in the LHC, to the search for exo-planets, the beginning of this century has amassed, utilized and analysed data on an unprecedented scale, due to the explosive growth and availability of computing power and artificial intelligence (AI).
Whilst it is well known that string theory has long been a Physical Beatrice to the Pure Mathematical Dante, it is perhaps less appreciated that over the past three decades they too have repeatedly provided benchmarks to computational mathematics, especially in computational algebraic geometry and combinatorial geometry. 

The story, as we will shall see, goes back to the late 1980s, reaching its first climax in the mid-1990s: interestingly, the now classic complete intersection Calabi-Yau (CICY) \cite{cicy} and Kreuzer-Skarke (KS) \cite{Avram:1997rs} datasets place ``big data'' in pure mathematics and mathematical physics to a time ordinarily thought to predate the practice of data-mining.
More recently, a steadily expanding group of researchers in mathematical physics, have been advocating to the physics community (q.v.~\cite{bookComp}) the importance of interacting with
{\sf Macaulay2} \cite{m2}, 
{\sf Singular} \cite{singular}, 
{\sf Bertini} \cite{bertini} in algebraic geometry, 
{\sf GAP} \cite{gap} in group theory, 
{\sf MAGMA} \cite{magma} in number theory etc., 
as well as the overarching {\sf SageMath} \cite{sage}. 
Indeed, the first sessions in theoretical high-energy physics \cite{siam} in the biannual ``Society for Industrial and Applied Mathematics'' (SIAM) meetings on applied algebraic geometry have emerged.

The protagonist of our chronicle is therefore engendered from the union of mathematics, physics and data science and it is this trinity that will be the theme of this book.
Over the years, various fruitful collaborations between theoretical physicists, pure and applied mathematicians, and now data scientists and computer scientists, have been constructing and compiling databases of objects which may {\it ab initio} seem to be only of abstract mathematical interest but instead links so many diverse areas of research: Calabi-Yau manifolds constitute the exemplar of this paradigm.
Thus, as much as our ethereal Zeitgeist being {\it Data}, it too should be {\it Interdisciplinarity}.

By now, there is a landscape of data and a plethora of analytic techniques as well as computer algorithms, freely available, but lacking a comprehensive roadmap. 
This is our {\bf Calabi-Yau Landscape}.
There are still countless goldmines buried in this landscape, with new mathematical conjectures laying hidden (e.g,~the vast majority of the manifolds obtainable from the KS dataset by triangulations  has been untouched \cite{Altman:2014bfa}) and new physics to be uncovered (e.g.~the unexpected emergence of Calabi-Yau structure in the scattering of ordinary $\phi^4$ QFTs \cite{Bourjaily:2018ycu,Bourjaily:2018yfy}).

Europe's obsession with Jerusalem in the Middle Ages led to the publication of one of the very first printed maps, based on Burchardus de Monte Sion's thirteenth century ``Prologue of the Holy Land'' \cite{prologus}.
We have thus named our prologue in homage to this early cartographical work, drawing upon the analogy that we too will be mapping out the landscape of an important terrain.

\section{A Geometrical Tradition}\label{s:geo}

\begin{figure}[h!!!]
\[
\begin{array}{|c|c|c|}\hline
\begin{array}{l}\includegraphics[width=0.5in,angle=0]{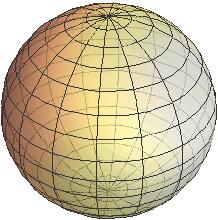}\end{array}
&
\begin{array}{l}\includegraphics[width=1in,angle=0]{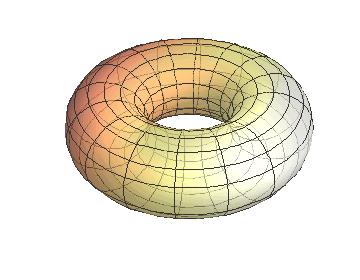}\end{array}
&
\begin{array}{cc}
&\\
\begin{array}{l}\includegraphics[width=1.2in,angle=0]{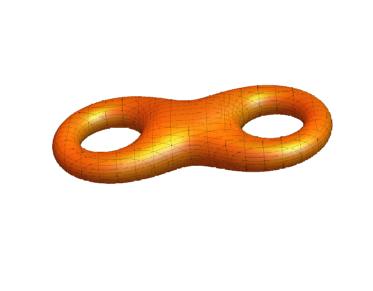}\end{array}&
\begin{array}{l}\includegraphics[width=1.2in,angle=0]{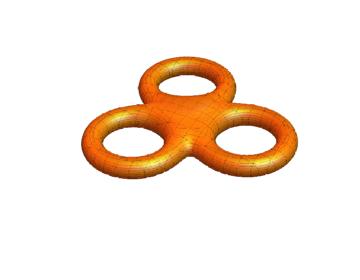}\end{array}\\
\begin{array}{l}\includegraphics[width=1.2in,angle=0]{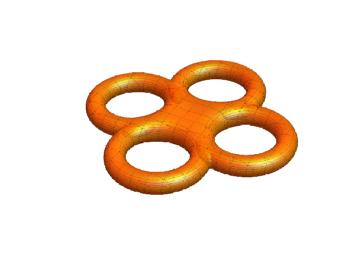}\end{array}&
\ldots \ldots
\end{array}
\\ \hline
g(\Sigma) = 0 & g(\Sigma) = 1 & g(\Sigma) > 1\\ \hline
\chi(\Sigma) = 2 & \chi(\Sigma) = 0 & \chi(\Sigma) < 0\\ \hline
\mbox{Spherical} & \mbox{Ricci-Flat: CY$_1$} & \mbox{Hyperbolic} \\ \hline
+ \mbox{ curvature} & 0 \mbox{ curvature} & - \mbox{ curvature}
\\ \hline
\end{array}
\]
\caption{{\sf
The trichotomy of Riemann surfaces $\Sigma$ of genus $g(\Sigma)$, in terms of its Euler number $\chi(\Sigma) = 2-2g(\Sigma)$.
}
\label{f:RiemannSurface}}
\end{figure}
\comment{
building g > 1

torusImplicit[{x_, y_, z_}, R_, r_] = (x^2 + y^2 + z^2)^2 - 
   2 (R^2 + r^2) (x^2 + y^2) + 2 (R^2 - r^2) z^2 + (R^2 - r^2)^2;

build[n_] := 
  Module[{f, cp, polys, cartPolys, cartPolys1},(*implicit polynomial*)
   f = Product[
      torusImplicit[{x - 1.5 Cos[i 2 Pi/n], y - 1.5 Sin[i 2 Pi/n], z},
        1, 1/4], {i, 0, n - 1}] - 10;
   cp = ContourPlot3D[
     Evaluate[f == 0], {x, -3, 3}, {y, -3, 3}, {z, -1, 1}, 
     BoxRatios -> Automatic, PlotPoints -> 35, 
     MeshStyle -> Opacity[.2], 
     ContourStyle -> 
      Directive[Orange, Opacity[0.8], Specularity[White, 30]], 
     Boxed -> False, Axes -> False]];
}

Our story, as many in mathematics, begins with Euler, Gau\ss, and Riemann.
The classic classification theorem of (compact, smooth, orientable) surfaces $\Sigma$ (without boundary and punctures) is well known.
These two-dimensional objects - smooth manifolds if you will - are topologically characterized by a single non-negative integer, the {\red genus}, which counts the number of holes.
This is shown in the top line of Figure \ref{f:RiemannSurface}.
The sphere $S^2$ is genus 0, the torus $T^2 \simeq S^1 \times S^1$ is genus 1, etc.
One says that this integer genus distinguishes the {\blue Topological type} of the surface.

The {\bf {\red Euler characteristic}} (or simply, Euler number) $\chi$ relates topology to combinatorics: take any convex polyhedron drawn on $\Sigma$, and find the signed alternating sum of the number of independent objects in each dimension, this should yield
\begin{equation}
\chi(\Sigma) = 2 - 2g(\Sigma) \ .
\end{equation}
For example, on $S^2$ where $g=0$, we can draw a cube, there are 8 vertices, 12 edges and 6 faces, and $8 - 12 + 6 = 2$;
this should be true for any convex polyhedron, a result dating at least to Euler, which is a precursor to the index theorem.
More strictly speaking, ``number of independent objects in each dimension'' is given by the {\blue Betti number} $b_i$ which counts the number of independent algebraic cycles in dimension $i$, which we will later see as the rank of the $i$-th (co-)homology group (suitably defined). Thus,
\begin{equation}
\chi(\Sigma) = \sum\limits_{i=0}^2 b_i \ , \quad b_i = \rk(H_i(\Sigma)) \ .
\end{equation}

Next, Gau\ss-Bonnet relates topology to differential geometry: $\chi$ should be the integral of (an appropriate) curvature $R$, here the Ricci curvature:
\begin{equation}\label{GB}
\chi(\Sigma) = \frac{1}{2\pi} \int_{\Sigma} R \ .
\end{equation}
Riemann Uniformization Theorem then gives us a wonderful trichotomy:
\begin{itemize}
\item $\chi(\Sigma) > 0$: spherical geometry $g(\Sigma)=0$ where the metric is of (constant) positive curvature;
\item $\chi(\Sigma) = 0$: flat/torus geometry $g(\Sigma)=1$ where the metric is Ricci flat;
\item $\chi(\Sigma) < 0$: hyperbolic geometry $g(\Sigma)>1$ where the metric is of negative curvature.
\end{itemize}
In particular, every $\Sigma$ admits one such constant curvature by (conformal) scaling.

We summarize the above discussion into a chain of equalities:
\begin{equation}
\chi(\Sigma) = 2-2g(\Sigma) =
\frac{1}{2\pi}\int_\Sigma R = 
 [c_1(\Sigma)] \cdot [\Sigma] = 
\sum\limits_{i = 0}^{\dim_{\IR} \Sigma = 2} (-1)^i b^i (\Sigma) \ .
\end{equation}
As we proceed from left to right, the first equality is combinatorics and topology; the second relates $\chi$ to analysis and differential geometry; the third rewrites the integral as intersection (of cycles) in algebraic geometry; and the last, to (co-)homological algebra.

It is fair to say that a significant portion of modern geometry is devoted to the generalization of this beautiful story in dimension 2.
This is, of course, extremely difficult; high dimensional geometry and topology often have unpredicted properties.
The Hopf Conjecture -- that for a compact Riemannian manifold $M$ of dimension $2n$,  positive (sectional) curvature means positive $\chi$ and negative (sectional) curvature means Sign$(\chi) = (-1)^n$ -- is one of the immediate generalizations to Figure \ref{f:RiemannSurface}.
For $n=1$, we are back to the case of $\Sigma$ and for $n=2$ it has also been settled by Chern-Gau\ss-Bonnet, which generalizes \eqref{GB} to $\chi(M) = \frac{1}{(2\pi)^n} \int_M \mbox{Pf}(\Omega)$, the integral of the Pfaffian of the curvature form $\Omega$ to the Levi-Civita connection (and for $n=2$ reducing to $\chi = \frac{1}{128 \pi^2}\int R_{abcd} R_{efgh} \epsilon^{abdf} \epsilon^{cdgh}$ for the Riemann curvature tensor $R_{abcd}$).
However, for $n>2$, this is still an open problem \cite{Hopf}!

By imposing extra symmetries problems in mathematics and physics typically become easier to manage.
If arbitrary high-dimensional manifolds are difficult to tame, what about sub-classes with further structures?
Indeed, for orientable surfaces such as $\Sigma$, one can impose a complex structure, much like rendering $\IR^2$ into the complex plane $\IC$ by adding $i = \sqrt{-1}$. 
Thus equipped, $\Sigma$ becomes a complex curve (real dimension 2 becomes complex dimension 1).
In other words, 2-manifolds can be complexified to 1-folds: these are the {\blue Riemann Surface} $\Sigma$ considered as complex curves.
Note that the standard though perhaps confusing appellation is that (complex) curves are manifolds of real dimension 2, i.e., surfaces; likewise, a (complex) surface is a real 4-dimensional manifold.
In order to avoid confusion, the widely used convention is that
\begin{itemize}
\item an {\red n-manifold} refers to a manifold of real dimension $n$;
\item an {\red n-fold} refers to one of complex dimension $n$.
\end{itemize}
A cultural distinction is that differential geometers use (1) more widely, and algebraic geometers, (2).

In fact, $\Sigma$ is furthermore K\''ahler, in that its local metric is the double derivative of a scalar potential function.
We leave a brief recapitulation, to Appendix \ref{ap:geo}, of the rudiments and notation of a successive specialization from real manifolds to complex, to Hermitian, and to K\"ahler geometry, as well as some key results from bundle and (co-)homology which we will use.
One advantage of complexification, as we will see in the next chapter, is that we can realize the manifold as vanishing loci of polynomials in complex projective space.
In this land of (complex) {\it algebraic} geometry (algebraic because we are dealing with polynomials) in projective space, everything is automatically compact and K\''ahler.

\subsection{A Modern Breakthrough}
Within the world of K\''ahler geometry, there are sufficient symmetries that a powerful statement exists for all dimensions.
In the mid-1950s, E.~Calabi made his famous conjecture \cite{calabi}
\begin{conjecture} [Calabi Conjecture]
Let $(M,g,\omega)$ be a compact K\"ahler manifold and $R$ a $(1,1)$-form such that $[R] = [c_1(M)]$.
Then $\exists \ !$ K\"ahler metric $\tilde{g}$ and K\"ahler form $\tilde{\omega}$ such that $[\omega] = [\tilde{\omega}] \in H^2(M; \IR)$ and that
$R = Ric(\tilde{\omega})$, the Ricci form of $\tilde{\omega}$.  
\end{conjecture}

This is quite a mouthful! And indeed, it is not the purpose of this prologue, nor of this book, to delve into the subtleties.
Let us first slightly unpackage the statement.
As mentioned earlier for complex dimension 1, the K\"ahler $n$-fold $M$ has a metric of the form
\begin{equation}
g_{\mu\bar{\nu}} = \partial \bar{\partial} K(z^\mu,\bar{z}^{\bar{\nu}})
\end{equation}
for local coordinates $z^{\mu = 1, \ldots, n}$ where $K$ is a scalar function called the  K\"ahler potential.
The  K\"ahler form is a 2-form $\omega := i g_{\mu\bar{\nu}} dz^{\mu} \wedge dz^{\bar{\nu}}$ and the existence of the potential is is equivalent to the closedness of $\omega$, i.e., $d \omega = 0$.
Immediately, we see why complex 1-folds are K\"ahler because the differential of a 2-form on a 2-dimensional manifold automatically vanishes.

The Chern classes $c_{k}(M) = c_k(T_M) \in H^{2k}(M)$, for K\"ahler manifolds, have an important property: the Ricci $(1,1)$-form, the 
K\"ahler $(1,1)$-form and $c_1(M)$, as real 2-forms, all reside in $H^2(M;\IR)$.
The content of the conjecture is that the Chern class determines, in a unique way, the behaviour of the curvature, much like Chern-Gau\ss-Bonnet mentioned above.
In fact, uniqueness was already shown by Calabi and it was the existence of $(\tilde{g}, \tilde{\omega})$ that remained the challenge.

Furthermore, we saw from Figure \ref{f:RiemannSurface}, that the genus 1 case acts as a boundary case in the trichotomy and this pattern of finite number of topologies for positive curvature and zero curvature is part of a bigger conjecture which we will discuss later in the book.
Here, in the Calabi conjecture, the case when $R = 0$ (Ricci-flatness) is also obviously important, where it means that for K\"ahler $M$, unique existence of Ricci-flat K\"ahler metric is equivalent to the vanishing of the first Chern class $c_1(M)$.
The special case of {\blue K\"ahler-Einstein} manifolds, i.e., when $R = \lambda g$, where the Ricci form is proportional to the metric form for some $\lambda \in \IR$ best illustrates the situation.
Here, the natural trichotomy: $\lambda >0$ (ample, Fano), $\lambda = 0$ (Ricci-flat) and $\lambda < 0$ (anti-ample, or general type) is the desire generalization to Figure \ref{f:RiemannSurface}.

From a computational perspective, to appreciate the power of the statement, suppose we wished to go hardcore and tried to find an explicit metric in some local coordinate. Even with the help of extra properties such as K\"ahler, the Ricc-flat equations are a set of complicated non-linear PDEs in the components of the metric for which there is little chance of a hope for analytic solution.
We are now at liberty to only check an algebraic/topological quantity, the first Chern class, many examples of which we shall compute throughout the book, would govern the Ricci curvature. In particular, its vanishing would guarantee the uniqueness and existence of a Ricci-flat metric.

The existence part of the Calabi Conjecture was finally settled by S.-T.~Yau \cite{yau}, some twenty years later:
\begin{theorem} [Yau]
The Calabi Conjecture holds.
\end{theorem}
In particular, the case of $R=0$ is now called {\red Calabi-Yau} manifolds in honour of this achievement, which uses ingenious existence methods for the solution of a Monge-Amp\`ere PDE and for which Yau received the Fields Medal in 1982.

It should be pointed out the proof is an existence proof and to date {\it no explicit Ricci-flat K\"ahler metric has been found on any compact Calabi-Yau manifold} (beyond the trivial case of products of tori) and this remains an important open problem.
Importantly, before proceeding, the reader is pointed to the survey \cite{yaubook2}, which gives a fantastic bird's-eye-view of the historical developments and well most of the active areas of research related to Calabi-Yau manifolds.

Henceforth, we will use the acronym {\red CY$_n$} to mean a Calabi-Yau $n$-fold, a Ricci-flat K\"ahler manifold of complex dimension $n$.

\subsection{Preliminary Examples: $1,2,? \ldots$}
We will give many explicit examples later in the book, but for now, in order to reify the above concepts, let us present some preliminary though extremely important examples of Calabi-Yau manifolds.
We saw that Riemann surfaces of real dimension 2 are complexifiable to K\"ahler 1-folds.
Indeed, the torus $T^2 \simeq (S^1)^2$ case of $g(\Sigma)=1$ is the {\it only} compact, smooth, Ricci-flat, K\"ahler manifold of complex dimension one.
That is,
\begin{proposition}\label{cy1}
A (compact, smooth) Calabi-Yau 1-fold is the torus $T^2$ as a Riemann surface/complex algebraic curve.
\end{proposition}

We are therefore reassured that CY$_1$  is none other than one of the most familiar objects in mathematics.
We can see this by recalling that $T^2$ is $\IC / L$, the quotient of the complex plane by a lattice (as we remember from our first lesson in topology that identifying the opposite sides of a sheet of paper gives a doughnut), and thus a metric on $T^2$ can be inherited from the flat metric on the plane.
Algebraically, a torus can be realized as an {\blue elliptic curve}, the central object to geometry and number theory; there is thus a programme in understanding the arithmetic, in addition to the geometry, of Calabi-Yau manifolds \cite{yuibook,Candelas:2000fq}.

In complex dimension 2, the 4-torus $T^4 \simeq (S^1)^4$ is analogously Calabi-Yau.
There is another, called the {\blue K3-surface} which deserves an entire treatise to itself.
Unfortunately, we do not have enough space in this book to discuss K3 surfaces and will trust the reader to the great texts of  \cite{beauville,schuett,Aspinwall:1996mn}.

So far, we have one CY$_1$ and two CY$_2$, at least topologically.
However, this seemingly innocent sequence of numbers 1,2, \ldots of Calabi-Yau manifolds in complex dimension 1,2, \ldots does {\it not} persist.
In fact, the next number in this sequence - which is perhaps the most pertinent to physics as we shall see next - for the number of inequivalent Calabi-Yau 3-folds, is already {\it unknown} (though it is conjectured by Yau to be finite in topological type).
What we do know is that there is a vast landscape thereof and much of this book will be a status report on how much we do know of this landscape.

\section{$10 = 4 + 2 \times 3$: a Physical Motivation}\label{s:compactification}
While the aforementioned advances in geometry were being made, theoretical physicists were working on the ultimate Theory of Everything.
By the late 1970s, a theory of strings, born from an initial attempt to explain the mass-spin correlation of the zoo of elementary particles in the 1960s, showed great promise in unifying gravity with quantum field theory.
As this book is neither on string theory nor really on algebraic geometry, but rather on the algorithms and data which emerge,
the reader is referred to the canon \cite{gsw}, or to \cite{dean} for technical and historical accounts, and encouraged to peruse this section as quickly as possible.

As far as Calabi-Yau manifolds are concerned, the physics side of the story began in 1985, with the paper which began what is now called {\blue ``string phenomenology''} \cite{Candelas:1985en}, a field aimed at constructing the Standard Model as a low-energy limit of string theory (the gravity side is called ``string cosmology'' and the conjunction of the two should, in principle, encompass all that is within the universe).
The paper immediately followed the discovery of the {Heterotic string} \cite{Gross:1984dd}, which, being a unified theory of gravity and a QFT with a Lie algebra as large as $E_8 \times E_8$ or $SO(32)$, caused much excitement\footnote{
By fusing the bosonic string (whose critical dimension is 26) with the superstring (whose critical dimension is 10) in the process of ``heterosis'' by assigning them to be respectively left and right moving modes of the string, the fact that $26 - 10 = 16$ gives 16 internal degrees of freedom to furnish a gauge symmetry.
Beautifully, in 16 dimensions there are only two even self-dual integral lattices in which quantized momenta could take value, viz., the root lattices of $E_8 \times E_8$ and of $D_{16} = \mathfrak{so}(32)$.
}.

The primary reason for this optimism is that the {\blue Standard Model} (SM) gauge group $G_{SM} = SU(3) \times SU(2) \times U(1) \subset E_8$
(recall that the $SU(3)$ factor for QCD governs the dynamics of baryons and the $SU(2) \times U(1)$ factor for QED, the leptons).
It has been an old question in particle physics as to why the Standard Model gauge group is of the structure of non-semisimple Lie group but is rather such a seemingly arbitrary product of three unitary groups.

The fact that $G_{SM}$ is not a (semi-)simple Lie group has troubled many physicists since the early days: it would be more pleasant to place the baryons and leptons in the same footing by allowing them to be in the same representation of a larger simple gauge group.
This is the motivation for the sequence in \eqref{embed} below: starting from $SU(5)$, theories whose gauge groups are simple are called {\bf grand unified theories} (GUTs), the most popular historically had been $SU(5)$, $SO(10)$ and $E_6$, long before string theory came onto the scene in theoretical physics.
Various attempts have been made in harnessing the following inclusion of Lie groups
\begin{equation}\label{embed}
SU(3) \times SU(2) \times U(1) \subset SU(5) \subset SO(10) \subset E_6 \subset E_7 \subset E_8 \ ,
\end{equation}
with all the various semisimple groups to the right of $G_{SM}$ furnishing different candidate GUT theories.
Thus, a unified theory with a natural $E_8$ gauge group was most welcome 
\footnote{According to Witten, in the sense of string phenomenology, ``heterotic compactification is still the best hope for the real world.''}
.
Oftentimes, we add one more $U(1)$ factor to $G_{SM}$, denoted as $U(1)_{B-L}$, to record the difference between baryon and lepton number, in which case 
\begin{equation}\label{GSM}
G_{SM}' = SU(3) \times SU(2) \times U(1) \times U(1)_{B-L}
\end{equation}
and the above sequence of embeddings skips $SU(5)$.

The downside, of course, is that (standard, critical, supersymmetric) string theory lives in $\IR^{1,9}$ and compared to our $\IR^{1.3}$ there must be 6 extra dimensions.
This remains a major challenge to the theory even today.
There are two philosophies in addressing $10 = 4 + 6$, to (1) take the extra dimension to be small in the spirit of Kaluza-Klein or to (2) take them to be large in the spirit of holography.
The former is called {\bf compactification} and the latter {\bf brane-world} and conveniently correspond mathematically to compact and non-compact Calabi-Yau varieties, as we shall see in the first two chapters of this book.

In brief, \cite{Candelas:1985en} gave the conditions for which the heterotic string, when compactified would give a supersymmetric gauge theory in 
$\IR^{1,3}$ with potentially realistic particle spectrum.
Here, compactification means that the 10-dimensional background is taken to be of the form $\IR^{1,3} \times M_6$ with $\IR^{1,3}$ our familiar space-time and $M_6$ some small (Planck-scale) curled up 6-manifold, endowed with a vector bundle $V$at the Planck scale too small to be currently observed directly.
Specifically, with more generality, the set of conditions, known as the {\it Hull-Strominger System} \cite{Hull,Strominger:1986uh}, for the low energy low-dimensional theory on $\IR^{1,3}$ to be a supersymmetric gauge theory are
\begin{enumerate}
\item $M_6$ is complex;
\item The Hermitian metric $g$ on $M_6$ and $h$ on $V$ satisfy 
\begin{enumerate}
\item $\partial \overline{\partial} g = i \tr F \wedge F - i \tr R \wedge R$ where $F$ is the curvature (field strength) 2-form for $h$ and $R$ the curvature 2-form for $g$;
\item $d^\dagger g = i (\partial - \overline{\partial}) \ln || \Omega ||$, where $\Omega$ is a holomorphic 3-form on $M_6$ which exists \footnote{
Recently, Li-Yau \cite{liyau} showed that this is equivalent to $\omega$ being balanced, i.e., 
$d\left( ||\Omega|| g^2 \right) = 0$.
};
\end{enumerate}
\item $F$ satisfies the Hermitian Yang-Mills equations
\begin{equation}\label{hym}
\omega^{a\overline{b}} F_{a\overline{b}} = 0 \ , \quad
F_{ab} = F_{\overline{a}\overline{b}} = 0 \ .
\end{equation}
\end{enumerate}

A sequence of arguments (cf.~Appendix \ref{ap:geoSpinor}) then leads to the fact that the simplest solution to the above conditions is that
\begin{proposition}
$M_6$ is K\"ahler, complex dimension 3, and of $SU(3)$ holonomy.
\end{proposition}
Furthermore, one can take the vector bundle $V$ to be simply the tangent bundle $T_M$.
We can now refer to Berger's famous  holonomy classification \cite{berger} as to what such a manifold is.
\begin{theorem} [Berger]
For $M$ a Riemannian manifold of real dimension $d$ 
which locally is not a product space nor a symmetric space, then special holonomy and manifold type obey:
\begin{center}
\begin{tabular}{|c|c|} \hline 
Holonomy ${\cal H} \subset$ & Manifold Type \\ \hline 
$SO(d)$ & Orientable \\
$U(d/2)$ & K\"ahler \\
$SU(d/2)$ & {\red Calabi-Yau} \\
$Sp(d/4)$ & Hyper-K\"ahler \\
$Sp(d/4) \times Sp(1)$ & Quaternionic-K\"ahler \\
\hline
\end{tabular}
\end{center}
Moreover, for $d=7,8$, there are two more possibilities, viz., $G_2$ and Spin$(7)$.
These two, together with the $SU(d/2)$ and $Sp(d/4)$ cases, are Ricci flat.  
\end{theorem}
Thus, {\em our internal 6-manifold $M$ is, in fact, a Calabi-Yau 3-fold} \footnote{
The reader might wonder how curious it is that two utterly different developments, one in mathematics, and another in physics, should both lead to Rome.
Philip Candelas recounts some of the story.
The paper \cite{Candelas:1985en} started life as two distinct manuscripts, one by Witten, and the other, by Candelas-Horowitz-Strominger, the latter began during a hike to Gibraltar reservoir, near Santa Barbara, in the fall of 1985.
Both independently came to the conclusion that supersymmetry required a covariantly constant spinor and thus vanishing Ricci curvature for the internal manifold $M$.
It was most timely that Horowitz had been a postdoctoral researcher of Yau, and Strominger also coincided with Yau when both were at the IAS,  thus a phone-call to Yau settled the issue about the relation to $SU(3)$ holonomy and hence Calabi-Yau.
Thus the various pieces fell, rather quickly, into place, the two manuscripts were combined, and the rest, was history.
In fact, it was the physicists who named these Ricci-flat K\"ahler manifolds as ``Calabi-Yau''.
}
and our two strands of thought, in the \S\ref{s:geo} and in the present, naturally meet.
This is another golden example of a magical aspect of string theory: it consistently infringes, almost always unexpectedly rather than forcibly, upon the most profound mathematics of paramount concern, and then quickly proceeds to contribute to it.

\subsection{Triadophilia}
As seen from the above, string phenomenology aims to obtain the particle content and interactions of the standard model as a low-energy limit.
In terms of the representation of $G_{SM}'$ in \eqref{GSM}, denoted as $({\bf a}, {\bf b})_{(c,d)}$ where ${\bf a}$ is a representation of $SU(3)$, ${\bf b}$, that of $SU(2)$, and $(a,b)$ are the charges of the two Abelian $U(1)$ groups, the SM elementary particles (all are fermions except the scalar Higgs) are as follows
\begin{equation}\label{SM}
\mbox{
\begin{tabular}{|l|l|l|}\hline 
$SU(3) \times SU(2) \times U(1) \times U(1)_{B-L}$ & Multiplicity & Particle
\\ \hline \hline
$({\bf 3,2})_{1,1}$ & $3$ & left-handed quark\\\hline

$({\bf 1,1})_{6,3}$ &$3$&left-handed anti-lepton\\\hline

$({\bf \overline{3},1})_{-4,-1}$&$3$&left-handed  anti-up\\ \hline

$({\bf \overline{3},1})_{2,-1}$&$3$&left-handed anti-down\\ \hline

$({\bf 1,2})_{-3,-3}$& $3$& left-handed lepton\\ \hline

$({\bf 1,1})_{0,3}$&$3$&left-handed anti-neutrino\\ \hline

$({\bf 1,2})_{3,0}$&$1$&up Higgs\\ \hline

$({\bf 1,2})_{-3,0}$&$1$& down Higgs\\\hline
\end{tabular}
}
\end{equation}

In addition to these are vector bosons: (I) the connection associated to the group $SU(3)$, called the gluons, of which there are 8, corresponding to the dimension of $SU(3)$, and (II) the connection associated to $SU(2) \times U(1)$, called $W^{\pm}$, $Z$, and the photon; a total of 4.
We point out that in this book the Standard Model - and indeed likewise for all ensuing gauge theories - we shall actually mean the (minimal) supersymmetric extension thereof, dubbed the {\bf MSSM}, and to each of the fermions above there is a bosonic partner and vice versa.

Of note in the table is the number 3, signifying that the particles replicate themselves in three families, or {\bf generations}, except for the recently discovered Higgs boson, which is only a single doublet under $SU(2)$.
That there should be 3 and only 3 generations, with vastly disparate masses, is an experimental fact with confidence \cite{pdg} level $\sigma = 5.3$ and has {\em no satisfactory theoretical explanation to date}.
The possible symmetry amongst them, called flavour symmetry, is independent of the gauge symmetry of $G_{SM'}$.

Upon compactification of the $E_8$ heterotic string on a Calabi-Yau 3-fold $M_6$, a low-energy supersymmetric QFT is attained.
Thus, our paradigm is simply
\[{\red
\mbox{Geometry of $M_6$ $\longleftrightarrow$ physics of $\IR^{1,3}$}.
}\]
What a marvellous realization of Kepler's old adage ``Ubi materia, ibi geometria''!

Now, the tangent bundle $T_M$ is of $SU(3)$ holonomy, thus $E_8$ is broken to $E_6$ since $SU(3)$ is the commutant of $E_6$ within $E_8$.
In particular, the fundamental 248 representation of $E_8$ branches as:
\begin{align}
\nonumber
E_8 \rightarrow & SU(3) \times E_6 \\
\label{E8E6}
248 \rightarrow & (1,78) \oplus (3,27) \oplus (\overline{3},\overline{27}) \oplus (8,1) \ .
\end{align} 
It is possible to package all of the SM particles in \eqref{SM} into the $27$ representation of $E_6$, in a SUSY $E_6$-GUT theory.
From \eqref{E8E6}, this is associated with the fundamental $3$ of $SU(3)$. 
The 27 representation is thus associated to $H^1(T_M)$ and the conjugate $\overline{27}$ to $H^1(T_M^\vee)$.
Similarly, the 1 representation of $E_6$ is associated with the 8 of $SU(3)$, and thus to $H^1(T_M \otimes T_M^\vee)$.
Thus, we have that
\begin{align}
\nn
\mbox{generations of particles} & \sim H^1(T_M) \ , \\
\nn
\mbox{anti-generations of particles} & \sim H^1(T_M^\vee) \ .
\end{align}
In general, even when taking an arbitrary bundle $V$ instead of just $T_M$, the lesson is that
\[
{\blue \mbox{Particle content in $\IR^{1,3} \longleftrightarrow$ \begin{minipage}{3in} cohomology groups of $V$, $V^\vee$ \\
			and of  their exterior/tensor powers \end{minipage}
}}
\]
The interactions, i.e., cubic Yukawa couplings in the Lagrangian, constituted by these particles (fermion-fermion-Higgs) are tri-linear maps taking the cohomology groups to $\IC$; this works out perfectly for a Calabi-Yau 3-fold: for example,
\begin{equation}
H^1(M,V) \times H^1(M,V) \times H^1(M,V) \rightarrow H^3(M,\cO_M) \simeq \IC \ .
\end{equation}

An immediate {\it constraint} is, as mentioned above, that there be 3 net generations, meaning that
\begin{equation}\label{TX3}
\left| h^1(X,T_M) - h^1(X,T_M^\vee) \right| = 3 \ .
\end{equation}
Thus, the endeavour of finding Calabi-Yau 3-folds with the property \eqref{TX3} began in 1986.
We will see in the following section that the difference on the left-hand-side is half of the Euler number $\chi$.

Thus, one of the first problems posed by physicists to what Candelas calls ``card-carrying'' 
\comment{
\footnote{
I have always interpreted  - as I have always perceived him to be a gentleman in the classic sense -   to mean ``visiting-card'' carrying, as in a passage from, say, Jane Austen.
However, when it finally occurred to me to check with him after misquoting him over the years, he assured me that he actually meant ``Party-membership-card'' carrying, as in a passage from, say, Solzhenitsyn.
In any event, whatever the card, one who carries it must be quite serious about it indeed. 
}} 
algebraic geometers was made specific:
\begin{question}
Are there smooth, compact, Calabi-Yau 3-folds, with Euler number $\chi = \pm 6$ ?
\end{question}

This geometrical ``love for threeness'', much in the same spirit as triskaidekaphobia, has been dubbed by Candelas et al.~as {\bf Triadophilia} \cite{Candelas:2007ac}.
More recently, independent of string theory or any unified theories, why there might be geometrical reasons for three generations to be built into the very geometry of the Standard Model has been explored \cite{He:2014oha}.

\subsection{Caveat \& Apology: Re the Title}\label{s:landscape}
As stated in the outset, the foregoing discussions of this section was meant to be a rapid summary of how the study of Calabi-Yau manifolds first came into physics in the mid-1980s; they are neither meant to be self-contained nor complete and chapters 12-15 of \cite{gsw}, Volume 2, is a far more pedagogical and enlightening.
More importantly, we have use the word ``{\blue Landscape}'' rather loosely, throughout and in the title.
In physics, the word has taken a more specific meaning on which we now briefly elaborate.

The reader should not be deceived into thinking that Equations \eqref{E8E6} - \eqref{TX3} apply universally in string theory.
The number of generations of particles in \eqref{TX3}, in particular, is only applicable to the so-called ``standard embedding'' of heterotic string compactification described above, as the first appearance of Calabi-Yau manifolds in physics.
Today, even within the heterotic community, people are no longer searching for Calabi-Yau threefolds with Euler number\footnote{
Serious effort, however, has been made by various groups, e.g., the UPenn-Oxford collaboration, to use stable holomorphic vector bundles $V$ on CY$_3$, whose index gives the number of particle generations.
The standard embedding is the special case where $V$ is the tangent bundle.
A review of some of this direction in recent times is in \cite{He:2010uj}.
} $\pm 6$.

With the advent of string dualities in the mid-1990s, from D-branes \cite{Polchinski:1995mt} to AdS/CFT \cite{Maldacena:1997re}, from M-theory on $G_2$-manifolds \cite{Atiyah:2001qf} to F-theory on 4-folds \cite{Morrison:1996na,F4fold}, etc., 
the space of possible vacuum solutions of string theory that lead to our $3+1$-dimensional world exploded to far beyond Calabi-Yau manifolds.
The reader is referred to the classic textbooks \cite{gsw,polchinski},  to \cite{BBS} for a modern treatment, and especially to
\cite{dine,ibanezuranga,stringdata10} for those interested in phenomenology and the standard model.
Since the focus of the book is to use the plethora (to which we vaguely refer as landscape) of Calabi-Yau manifolds as a playground for computational geometry and data-science, rather than an introductory text on string theory, we will not delve into the vastness of the subject of the string landscape.
For the benefit of the interested, however, we will attempt a precis of the major directions of reaching the standard model of particle physics from string theory, accentuating on where Calabi-Yau spaces appear.

\paragraph{CFT Constructions: }
The context,  viz., heterotic compactifications, wherein CY$_3$ first arose was presented above, with the emphasis on translating between the particle physics and the algebraic geometry (and for this reason we have adhere to this train of thought). However, from the point of view of the world-sheet conformal field theory (CFT), this scenario is difficult and not exactly solvable. Whilst the 6-torus $T^6$ is a trivial Calabi-Yau, one could take {\it orbifolds} thereof, which means we take the quotient $T^6 / \Gamma$, typically for a product of cyclic groups $\Gamma = \IZ/(n\IZ) \times \IZ/(m\IZ) \subset SU(3)$, and consider this to be our compactification space.

Alternatively, one could start with a solvable non-trivial 2-dimensional CFT, notably minimal models with specific central charges, and tensor them together; this is the {\it Gepner Model}.
There is a beautiful story here, that the Laudau-Ginzberg (LG) realization of the minimal model CFT is given by a polynomial Lagrangian governed by a (super-)potential $W$, which is precisely the defining hypersurface in a weighted projective space for a CY$_3$. This is the LG/CY Correspondence.
Both these directions have led to exact MSSM particle spectrum.
In fact, the sheer number of such CFT constructions, coming from the different types of lattices and orbifold groups, first gave physicists a hint of the vastness of possible string vacua, for which the word ``landscape'' was applied in the mid-1980s \cite{landscape1}.
Indeed, the idea of the computational complexity of the {\it string landscape} was already contemplated back in
\cite{Denef:2006ad} (q.v.~\cite{Halverson:2018cio} for recent studies).

\paragraph{Type II Constructions: }
In type II superstring theory, there are D-branes, which are extended dynamical objects of spatial dimensions $-1,1,3,5,7,9$ for type IIB and $0,2,4,6,8$ for type IIA.
The most promising idea of obtaining the standard model here is to place configurations of D6-branes at angles to each other, within a CY$_3$ background so that the D6-branes wrap 3-cycles (elements of $H_3$) within the compact CY.
This is the {\it intersecting brane-world} construction and the standard model particles are given as modes of the open string stretched between the D6 branes.

One can apply thrice T-duality (mirror symmetry) on the above to D3-brane in type IIB.
In Chapter 2 of this book, we will focus specifically on D3-branes transverse to non-compact CY$_3$ spaces, a prototype of AdS/CFT. We will do so not because the world-volume physics therein are proficient in obtaining the standard model, nor because of the profound holographic ideas of quantum gravity which arises,
but because they afford quiver realizations that illustrate the inter-play between representation theory and algebraic geometry, often in a computational and algorithmic way.

\paragraph{F-theory \& CY$_4$: }
Type IIA can be geometrized to 11-dimension M-theory, which is an $S^1$-fibration over the former.
Thus, intersecting brane models can be mapped to M-theory on 7-manifolds of $G_2$ holonomy (much like these latter manifolds can be constructed from a CY$_3$ $M$ by direct product  $M \times S^1$, and then quotienting by certain actions on special Langrangian sub-manifolds of $M$).
If we go up one more dimension, we are in 12-dimensional F-theory, which is an elliptic fibration over IIB.
Here, we can compactify on CY$_4$ (the analogue of $10=4+2\times3$ is $12=4+2\times4$) and all the constructions in this book can be directly generalized. In terms of standard-model building, F-theory has an advantage of being purely geometric, it does not require bundle information over CY$_3$ as in (non-standard) heterotic compactification.
Moreover, CY$_4$, being complex, is easier to handle than $G_2$ manifolds, which rests on the much more difficult mathematics of real algebraic geometry (real numbers are not algebraically closed and over them we do not have the fundamental theorem of algebra).

\paragraph{Flux Compactification \& the famous $10^{500}$: }
A key ingredient of a string theory is the presence of higher-form fields, i.e., multi-indexed tensors (generalizing the photon/1-index, graviton/2-index, B-filed/2-index, etc.).
In type II, the branes mentioned above are precisely the objects which ``couple'' (i.e., provide the support of region of integration of) to these form-fields.
Specifically, these couplings all contribute to the low-energy effective Langrangian.
This is the {\it flux compactification} scenario.

An archetypal situation is type IIB compactification on CY$_3$, but with 3-form field strengths (e.g., the NSNS $H_3$ and the RR $F_3$) contributing, and satisfying a quantization condition of their integrals over 3-cycles in the CY$_3$ being integers. This is a scenario which works well in producing low-energy effective theories with {\it moduli stablization} (which ensures that the compact dimensions remain so under the dynamics).

The rough estimate of the possible string vacua from flux compactification proceeds as follows \cite{landscape2}.
The typical flux (i.e., as the integer integral of the 3-forms around the 3-cycles) takes about 10 values, the typical number of 3-cycles in a compact smooth CY$_3$ is about 500 (we will see this estimate of the third Betti number $b_3$ against the actual datasets in the next chapter).
Hence, there are around $10^{500}$ possible (flux) vacua, a number which has at the turn of the millennium very much entered into the pubic psyche as the ``landscape''.
 
As one can see, this is a back-of-the-envelope assessment and says nothing about how many of such vacua even remotely resemble our observed 4-dimensional universe. 
Some recent estimates by explicit computer scan in type II and heterotic context separately \cite{Gmeiner:2005vz,hetLine,Deen:2020dlf} showed that about 1 in a billion stable vacua has an exact standard model particle content. It would be interesting to see, once numerical metrics can be more efficiently computed \cite{metricDon1,metricDon2,Ashmore:2019wzb}, how correct particle masses would further reduce the possibilities.

~\\
~\\

\section{Topological Rudiments}\label{s:cyintro}
From the previous two sections, we have seen two parallel strands of development, in mathematics and in theoretical physics, converging on the subject of Calabi-Yau manifolds, a vast subject on which this book can only touch upon a specialized though useful portion; for such marvelous topics as mirror symmetry, enumerative geometry, etc., the reader is referred to the classics \cite{mirror,katz}. 

For now, we nevertheless have a working definition of a (smooth, compact) Calabi-Yau $n$-fold, it is a manifold $M$ of complex dimension $n$, which furthermore obeys one of the following {\it equivalent} conditions:
\begin{itemize}
\item K\"ahler with $c_1(T_M)=0$;
\item K\"ahler $(M, g, \omega)$ with vanishing Ricci curvature $R(g, \omega) = 0$;
\item K\"ahler metric with global holonomy $\subset  SU(n)$;
\item K\"ahler and admitting nowhere vanishing global holomorphic $n$-form $\Omega^{(n,0)}$;
\item Admits a covariantly constant spinor;
\item $M$ is a projective manifold with (1) Canonical bundle (sheaf), the top wedge power of the cotangent bundle (sheaf),
\[
K_M := \bigwedge^n T_M^\vee
\] 
being trivial, i.e., $K_M \simeq \cO_M$ and 
(2) $H^i(M, K_M) = 0$ for all $0 < i < n$. 
\end{itemize}
We remark that the global form $\Omega^{(n,0)}$ prescribes a volume-form $V = \Omega \wedge \overline{\Omega}$.
When $n=3$,   it can be written in terms of the Gamma matrices $\gamma$ and the covariant constant spinor $\eta$ 
(cf.~Appendix \ref{ap:geo}) as $\Omega^{(3,0)} = \frac{1}{3!} \Omega_{mnp} dz^m\wedge dz^n\wedge dz^p$ with
$\Omega_{mnp} := \eta_-^T \ \gamma^{[m} \gamma^n \gamma^{p]} \ \eta_-$.

It is clear that in the above, the last definition is the most general.
So long as $M$ is a projective variety it is not even required that it be smooth and $K_M$ could be a sheaf obeying the said conditions.
Throughout this book, because the emphasis is on the algebraic, liberal usage of the first and last definitions will be made, and we will see how they translate to precise algebraic conditions from which we will make explicit construction of examples.

\subsection{The Hodge Diamond}\label{s:diamond}
We now present the most key topological quantities of a Calabi-Yau 3-fold.
Again, a lightning introduction of key concepts such as (co-)homology, bundles and complex geometry, in as self-contained a manner as possible, is given in Appendix \ref{ap:geo}.
From the Hodge decomposition, Eq.~\eqref{hodgedecomp}, the Betti numbers of a K\"ahler $n$-fold $M$ splits as
\begin{equation}\label{b2h}
b_k = \sum\limits_{\substack{p,q=0 \\ p+q=k}}^{n} h^{p,q}(M) \ , \qquad k = 0, \ldots, 2n \ .
\end{equation}
Now, complex conjugation (since $M$ has at least complex structure) implies that $h^{p,q} = h^{q,p}$.
Moreover, Poincar\'e duality implies that $h^{p,q} = h^{n-p, n-q}$.
Therefore, the $(n+1)^2$ {\red Hodge numbers} $h^{p,q}$ automatically reduce to only $((n+1)(n+2)/2 - (n-1))/2 + (n-1)$.
The clearest way to present this information is via a so-called {\red Hodge Diamond} (a rhombus, really).

For $n=3$, for example, the diamond looks like
\begin{equation}
\begin{array}{ccccccc}
& & & h^{0,0} & & &  \\
& & h^{0,1} &  & h^{0,1} & & \\
& h^{0,2} & &  h^{1,1} & & h^{0,2} &  \\
h^{0,3} & & h^{2,1} & & h^{2,1} & &  h^{0,3} \\
& h^{0,2} & &  h^{1,1} & & h^{0,2} &   \\
& & h^{0,1} &  & h^{0,1} & &   \\
& & & h^{0,0} & & &   \\
\end{array} \ .
\end{equation}

Further simplifications to the diamond can be made.
Because there is a unique non-vanishing holomorphic $n$-form,  $h^{n,0} = h^{0,n} = 1$.
Next, we can contract any $(p,0)$-form with $\overline{\Omega}$ to give a $(p,n)$-form, and a subsequent Poincar\'e duality gives a $(n-p,0)$-form.
This thus implies that $h^{p,0} = h^{n-p,0}$.

Next, we consider some special impositions.
In the case of $M$ being compact and connected, it is standard that $b_0 = 1$. Therefore, the top and bottom tip $h^{0,0} = 1$.
Furthermore, if $M$ is simply-connected, i.e., the first fundamental group $\pi_1(M) = 0$, then $H_1(M)$, being the Abelianization of $\pi_1(M)$, also vanishes.
This means that $h^{1,0} = h^{0,1} = 0$.

Putting the above together, the Hodge diamond of a compact, connected, simply connected Calabi-Yau 3-fold becomes:
\begin{equation}\label{diamond}
\begin{array}{cccccccccccc}
& & & 1 & & &    &&b^0 &&1 \\
& & 0 &  & 0 & & &&b^1 &&0\\
& 0 & &  h^{1,1} & & 0 &  &&b^2 &&h^{1,1}\\
1 & & h^{2,1} & & h^{2,1} & &  1 &\qquad \longrightarrow \qquad\qquad &b^3 &&2 + 2 h^{2,1}\\
& 0 & &  h^{1,1} & & 0 &  &&b^4 && h^{1,1} \\
& & 0 &  & 0 & &  &&b^5 && 0\\
& & & 1 & & & &&b^6  && 1  \\
\end{array} \quad \ .
\end{equation}
In other words, the Calabi-Yau 3-folds of our primary concern are actually governed by only 2 degrees of freedom
\begin{enumerate}
\item The {\blue K\"ahler parametres}, of which there are $h^{1,1}(M)$;
\item The {\blue complex structure parametres}, of which there are $h^{2,1}(M)$. 
\end{enumerate}

More precisely, by Hodge decomposition $H^{p,q}(M) = H^{q}(M, \bigwedge^p T_M^\vee)$ (cf.~\eqref{hodgedecomp}), we can express our two key Dolbeault cohomology groups as those valued in the tangent bundle $T_M$ and its dual, $T_M^\vee$, the cotangent bundle: 
\begin{equation}
H^{1,1}(M) = H^1(M, T^\vee_X), \qquad
H^{2,1}(M) = H^{1,2}(M) = H^2(X, T^\vee_X) \simeq H^1(X, T_X) \ ,
\end{equation}
where in the last step we have used {\blue Serre Duality}, that for a vector bundle $V$ on a compact smooth K\"ahler manifold $M$ of complex dimension $n$, we have 
\begin{equation}
H^i(M,V) \simeq H^{n-i}(M, V^\vee \otimes K_M) \ ,
\end{equation}
 and since $M$ is Calabi-Yau, the canonical bundle $K_M \simeq \cO_M$.
This pair of non-negative\footnote{
In almost all cases, they are both positive integers. The case of $h^{2,1}=0$ is called {\it rigid} because here the manifold would afford no complex deformations. The Hodge number $h^{1,1}$, on the other hand, is at least 1 because $M$ is at least K\"ahler.
} integers $(h^{1,1}, h^{2,1})$, the {\red Hodge pair}, will play a crucial r\^ole in the ensuing.

It is well known that the Euler number of a compact smooth manifold can be expressed as an alternating sum of Betti numbers:
\begin{equation}
\chi(M) = \sum\limits_{i=0}^{\dim_{\IR}(M)} (-1)^i b^i \ .
\end{equation}
Thus, combining with \eqref{b2h} and for our shape of the Hodge diamond, we arrive at
\begin{align}
\nn
\chi(M) &= 1 - 0 + h^{1,1} - (2 + 2 h^{2,1}) + h^{1,1} - 0 + 1 \\
\label{euler}
& = 2(h^{1,1} - h^{2,1}) \ 
\end{align}
for our Calabi-Yau 3-fold.

\setcounter{part}{1}
\chapter{The Compact Calabi-Yau Landscape}

\begin{chapquote}{Shing-Tung Yau}
``In the end, the search for a single, all-encompassing theory of nature amounts, in essence, to the search for the symmetry of the universe.''
\end{chapquote}

By now we hope the reader would be intrigued by the richness of Calabi-Yau manifolds and thus motivated, would be impatient for explicit examples with which to befriend.
Acquainting ourselves with a ``bestiary'' (in the spirit of \cite{hubschbook}) of Calabi-Yau manifolds will indeed constitute much of the ensuing.
As the approach of this book is algebraic, we will construct all our manifolds as {\bf {\red Algebraic Varieties}}.

The neophyte might be alarmed at the seeming sophistication of our choice, since it has become customary for the beginning student to learn the rudiments of differential geometry before algebraic geometry, and thereby be familiar with charts and transition functions, {\it before} polynomial rings and ideals.
This is perhaps ironic, since from a practical, and certainly computational, point of view,  algebraic varieties are simply vanishing loci of multi-variate polynomials, well known to school mathematics after an entr\'ee to Cartesian geometry.
Luckily, powerful theorems such as that of Chow \cite{GH} allow us to realize (complex analytic) manifolds as algebraic varieties and vice versa.
In case the reader needs a quick introduction to polynomial rings and how ideals correspond to varieties, we leave a somewhat self-contained introduction to Appendix \ref{ap:spec}.

Since we are only considering complex manifolds, all our examples will be constituted by the intersection of polynomials in complex variables.
Immediately, our Calabi-Yau 1-folds as algebraic tori described in Propostion \ref{cy1} find their incarnation: we recall from any second term course in complex analysis that the zero-locus of a cubic polynomial in two complex variables, say $(x,y) \in \IC^2$, is a torus.
Thus, a Calabi-Yau 1-fold as an elliptic curve is a cubic in $\IC^2$; this is an affine variety in $\IC[x,y]$.

While in the next chapter focus will be made on affine varieties, for the present chapter we will exclusively consider {\bf {\red projective varieties}}.
This is for the sake of compactness, as it is well-known that a projective space is attained by adding the ``point-at-infinity'' to an affine space.
Hence, affine complex coordinates $(x,y) \in \IC^2$ are promoted to projective  coordinates $(x,y,z) \in \IC^3$ with the extra identification that $(x,y,z) \sim \lambda (x,y,z)$ for any non-zero $\lambda \in \IC^{\times}$.
In other words, the elliptic curve is a cubic in $\IC\IP^2$ with homogeneous coordinates $[x:y:z]$.
We will adopt the convention in complex algebraic geometry that $\IP^n$ is understood to be $\IC\IP^n$, and define it as
\begin{equation}\label{Pn}
\IP^n_{[z_1: z_2 : \ldots : z_n]} := \IC^{n+1}_{(x_0, x_1, \ldots, x_n)} \slash \sim \ ;  \qquad (x_0, x_1, \ldots, x_n) \sim \lambda (x_0, x_1, \ldots, x_n) \ , \quad
\lambda \in \IC^{\times} \ .
\end{equation}
We will almost exclusively work with polynomials embedded in \eqref{Pn} as well as some of its natural generalizations.
Again, we leave the reader to Appendix \ref{ap:geo} for an attempt of a self-contained initiation to all the requisite knowledge of complex algebraic geometry.

\section{The Quintic}\label{s:Q}
In the above, we constructed CY$_1$ as a cubic ($d=3$) in $\IP^2$ ($n=2$), algebraically furnishing a torus into an elliptic curve \footnote{
Here is a wonderful quote and homage to Kronecker, from Tristan H\"ubsch, ``Algebraic curves were created by God, algebraic surfaces by the Devil; it is then only human to create algebraic 3-folds.''
}.
Two simple numerologies should be noted: 
\begin{itemize}
\item $1 = 2 - 1$: the dimension of the CY is 1 less than that of the ambient space since it is defined by a single polynomial; this is an example of a {\blue {\bf hypersurface}}.
\item $3 = 2 + 1$: the degree of the polynomial exceeds the dimension of the ambient projective space by 1, and is thus equal to the number of homogeneous coordinates; this we will see to be precisely the Calabi-Yau condition.
\end{itemize}

We wish to construct a Calabi-Yau $(n-1)$-fold $M$ realized as a {\red hypersurface} in $\IP^{n}$, i.e., a single homogeneous polynomial, of degree $d$ in $n+1$ projective coordinates.
The formal way of doing this is to first write down the short exact sequence \cite{hartshorne} (q.v.~Appendix \ref{ap:geo})
\begin{equation}\label{eulerES}
0 \to T_M \to \left. T_{\IP^n}\right|_M \to N_{M / \IP^n} \to 0 \ .
\end{equation}
The sequence essentially states that the tangent bundle of the ambient space $A = \IP^{n}$, when restricted to the hypersurface $M$, breaks up into the tangent bundle $T_M$ of $M$ and the normal bundle $N_{M / \IP^n}$ as $M$ embeds into $\IP^n$.
However, $T_A$ does not quite {\it split} into the direct (Whitney) sum of the two, but is a non-trivial {\it extension} of the two.
One consequence of \eqref{eulerES} is the Adjunction \cite{hartshorne} formula $K_M = \left. \left( K_{\IP^n} \otimes N_{M / \IP^n}^\vee \right) \right|_M$ for the {\red Canonical Bundle} $K_M$.
Now, because $M$ is defined by a single polynomial of degree $d$, this normal bundle is simply $\cO_{\IP^n}(d)$.

We now appeal to the axioms of the (total) {\red Chern Class} (cf.~Definition \ref{chern} in Appendix \ref{ap:geo}).
First, normalization axiom $c(\cO_{\IP^n}(1)) =  1 + H$ gives us the total Chern class of the tangent bundle \footnote{
This can itself be derived from the Euler sequence of bundles on projective space:\\
$0 \to \cO_{\IP^n} \to \cO_{\IP^n}(1)^{\oplus (n+1)} \to T_{\IP^n} \to 0 \ .$
}: 
\begin{equation}
c\big(\left. T_{\IP^n}\right|_M \big) =  (1 + H)^{n+1} \ ,
\end{equation}
where $H$ is the hyperplane (divisor) class of $\IP^{n-1} \hookrightarrow \IP^{n}$.
Next, the normal bundle, being $\cO_{\IP^n}(d)$, is a line bundle of degree $d$ and has only two contributions to the total Chern class
\begin{equation}
c\big(\cO_{\IP^n}(d) \big) = c_0\big(\cO_{\IP^n}(d) \big) + c_1\big(\cO_{\IP^n}(d) \big) = 1 + d\ H \ .
\end{equation}
The short exact sequence \eqref{eulerES} (cf.~\eqref{ssChern}) implies that
\begin{equation}
c(T_M) c\big(N_{M / \IP^n}\big) = c\left( \left. T_{\IP^n}\right|_M \right) \ , 
\end{equation}
whence, recalling that $c_k(T_M) \in H^{2k}(M; \IR)$, we arrive at the expression for our desired Chern class for $M$:
\begin{align}
\nn
c(T_M) & = (1+H)^{n+1} (1+d \ H)^{-1} \\
\nn
& = (1 + (n+1) \ H + {n+1 \choose 2} H^2 + \ldots ) (1 - d\ H + d^2H^2 - \ldots) 
\\ \label{cQ}
&= 1 + (n+1-d) H + \left( {n+1 \choose 2} + d^2 - d(n+1) \right) H^2 + \ldots
\end{align}
In the above, the order $H^k$ term is $c_k(T_M)$ so we expand only to order $H^{n-1}$ since $M$ is of complex dimension $n-1$ and cannot afford any Chern classes above this dimension.

Immediately, we see that $c_1(T_M) = (n+1-d) H$.
Using the vanishing $c_1$ condition in the definition of Calabi-Yau manifolds from \S\ref{s:cyintro} (because $\IP^n$ is K\"ahler, its holomorphic subvarieties as defined by polynomials in the projective coordinates are automatically K\"ahler), we arrive at our suspected result:
\begin{proposition}\label{CYn-1}
A homogeneous degree $d = n+1$ polynomial (in the $n+1$ projective coordinates) as a hypersurface in $\IP^n$ defines a Calabi-Yau $(n-1)$-fold.
\end{proposition}
In particular, we encounter our first Calabi-Yau 3-fold: the degree 5 polynomial in $\IP^4$, viz., the {\red Quintic}, $Q$.

In fact, we have done more.
In \eqref{cQ}, settig $n=4, d=5$, we have all the Chern classes for $Q$:
\begin{equation}\label{chernQ}
c(T_Q) = 1 + 10 H^2 - 40 H^3 \ .
\end{equation}
The top Chern class, here $c_3$, is the {\bf Euler class}. Its integral over $M$ should give the Euler number:
\begin{equation}\label{dQ}
\chi(Q)= -40 \int_Q H^3 = -40 \cdot 5 = -200 \ ,
\end{equation}
where the integral of $H^3$ over $M$ is a consequence of B\'ezout's theorem: $H^3$, the intersection of 3 hyperplanes, is the class of 5 points.
Another way to think of this is to pull back the integral from $M$ to one on the ambient space $A$.
For $\omega \in H^{n,n}(M)$ and $\mu \in H^{k,k}(A)$ where $M$ is co-dimension $k$ in $A$ so that the normal bundle is rank $k$, we have $\int_M \omega = \int_A \mu \wedge \omega$. For our present case $\mu = d \ H$.

\subsection{Topological Quantities: Exact Sequences}
We now have our favourite Calabi-Yau 3-fold, the Quintic, of Euler number $-200$.
From \eqref{euler}, we immediately see that
\begin{equation}\label{eulerQ}
h^{1,1} - h^{2,1} = -100 \ .
\end{equation}
Obtaining each individual Hodge number, on the other hand, is an entirely different kettle of fish.
This perfectly illustrates the principle of index theorems: the alternating sum is given by some topological quantity whilst individual terms require more sophisticated and often computationally involved methods.
Within our context, the specific index theorem is a version of Atiyah-Singer \cite{GH},  stating that
\begin{theorem}[Atiyah-Singer, Hirzebruch-Riemann-Roch]\label{HRR}
For a holomorphic vector bundle $V$ on a smooth compact complex manifold $M$ of complex dimension $n$, the alternating sum (index) in bundle-valued cohomology is given by
\[
{\rm ind}(V) = \sum\limits_{i=0}^n (-1)^i \rk H^i(M, V) = \int_M \ch(V) \wedge \td(T_M) \ ,
\]
where $\ch(V)$ is the Chern character for $V$ and $\td(T_M)$ is the Todd class for the tangent bundle of $M$.
\end{theorem}
If we took the simple case of $V = T_M$, and use \eqref{Ch} and \eqref{Td} from the appendix, we would obtain the {\red Euler Number} as a special case of the index theorem \footnote{
In detail, we proceed as follows. Take $V=T_M$, of rank $r = n = \dim_{\IC} M$, we have
\begin{equation*}\begin{array}{lll}
\int_M c_n(T_M) & = \sum\limits_{i=0}^r (-1)^i \ch(\bigwedge^i T_M^\vee)  \td(T_M) & \quad \mbox{Borel-Serre, \eqref{BS}} \\
& = \sum\limits_{i=0}^n (-1)^i \chi (M, \bigwedge^i T_M^\vee)  & \quad \mbox{Hirzebruch-Riemann-Roch, Theorem \ref{HRR}} \\
& = \sum\limits_{p,q=0}^n (-1)^{p+q} \rk H^q (M, \bigwedge^p T_M^\vee)  & \quad \mbox{definition of $\chi$ as index} \\
& = \sum\limits_{i=0}^n (-1)^{i} h^i (M, \IC) = \chi(M) & \quad \mbox{Hodge Decomposition \ref{hodgedecomp}} \ .
\end{array}\end{equation*}
} so that for 3-folds,
\begin{equation}\label{topChern}
\chi(M) = \chi(T_M) = \int_M c_3(T_M) \ .
\end{equation}.

Luckily, we do not need to do much in this very case to obtain the individual terms.
First, there is only one K\"ahler class, that inherited from the ambient $\IP^4$ -- indeed, $H^2(\IP^n; \IZ) \simeq \IZ$ for any $n$, corresponding to the hyperplane class, so that $h^{1,1}(\IP^n) = 1$ upon Hodge decomposition. 
Hence,
\begin{equation}
h^{1,1}(Q) = 1 \ .
\end{equation}
In general, however, it must be emphasized that ambient classes {\em do not} descend 1-1 to projective varieties and other measures need to be taken, as we shall see.

Alternatively, we can try to obtain $h^{2,1}$.
As mentioned in \S\ref{s:diamond}, these correspond to complex structure. What this means is the following.
A generic quintic in $\IP^4_{[z_0:\ldots:z_4]}$ can be written as a sum over all possible degree 5 monomials in the 5 projective variables $z_i$; each monomial has a complex coefficient.
Perhaps the most familiar form of $Q$ is 
\begin{equation}\label{specialQ}
Q := \sum\limits_{i=0}^4 z_i^5 - \psi \prod\limits_{i=0}^4 z_i \ , \quad \psi \in \IC \ ,
\end{equation}
consisting of the sum of the 5-th powers, usually called the {\blue Fermat form}, together with a deformation by the product, with $\psi$ as an explicit complex parameter.
The question of complex structure is then to ascertain how many such coefficients, up to variable redefinitions, there are. 
Again, what we shall do below is a back-of-the-envelope calculation, which {\em does not} work in general. Nevertheless, its simplicity is illustrative.

First, how many possible degree 5 monomials can be written in 5 variables?
This is a standard combinatorial problem and one can quickly convince oneself that the number of  degree $d$ monomials in $k$ variables
is the so-called multiset coefficient ${n+d-1 \choose d}$.
Next, we need to subtract the reparametrizations coming from linear transformations, i.e., from $PSL(5; \IZ)$, of which there are $5^2 - 1$.
Finally, we are at liberty to rescale the polynomial by an overall constant.
Therefore, in all, there are
\begin{equation}\label{h21Q}
h^{2,1}(Q) = {5+5-1 \choose 5} - (5^2 - 1) - 1 = 126 - 24  - 1 = 101
\end{equation}
complex deformation parameters.
We have therefore rather fortuitously obtained the correct values of the Hodge pair $(1,101)$ for the quintic and we can check that indeed their difference is -100, as is required by the Euler number in \eqref{eulerQ}.

The proper way to perform this above computation is, as might be expected, by sequence chasing.
We once more appeal to the Euler sequence \eqref{eulerES}, which induces a long {\red Exact Sequence} in cohomology (the sexy way of saying this is that the cohomological functor $H^\bullet$ is covariant on short exact sequences):
\begin{equation}\label{ssHQ}
\begin{array}{cccccccc}
0 
& \to & \cancelto{0}{H^0(Q, T_Q)} & \to & H^0(Q, T_{\IP^4}|_Q) & \to & H^0(Q, N_{Q/\IP^4}) & \to
\\
& \to & \fbox{$H^1(Q, T_Q)$} & \stackrel{d}{\to} 
  & H^1(Q, T_{\IP^4}|_Q) & \to & H^1(Q, N_{Q/\IP^4}) & \to \\
& \to & H^2(Q, T_Q) & \to & \ldots &&&
\end{array}
\end{equation}
In the above, we have boxed the term which we will wish to compute, viz., $H^{2,1}(Q) \simeq H^1(Q, T_Q)$.
The first term $H^0(Q, T_Q) = H^{1,3}(Q)$ vanishes because $Q$ is Calabi-Yau.
The boundary map $d$ actually has 0 rank (q.v.~\cite{hubschbook} and Eq (6.1) of \cite{Green:1987cr}) and we thus have a short exact sequence 
$0 \to H^0(Q, T_{\IP^4}|_Q) \to H^0(Q, N_{Q/\IP^4}) \to H^{2,1}(X) \to 0$ so that
\begin{equation}
h^{2,1}(X) = h^0(X, N_Q) - h^0(X,T_{\IP^4}|_Q) \ .
\end{equation}
Thus, we have reduced the problem to counting global holomorphic sections.
Each of the two terms can be obtained by Leray tableaux \cite{GH} (cf.~summary in Appendix C.2 of \cite{Anderson:2008uw} for the present case).

\subsection{Topological Quantities: Computer Algebra}
If a single manifold as simple as the quintic requires so extensive a set of techniques, one can only imagine how involved the generic computation in algebraic geometry is.
The beginning student may not appreciate how {\em difficult} dealing with systems of polynomials (a concept introduced at primary school) really is: the system is highly non-linear, especially when the degrees are high. 
Whilst non-linearity typically invokes the thought of transcendental functions,  the field of non-linear algebra is an entire enterprise by itself. 

We are fortunate that we live within the age of computers, where at least anything algorithmic such as exact or Leray sequences need not be performed by hand.
The discipline of {\bf computational algebraic geometry} is vast and it is not our place to give an introduction thereto here.
The keen reader is referred to the fantastic textbook \cite{m2book} for a pedagogical treatment (cf.~also Appendix A.3 of \cite{Anderson:2007nc} for a tutorial in the present context of computing bundle cohomology).
Nevertheless, the key idea to the matter, viz, the {\bf Gr\"obner Basis}, is so vital we shall take an immediate  foray in Appendix \ref{ap:gb}.

The take-home message is a matter of nomenclature.
In differential geometry, we think of a manifold in terms of local patches at the intersection of which are smooth transition functions.
In algebraic geometry, we think of it as an {\red algebraic variety}, or the vanishing locus of a system of multivariable polynomials 
$\{ f_i(z_k) \}$ in the coordinates $\{ z_k \}$ of the ambient space, typically $\IC\IP^n$.
The fancier, purely algebraic way, of saying this is that we start with \footnote{
The set of polynomials in a fixed number of variables clearly form a ring under ordinary $+, \times$.
} a polynomial ring  $\IC[z_k]$  and any vanishing system of polynomials therein forms an ideal \footnote{
It is an {\red ideal} since  the vanishing locus of a system of polynomials is the same as that the system multiplied by arbitrary polynomials.
}.
Note that we use standard notation that  $\IC[z_k]$ means the set of polynomials in the variables $z_k$ and with coefficients in $\IC$.
Hence, the vanishing locus defining the manifold corresponds to ideals in $\IC[z_k]$.
This is the starting point of any code in a computer algebra system for defining a variety, as we will now see.

If the reader only wishes for a black-box, much as one wishes to evaluate a complicated integral on {\sf Mathematica} without the desire for the detailed steps, then perhaps the most intuitive software out there is {\sf Macaulay2} \cite{m2} a glimpse of whose powers we will now see.
The advantages of {\sf Macaulay2} are manifold:
\begin{itemize}
\item it is free (downloadable from\\
 \url{http://www2.macaulay2.com/Macaulay2/}
 \\
and every student of algebraic geometry is encouraged to do so);
\item it is included in  latest distributions of linux/MacOS/Win10 Ubuntu;
\item  it has been incorporated into SageMath \cite{sage}
(cf.\\
\url{http://doc.sagemath.org/html/en/reference/interfaces/sage/interfaces/macaulay2.html}
\\
for calling {\sf Macaulay2} from within {\sf SageMath});
\item the aforementioned book \cite{m2book}  uses it as a companion in instruction; and 
\item for a quick online experiment without the need to download, one could simply access it ``in the cloud'' at
\url{web.macaulay2.com}. 
\end{itemize}

Let us now redo the above Hodge computations for the Quintic in {\sf Maucalay2}.
All commands in {\sf Macaulay2} are case-sensitive and there is a habit that the first letter is in lower case and in the case of multiple words concatenated into one, the second starts with a capital. 
First, we define the polynomial ring $R$ in which we work, consisting of 5 complex variables $z_i$.
The coefficient field \footnote{
One must not confuse the coefficient ring/field with the ground ring/field: the former is where the coefficients take value and can be varied quite liberally in all our computations, the latter is where the variables take value, which for us will always be $\IC$.
} is here taken to be $\IZ/1979\IZ$ (or any other large-ish prime of one's choice); this is the case with most packages in algebraic geometry as the coefficients can grow very quickly in manipulations such as Gr\"obner bases so we take them modulo some prime.
One could always try a few different primes in case the variety becomes singular over particular primes of bad reduction.
Thence, we define
\begin{verbatim}
R = ZZ/1979[z_0, z_1, z_2, z_3, z_4] ;
\end{verbatim}
A shorthand for writing this is \verb|R = ZZ/1979[z_0 .. z_4]|.
Next, we define a specific quintic polynomial ($\psi=2$ in \eqref{specialQ})
\begin{verbatim}
polyQ = z_0^5 + z_1^5 + z_2^5 + z_3^5 + z_4^5 + 2*z_0*z_1*z_2*z_3*z_4 ;
\end{verbatim}
A shorthand for this is \verb|poly = (sum for i from 0 to 4 list x_i^5)| \\
\verb|+ 2 * (product for i from 0 to 4 list z_i) ; |\\
we remark that we have chosen a particular homogeneous degree 5 polynomial for convenience. Final topological quantities should neither depend on the choice of prime for the coefficient ring nor on the particular form of the polynomial (we remember from Cartesian geometry that ellipses are simply deformed circles).

We now enter the important step of defining the algebraic variety:
\begin{verbatim}
quintic = Proj( R/ideal(polyQ) ) ; 
\end{verbatim}
Here, we define $polyQ$ as an ideal of the ring $R$ and form the quotient ring $R / \left< polyQ \right>$ first \footnote{
Indeed, unlike {\sf Mathematica}, whose primary user is the theoretical physicist, to whom notational flexibility is more important than rigour, all the software mentioned in the preface, and {\sf Macaulay2} in particular, are written for the mathematician, where clarity and strictness cannot be relaxed.
}.
From this quotient ring we form the Proj, which is the projective version of Spec, the maximal spectrum (i.e., set of maximal ideals) of a ring, to which affine varieties correspond (again, Spec, coordinate rings, etc. are described in an introductory manner in Appendix \S\ref{ap:spec}).
Roughly, Proj turns quotient rings to projective algebraic varieties, furnishing the explicit map in the ring-variety correspondence \cite{CLO}.
In the case of multiple defining polynomials, say $f_1, f_2, \ldots$, we can simply use \verb|ideal( {f_1, f_2, ... }| )\ , which is the most general situation.

Thus, we have \verb|quintic| as a projective variety. We can check \footnote{
One thing about Macaylay2 commands is that unlike most programming languages one could in fact dispense with the brackets ``(''  and ``)''. Therefore, commands such as {\sf ideal(polyQ|} or {\sf |dim quintic} are perfectly legitimate.
Dan Grayson, one of the co-founders of the project, tells me that he has spent an enormous effort to make this syntactically possible.
However, to avoid confusing the reader, we will not take advantage of this liberty in this book.
} its dimension using 
\begin{verbatim}
dim(quintic)
\end{verbatim}
which returns \verb|3|, as required.
Note that without the \verb|Proj|, the dimension would be \verb|4| since we have not yet projectivized.
Now, we can check that this quintic is smooth with the built-in command 
\begin{verbatim}
singularLocus(quintic)
\end{verbatim}
which returns \verb|Proj(R/1)|, meaning that the smooth locus corresponds to an ideal defined by 1, which of course is not  possible since ideals correspond to the vanishing set of polynomials.
In other words, there are no singular loci and our quintic is smooth. By convention, the dimension of \verb|Proj(R/1)| is recorded as $-\infty$, as can be checked by the \verb|dim| command.
This is indeed in congruence with our conclusions from \eqref{smoothQ} from the Appendix, where $\psi = 2$ is a smooth point in complex structure moduli space.

Now, we move into the heart of the matter by establishing the cotangent bundle:
\begin{verbatim}
cotan = prune( cotangentSheaf(quintic) ); 
\end{verbatim}
First, in any computational algebraic geometry package, no assumption about smoothness is a priori made, nor is the assumption about the existence of (holomorphic) bundles.
In Appendix \ref{ap:geo}, we made no mention of sheaves in order to simplify the discussion.
Indeed, we will make no use of sheaf theory in this book. The reader could roughly bear in mind that a sheaf is just a bundle which could change rank at sub-regions of the variety, just like a singular variety is one which could change dimension at sub-regions of the manifold.
Hence, there is no command ``cotangentBundle'' in Macaulay2, but rather ``cotangentSheaf'', which makes assumption of smoothness.
Moreover, \verb|prune| is a built-in simplifies the sheaf by ``pruning'' (a play on the word ``sheaf'').

One can check that the cotangent bundle has 3 (as is equal to the dimension of the base manifold) simply with 
\begin{verbatim}
rank(cotan)
\end{verbatim}
Moreover, one can find the canonical bundle as (the exterior power is the wedge product $\wedge^3$)
\begin{verbatim}
KQ = prune( exteriorPower(3, cotan) ) ; 
\end{verbatim}
At this point, we appreciate that whilst the software is extremely convenient, even the state-of-art computers are beginning to struggle a bit: all commands so far take milliseconds to execute on a simple laptop, the exterior power, however, will take quite some time and memory and one might even need to set the prime 1979 to something smaller to enable memory allocation to store the coefficients generated in due course.

We should obtain the result \verb|OO_quintic^1| for \verb|KQ|, meaning that $K_Q = \cO_Q$, the trivial structure bundle
\footnote{
In computational algebraic geometry, sheafs/bundles are denoted by modules via the {\blue sheaf-module correspondence},
a quick discussion of this in Appendix A.1 of \cite{Anderson:2007nc}.
}
discussed in Appendix \ref{ap:geo}, whereby computationally proving that $Q$ is indeed a Calabi-Yau 3-fold.

Finally, we can use Hodge decomposition \eqref{hodgedecomp} to compute the requisite topological invariants:
\begin{verbatim}
H11Q = HH^1 ( cotan );
H21Q = HH^1 ( exteriorPower( 2, cotan ) );
\end{verbatim}
In the above, \verb|HH| is the sheaf-cohomology operator, and we readily obtain the results \verb|(ZZ/1979)^1| and \verb|(ZZ/1979)^101| as free-modules respectively for \verb|H11Q| and \verb|H21Q| (one can use the \verb|rank| command to obtain the numbers 1 and 101 respectively.

We conclude this section with a few lessons:
(1) algebraic geometry is algorithmic. Finding geometric quantities ultimately reduces to manipulating systems of polynomials, be they reduction or finding (co)kernels of matrices with polynomial components;
(2) software such as {\sf Macaulay2} have very conveniently implemented the cruces such as finding Gr\"obner bases or syzygies, and for simple cases are our indispensable aide; 
(3) Sadly, computing  Gr\"obner bases is a doubly-exponential running time process, for example, an estimate on the complexity of the Buchberger algorithm has an upper-bound on the degrees of the elements as $\sim (d^2/2 + d)^{2^{n-1}}$ where $d$ is the maximal total degree of the input polynomials and $n$ is the number of variables \cite{dube}.

\comment{
R = ZZ/1979[z_0, z_1, z_2, z_3, z_4] ;
polyQ = z_0^5 + z_1^5 + z_2^5 + z_3^5 + z_4^5 + + 2*z_0*z_1*z_2*z_3*z_4 ;
quintic = Proj( R/ideal(polyQ) ) ; 
cotan = prune( cotangentSheaf(quintic) ); 
KQ = prune( exteriorPower(3, cotan) ) ; 
}

\section{CICY: Complete Intersection Calabi-Yau}
While we had enjoyed our promenade in the land of exact sequences and of computational geometry, the quintic - the wealth of beautiful mathematics and physics we had only a tiny glimpse - was less than desirable as far as \eqref{TX3} is concerned: the Euler number \eqref{eulerQ} of -200 is far from $\pm 6$, it is not even divisible by 3.

An immediate generalization presents itself: instead of a single {\em hypersurface}, what about a set of polynomials in $\IP^n$; after all, any projective variety can be so realized.
This is clearly too much.
The first generalization of a hypersurface is a {\bf {\blue complete intersection}}, where the codimension of the variety is equal to the number of polynomials.
This is a case of the ``best possible'' intersection where each defining polynomial slices out one dimension.
It should be emphasized that in general if one has $k$ random polynomials intersecting in $\IP^n$, it is {\em not} guaranteed by any means to give a variety of dimension $n-k$. 

\paragraph{Cyclic Manifolds} How many complete intersection Calabi-Yau 3-folds can be written inside $\IP^n$.
Again, this is a simple combinatorial problem: partition $n + 1= \sum\limits_{\substack{n_i \in \IZ_{>0}\\ i=1}}^k n_i$ (the generalization of Proposition \ref{CYn-1}), together with $n-3 = k$ (complete intersection condition).
There are easily seen to be 5 solutions in total, of which $n=4, k=1, n_1 = 5$ is the quintic.
By similar considerations as in \S\ref{s:Q}, we can compute their Hodge numbers.
They all have $h^{1,1} = 1$, as descended from the ambient $\IP^n$.
These are consequently called {\bf cyclic} because 
$H^2(M;\IZ) \simeq \IZ$, the infinite cyclic group.
Adopting the notation $[n | n_1, n_2, \ldots, n_k]$ to mean the complete intersection of homogeneous polynomials $f_1$ of degree $n_1$, $f_2$ of degree $n_2$, etc., the 5 solutions are 
\begin{equation}\label{5cyclic}
\mbox{$[4|5]$, $[5 |2,4]$, $[5|3,3]$, $[6|3,2,2]$ and $[7|2,2,2,2]$} \ .
\end{equation}
Respectively, these are the quintic in $\IC\IP^4$, the intersection of a quadric and a quartic in $\IC\IP^5$,
that of 2 cubics in $\IC\IP^5$, that of a cubic with 2 quadrics in $\IC\IP^6$, and that of 4 quadric in $\IC\IP^7$.
Again, every polynomial is considered {\it generic} is that the coefficients are sufficiently random so that the variety is smooth.
Another thing is that algebraic geometers refer to quadratic polynomials as {\it quadrics}.
This is due to historical reasons when quadratic surfaces (like conic sections) were called quadrics.

In fact, the first column is redundant as it is one less than the sum of the rest.
By adjunction formula/Euler sequence as before, we can also compute the second Chern class (the first is, of course, 0, and the third integrates to the Euler number) as a multiple of the class $H^2$ for the hyperplane class $H \subset \IP^n$ (for $Q$, we recall from \eqref{chernQ}, this is 10).
By B\'ezout, we can also find the volume normalization (triple intersection) $d(M) = \int_M H^3$ (for $Q$, we recall from \eqref{dQ} that this is 5).
We summarize the data for all five in Table \ref{t:cyclic}.

\begin{table}[h!]
\begin{tabular}{|c|c|c|c|c|c|c|}\hline
 $A$ &Configuration& $\chi (M)$ & $h^{1,1}(M)$ &
 $h^{2,1}(M)$ & $d(M)$ & $c_2(T_M)$ \\ \hline
$\IP^4$& $[4|5]$ & $-200$ & $1$ & $101$&$5$ & $10$ \\
$\IP^5$ & $[5|2 \ 4]$ & $-176$ & $1$ & $89$ &$8$&
  $7$ \\
$\IP^5$ & $[5|3 \ 3]$ &$-144$ & $1$ &$73$ &$9$& $6$ \\
$\IP^6$ & $[6|3 \ 2 \ 2]$ & $-144$ & $1$ &
  $73$ & $12$ & $5$ \\
$\IP^7$ & $[7|2 \ 2 \ 2 \ 2]$ & $-128$ & $1$ & $65$ & $16$&
  $4$ \\ \hline
\end{tabular}
\caption{{\sf The 5 cyclic Calabi-Yau 3-folds as complete intersections in an ambient space which is a single $\IP^n$.
The configuration, Hodge numbers, Euler number are given, together with the second Chern class $c_2(T_M)$, as the coefficient of $H^2$ for the hyperplane $H \in \IP^n$, as well as the volume $d(M) = \int_M H^3$.
}
\label{t:cyclic}}
\end{table}

Our bestiary of Calabi-Yau 3-folds has now increased from 1 to 5.
Again, none has $\chi = \pm 6$, though 2 have Euler number divisible by $6$.
Can we proceed further?
What about generalizing the ambient space to a product of projective spaces instead of a single one?
This was considered by \cite{cicy,cicy2,cicy3,Green:1987cr} in the second lustrum of the 1980s.
As with the partition problem of \eqref{5cyclic}, this amounts to writing the configuration
\begin{equation}\label{cicy}
\begin{array}{ccc}
    M = 
      \left[\begin{array}{c|cccc}
      n_1 & q_{1}^{1} & q_{1}^{2} & \ldots & q_{1}^{K} \\
      n_2 & q_{2}^{1} & q_{2}^{2} & \ldots & q_{2}^{K} \\
      \vdots & \vdots & \vdots & \ddots & \vdots \\
      n_m & q_{m}^{1} & q_{m}^{2} & \ldots & q_{m}^{K} \\
      \end{array}\right]_{m \times K} \ ,
      &
      &
      \begin{array}{l}
      \sum\limits_{r=1}^m n_r = K+3\\
  	\sum\limits_{j=1}^K q_{i}^{j} = n_i + 1   \ \forall i = 1, \ldots, m
      \end{array}
      \end{array}
\end{equation}
as a set of $K$ polynomials of multi-degree $q_j^i \in \IZ_{\geq 0}$ in the ambient space $A = \IP^{n_1} \times \ldots \times \IP^{n_m}$.
The complete intersection condition $\dim(A) - \dim(M) = K$ translates to $\sum\limits_{r=1}^m n_r = K+3$, and the $c_1(T_M)=0$ condition, to
$\sum\limits_{j=1}^K q_{r}^{j} = n_i + 1$ for each $r = 1, \ldots, m$.
Again, the left-most column is redundant and may be omitted without ambiguity.
The Calabi-Yau 3-fold corresponding to such a configuration as \eqref{cicy} is called with the affectionate acronym 
{\red ``CICY''}, for Complete Intersection CY$_3$.

\paragraph{CICY} 
Classifying CICY matrices, checking redundancies and equivalence, is already a highly non-trivial task, and in some sense constituted one of the very first emergence of (then) big-data to pure mathematics and theoretical physics.
Candelas and Lutken recount the fascinating tale of how this was achieved on the then CERN super-computer, involving
dot-matrix printers and magnetic tapes, or Schimmrigk, on staring at VAX machines in
Austin over long nights.

The result has persisted in two amusing forms, in the corner of Philip's office as a rather large pile of perforated print-out and in Andy's, on a magnetic  tape, should the reader even belong to the generation to know what these mean.
Luckily, the data has been resurrected \cite{Anderson:2008uw} and is now accessible at \\
\url{http://www-thphys.physics.ox.ac.uk/projects/CalabiYau/cicylist/} \\
in text and {\sf Mathematica} \cite{mathematica} format (indeed, the webpage of Andr\'e Lukas has a very useful collection of CICY and other related matter \cite{andrepage}).
All requisite data, viz., the Hodge pair, the second Chern class and the triple intersection $d(M)$, have also been computed by \cite{cicy,cicy2,cicy3,Green:1987cr,hubschbook}.
Hence, though we reside firmly within the Age of Data and AI and nowadays even the most abstract branches of mathematics have enjoyed information explosion, data mining in algebraic geometry and theoretical physics dates to as early as the late 1980s.
One is perhaps reminded of a popular internet meme that technically Moses was the first person to download from the Cloud onto a tablet.

The presentation of the topological data in addition to the Hodge pair require the following.
One first fixes a basis $\{J^r\}_{r = 1, \ldots, h^{1,1}}$  of $H^2(M, \IZ)$ so that the K\"ahler cone is $\cK = \{t_r J^r | t_r \in \IR_{>0} \}$.
For the cyclic case, there is only one $J = H$.
In general, $J$ will not descend 1-1 from the $m$ hyperplane classes of the ambient product of projective spaces, and when they do, the CICY is called {\blue {\it favourable}}.
The triple intersection form in this basis is written as
\begin{equation}\label{drst}
d_{rst} := \int_M J^r \wedge J^s \wedge J^t \ ,
\end{equation}
which can be computed by pulling back integration from the ambient product of projective space where integration is standard, viz.,
\begin{equation}
\int_M \bullet = \int_A \mu \wedge \bullet\ , \qquad  \mu := \bigwedge_{j=1}^K \left( \sum_{r=1}^m q^j_r J_r \right) \ .
\end{equation}
The form $d_{rst}$ is a {\em completely symmetric tensor} with respect to the three indices (note that $J^r$ is a real 2-form so that commuting them across $\wedge$ generates 2 minus signs that cancel).

The total Chern class of a K\"ahler 3-fold can be written as 
\begin{equation}
c(T_M) = \sum_r [c_1(T_M)]_r J^r + \sum_{r,s} [c_2(T_M)]_{rs} J^r J^s + \sum_{r,s,t} [c_3(T_M)]_{rst} J^r J^s J^t \ .
\end{equation}
For us, $c_1 = 0$ and moreover  \eqref{topChern} reads
\begin{equation}
\chi(M) = \sum\limits_{r,s,t} d_{rst} [c_3(T_M)]_{rst} \ ,
\end{equation}
leaving us with $c_2$ as an independent set of quantities to be determined.

\paragraph{Compact CY$_3$ Data: }
We remark that the above data is in some sense complete because of an important result of Wall \cite{wall}
\begin{theorem}[Wall]\label{thm:wall}
The topological type of a compact K\"ahler 3-fold is completely determined by 
(1) the Hodge numbers $h^{p,q}$; 
(2) the triple intersection number $d_{rst}$; and
(3) the first Pontrjagin class $p_1(T_M) = c_1(T_M)^2 - 2 c_2(T_M)$.
\end{theorem}
Therefore, for all the Calabi-Yau databases out there, it suffices to record the Hodge numbers, the intersection 3-tensor $d$ and $c_2$ (often written as a vector in the dual basis to $J^r$ by contracting with $d$, i.e., as $[c_2(T_M)]_r := \sum\limits_{s,t}  [c_2(T_M)]_{rs} d_{rst}$.
This was indeed the data shown, for example, for the 5 cyclics in Table \ref{t:cyclic}.
In this sense, our Calabi-Yau 3-fold data is a list of integers
\begin{equation}\label{CYdata}
{\red \fbox{$ \big\{ (h^{1,1}, h^{2,1}) \ ; \quad \ [c_2]_r \ ; \quad d_{rst} \big\} $}  \ , \qquad r,s,t = 1, \ldots, h^{1,1} } \ .
\end{equation}

Theorem \ref{thm:wall} should also be contrasted with the complex dimension 1 case: there, as we recall, a single integer (the genus or the Euler number) completely characterizes the topological type of a Riemann surface. In complex dimension 3, we essentially need three sets of integers.
In complex dimension 2, the situation is actually quite complicated and is intimately related to the celebrated Poincar\'e Conjecture for 4-manifolds.
At least for simply-connected cases, a classical result of Milnor \cite{milnor} states that the intersection form in middle cohomology suffices to classify the topology.

While we are on the subject of generalities, an important result should be borne\footnote{
I am grateful to Mark Gross for reminding me of this.
The argument uses the fact that the  secant variety to a smooth 3-fold $M$ is at most of complex dimension $2\cdot3+1=7$.
Embed this into $\IP^r$. If $r > 7$, then there is a point $p \in \IP^r$ not on $M$, so that there is a projection $\pi_p$ from $p$ to a hyperplane $H$ , giving us $\pi_p : M \hookleftarrow H \simeq \IP^{r-1}$. We can inductively proceed until $r = 7$.
} in mind (q.v.~e.g., \cite{CGKK}):
\begin{proposition}
All compact smooth 3-folds can be smoothly embedded into $\IP^7$.
\end{proposition}
In other words, {\it in principle}, to classify all Calabi-Yau 3-folds, we only need to systematically write all possible polynomials, degree by degree, hypersurface by hypersurface (indeed, we are by no means restricted to complete intersection - of which there is only one, as we saw in Table \ref{t:cyclic} - there can be any number of defining polynomials).
As deceptively simple as this may seem, it is evidently far too difficult a task upon further reflection!

\subsection{Topological Quantities: Statistics}
Let us return to CICYs.
The configuration matrix \eqref{cicy}, as might be expected, readily gives almost all the data in \eqref{CYdata}.
One can show \cite{hubschbook} that, with $\mu := \bigwedge_{j=1}^K \left( \sum\limits_{r=1}^m q^j_r J_r \right)$ and with normalization
$\int_{\IP^{n_r}} J^s = \delta_{rs}$ upon integrating the ambient $A$ as a product of projective spaces: 
\begin{align}
\nn
c_1^r(T_M) &= 0 \\
\nn
c_2^{rs}(T_M) &= \frac12 \left[ -\delta^{rs}(n_r + 1) + 
  \sum\limits_{j=1}^K q^r_j q^s_j \right] \\
\nn
c_3^{rst}(T_M) &= \frac13 \left[\delta^{rst}(n_r + 1) - \sum\limits_{j=1}^K q^r_j q^s_j q^t_j \right] \\
\nn  
d_{rst} &= \int_A \mu\wedge J^r \wedge J^s \wedge J^t \\
\chi &=   \sum\limits_{r,s,t} d_{rst} c_3^{rst} = 
	\mbox{Coefficient}(c_3^{rst}J_rJ_sJ_t \mu, \quad \prod_{r=1}^m J_r^{n_r})  \ .
\end{align}
We said `almost', because as always, it is a difficult task to find the individual Hodge numbers for which there is no short-cut, nor explicit formulae as above.

Nevertheless, the full Hodge list was computed in \cite{Green:1987cr} and all the inequivalent configurations, classified in \cite{cicy,cicy2,cicy3}.
We summarize some key results as follows
\begin{itemize}
\item There are {\bf 7890} matrices (dropping the first redundant column in \eqref{cicy}) from $1 \times 1$ (the Quintic, denoted as $[5]^{1,101}_{-200}$) to a maximum of 12 rows, or a maximum of 15 columns;
\item All entries $q_j^r \in [0,5]$;
\item The transpose of any CICY matrix is also a CICY;
\item The 5 cyclics in Table \ref{t:cyclic} are the only ones with a single row, and their transposes give 4 more.
\item There are 266 distinct Hodge pairs $(h^{1,1},h^{2,1}) = (1,65), \ldots, (19,19)$;
\item There are 70 distinct Euler numbers $\chi \in [-200,0]$ (all non-positive and none has $|\chi|=6$).
\end{itemize}
Recently, it was shown \cite{Braun:2010vc} by explicit computer-check using \cite{gap} that 195 have freely-acting symmetries (whereby admitting smooth quotients) with 37 different finite groups, from the simplest $\IZ/2\IZ$, to $(\IZ/8\IZ) \rtimes H_8$, of order 64.

Other famous CICYs include the Tian-Yau manifold $TY =  \left[\begin{array}{ccc} 1&3&0\\ 1&0&3\\  \end{array}\right]^{14,23}_{-18}$ as well as its transpose, the Sch\"on manifold $S = \left[\begin{array}{cc} 1&1 \\ 3&0\\ 0&3\\ \end{array}\right]^{19,19}_0$, both of which have been central in string phenomenology.
It was found by Yau in the early days that $TY$ had a freely-acting $\IZ/3\IZ$ symmetry, so that its quotient  (which is not a CICY) $\left[TY/(\IZ/3\IZ) \right]_{-6}^{6,9}$ has $\chi = -6$.
At this time, this quotient was {\it the} answer to triadophilia. Whilst it had problems with too much extra matter and a less-than-ideal $E_6$ GUT group, it was taken seriously as string theory's potential solution to the universe \cite{Greene:1986ar}.

The manifold $S$ is also interesting, it already existed in the mathematics literature \cite{schoen} as a prime example of elliptic fibration (to which we shall turn later) and a paragon of a self-mirror ($h^{1,1} = h^{2,1}$) manifold.
On $S$, one could find a $(\IZ/3\IZ)^2$ discrete symmetry so that the quotient is a very beautiful (again, non-CICY) self-mirror manifold of Hodge pair $(3,3)$.
In \cite{Candelas:2007ac}, a discussion of these two manifolds, their related cousins as well as quotients, is presented in pedagogical detail.

\begin{figure}[h!!!]
\begin{center}
\begin{tabular}{cc}
\begin{tabular}{c}
$h^{1,1} \begin{array}{l}\includegraphics[width=1.5in,angle=0]{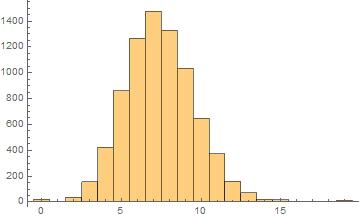}\end{array}$
\\
$h^{2,1} \begin{array}{l}\includegraphics[width=1.5in,angle=0]{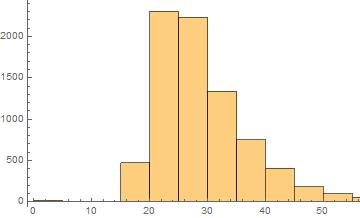}\end{array}$
\\
$\chi \begin{array}{l}\includegraphics[width=1.5in,angle=0]{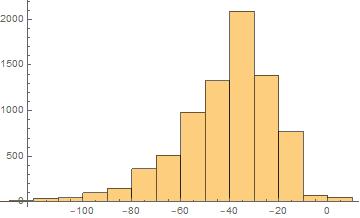}\end{array}$
\end{tabular}
&
$\begin{array}{l}\includegraphics[width=3in,angle=0]{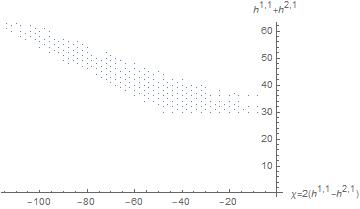}\end{array}$
\end{tabular}
\end{center}
\caption{
{\sf 
The distribution of the Hodge numbers and the Euler numbers for CICYs; 
a plot of the distinct Hodge pairs, organized as $h^{1,1}+h^{2,1}$ versus Euler number 
$\chi = 2(h^{1,1}-h^{2,1})$ .}
}
\label{f:cicydis}
\end{figure}

It is expedient to present the distribution of the {\red Hodge numbers} over the CICY dataset.
In Figure \ref{f:cicydis} we show the frequency plot/histogram for $h^{1,1}$ (somewhat Gaussian), $h^{2,1}$ (somewhat Poisson) and $\chi$ respectively on the left.
On the right of the said figure, we plot the distinct values of the Hodge pair (there are only 266 points since there is much multiplicity).
As is customary we plot twice the difference, i.e., the Euler number on the abscissa and the sum, on the ordinate. This way, mirror symmetry will be very apparent as a flip along the vertical: a mirror pair of Calabi-Yau 3-folds will be two points symmetric about the y-axis.
Moreover, the line $y = |x|$ serves as a natural lower perimeter.

Note the peculiarity of the dataset, that all Euler numbers are non-positive.
This is perfect case where one should be constantly vigilant when confronting data (by the standards of the 1990s, rather ``big'' data), 8 thousand points might have led some to falsely speculate that Calabi-Yau Euler numbers cannot be positive and mirror symmetry would not exist!

\section{Other Datasets}
Prompted by simultaneous developments in physics and mathematics, the early 1990s saw further Calabi-Yau datasets, the prominent ones of which we now present.

\subsection{Hypersurfaces in Weighted $\IC\IP^4$}\label{s:WP4}
Perhaps puzzled by the skewness in the CICY $\chi$ distribution, Philip Candelas - who, being one of the forefathers of mirror symmetry, was surely aware of this asymmetry - looked to beyond CICYs with his friends. 
This led to a series of works shortly after the CICYs in the early 1990s \cite{wp4,Candelas:1994bu}.
Now, CICYs generalized the quintic by extending a single $\IP^4$ to a product of $\IP^n$'s, another natural extension is to consider weights.
Consider the ambient space $A$ as {\it weighted} projective $\IP^4$
\begin{align}\label{wp4}
\nn
& A = W\IP^4_{[w_0 : w_1 : w_2 : w_3 : w_4]} := (\IC^5 - \{0\}) / \sim \ , \qquad  \mbox{ \qquad where } \\
& \qquad \qquad
( x_0, x_1, x_2, x_3, x_4) \sim 
		(\lambda^{w_0}x_0, \lambda^{w_1}x_1, \lambda^{w_2}x_2, \lambda^{w_3}x_3, \lambda^{w_4}x_4) \ .
\end{align}
Indeed, when all $w_i = 1$, this is simply $\IP^4$.
As one might expect, a homogeneous polynomial of degree $\sum\limits_{i=0}^4 w_i$ embedded in $A$ as a hypersurface is Calabi-Yau.

There is a serious complication however: $W\IP^4$ is in general singular and resolution is needed in addition to choosing a generic enough hypersurface which avoids any singularities.
All this was performed in \cite{wp4} and we will not delve too much into the details here as we shall later introduce a dataset into which these $W\IP^4$ hypersurfaces are subsumed.
Nevertheless, this dataset is very convenient in presentation - a single 5-vector of positive integers and, as we see below, is very illustrative in many regards.

\begin{figure}[h!!!]
\begin{center}
\begin{tabular}{cc}
\begin{tabular}{c}
$h^{1,1} \begin{array}{l}\includegraphics[width=1.5in,angle=0]{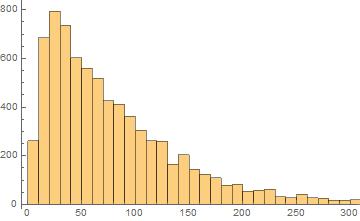}\end{array}$\\
$h^{2,1} \begin{array}{l}\includegraphics[width=1.5in,angle=0]{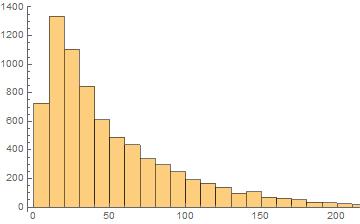}\end{array}$\\
$\chi \begin{array}{l}\includegraphics[width=1.5in,angle=0]{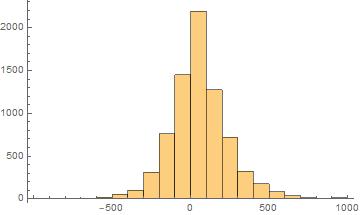}\end{array}$
\end{tabular}
$\begin{array}{l}\includegraphics[width=3in,angle=0]{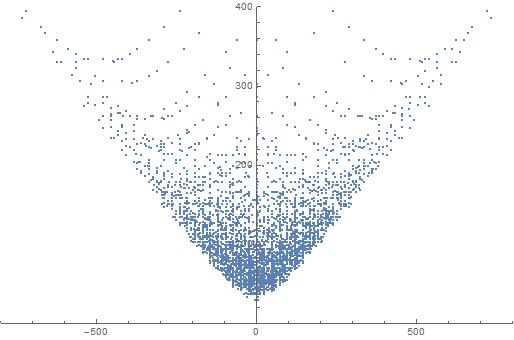}\end{array}$
\end{tabular}
\end{center}
\caption{
{\sf 
The distribution of the Hodge and the Euler numbers for hypersurfaces in $W\IP^4$; 
a plot of the distinct Hodge pairs, organized as $h^{1,1}+h^{2,1}$ versus  
$\chi = 2(h^{1,1}-h^{2,1})$ .}
}
\label{f:wp4dis}
\end{figure}

In all, \cite{wp4} (also independently Klemm, Schimmrigk, Kreuzer-Skarke) found {\red 7555} inequivalent Calabi-Yau 3-folds as 5-vectors, giving us 2780 distinct Hodge pairs and with 
\begin{equation}
\chi \in [-960, 960] \ ; \qquad
h^{1,1} \ , \ \ h^{2,1} \in [1, 491] \ .
\end{equation}
The value 960 is interesting \footnote{
I have bet a bottle of port, as it is a postprandial tradition fortified by drink in some Oxbridge colleges, against Andrew Dancer of Jesus, Oxford, in his College betting book - the volume into which we signed starting from when Napoleon matched into Prussia - that the 960 could never be exceeded.
In retrospect, it was realized that the terms of the bet - neither of us being legally trained - were not entirely unambiguous: should some $|\chi|>960$ 3-fold ever be found after our lifetimes, who would be there to collect or offer the prize? 

Richard Eager has also pointed out that the extremal manifold of Euler number $-960$, is the hypersurface in weighted projective space $W\IP^4_{[1:1:12:28:42]}$, so that after all, 42, the answer to the meaning of life, the universe, and everything \cite{hitchhiker} is involved \cite{DegaWend,He:2014uma}.
}: it is in fact the largest in magnitude for $\chi$ of any Calabi-Yau 3-fold known to date, despite all the plethora of constructions over the decades.
In Figure \ref{f:wp4dis}, we show the histograms for $h^{1,1}$, $h^{2,1}$ (somewhat Poisson) and $\chi$ (somewhat Gaussian), as well as a plot of the distinct Hodge numbers. Now, the Euler number has both positive and negative values, with most concentration on the vertical: the left-right symmetry is mirror symmetry, with apparently self-mirror manifolds dominating in number.
The full list of distinct Hodge paris can be downloaded from  \\
\url{http://hep.itp.tuwien.ac.at/~kreuzer/pub/misc/wp4.spec.gz} \\
as part of the legacy database of \\
\url{http://hep.itp.tuwien.ac.at/~kreuzer/CY/}\\
 upon which we shall expound in depth later in the chapter.

\subsection{Elliptic Fibrations}\label{s:ellipticCY3}
The 1990s saw a surge of interest in elliptic fibrations, in the mathematics because of the minimal model programme \cite{grassiell,grossell,grossdolga}, and and systematic construction of bundles over threefolds \cite{donagi,FMW} (generalizing Atiyah's classical result on vector bundles over the elliptic curve \cite{atiyahEll}), and in the physics because of the emergence of so-called F-theory, which is an elliptic-fibration over type IIb string theory \cite{Friedman:1997yq,Weigand:2018rez}.

The idea of an elliptic fibration is straight-forward. Recall that an elliptic curve - an algebraic realization of CY$_1$ -  is a cubic in $\IP^2$.
This can be brought to Weierstra\ss\ form
\begin{equation}\label{weierstrass}
zy^2 = 4x^3 - g_2 x z^2 - g_3 z^3 \ ,
\end{equation}
where $x,y,z$ are projective coordinates on $\IP^2$ and $g_2, g_3$ are complex parameters \footnote{
The version with which we are perhaps more familiar from our undergraduate days is when $z=1$, so that we are dealing with the non-compact version embedded in $\IC[x,y]$.
} which in terms of the modular parameter $\tau$ are the celebrated Eisenstein series.

Now, fibration simply means that we promote $g_2$ and $g_3$ to specific functions (sections) of chosen coordinates of a base $B$.
Being a 3-fold and with fibres elliptic curves ($dim_{\IC} = 1$), it means that the base $B$ must be a complex surface ($\dim_{\IC}=2$).
Thus one only needs to modify our variables so that they become sections of appropriate bundles over $B$.
Specifically, it turns out taking the anti-canonical line bundle $L := K_B^{-1} = (\wedge^2 T_B^\vee)^{-1}$ of $B$, it suffices to take $(x,y,z,g_2, g_3)$ as global sections of $(L^{\oplus 2}, \ L^{\oplus 3}, L^{\oplus 0}=\cO_B, \ L^{\oplus 4}, \ L^{\oplus 6})$ respectively (one can check that the equation become homogeneous degree 6 in terms of the sections of $L$).
This is a fancier way of saying that we consider the fiber \eqref{weierstrass}  itself  as being a hypersurface in the weighted projective space $W\IP^2_{[2:3:1]}$.

The situation is simplest, for instance, when the base is $\IP^2$, say with homogeneous coordinates $r,s,t$.
We can directly write the variables as homogeneous polynomials of the specified degrees in terms of $(r,s,t)$ and we are back to writing the Calabi-Yau 3-fold as a hypersurface.
The general case is more involved as one needs to find the right projective coordinates to embed $K_B^{-1}$ so that the 3-fold can be written as a projective variety.

There is a common belief that most Calabi-Yau 3-folds are elliptically fibered (numbers such as 80\%, if not more, have been in the air) and it is still very much an active area of research to identify which CY$_3$ is in fact an elliptic fibration.
For CICYs, this was done in \cite{Anderson:2017aux} and for the largest set of toric hypersurfaces (which we shall shortly address), systematic studies were carried out in \cite{Huang:2018esr}, especially for the extremal values.
Some explorations in the landscape of elliptic Calai-Yau 3-folds have been nicely summarized on Wati Taylor's webpage at \\
\url{http://ctp.lns.mit.edu/wati/data.html}.

Whilst there is, to our knowledge, no complete database of elliptic fibration yet, there is a classification \cite{Morrison:1996na} of the possible bases $B$ as well as a computation of Hodge numbers \cite{ellhodge}.
In brief, the Chern classes of the 3-fold can be written in terms of those of the base as \cite{FMW}
\begin{align}
\nn c_1(X) &= 0, \\
\nn c_2(X) &= c_2(B) + 11 c_1(B)^2 + 12 \sigma c_1(B),\\
c_3(X) &= -60 c_1(B)^2 \ .
\end{align}
The base itself can only be of 4 types:
\begin{enumerate}
\item del Pezzo surface { $dP_{r = 0, 1, \ldots, 9}$}, i.e.,
    $\IP^2$ blown up at $r$ points;
\item Hirzebruch surface {$\IF_{r = 0, \ldots 12}$}, i.e.,
    $\IP^1$-bundle over $\IP^1$;
\item Enriques surface {$E$}, i.e, a particular involution of K3 surface;
\item Blowups of $\IF_r$;
\end{enumerate}
The del Pezzo and Hirzebruch surfaces are so central both to complex geometry and to today's theoretical physics that we shall devote an appendix introducing some of their properties in \ref{ap:dPF}.
Furthermore, a beautiful fact about elliptic fibrations is the classification by Kodaira of possible singularity types, for which we also leave the reader to appendix in \ref{ap:Kodaira}.

\section{An Explosion: The Kreuzer-Skarke Dataset}
The mid to late 1990s saw the creation of what still is, by several orders of magnitude, the largest database in Calabi-Yau 3-folds (about $10^{10}$, as we shall soon see), or, for that matter, the largest dataset in pure mathematics \footnote{
Comparable dataset include the GAP project \cite{gap} and related atlas \cite{atlas} on finite groups (about $10^7$), the graded rings project \cite{grdb} of algebraic varieties, especially Fano 3-folds (about $10^8$), the knots \cite{knots} database (about $10^6$) and the L-functions and modular forms database \cite{lmfdb} (about $10^6$).
}.
The idea is as before: how does one generalize $\IP^4$?
We have seen weighting and taking products as ambient spaces \footnote{
Another intermediate is to take products of weighted projective spaces, combining the ideas of the previous two datasets, this was performed in \cite{Kim:1989dq}.
}, a natural next generalization is to take $A$ a toric variety.

Space-time certainly does not permit an introduction to this vast subject of toric varieties, and the reader is referred to the classic of \cite{fultontoric}, the modern tome of \cite{schencktoric}, or such preludes as the pertinent chapter in \cite{mirror} and the brief introduction of \cite{Bouchard:2007ik,Closset:2009sv}, as well as the tutorial in the appendix of \cite{He:2009wi} in our present context.
We also leave the readers to Appendix \ref{ap:toric} to fresh their memory on some notations.

Suffice it to say here that whilst a (weighted) projective space of complex dimension $n$ is $\IC^{n-1}$ (minus the origin), modulo the equivalence relation of the form \eqref{wp4}, a toric variety of complex dimension $n$ is $\IC^{n+k}$ (minus a point set furnished by the so-called Stanley-Reisner ideal), modulo a set of $k$ equivalence relations (encoded by a charge matrix).
All this data can be conveniently repackaged into lattice cones and polytopes in $\IR^n$, and in particular the concept of reflexive polytopes, on the fundamentals of which we now take a lightning review.

\subsection{Reflexive Polytopes}
We begin by recalling
\begin{definition}
A {\blue Convex Polytope} $\Delta$ has 2 equivalent definitions:
\begin{enumerate}
    \item (The Vertex Representation) Convex hull of set $S$ of $k$ points (vertices)
      $p_i \in \IR^n$
      \[ {\rm Conv}(S) = \left\{
      \sum\limits_{i=1}^k \alpha_i p_i \quad : \quad
      \alpha_i \ge 0, \ \sum\limits_{i=1}^k \alpha_i = 1
      \right\} \] 
    \item  (The Half-hyperplane Representation):
      Intersection of linear inequalities (hyperplanes)
      $H \cdot \underline{x} \geq \underline{b}$, where $\underline{b}$ and $\underline{x}$ are real $n$-vector and $H$ is some $k \times n$ matrix. 
\end{enumerate}
\end{definition}
On $\Delta$, the extremal points are called vertices, extremal lines, edges, and then 2-faces, 3-faces, etc, and the $(n-1)$-faces of codimension 1 are called facets.

Next, we henceforth focus only on {\it lattice} or {\it integral} polytopes
\begin{definition}
A convex {\blue lattice} polytope $\Delta_n$ is a convex polytope whose vertices are integral (lattice points), i.e., $p_i \in \IZ^n$.
\end{definition}
Often we use subscript $\Delta_n$ to emphasize the dimension $n$, and we also often drop the adjectives ``convex'' as it is understood to be so.
In low dimension, we are very familiar with $\Delta_n$: $n=2$ give lattice polygons (i.e., polygons whose vertices are given by a pair of integer Cartesian coordinates), $n=3$ give lattice polyhedra, etc.

Finally, in the land of lattice convex bodies, the concept of duality is central:
\begin{definition}
Given a lattice polytope $\Delta$, the {\blue Polar Dual} is the polytope
\[
{\red \Delta^\circ} := \{ \underline{v} \in \mathbb{R}^n ~|~ \underline{m} \cdot \underline{v} \geq -1\;\; \forall \underline{m} \in \Delta \} \ .
\]
\end{definition}
Indeed, this is a duality in the sense that $(\Delta^\circ)^\circ = \Delta$.
Note that as defined, $\Delta^\circ$ is not necessarily a lattice polytope since upon solving the inequality in the definition, the vertices of $\Delta^\circ$ are not guaranteed to be integer, but will be rational in general.
However, in the special case that the polar dual is a lattice polytope, we have that
\begin{definition}
If $\Delta^\circ$ is also a (convex) lattice polytope, then $\Delta$ (and also $\Delta^\circ$ by duality) is called {\red reflexive}. 
\end{definition}
 In the even more special case that $\Delta = \Delta^\circ$, they are call self-dual or self-reflexive.

To illustrate the above definitions, we give a pair of examples in Figure \ref{f:DDual}, which we might have seen from school days.
Here, $\Delta_2$ and its polar dual $\Delta_2^\circ$, both lattice polygons, are given in vertex and half-plane representations.
One can check the duality between them and the fact that both enjoy integer vertices.

\begin{figure}[t!]
\[
  \begin{array}{l|c|l}
    {\red     \Delta_2 }
    &
    \begin{array}{c}
      \includegraphics[width=1.2in,angle=0]{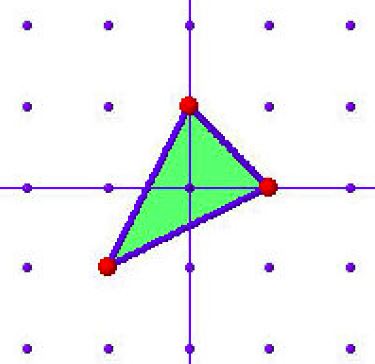}\\
    \end{array}
    &
    \begin{array}{l}
      {\rm Vertices: } \quad (1,0), \ (0,1), \ (-1,-1)
      \\
        {\rm Facets: } \left\{
        \begin{array}{l}
        -x-y \geq -1 \\
        2x-y \geq -1 \\
        -x+2y \geq -1
        \end{array} \right.
    \end{array}
    \\ \hline
    {\red   \Delta^\circ_2 }
    &
    \begin{array}{c}\\
      \includegraphics[width=1.2in,angle=0]{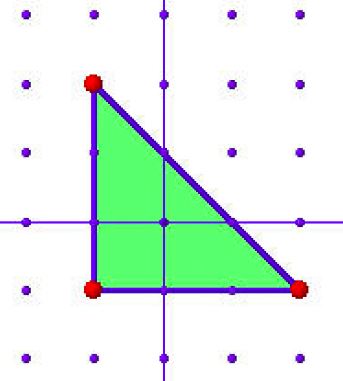}
    \end{array}
    &
    \begin{array}{l}
      {\rm Vertices: } \quad (-1,2), \ (-1,-1), \ (2,-1)
      \\
        {\rm Facets: } \left\{
        \begin{array}{l}
        -x-y \geq -1 \\
        x \geq -1 \\
        y \geq -1
        \end{array} \right.
    \end{array}
  \end{array}
\]
\caption{
{\sf 
A pair of reflexive polytopes (here, polygons), in vertex and hyper-plane representations.
}
}
\label{f:DDdual}
\end{figure}

One important observation we make is the origin is the central point. This is, in fact, a general statement (cf.~\cite{doran} for a popular account):
\begin{theorem}
$\Delta$ is a reflexive polytope $\Leftrightarrow$ $\Delta$ has a single interior lattice point (which can be shifted to be the origin) such that the facets of $\Delta$ are distance 1 hyperplanes therefrom. 
\end{theorem}
We remark that in dimension 2, i.e., for reflexive polygons, we do not even need the second half of the condition: 
reflexivity for convex lattice polygons is the same as having a single interior lattice point \footnote{
I am grateful to Alexander Kaspryzyk for pointing this out.
}.
In dimension greater than 2, however, the situation is more subtle and we need the condition that all facets are distance 1 from the single interior lattice point.

\begin{figure}[h!!]
\[
\Delta_2 =
    \begin{array}{c}
      \includegraphics[width=1in,angle=0]{./PICTS/dP0polygon.jpg}
    \end{array}
    \quad
    \Rightarrow
    \Sigma(\Delta_2) = 
    \begin{array}{c}
      \includegraphics[width=1.2in,angle=0]{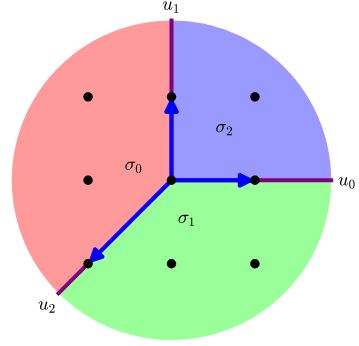}
    \end{array}
    \quad
    \Rightarrow
    X(\Sigma(\Delta_2)) = \IP^2 \ .
\]
\caption{
{\sf 
The (reflexive) polytope giving the face fan for $\IC\IP^2$.
}
}
\label{f:toricP2}
\end{figure}

Having briefly refreshed our minds on reflexive polytopes, the key fact we now use is that they allow us to construct compact toric varieties (the reader is referred to the canonical textbooks \cite{fultontoric,schencktoric} as well as Appendix {ap:toric} for a quick overview).
This is done so via the so-called {\blue Face Fan}
    ${ \Sigma(\Delta)} \equiv \{
    \sigma = \text{pos}(F) \big| F \in \text{Faces}(\Delta)
    \}$ \ where
    $\text{pos}(F) \equiv
    \left\{
    \sum_i \lambda_i \underline{v}_i \big|
    \underline{v}_i \in F ~,~ \lambda_i \geq 0
    \right\}$.
    In other words, as we have a single interior point, we can subtend cones therefrom, joining the various vertices.
    Once we have the fan, we can can obtain a compact toric variety $X(\Sigma)$.
For our $\Delta_2$ example above, we see the standard fan for $\IP^2$, as shown in Figure \ref{f:toricP2}.
This is a nice way to think of $\IP^2$, as encoded by the lattice triangle with vertices $\{(1,0), \ (0,1), \ (-1,-1)\}$.

In general, a reflexive polytope $\Delta_n$ will define a compact (complex) $n$-fold, which is, however, not guaranteed to be smooth.
They are called Gorenstein Fano, in that the anti-canonical sheaf (note that we use sheaf here since the underlying variety is not guaranteed to be smooth) is ample \cite{gorentoric}.
Indeed, as with toric varieties \cite{schencktoric}, $X(\Delta)$ is smooth iff the generators of all the cones $\sigma \subset \Sigma$ are part of a $\IZ$-basis (i.e., 
$\det({\rm gens}(\sigma)) = \pm 1$).
In such a smooth case, $\Delta$ is called {\red regular}.

\subsection{CY Hypersurfaces: Gradus ad Parnassum}\label{s:CYgradus}
Once we have a reflexive polytope $\Delta_n$ and its associated compact {\red Toric Variety} $X(\Delta_n)$, a beautiful construction gives us
\begin{theorem} [Batyrev-Borisov \cite{BB}]\label{thm:BB}
The anti-canonical divisor in $X(\Delta_n)$ gives a smooth Calabi-Yau $(n-1)$-fold as a hypersurface:
	\[
	0 = \sum\limits_{\underline{m} \in \Delta} C_{\underline{m}} \prod\limits_{\rho =1}^k x_\rho^{\langle \underline{m}, \underline{v}_\rho \rangle + 1} \ ,
	\]
where $\underline{m}$ (over which the sum is performed) are all the lattice points inside and on $\Delta$ while $\underline{v}_\rho$ are the vertices of $\Delta^\circ$. The coefficients $C_{\underline{m}}$ are complex numbers specifying, as for projective varieties, the complex structure.
\end{theorem}
In other words, if we have a reflexive $\Delta_4$, then we can easily obtain a hypersurface therein according to the recipe above, which is CY$_3$.
We have not defined divisors in this book because we will not really need this concept.
We can think of it as a codimension 1 sub-manifold (e.g., hypersurface) and the anti-canonical divisor is one of a specific degree.

The simplest {\blue Fano} (toric) 4-fold is $\IP^4$ (all $\IP^n$ are Fano because they have positive curvature), it corresponds to a $\Delta_4$ much like how $\IP^2$ is a toric variety in Figure \ref{f:toricP2}.
Here, the vertices of the polytope and its polar dual are easily checked to be
\begin{equation}\label{P4polytope}
	\begin{array}{cc}
	\Delta:
	\begin{array}{rcl}
	\underline{m}_1 &=& (-1,-1,-1,-1), \\
	\underline{m}_2 &=& (~4,-1,-1,-1) , \\	
	\underline{m}_3 &=& (-1,~4,-1,-1), \\
	\underline{m}_4 &=& (-1,-1,~4,-1), \\
	\underline{m}_5 &=& (-1,-1,-1,~4) \ ,
	\end{array}
	&
	\Delta^\circ:
	\begin{array}{rcl}
	\underline{v}_1 &=& (1, 0,0,0), \\
	\underline{v}_2 &=& (0,1,0,0), \\
	\underline{v}_3 &=& (0,0,1,0), \\
	\underline{v}_4 &=& (0,0,0,1), \\
	\underline{v}_5 &=& (-1,-1,-1,-1) \ .
	\end{array}
	\end{array}
\end{equation}
We can find algorithmically (shortly we will see how this is done on the computer) all the lattice points in $\Delta$ of which there are 126 (reminiscent of \eqref{h21Q}), giving us 126 monomials of degree 5 upon taking the dot product in the exponent.
We have thus retrieved our favourite quintic 3-fold $Q$.

\paragraph{A SageMath Digression: }
It is expedient to take a slight digression on the details of the above computation, as a means to familiarize the reader with SageMath \cite{sage} (luckily the software Polymake \cite{polyDB}, like Macaulay2 and Singular, have been incorporated, and we leave a somewhat combined treatment to Appendix \ref{ap:sage}).
Indeed, the Python-style environment of SageMath and its over-arching vision has rendered it an almost indispensable tool to many contemporary researchers of mathematics and the student versed therein would be very much at an advantage.
Other than downloading the freely available software from \\
\url{http://www.sagemath.org/}, \\
an extremely convenient way to run SageMath is to do so ``via the cloud'', which also allows collaborations, at \\
\url{https://cocalc.com/}.

We begin with defining the polytope for $\IP^4$.
For convenience, we define the dual $\Delta^\circ$ from $\underline{v}$ in \eqref{P4polytope}:
\begin{verbatim}
      P4dual = LatticePolytope([[1,0,0,0],[0,1,0,0],[0,0,1,0],
                         [0,0,0,1],[-1,-1,-1,-1]]);
      P4 = P4dual.polar();
\end{verbatim}
One now checks that the vertices of $\Delta$ are indeed as given by $\underline{m}$ in \eqref{P4polytope}:
\begin{verbatim}
      P4.vertices()
\end{verbatim}
All lattice points on and inside $\Delta$ can now be readily found
\begin{verbatim}
     pts = P4.points()
\end{verbatim}
This returns a long list of 4-vectors, and \verb|len(pts)| checks that indeed there are 126 of them.
Moreover, we can also check polar duality, that $(\Delta^\circ)^\circ = \Delta$ by
\verb|LatticePolytope(pts).polar().vertices()|, giving us back the vertices $\bold v$ of $\Delta^\circ$. 

\paragraph{Increasing Sophistication: } 
Returning to our question of Calabi-Yau 3-folds, one would instantly ask whether there are any more than our old friend $Q$ in $\IP^4$.
Let us re-examine the 5 cyclic 3-folds in \eqref{5cyclic}, their transposes, being CICYs, are also CICY, and thus Calabi-Yau, though the ambient space $A$ is more involved:
\begin{equation}\label{cyclicT}
[4|5]^{1,101}_{-200}
\ , \quad
\left[
\begin{array}{c|c}
1 & 2 \\
3 & 4
\end{array}
\right]^{2,86}_{-168}
\ , \quad
\left[
\begin{array}{c|c}
2 & 3 \\
2 & 3
\end{array}
\right]^{2,83}_{-162}
\ , \quad
\left[
\begin{array}{c|c}
2 & 3 \\
1 & 2 \\
1 & 2
\end{array}
\right]^{3,75}_{-144}
\ , \quad
\left[
\begin{array}{c|c}
1 & 2 \\
1 & 2 \\
1 & 2 \\
1 & 2
\end{array}
\right]^{4,68}_{-128}
\end{equation}
These are all hypersurfaces in (smooth, toric) Fano 4-folds, with the ambient space being, respectively, $\IP^4$, $\IP^1 \times \IP^3$, $\IP^2 \times \IP^2$, $\IP^2 \times \IP^1 \times \IP^1$ and $(\IP^1)^4$, the polytope data for which can also be readily written down.
In a way, proceeding from these 5, to the weighted projective hypersurfaces, to the hypersurfaces in Gorenstein Fano toric 4-folds, constitute a sequential generalization and a slow climb in sophistication, a {\it gradus ad Parnassum}, as it were.

\subsection{$1, 16, 4319, 473800776 \ldots$}
\begin{figure}[h!!!]
\begin{center}
  $\begin{array}{c}\includegraphics[trim=1mm 1mm 1mm 6mm, clip, width=5.3in]{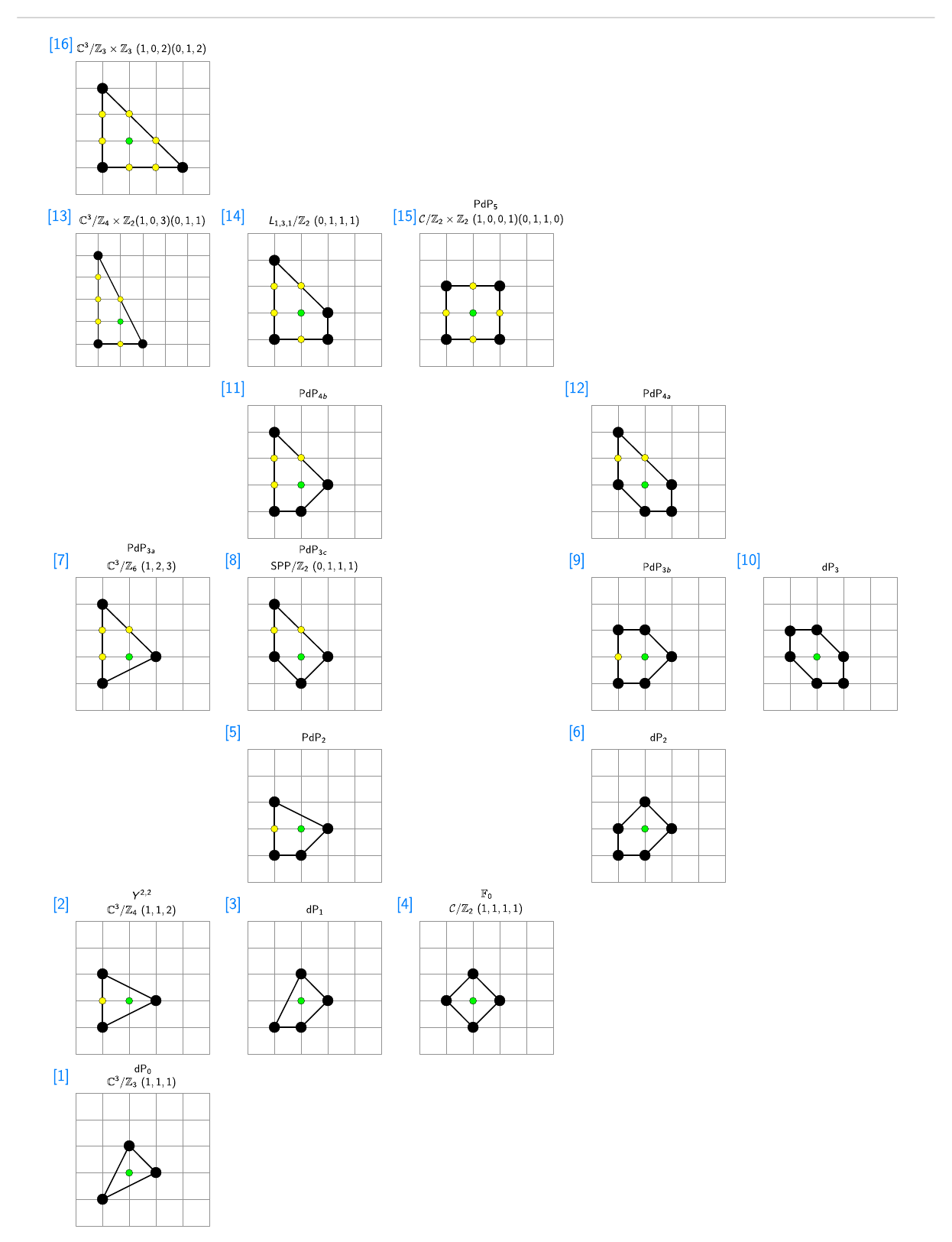}\end{array}$
\end{center}
\caption{
{\sf 
The 16 reflexive polygons $\Delta_2$. The single interior point (origin) is marked green, the vertices, black, and lattice points on the facets but are not vertices, in yellow. Figure taken from \cite{tilings+5,He:2017gam}.
}
}
\label{f:16delta2}
\end{figure}
How many reflexive polytopes $\Delta_n$ are there one might ask.
That is, how many $\Delta_n$ are there in each dimension, up to $GL(n;\IZ)$ equivalence, an equivalence up to which toric varieties are defined.
For $n=1$, there is clearly just 1, the point.
For $n=2$, this is already non-trivial and it turns out there are exactly 16, a classical result dating at least to the early C20th (cf.~\cite{bjorn}  for an interesting account).
For reference, we present these 16 in Figure \ref{f:16delta2}.
The nomenclature may seem mysterious, and to that we shall return in the next chapter.

The diagram is organized so that the number of vertices increases from 3 to 6 from left to right and that of the the area (total number of lattice points) increases from 4 to 10 from bottom to top. Polar duality is reflection along the middle horizontal, on which there are 4 self-reflexive ones.
We have already seen the pair (1 - 16) in Figure \ref{f:DDdual}.
Notable cases are toric del Pezzo surfaces $dP_{0,1,2,3}$ and the zeroth Hirzebruch surface $\IF_0$  (cf.~Appendix \ref{ap:dPF}), these are the 5 smooth Fano varieties of dimension 2.
Hypersurfaces of the form in Theorem \ref{thm:BB} would give CY$_1$, or 16 special elliptic curves.

At $n=3$, the number of reflexive polyhedra was already unknown to the mathematics community until our present story.
Inspired by \cite{BB}, M.~Kreuzer and H.~Skarke [KS] undertook the tremendous task to use the computing technology of the mid 1990s to address the $n=4$ case (which would give a class of desired Calabi-Yau 3-folds) \cite{Kreuzer:1995cd,Kreuzer:2000xy}, and in passing, solved the $n=3$ case as a stepping stone \cite{Kreuzer:1998vb}.
To confront so formidable a problem by a brute-force computer search (for which the package PALP \cite{Kreuzer:2002uu}, one of the first softwares for combinatorial geometry, was developed \footnote{
A recent update and useful manual have been provided in \cite{Braun:2011ik}.
}) indeed required the courage of physicists more than the sublimity of mathematicians.

KS found that there are 4319 $\Delta_3$ upto $GL(3;\IZ)$, of which 18 are regular (and correspond to smooth toric Fano 3-folds).
Hypersurfaces of the form in Theorem \ref{thm:BB} would give CY$_2$, or 4319 algebraic K3 surfaces.
The {\it ne plus ultra} of computer work for mathematics at the turn of the millennium was the tour-de-force computation for $n=4$, taking more than half a year on two dual Pentium III/600MHz PCs and between 10 and 30 processors on a 64 processor SGI Origin 2000 (if the younger members of the readership even know what these are) \cite{Kreuzer:2000xy}, giving us $473,800,652$ $\Delta_4$ up to $GL(4;\IZ)$, of which 124 are regular.

This is ``big data'' even by today's standards, let alone at the turn of the century. 
Thus, our bestiary of Calabi-Yau 3-folds, grew steadily to about $10^4$ between the late 1980s to the mid 1990s, and suddenly exploded to $10^{10}$ (the actual number exceeds even this by many orders of magnitude as we shall see shortly) by the end of last decade of C20th.

In summary, we have two fascinating sequences:
\begin{equation}\label{reflexives}
    \begin{array}{|c||c|c|c|c|c|}\hline
    \mbox{dimension} & 1 & 2 & 3 & 4 & \ldots \\ \hline
    \mbox{\# Reflexive Polytopes} & 1 & 16 & 4319 & 473,800,776 & \ldots \\
    \hline
    \mbox{\# Regular/Smooth} & 1 & 5  & 18 & 124 & \ldots \\ \hline
    \end{array}
\end{equation}

We have no idea what the next number is for the top row and for the bottom row, which is significantly fewer, at least a handful more have been found by exhaustive search \cite{regDelta}, viz., $\{ 1, 5, 18, 124, 866, 7622, 72256, 749892, 8229721 \ldots \}$.
It seems possible that there might be a generating function for this, though so far there is no progress on this front.
The reader is also referred to an excellent recent  account of polytope classification and databases in \cite{polyDB}.
In any event, the relevant data is available from KS's Calabi-Yau page\\
 \url{http://hep.itp.tuwien.ac.at/~kreuzer/CY/} \\
 (the some 473 million requires about 5 Gb of storage, which is not so astronomical for today's laptop) as well as the polytope database\\
\url{https://polymake.org/polytopes/paffenholz/www/fano.html}.

Thus the status stands, partially impeded by the most untimely death of Max Kreuzer
\footnote{
I have a profound respect for Max. It was not long after his visit to Oxford in 2009 - a very productive
and convivial period from which I still vividly remember his distinctive and infectious laughter -  
that Philip, Andre and I received the shocking email that his doctors gave him only a few
months. During the last weeks on his deathbed as cancer rapidly took hold of him, Max emailed
us regularly and our many discussions continued as normal. His several posthumous papers on the
ArXiv are testimonies to his dedication to science. I am honoured and humbled that he, one of the great pioneers of Calabi-Yau data,
should write his last journal paper with us \cite{He:2011rs}. {\it In pace requiescat.}
}
until recently when Harald Skarke carried the torch and produced the remarkable estimate on the next number \cite{Scholler:2018apc}, a staggering 
$2^{{2^6}-4} \simeq 1.15 \times 10^{18}$ of which $185,269,499,015$ are explicitly found.

\paragraph{Topological Data: }
In dimension 3, we saw in \eqref{CYdata} that the CY data is specified by the Hodge numbers, the 2nd Chern class and the intersection numbers.
There is a beautiful formula \cite{Baty} which gave the Hodge numbers in terms of the polytope data
\begin{align}
\nn
	h^{1,1}(X) &= \ell(\Delta^\circ) - \sum\limits_{{\rm codim} \theta^\circ = 1} \ell^\star (\theta^\circ) +
  	\sum\limits_{{\rm codim} \theta^\circ = 2} \ell^*(\theta^\circ) \ell^\star (\theta) - 5 ;
  	\\
\label{hKS}  	
 	h^{1,2}(X) &= \ell(\Delta) - \sum\limits_{{\rm codim} \theta = 1} \ell^\star (\theta) +
  	\sum\limits_{{\rm codim} \theta = 2} \ell^*(\theta) \ell^\star (\theta^\circ)- 5\ .
\end{align}	
In the above, $\Delta$ is the defining polytope for the Calabi-Yau hypersurface, $\Delta^\circ$, its dual; $\theta$ and $\theta^\circ$
are the faces of specified codimension of these polytopes respectively. 
Moreover, $\ell (~)$ is the number of integer points of a polytope while $\ell^\star (~)$ is the number of interior integer points.
From the symmetry between the two expressions, we see immediately that
\begin{corollary}
Polar duality $\Delta \leftrightarrow \Delta^\circ$ for the ambient toric variety  is mirror symmetry for the CY$_3$ hypersurface.
\end{corollary}

However, in order to compute the second Chern class and the intersection numbers, we again need to establish the sequence-chasing as in \S\ref{ssHQ}, requiring that, in particular, the ambient space be smooth.
As we saw in \eqref{reflexives}, the number of regular polytopes (smooth varieties) are very rare and almost all the $\Delta$ require desingularization, which on the level of the polytope corresponds to ``maximal triangulations'' \cite{Kreuzer:1995cd,Kreuzer:2002uu} (q.v., \cite{Altman:2014bfa} for a tutorial on how this is done using {\sf PALP}, and how to extract the Chern classes and intersection numbers).
It should thus be emphasized that the actual CY hypersurfaces are much, much more than $473,800,776$: even though the Hodge pair does not depend on triangulation, the Chern class $c_2$ and the intersection form $d_{rst}$ do, so the topological type of the CY$_3$ crucially depend on different triangulations.

Unfortunately, triangulating polytopes is an exponentially expensive process.
For small $h^{1,1}$ (up to 6), the full triangulations can be done on the computer while for large $h^{1,1}$ (between 240 and 491), methods were also devised to do so (q.v.~Table 1 of \cite{Demirtas:2018akl} for the number of polytopes organized by $h^{1,1}$).
To give an idea of the growth rate, the results from  \cite{Altman:2014bfa} on the low K\"ahler parameters are
\begin{equation}
\begin{array}{|c||c|c|c|c|c|c|c|}\hline
h^{1,1} & 1 & 2 & 3 & 4 & 5 & 6 & \ldots \\ \hline
\mbox{\# of Polytopes} & 5 & 36 & 244 & 1197 & 4990 & 17101 & \ldots \\
\mbox{\# of Triangulations} & 5 & 48 &526 &5348 &57050 &590085 & \ldots \\
\mbox{\# of CY} & 5 &39 &306 &2014 &13635& 85682 & \ldots \\ \hline
\end{array}
\end{equation}
where after triangulation, the number of unique Calabi-Yau hypersurfaces (having distinct data in the sense of \eqref{CYdata}) is also checked.
Note that the 5 $h^{1,1}$ cases are the 5 transposes of the cyclics as CICYs, the ambient spaces for which are of course smooth and require no triangulation.

The motivated readers, especially when led by a curiosity induced by insomnia to befriend some Calabi-Yau 3-folds -  is highly encouraged to play with the interactive webpages (cf.~accompanying papers \cite{Gray:2012jy,Altman:2014bfa})  of 
Benjamin Jurke \\
\url{https://benjaminjurke.com/academia-and-research/calabi-yau-explorer/}
\\
and
Ross Altman \\
\url{http://www.rossealtman.com/} \ .
 
Furthermore, the reader will be happy to learn that the 16 and the 4319 are built into SageMath \cite{sage} using the command \verb|ReflexivePolytope(d,n)| where $d$ is the dimension and $n$ is the $n$-th $\Delta$.
The $473,800,652$, however, are currently too large to be built-in, but can be downloaded from \\
\url{http://hep.itp.tuwien.ac.at/~kreuzer/CY/} \\
in {\sf PALP} format.
We mention that in the KS dataset, the simplest case of $\IP^4$ is given in a different but $SL(4; \IZ)$ equivalent form, with $\Delta$ given as
\begin{verbatim}
 4 5  M:126 5 N:6 5 H:1,101 [-200]
     1   1   1   1  -4
     0   5   0   0  -5
     0   0   5   0  -5
     0   0   0   5  -5
\end{verbatim}
Here, the first row records that the ensuing matrix for $\Delta^\circ$ will be $4 \times 5$, that there are 126 lattice points and 5 vertices for $\Delta$, 6 lattice points and 5 vertices for $\Delta^\circ$ (the one more being the origin in the interior).
Consequently the Hodge numbers are $(1,101)$ and $\chi = -200$ for the CY hypersurface.

\paragraph{A Georgia O'Keefe Plot: }

\begin{figure}[h!]
\begin{center}
      \begin{tabular}{cc}     
      (a) $\begin{array}{l}\includegraphics[width=2.6in,angle=0]{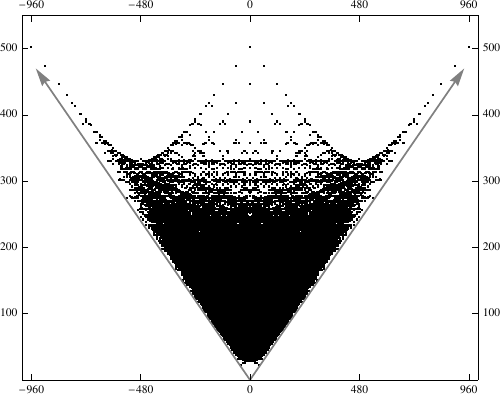}\end{array}$
      &
      (b) $\begin{array}{l}\includegraphics[width=2.4in,angle=0]{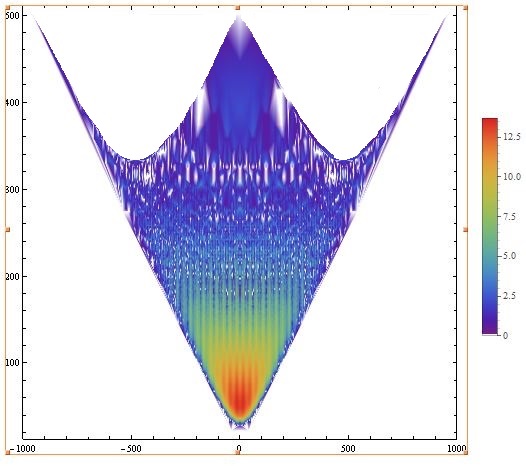}\end{array}$
    \end{tabular}
\end{center}
\caption{
{\sf 
(a) A plot of the 30,108 distinct Hodge pairs for the Kreuzer-Skarke dataset of Calabi-Yau  hypersurfaces in Gorenstein Fano toric 4-folds;
(b) the same but with multiplicity over the log-density of the $473,800,652$ over these distinct values.
}
}
\label{f:KSplot}
\end{figure}

The KS dataset  produced 30,108 distinct Hodge pairs, $\chi \in [-960, 960]$ (note that since the Hodge numbers are triangulation independent, even getting the full list of CY hypersurfaces someday when there is computing power to do all triangulations will not change this).
The extremal values of $\pm 960$, as mentioned in the footnote of \S\ref{s:WP4}, are actually two hypersurfaces in weighted $\IP^4$, corresponding to the mirror pair of $(11, 491)$ (for weights $w_i = [1: 1: 12: 28: 42]$) and $(491,11)$ (for weights $w_i = [21: 41: 249: 581: 851]$).
As always, we can plot $h^{1,1}+h^{2,1}$ versus $\chi$, as was done in \cite{Kreuzer:2000xy}.

\begin{figure}[h!!]
(a)		\includegraphics[width=2.3in]{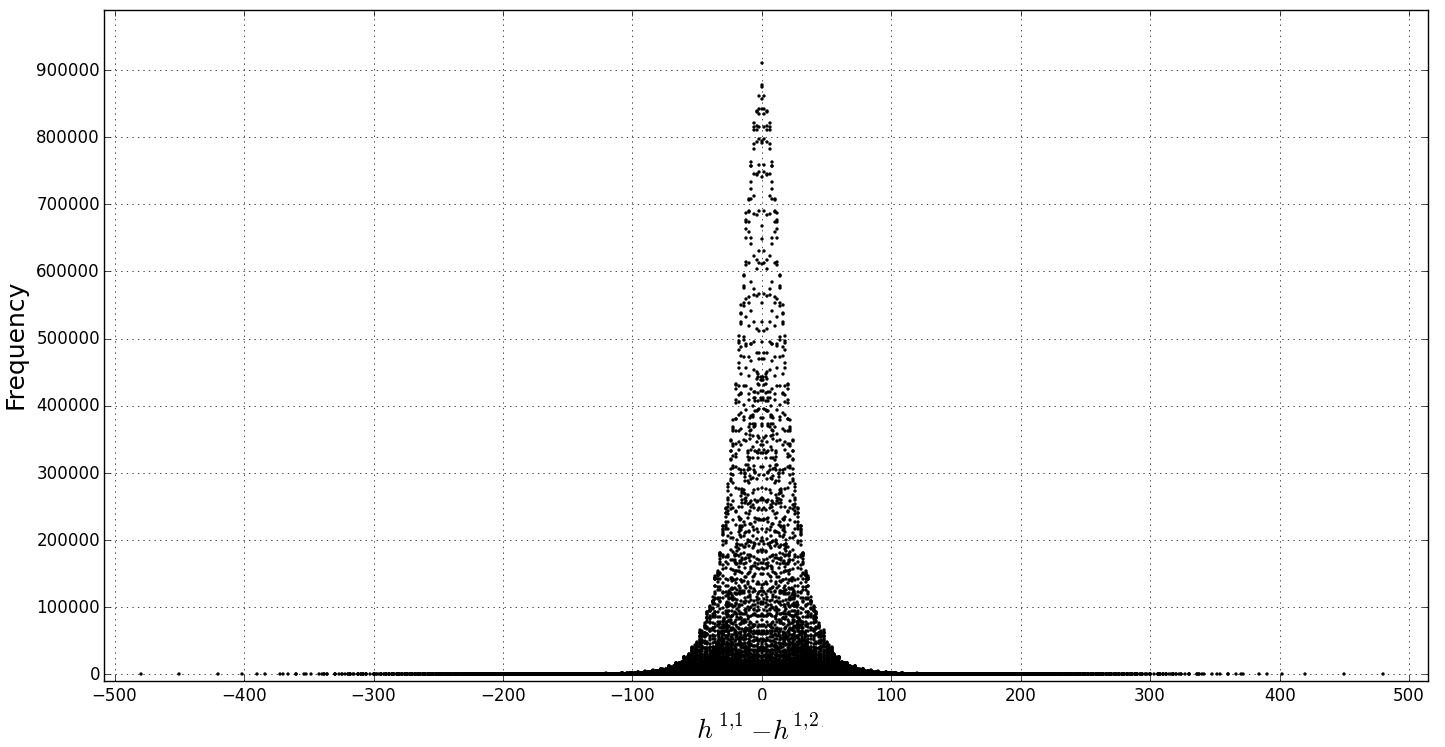}
(b)		\includegraphics[width=2.3in]{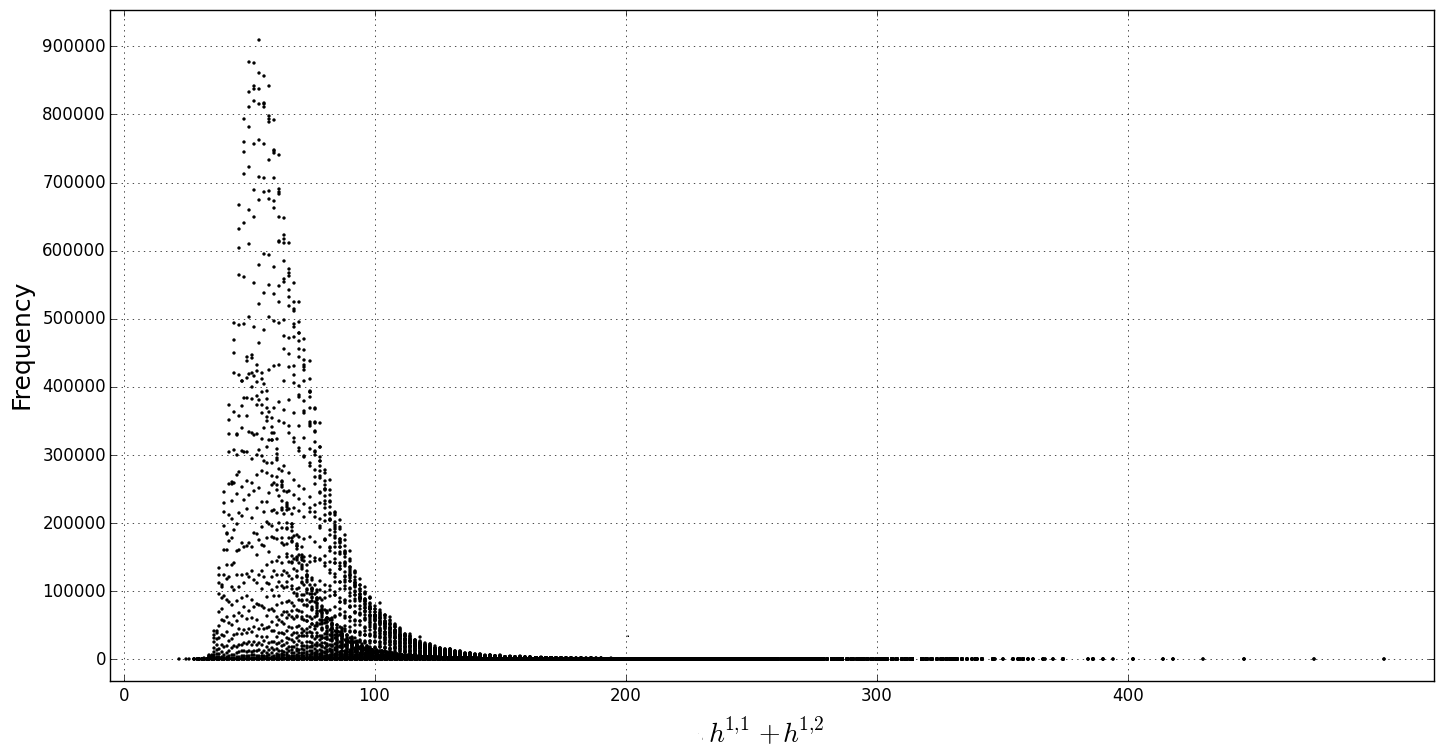}
\caption{
{\sf
  (a) Frequency $f$ plotted against $\frac12 \chi = (h^{1,1} - h^{1,2})$; 	
  (b) Frequency $f$ plotted against the sum of Hodge numbers $(h^{1,1} + h^{1,2})$.}  	
\label{f:histogramKS}
}
\end{figure}

This has become of the most iconic plots in string theory, as it is, {\it inter alia}, the best experimental evidence for mirror symmetry: every point has its mirror image along the vertical.
We reproduce this plot in part (a) of Figure \ref{f:KSplot}, which is framed in Philip Candelas' office.
My only contribution - undoubtedly inspired by my 4-year-old daughter - was to colour it in \cite{He:2013epn}:
in part (b) of the said figure, a heat plot of the log-density (i.e., log of the multiplicity of how many different polytopes per point of Hodge pair: we have  $473,800,652$ $\Delta_4$ but only  30,108 distinct values using \eqref{hKS}) is presented.
It is also curious to note that $(27,27)$ is the most occupied point: with a multiplicity of $910113$.
The distribution of the Hodge numbers follows pseudo-Voigt/Planickian curves and is the subject of \cite{He:2015fif}.
For given an idea of the sharp peak around $(27,27)$, we show a histogram of the number of manifolds versus the two Hodge numbers in Figure \ref{f:histogramKS}.

There are numerous features of the plot in Figure \ref{f:KSplot}, most of which are still unexplained.
Other than the extremals of $\pm 960$, why is there a boundary on top (the two bounding straight-lines of funnel shape is just by definition of the plot, that the Hodge numbers are non-negative), why do they appear as parabolae?
The papers of \cite{Johnson:2014xpa,Demirtas:2018akl} identify these as elliptc fibrations while \cite{Candelas:2012uu} find intriguing $E_n$ (del Pezzo) structure.

Independent of the KS dataset, no Calabi-Yau from {\it any} construction to date has ever produced a Hodge pair above those two puzzling parabolae though there is no theoretical reason why this is so.
On the other hand, at the bottom tip, there is also a paucity of manifolds, and this is also true for CY databases in general.
This zoo of manifolds of small Hodge numbers \footnote{
This tip of the plot, where Hodge number are small, is what Philip \cite{Candelas:2007ac} calls a ``des res'', or a ``desired residence'', in reference to newspaper advertisements back in the day before social media.
}, is also much investigated \cite{Candelas:2007ac,Candelas:2016fdy,Candelas:2008wb}.
Recently, there is systematic construction \cite{HalMikeSmall} for (non-CI) CY in $\IC\IP^n$ which gives small Hodge numbers.

Of course, in addition to all the datasets mentioned above, the CICYs, the elliptic fibrations, the KS, etc, there have been many other constructions, such as using Grassmannians or flag varieties as ambient space \cite{grassmannian} or various quotients, etc; and a good compendium is in Table 9 of \cite{Candelas:2008wb}.
Since these have not produced large or available databases online, nor are comparable in number to the KS set, we shall not address them here.
Finally, we mention that shortly before Max Kreuzer's passing, he produced partial results for some $10^{10}$ double-hypersurfaces in toric Fano varieties coming from reflexive polyhedra in dimension 5. Amazingly, none of their Hodge pairs reside outside the bounds of Figure \ref{f:KSplot}.

\paragraph{Caveat Emptor: }
We need to warn the reader that the ``explosion'' in the title of this section simply refers to the vastness of the Kreuzer-Skarke dataset in contrast to the others, especially those emerging from the 1990s-2000s (even the graded rings database with its impressive list of Fano varieties \cite{grdb} is not of the same order of magnitude).
We will shortly see an important conjecture on the finiteness of topology types for Calabi-Yau manifolds. However, one could still have infinite families of (compact, smooth, connected) Calabi-Yau manifolds distinguished by properties more refined than mere topology, such as Gromow-Witten (GW) invariants.
For instance, an ongoing  programme \cite{GWinf} to understand the infinite possibilities of GW invariants from so-called generalized CICYs \cite{Anderson:2015iia}, which relax the non-negativity condition in the configurations for CICYs.
Likewise, \cite{gasparim} has recently shown infinite family of deformations.

\section{Cartography of the Compact Landscape}\label{s:compVenn}
Thus we have taken a promenade in the landscape of Calabi-Yau 3-folds, a contiguous development spanning a decade, from the late 1980s till the turn of the millenium.
With the discovery of D-branes \cite{Polchinski:1995mt}, M-theory on $G_2$ manifold \cite{Atiyah:2001qf}, F-theory on 4-folds \cite{Morrison:1996na,Friedman:1997yq,F4fold}, and, of course, the holographic correspondence \cite{Maldacena:1997re} all approximately within the final lustrum of the century, coupled with the computational limits of the time, the fervour of constructing datasets of compact smooth Calabi-Yau 3-folds by theoretical physicists was relatively cooled.
Meanwhile, pure mathematicians interested in Calabi-Yau manifolds had too great a wealth of ideas ranging from enumerative geometry to homological mirror symmetry to preoccupy themselves with the mundanity of data-mining.

\begin{figure}[t!]
	\centering
	\includegraphics[width=4in,angle=0]{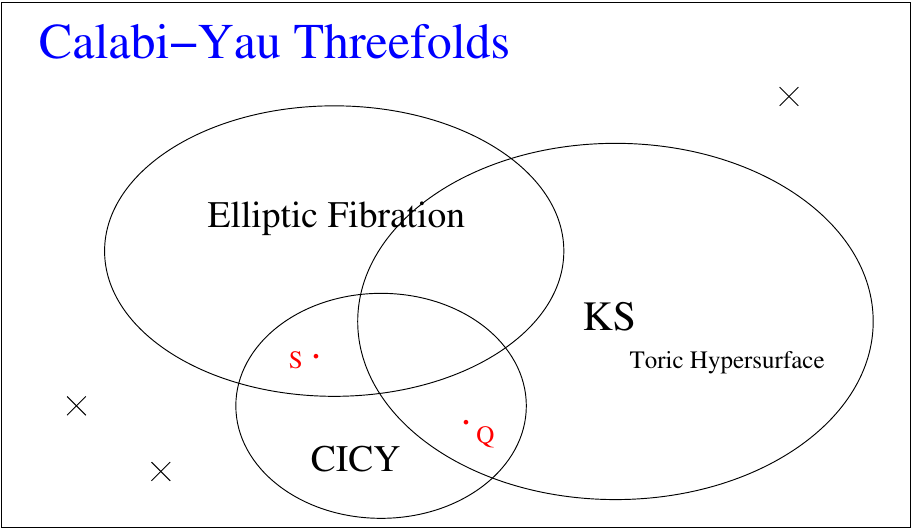}
	\caption{{\sf The landscape of compact smooth Calabi-Yau 3-folds.}}
	\label{f:landCompact}
\end{figure}

We saw above, and will shortly see and  shall see again in Chapter \ref{s:learn} how there has been renewed interest in the various datasets.
For now, we summarize the {\bf landscape} of compact smooth Calabi-Yau 3-folds in the Venn diagram in Figure \ref{f:landCompact}.
The three major datasets are shown with size of the bubbles not to scale. The crosses are supposed to signify various other constructions mentioned above. Our most familar friends, the quintic $Q$ and the Schoen $S$, are marked.
With the continual growth in number of the Calabi-Yau data, one might be led to wander whether there might be an infinite number.
While this is unsettled, there is an important conjecture of Yau \cite{yaubook2} that
\begin{conjecture}[Yau]\label{yauconj}
There is a finite number of {\blue Topological Types} of (compact, smooth, connected) Calabi-Yau threefolds in the sense that there is a finite number of possible values to the data \eqref{CYdata}.
\end{conjecture}
So far, this full data, viz.,
\begin{equation}
\big\{ (h^{1,1}, h^{2,1}) \ ; \quad \ [c_2]_r \ ; \quad d_{rst} \big\}  \ , \qquad r,s,t = 1, \ldots, h^{1,1} \ ,
\end{equation}
is really only known for the CICYs, \\
\url{http://www-thphys.physics.ox.ac.uk/projects/CalabiYau/cicylist/} \ , \\
and very partially, for the KS data:\\
\url{http://www.rossealtman.com/} \\
and it would be interesting to study the statistics thereof.
For known boundedness results, q.v.~\cite{wilson} and the Reid's Fantasy \cite{reidfantasy}

Conjecture \ref{yauconj} is even expected to hold for {\it any dimension}, in that the topological types of the boundary case of Ricci-flatness (Calabi-Yau) is finite.
In complex dimension one, we have already seen the validity of the conjecture from Figure \ref{f:RiemannSurface}: the genus 1 torus is the only Ricci-flat (compact, connected, complex) 1-fold. Negative Ricci curvature, on the other hand, has an infinite number, viz., all $g > 1$.
In complex dimension two, Ricci-flatness only comprises of the topology of the 4-torus $T^4$ and the K3-surface, whilst there is an infinity of surfaces of general type, with negative curvature.

\section{Epilogue: Recent Developments}
Whereas the original triadophilia, which we recall to be the physicists' search for smooth compact Calabi-Yau 3-folds with $\chi = \pm 6$, saw its {\it fin-de-si\`ecle} ebb by the  aforementioned  explosion of other string vacua and by the enormity of the KS data (there are more than $10^6$ in the list with $\chi = \pm6$), there had been renewed interest over the past decade in an extension of more physical importance.

Again, we emphasize that by ``recent developments'' we only refer to the Heterotic programme which was the original one that brought Calabi-Yau manifolds into physics.
As listed in \S\ref{s:landscape}, there is a host of ongoing activities which study the ``string landscape'', and which 
study a myriad of interesting geometrical and combinatorial problems in Calabi-Yau varieties and beyond.
In some sense, string theory has traded one very hard problem - the quantization of gravity - by another very hard problem - the selection of the right vacuum. Perhaps to the mathematician, the latter is even more interesting as it compels us to probe the landscape of geometric and algebraic structures.

Returning to the heterotic string, we saw that
using the tangent bundle $T_M$ in \eqref{hym} gave, by \eqref{E8E6}, GUT theories with $E_6$ gauge group.
By taking $V$ not being $T_M$, but, for example, a stable $SU(4)$ or $SU(5)$ bundle, one could obtain the more interesting commutant $SU(10)$ or $SU(5)$ GUTs. This has come to be known as ``non-standard'' embedding and has with the developments in the theory of stable bundles on Calabi-Yau manifolds become an industry of realistic model building (cf.~e.g., \cite{Donagi:2004ia}).

One can do even better.
With the incorporation of Wilson lines, i.e., representations of the fundamental group $\pi_1(M)$, should it be non-trivial, $SO(10)$ and $SU(5)$ can be further broken to the exact standard model group and instead of Hodge numbers (which are cohomologies valued in $T_M$ its duals and exterior powers), one computes the more general bundle-cohomologies valued in $V$, projected by the $\pi_1(M)$-representations.

\begin{question}
In summary, the mathematical problem of general heterotic string compactification becomes
\begin{itemize}
\item Find a smooth, compact Calabi-Yau 3-fold $M$ with non-trivial fundamental group $\Gamma = \pi_1(M)$.
	This is usually done by finding an ``upstairs'' CY 3-fold $\widetilde{M}$ with a freely acting discrete symmetry $\Gamma$ so that 
	``downstairs'' $M \simeq \tilde{M} / \Gamma$;
\item Find a stable holomorphic $SU(4)$ or $SU(5)$ vector bundle $V$ with index $\pm3$ on $M$ (cf.~Theorem \ref{HRR});
\item Compute the cohomology groups $H^*(M, \wedge^i V)$ which must match the table in \eqref{SM}. 
This done by constructing $\widetilde{V}$ (of index $\pm 3 |\Gamma|$) on $\widetilde{M}$ equivariant under $\Gamma$ and computing the equivariant cohomologies  
$H^*(\widetilde{M}, \wedge^i \widetilde{V})^\Gamma$ together with the projection of a chosen representation of $\Gamma$.
\end{itemize}
\end{question}
This is a rich and interesting story in algebraic geometry and in physics and we do apologize that a subject deserving an entire book by itself - a feeble attempt was made in \cite{He:2010uj} as a quick introduction - has been relegated to an epilogue.
However, venturing to the landscape of the extra data of bundles over Calabi-Yau manifolds will clearly be incompatible with the limitations of space and energy.

The problem, due to its physical significance, galvanized Burt Ovrut \footnote{
The relentlessness with which Burt pursues this ultimate Triadophilia -- of not only 3 generations but the exact spectrum and precise properties of the standard model -- is truly inspiring. I deeply appreciate the guidance he has given me over the years.
} to launch an ongoing programme by teaming up with card-carrying geometers Ron Donagi and Tony Pantev at UPenn during the early years of the millennium.
It was supported by one of the earliest inter-disciplinary grants for algebraic geometry/theoretical physics by the USNSF \cite{PennProg}, a collaboration at whose fruiting stages I was fortunate to join.
The first answer to the above question, called the ``Penn Model'',  was found in \cite{HetMSSM}, built on a particular bundle of extension type on a quotient of the Schoen manifold, by tuning parameters in constituent line bundles in the exact sequence and exploiting the double elliptic-fibration structure of $S$.

Again, with the advances in computer algebra, the above problem can be algorithmized \cite{Anderson:2007nc} over the various datasets (CICYs \cite{Anderson:2008uw}, elliptic fibration \cite{Gabella:2008id} and some preliminary cases of the KS set \cite{He:2009wi,He:2011rs}), with a few more exact solutions produced \cite{newExact}, before a tour-de-force scan over Whitney sums of line-bundles within the regions of stability inside the K\"ahler cone was performed over the CICY list \cite{hetLine}.
Over an impressive consortium of some $10^{10}$ candidates which would have made KS proud, \cite{hetLine} found about 200 exact solutions to the problem, whereby giving 200 possibilities of (heterotic) string compactifications on smooth compact Calabi-Yau 3-folds endowed with (poly-)stable holomorphic vector bundles whose cohomologies give the {\it exact} charged matter content of the MSSM \footnote{
As always with these constructions, the uncharged matter corresponds to $H^*(M, V \otimes V^\vee)$, which give scalar moduli in the field theory.
}.

It is interesting that this 1 in a billion factor of reduction is very much in line with independent model-building results from type II strings on branes \cite{Gmeiner:2005vz}. Recently it has been suggested that even within the CICY list alone, there might be $10^{23}$ bundles \footnote{
Tristan H\"ubsch very charmingly calls it ``a mole of models'' and we almost entitled the paper as such.
} giving the exact standard model matter content \cite{Constantin:2018xkj}.
Indeed, the string landscape is far more vast \cite{complexity} that the initial quick estimates of $10^{500}$ \cite{Kachru:2003aw}.
The statistics of string vacua is very much an active field \cite{stats} and one could, as with fields of cosmogony or exo-planets, take either the anthropocentric view that there is a fundamental selection rule rendering our universe ``special'' or the more existentialist  one that we are simply a stochastic point in a multitude of universes.

\section{Post Scriptum: Calabi-Yau Metrics}\label{s:metric}

We have now taken a long stroll in the garden of Calabi-Yau manifolds, one of whose key definitions is its Ricci-flat metric, without a single mention of the metric.
This, as mentioned from the outset, is not due to negligence but to the fact (and challenge!) that no analytic Ricci-flat K\"ahler metric has ever been found on a compact Calabi-Yau $n$-fold (other than the trivial case of tori $T^{2n} \simeq \IC^n / \Lambda$ for some lattice $\Lambda$ where the flat metric on $\IC^n$ is inherited from the discrete quotient).
One could try to pull-back the Fubini-Study metric on $\IP^4$ onto the quintic as a natural course of action, but would find the resulting metric to be positively curved in some patches and analytically finding a Ricci-flat metric globally has so far defied every attempt.

Once we dispense with ``compactness'', analytic metrics can indeed be found, as we will see in this next chapter.
As a postscriptum to this chapter, let us see some important directions in at least numerically computing the Calabi-Yau metric.
It will also serve to put on concrete footing some of the terminology introduced in the previous and present chapters, as well as Appendix \ref{ap:geo}, such as K\"ahler and volume forms.
There are two main approaches to this: the so-called Donaldson algorithm \cite{metricDon1,metricDon2,Douglas:2006rr,Braun:2007sn,Anderson:2010ke,Ashmore:2019wzb} and the functional method \cite{Headrick:2005ch,Headrick:2009jz}.
As the focus on this book is on the algebraic and algorithmic, we will give a brief exposition of the former.

\newcommand{\xb}{\bar{x}}
\newcommand{\del}{\partial}
\newcommand{\delb}{\bar{\partial}}
\newcommand{\bb}{\bar{\beta}}
\newcommand{\sbar}{\bar{s}}
\newcommand{\Ob}{\overline{\Omega}}
\newcommand{\gb}{\bar{\gamma}}
\newcommand{\gk}{g^{(k)}_{a\bar b}}
\newcommand{\db}{\bar{\delta}}
\newcommand{\zb}{\bar{z}}
\newcommand{\jb}{\bar{j}}
\newcommand{\hmat}{h^{\alpha\bar\beta}}
\newcommand{\vcy}[1]{\text{Vol}_{\text{CY}}}
\newcommand{\vk}[1]{\text{Vol}_{\text{K}}}

Let $X$ be a smooth, compact Calabi--Yau threefold, a compatible Hermitian metric $g_{a\bar b}$, with its associated
K\"ahler form $\omega =  i g_{a\bar{b}} dz^{a} \wedge dz^{\bar{b}}$
and a nowhere-vanishing complex (volume) three-form $\Omega$. 
The Calabi-Yau condition is then:
\begin{equation}
d \omega = 0 (\mbox{K\"ahler}),\qquad \Omega \stackrel{!}{\neq}  0, d \Omega = 0 (\mbox{Ricci-flat}) \ .
\end{equation}
Let $x^a$, $a=1,2,3$ be the three complex coordinates on $X$. 
Since $g_{a \bar b}$ is Hermitian, the pure holomorphic and anti-holomorphic components of the metric must vanish.
Only the mixed components survive, which are given as the mixed partial derivatives of a single real scalar function, the K\"ahler potential $K$:
\begin{equation}\label{gijbar}
g_{a b}(x, \xb) = g_{\bar a \bar b}(x, \xb) = 0 \ , \qquad
g_{a \bar b}(x, \xb) = \partial_a \partial_{\bar b} K(x, \xb).
\end{equation}
The K\"ahler form derived from the K\"ahler potential is
\begin{equation}
\omega = \frac{i}{2} \sum_{a, \bar b = 1}^3 g_{a \bar b}(x, \xb) d x^a \wedge d \bar{x}^{\bar b} = \frac{i}{2} \del \delb K(x, \xb),
\end{equation}
where $\del$ and $\bar{\del}$ are the Dolbeault operators (so that in terms of the ordinary differential
$d = \del + \delb$ and $\del^2 = \delb^2 = 0$).

Recall that $K$ is only {\it locally} defined -- globally one needs to glue together the local patches by finding appropriate transition functions $f$ (K\"ahler transformations) so that
\begin{equation}
K(x, \xb) \sim K(x, \xb) + f(x) + \bar{f}(\xb) .
\end{equation}
Since $X$ is K\"ahler, the Ricci tensor is given by
\begin{equation}
R_{a \bar b}=\partial_{a}\partial_{\bar{b}}\ln \det g_{a \bar b} .
\end{equation}
In practice, finding a Ricci-flat K\"ahler metric on $X$ reduces to finding the corresponding K\"ahler potential as a real function of $x$ and $\xb$. 
Yau's celebrated proof~\cite{yau} of the Calabi conjecture~\cite{calabi} guarantees that this Calabi--Yau metric is unique in each K\"ahler class.

The general idea of Donaldson's algorithm~\cite{metricDon1} is to approximate the K\"ahler potential of the Ricci-flat metric using a finite basis of degree-$k$ polynomials $\{s_\alpha\}$ on $X$, akin to a Fourier series representation. 
This ``algebraic'' K\"ahler potential is parametrized by a hermitian matrix $h^{\alpha\bar\beta}$ with constant entries. Different choices of $h^{\alpha\bar\beta}$ correspond to different metrics within the same K\"ahler class. 
One then iteratively adjusts the entries of $h^{\alpha\bar\beta}$ to find the ``best'' (in the sense of a \emph{balanced} metric at degree $k$ which we define shortly) degree-$k$ approximation to the unique Ricci-flat metric. 
As one increases $k$, the balanced metric becomes closer to Ricci-flat at the cost of exponentially increasing the size (and thus sadly increasing computational complexity) of the polynomial basis $\{s_\alpha\}$ and the matrix $h^{\alpha\bar\beta}$.
Specifically, we perform the following:

\begin{enumerate}
	
	\item Let the degree $k$ be a fixed positive integer and denote by $\{s_\alpha \}$ a basis of global sections of $\cO_X(k)$ (q.v.~Appendix \ref{ap:geo}):
	\begin{equation}
	H^0(X, \cO_X(k)) = \operatorname{span} \{ s_\alpha \} , \qquad \alpha = 1, \ldots, N_k.
	\end{equation}
	In other words, we choose an $N_k$-dimensional basis of degree-$k$ holomorphic polynomials $s_\alpha(x)$ on $X$. The values of $N_k$ grow factorially with $k$.
	
	\item Make an ansatz for the K\"ahler potential of the form
	
	\begin{equation}\label{eq:KP}
	K(x, \xb) = \frac{1}{k \pi} \ln \sum_{\alpha, \bb = 1}^{N_k} h^{\alpha \bb} s_{\alpha}(x) \sbar_{\bb}(\xb),  
	\end{equation} 
	
	where $h^{\alpha \bb}$ is some invertible hermitian matrix. We see that without the mixing by the matrix $h$, the shape of $K$ is that of the potential of the standard Fubini-Study metric on $\IC\IP^n$.

	\item The pairing $h^{\alpha \bb} s_{\alpha} \sbar_{\bb}$ defines a natural inner product on the space of global sections, so that $h$ gives a metric on $\cO_X(k)$. Consider the Hermitian matrix
	\begin{equation}\label{Hh}
	H_{\alpha \bb} \equiv  \frac{N_k}{\vcy{X}} \int_X d \vcy{X}
	\frac{s_\alpha \sbar_{\bb}}{ h^{\gamma \db} s_{\gamma} \sbar_{\db}},
	\end{equation}
	where $\vcy X$ is the integrated volume measure of $X$:
	\begin{equation}
	d \vcy{X} = \Omega \wedge \Ob.
	\end{equation}
	In general, $\hmat$ and $H_{\alpha\bar\beta}$ will be unrelated. However, if they are inverses of each other
	\begin{equation}\label{balanced}
	h^{\alpha \bb} = ( H_{\alpha \bb} )^{-1} ,
	\end{equation}
	the metric on $\cO_X(k)$ given by $\hmat$ is said to be {\bf balanced}. This balanced metric then defines a metric $g^{(k)}_{a \bar b}$ on $X$ via the K\"ahler potential \eqref{eq:KP}. We also refer to this metric on $X$ as balanced.
	
	\item Donaldson's theorem then states that for each $k \geq 1$ a balanced metric exists and is \emph{unique}. Moreover, as $k\to\infty$, the sequence of metrics $g^{(k)}_{a \bar b} = \partial_a \partial_{\bar b} K$ converges to the unique Ricci-flat K\"ahler (Calabi--Yau) metric on $X$.
	
	\item In principle, for each $k$, one could solve \eqref{balanced} for the $\hmat$ that gives the balanced metric using \eqref{Hh} as an integral equation. However, due to the highly non-linear nature of the equation, an analytic solution is not possible. Fortunately, for each integer $k$, one can solve for $\hmat$ iteratively as follows:
	\begin{enumerate}
		\item Define Donaldson's ``$T$-operator'' as
		\begin{equation}\label{eq:T_operator}
		T \colon h^{\alpha \bb}_{(n)} \mapsto   T(h_{(n)})_{\alpha \bar{\beta}} = \frac{N_k}{\vcy{X}} \int_X d \vcy{X}
		\frac{s_\alpha \sbar_{\bb}}{ h^{\gamma \db}_{(n)} s_{\gamma} \sbar_{\db}} .
		\end{equation}
		\item Let $h^{\alpha\bar\beta}_{(0)}$ be an initial invertible hermitian matrix.
		\item Then, starting with $h^{\alpha\bar\beta}_{(0)}$, the sequence
		\begin{equation}
		h_{(n+1)} = [ T(h_{(n)}) ]^{-1}
		\end{equation}
		converges to the desired balanced metric $\hmat$ as $n \to \infty$.
	\end{enumerate}
	The convergence is very fast in practice, with only a few iterations ($\lesssim 10$) necessary to give a good approximation to the balanced metric.
	
	\item There are several measures of convergence one could use; a standard one is defined as follows.
	Consider the top-form $\omega^3$ defined by the K\"ahler form. Since $X$ is a Calabi--Yau threefold with $\Omega  \wedge \overline{\Omega}$ the unique (up to scaling) non-vanishing $(3,3)$-form, these two must proportional, i.e., $\omega\wedge\omega\wedge\omega \propto \Omega  \wedge\overline{\Omega}$.
	 Thus, the ratio of these two top-forms, compared to the ratio of their respective integrals, must be equal:
	\begin{equation}
	\begin{array}{l}
	\vk{X} := \int_X \omega^3 \\
	\vcy{X} := \int_X \Omega  \wedge \Ob
	\end{array}
	\qquad \Rightarrow \qquad
	\frac{\omega^3}{\vk{X}} = \frac{ \Omega  \wedge \Ob}{\vcy{X}}  \ .
	\end{equation} 
	In other words the the quantity
	\begin{equation}\label{eq:sigma_int}
	\sigma_k \equiv \frac{1}{\vcy{X}} \int_X d {\vcy{X}}\, \left| 
	1 - \frac{\omega^3_k \slash \vk{X}}{\Omega  \wedge \overline{\Omega} \slash \vcy{X} }
	\right|
	\end{equation}
	is 0 if and only if $\omega_k$ converges to the the Ricci-flat K\"ahler (CY) metric, as we increase the degree $k$.
\end{enumerate}

\subsection{Numerical Metric on the Quintic}
As always, we default to the quintic CY$_3$ to illustrate the details.
Take the Fermat Quintic $Q$ from \eqref{specialQ} with the parameter $\psi = 0$:
\begin{equation}\label{defQ}
Q := \{  \sum_{i=0}^4 z_i^5 = 0 \} \subset \IP^4.
\end{equation}

Fixing a positive integer $k$, $\{s_\alpha\}$ can be chosen to be monomials of degree $k$ on $Q$. On $\IP^n$, finding all monomials of degree $k$ is a standard problem in combinatorics and amounts to choosing $k$ elements from $n+k-1$. Since the ambient space is $\IP^4$, there are $\binom{5+k-1}{k}$ ways of doing so.
Explicitly, for the first few values of $k$, the monomial bases are (the arguments here are a direct generalization of \eqref{h21Q}):
\begin{equation}
\begin{array}{c|c|c}
k & N_k & \{s_\alpha\} \\ \hline
1 & 5 & z_{i = 0, \ldots, 4} \\ \hline
2 & 15 & z_i z_j \ , 0 \leq i \leq j \leq 4 \\ \hline 
3 & 35 & z_i z_j z_k \ , 0 \leq i \leq j \leq k \leq 4 \\ \hline
4 & 70 & z_i z_j z_k z_\ell \ , 0 \leq i \leq j \leq k \leq \ell \leq 4
\end{array}
\end{equation}
On $Q$ we need to impose the defining quintic equation \eqref{defQ} when we encounter variables of powers greater than or equal to 5. For example, we can choose to replace $z_0^5 \to - \sum_{i=1}^4 z_i^5$.
This amounts to a reduction of the number of independent monomials of degree $k \geq 5$.
In general, we have that
\begin{equation}\label{Nk}
N_k = \left\{
\begin{array}{lcl}
\binom{5+k-1}{k},& 0 < k \leq 4 \\ 
 \binom{5+k-1}{k} -  \binom{k-1}{k-5} ,& k \geq 5 \\
\end{array}
\right.
\end{equation}

Since the metric is a local quantity, we need to focus on particular affine patches of $Q$. Suppose we are in the $z_0 = 1$ patch. We can eliminate one of the remaining four coordinates, say, $z_1 = (-1 - z_2^5 - \ldots -z_4^5)^{1/5}$,  so that the good local coordinates are $(z_2, z_3, z_4)$, which we then set to be $x^a$. The holomorphic volume form is then
\begin{equation}\label{VolForm}
\Omega = \int_Q \frac{d z_1 \wedge d z_2 \wedge d z_3 \wedge d z_4}{ 1 + z_1^5+ z_2^5+ z_3^5+ z_4^5} = 
\frac{d z_2 \wedge d z_3 \wedge d z_4}{5 z_1^4} ,
\end{equation}
where the (Griffith) residue theorem \cite{GH} is applied in the last equality upon integrating out $z_1$.

\begin{figure}[t!!!]
(a) \includegraphics[width=2.3in,angle=0]{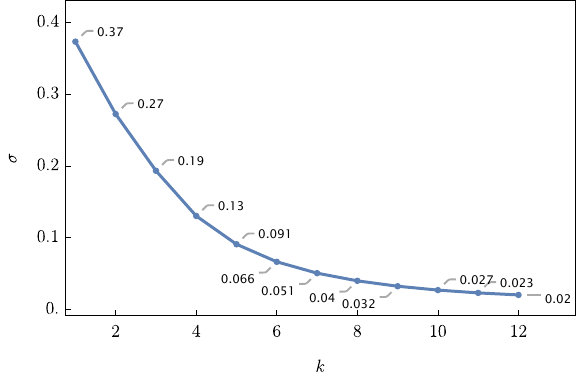}
(b) \includegraphics[width=2.3in,angle=0]{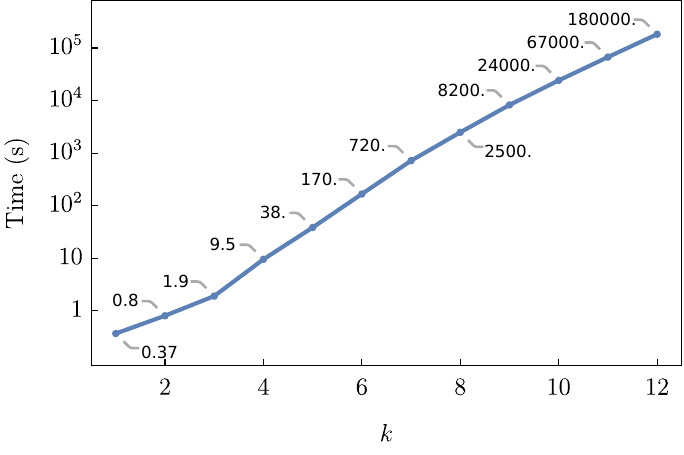}
\caption{
{\sf 
(a) The convergence of the $\sigma$ measure for the quintic as we increase $k$;
(b) time taken in seconds of the computation for the numerical metric on a log scale.
}
\label{f:donaldson}
}
\end{figure}

In general, we
\begin{enumerate}
	\item
	Work in the affine patch defined by $z_I$ for some $I=0, \ldots, 4$. For numerical stability, $z_I$ should have the largest norm of the $z_i$.\footnote{In other words, if one works in homogeneous coordinates, we take $z_I$ to be the coordinate with the largest norm and then divide the other coordinates by $z_I$.}
	\item
	Eliminate one of the four remaining variables $z_{J \neq I}$ by solving for $z_J$ using the defining equation of $Q$. The remaining three variables then constitute the ``good coordinates'' of the patch on $Q$ -- we denote these by $x^a$ for $a=1,2,3$. For numerical stability (cf.~\S 3.4.2 of \cite{Douglas:2006rr}), we eliminate the variable for which $|\partial Q/\partial z_J|$ is the largest.
In practice, we need to be careful about sampling random points on a manifold, with a weight measure; details can be found in \cite{Douglas:2006rr,Braun:2007sn} (see Appendix A of \cite{Ashmore:2019wzb} for a summary).

	\item 
	We can now evaluate elements of the monomial basis $\{s_\alpha\}$ at each point $p_M$ and obtain the $T$-operator. Choosing some initial invertible hermitian matrix $h^{\alpha \bb}$, we numerically perform the integration in \eqref{eq:T_operator} with respect to the weight measure of points.

	\item
	Set the new $h^{\alpha \bb}$ to be $(T_{\alpha \bb})^{-1}$ and iterate.
	The algorithm is insensitive to the initial choice of $\hmat$ and in practice fewer than 10 iterations are needed to converge to the balanced metric. 
	The actual computation for the metric $\gk$ is as follows.
	\begin{enumerate}
		\item 
		From $h^{\alpha \bb}$ we obtain the K\"ahler form in terms of the coordinates $(z_0,\ldots, z_4)$ 
		of the ``ambient'' affine patch (with, for example, $z_0=1$) on $\IP^4$:
		\begin{equation}
		K(z, \zb) = 
		\frac{1}{k \pi} \ln \sum_{\alpha, \bar{\beta} = 1}^{N_k} h^{\alpha \bb} s_{\alpha}(z) \sbar_{\bb}(\zb)\;
		\Rightarrow\;
		\tilde{g}_{i\bar j}(z, \zb ) = \partial_i \partial_{\jb} K .
		\end{equation}
		
		\item
		We pull-back this metric via the immersion of the Fermat quintic polynomial to find the metric $g_{a \bar b}$ on $Q$.
		Suppose, without loss of generality, that we are in the patch $z_0=1$ and the good coordinates on $Q$ are $x_a=(z_2, z_3, z_4)$ with $z_1$ to be eliminated via
		\begin{equation}
		z_1^5  = -1 - \sum_{i=2}^4 z_i^5 \quad
		\Rightarrow \quad
		\frac{\partial z_1}{\partial z_i} = - \frac{z_i^4}{z_1^4}\quad \ i = 2,3,4 .
		\end{equation}
		The other situations of different $I$ and $J$ are simply permutations of the following discussion. As the metric is a tensor, the pull-back is given simply by multiplying with Jacobian $J^i{}_a$. Explicitly, the Jacobian is
		$
		J^i{}_a = \frac{\partial z_i}{\partial x_a}=
		\frac{\partial z_{i=0,\ldots,4}}{\partial z_{k = 2,3,4}}
		=
		\left(
		\begin{array}{c|c|ccc}
		0&-z_2^4/z_1^4 & 1 & 0 & 0 \\
		0&-z_3^4/z_1^4 & 0 & 1 & 0 \\
		0&-z_4^4/z_1^4 & 0 & 0 & 1 \\
		\end{array}
		\right).
		$
		\item Finally, using the metric $g_{a \bar b}$ on $Q$, the K\"ahler form is
		\begin{align}
		\omega = \frac{i}{2} \sum_{a, \bar b}^3 g_{a \bar b}(x, \bar x) d x^a \wedge d \bar{x}^{\bar b}.
		\end{align}
	\end{enumerate}

	\item
	The error measure $\sigma$  in \eqref{eq:sigma_int} can also be obtained by numerical integration.
	In Figure \ref{f:donaldson} we show the relevant quantities.
	We sample 500K points random points on $Q$ for the Donaldson algorithm to obtain the measure $\sigma$. This is shown in part (a) for $k = 1, \ldots 12$. In part (b), we keep track of the time taken on Mathematica on , and see that the exponential growth of complexity
(q.v.\cite{Ashmore:2019wzb}, courtesy of A.~Ashmore for re-computing and re-drawing the figure).
\end{enumerate}

\setcounter{part}{1}
\chapter{The Non-Compact Calabi-Yau Landscape}

\begin{chapquote}{Gian-Carlo Rota, {\it Ten Lessons I wish I had given}}
``At the end of the lecture, an arcane dialogue took place between the speaker [Calabi] and some members of the audience, Ambrose and Singer if I remember correctly. There followed a period of tense silence. Professor Struik broke the ice. He raised his hand and said: `Give us something to take home!' Calabi obliged, and in the next five minutes he explained in beautiful simple terms the gist of his lecture. Everybody filed out with a feeling of satisfaction.''
\end{chapquote}


Relaxing the condition of compactness usually grants us liberty, as for instance in the familiar setting of Liouville's theorem on bounded entire functions.
In other words, one could think of local models of Calabi-Yau manifolds when one works on some affine patch.
In this sense, in the following we will use the words {\blue ``non-compact''}, {\blue ``affine''}, and {\blue ``local''} inter-changeably.

What is the simplest non-compact Calabi-Yau $n$-fold? Clearly, it is just $\IC^n$, which is not only Ricci flat, but completely flat.
Therefore, as the quintic was the {\it point d'appui} in the previous chapter, $\IC^3$ will be so for our present one.
In particular, $\IC^3 \simeq \IR^6$ is a cone over the round sphere $S^5$ (just like the more familiar case of $\IR^3$ being a cone over $S^2$) in that its flat metric \footnote{
In the few times when I met Calabi when I was a postdoc at the University of Pennsylvania he struck me as a gentleman of the Old School.
I remember when Serge Lang came to give a colloquium (in his 70s) and in his typically flamboyant style turned to Calabi in the middle of his talk when he came to a certain manifold - having repeatedly cross-examined various members of the audience throughout the lecture, and even jokingly told a poor chap to get out when he could not provide a correct answer - ``You, Calabi! is it Ricci flat?!'' Eugenio calmly answered with his usual charm and diplomacy, ``Sorry, I am afraid I was asleep.''  
} can be written as
\begin{equation}\label{C3S5}
ds^2({\IC^3}) = dr^2 + r^2 ds^2(S^5) \ ,
\end{equation}
where $r$ is the radial coordinate of the cone with $r=0$ being at the origin.
One might be misled to think all that this is too trivial but we will soon see a richness in both the mathematics and the physics.

\section{Another $10 = 4 + 2 \times 3$}
While local Calabi-Yau manifolds enjoy a wealth of usefulness in many contexts, in parallel  to \S\ref{s:compactification} from the Prologue, we take the point of view of its r\^ole in building the standard model from string theory.
Our non-compactness began with the late Joe Polchinski's discovery of Dirichlet branes, or {\red D-brane} as dynamical objects in string theory \cite{Polchinski:1995mt}.
In a nutshell, a Dp-brane is, by convention, a $(p+1)$-dimensional space-time object (one of whose dimension is time) whose world-volume supports a $(p+1)$ form so that a charge can be obtained by integration.
One can see the form as the connection of a $U(1)$-bundle on the Dp-brane.
Importantly a stack of $N$ D-branes (i.e., $N$ of them placed in parallel and with separation taken to the zero limit), the gauge group is ``enhanced'' from $U(1)^N$ to $U(N)$.
This is all we will need from the vast theory of D-branes.

It is therefore clear the importance which the D3-brane plays: its world-volume is $3+1$ dimensional, and, for our {\it local} purposes, is $\IR^{1,3}$ which gives us a {\it brane-world} \footnote{
The curvature of the brane and back-reactions with the background metric, for our algebro-geometric purposes, will be neglected throughout.
} scenario \cite{Randall:1999vf}. 
The 6 directions perpendicular to a stack of $N$ D3-branes, will again be Calabi-Yau, albeit non-compact.
In this set-up, in contrast to the {\it compactification} scenario of the previous chapter, the (supersymmetric) standard model lives on the $\IR^{1,3}$ of the D3-brane, interacting with the transverse or bulk dimensions only via gravity.

It should be pointed out that phenomenologically viable theories from brane constructions need to satisfy global consistency conditions; for these, the type IIA intersection brane mentioned in \S{s:landscape} are much more conducive to model building. In this chapter, we will take the mirror consideration of IIB D3-branes at singularities for several reasons.
First, in line with the theme of this book, they better illustrate singular and non-compact Calabi-Yau varieties and their resolutions.
Second, they were the original archetypes for the gauge-gravity holographic correspondence (aspects of which we will shortly attempt an explanation to the pure mathematician).
Finally, they offer an interesting forum for some beautiful dialogue between physics, geometry and representation theory.

\subsection{Quiver Representations \& a Geometer's AdS/CFT}\label{s:geoAdS}
While the setup of the D-brane's correspondence between gravity in the bulk and gauge theory on the world-volume goes under the rubric of {\blue holography} or {\blue AdS/CFT}, the challenge of captivating the interest of an algebraic geometer is the one which we will embrace here.
Once we phrase at least some of the subject purely in term of the mathematics, we will then see how there too is a plethora of combinatorial data.

In order to venture into this landscape of non-compact Calabi-Yau spaces. We need a few preliminary concepts from representation theory.
\begin{definition}
  A {\red Quiver} $\cQ = (\cQ_0, \cQ_1, W)$ is a finite directed graph with the set of vertices $\cQ_0$ and arrows $\cQ_1$, the cardinalities of which are $N_0$ and $N_1$ respectively:
  \begin{itemize}
  \item $\cQ$ is equipped with a {\em representation}, meaning that we attach vecto spaces $V_i \simeq \IC^{n_i}$ to each node for some positive integer $n_i$, whence each arrow $(X_{ij} \in \cQ_1) \in \hom(V_i, V_j)$ can be considered as an $n_j \times n_i$ matrix; we allow self-adjoining arrows, $\phi_i = X_{ii}$, as well as cycles which are closed loops $X_{i_1i_2} X_{i_2i_3} \ldots X_{i_ki_1}$ formed by the arrows.
  \item $\cQ$ is also furnished by {\em relations}; this is imposed by the {\blue superpotential} $W$ which is a polynomial in all the arrows treated as formal matrix variables:
    \begin{equation}\label{W}
    W = \sum_{k=1}^{N_2} c_k \tr(\prod X_{ij}) \ldots \tr(\prod X_{i'j'})
    \end{equation}
    summed over possible cycles or products therein with coefficients $c_k \in \IC$. The formal polynomial relations amongst the arrows are determined by the vanishing of the Jacobian $\partial_{X_{ij}} W$.
    We let the number of monomial terms in $W$ be $N_2$.
  \end{itemize}
\end{definition}
 The term quiver was coined by Gabriel \cite{gabriel} because its nodes, like the holster for the weapon, is a holder for arrows.

Indeed, even the pure mathematics community is using the term ``superpotential'', which originates in supersymmetric gauge theories and string theory \footnote{
There is an underlying 4-dimensional supersymmetric gauge theory, whose action, in $\cN=1$ superspace is
\[
S = \int d^4x\ [ \int d^2\theta d^2\bar{\theta}\ \Phi_i^\dagger e^V \Phi_i + 
  \left( \frac{1}{4g^2} \int d^2\theta\ \tr{\cW_\alpha \cW^\alpha} + 
  \int d^2\theta\ W(\Phi) + {\rm c.c.} \right)  \ ,
\]
where $V$ is the vector multiplet and $\Phi$, the hypermultiplet. The effective potential in terms of the scalars is
\[
V(\phi_i, \bar{\phi_i}) = \sum_i \left| \diff{W}{\phi_i} \right|^2 +
 \frac{g^2}{4}(\sum_i q_i |\phi_i|^2)^2
\]
whose vanishing is precisely the vacuum defined by the D- and F-terms being 0.
}.
The allowance for loops and cycles significantly complicates the representation theory of $\cQ$ but this is necessary for the physics.
The integers $N_0$, $N_1$ and $N_2$ will play an interesting combinatorial r\^ole for a wide class (almost all which we will consider) of quivers.
To the above quiver data, one associates a 4-dimensional supersymmetric gauge theory with gauge group $\cG = \prod\limits_{i=1}^{N_0} U(n_i)$ under the dictionary:
  \begin{description}
    \item [Node $i$: ] Factor $U(n_i)$ in $\cG$
    \item [Arrow $i \to j$: ]  a so-called bi-fundamental field $X_{ij}$ transforming as
    		$(\fund,\antifund)$ of $U(N_i)\times U(N_j)$
    \item [Loop $i \to i$: ] a so-called adjoint field $\varphi_i = X_{ii}$ of $U(N_i)$
    \item [Cycle $i_1 \to i_1 \to \ldots i_k \to i_1$: ] Gauge Invariant Operator (GIO) created by concatenating along a path, via matrix multiplication and 	
    	finishing with an overall trace:
    	$\tr(X_{i_1i_2} X_{i_2i_3} \ldots X_{i_ki_1})$; a term such as $\tr(\prod X_{ij})$ is called a Single-trace GIO while products thereof 
    	$\tr(\prod X_{ij}) \ldots \tr(\prod X_{i'j'})$ are Multi-trace GIO
    \item[2-Cycle $X_{ij}X_{jk}$: ] Mass term
    \item[Superpotential \eqref{W}: ] Superpotential in the Lagrangian with couplings $c_i$; and the set of polynomials $\left\{ \partial_{X_{ij}} W \right\}$ are the {\bf F-Terms}.
  \end{description}

The list of labels $\vec{n} = (n_1, n_2, \ldots, n_{N_0})$ is called the {\it dimension vector} and for the simplest case when all $n_i = 1$, the arrows are just complex numbers and the quantum field theory has gauge group $U(1)^{N_0}$.
Finally, recall 
\begin{definition}
The {\em incidence matrix} of $\cQ$ is is an $N_0 \times N_1$ integer matrix $d_{i\alpha}$ where $i = 1, \ldots , N_0$ indexes the nodes and $\alpha = 1, \ldots, N_1$, the arrows, such that each arrow $i \to j$ gives a new column in $d_{i\alpha}$, with $-1$ at row $i$ and $+1$ at row $j$, and 0 otherwise.
\end{definition}
 Then $\sum\limits_{\alpha} d_{j \alpha} |X_{ij}|^2 - \zeta_i$ are known as {\bf D-terms}, where $\zeta_i \in \IC$ are so-called Fayet-Iliopoulos (FI) parameters \footnote{
  In general, one assigns charges $q_\alpha$ to the fields and sum over $q_\alpha |X_{i\alpha}|^2$ but for our present purposes of $U(1)^{N_0}$ theories, the incidence matrix serves to encode the charges.
    Moreover, the FI-parametres exist only for the $U(1)$ gauge group factors.}.
  Note that unlike the F-terms, these are non-holomorphic, as they involve the complex conjugates of $X_{ij}$.
  
Of course, D-terms and F-terms have a long history in supersymmetric gauge theories and possess many subtle quantum field theoretical properties.
Continuing with the theme of the book being accessible to a general mathematically inclined readership, we will use no more than the following (rough) definition:
\begin{align}
\nn
\mbox{D-term $\sim$} & \mbox{the incidence matrix }  d_{j\alpha} \mbox{ of the quiver } \cQ \ ; \\
\nn
\mbox{F-term $\sim$} & \mbox{the partial derivatives (Jacobian variety) of a formal polynomial} \\ 
	& W(X_{ij}) \mbox{ (the superpotential) in the arrows} X_{ij} \ .
\end{align}

What does all of this have to do with geometry?
There is a key object in geometric representation theory \footnote{
Strictly speaking a {\it quiver variety} in the sense of \cite{naka} is a more general and decorated (in the sense of extra data) notion, 
and  $\cM(\cQ)$ really aught to be called the King variety, after \cite{king}.
I had suggested this to fellow numerologist Alastair King but he, in his humility, insisted that it be called the representation variety.
}, coming from a quiver  \cite{naka,king}.
\begin{definition}
  The representation variety $\cM(\cQ)$  is the GIT quotient of the representations $Rep(\cQ) = \bigoplus\limits_{i,j} \hom(\IC^{n_i}, \IC^{n_j})$, with relations from $W$, quotiented by the complexified group $\cG_{\IC} = \prod_i GL_n(\IC)$
  \[
  \cM(\cQ) = {\rm Spec} \IC[ Rep(\cQ) / \left< \partial_{X_{ij}} W \right>]^{\cG_{\cC}} \simeq \left< \partial_{X_{ij}} W \right> // \cG_{\IC} \ ,
  \]
where ${\rm Spec}$ is the maximal spectrum of the quotient ring.
\end{definition}
From the point of view of geometric representation, $\cM(\cQ)$ can be construed as the centre of the path algebra of $\cQ$.

As with many objects in mathematics, there is the {\it formal} definition and there is the {\it working} definition.
The above formal definition will mean absolutely nothing to the beginning student, nor will we digress too far here to delve into all the terms such as GIT (geometric invariant theory), the double slash notation, the vector space $Rep$, etc.
The only one on which we have spent some time is ``Spec'', since its usage appears in several places in the book; this, together with some other elements of algebraic geometry, is summarized in Appendix \S\ref{ap:geo} (especially \S\ref{ap:spec}).

What the above theorem computationally means, roughly, is just the following:
\begin{equation}
\cM(\cQ) \simeq
\mbox{set of minimal cycles in $\cQ$} / \mbox{F-term relations} \ .
\end{equation}
In other words, we find the (Eulerian) minimal cycles in the quiver, for each we can perform successive matrix multiplication (since each arrow is a matrix), and then finish off with matrix trace.
The result is a list of (very big) polynomials, one for each minimal cycle;
each is a polynomial in all the component variables of the arrow matrices.
If there are no F-terms (i.e., no superpotential), then finding the relations (maximal Spec) amongst these polynomials will give the defining equation of $\cM(\cQ)$.
If there are non-trivial F-terms, then we need to find the relations, up to these F-terms.
We will illustrate this will many upcoming examples, illustrating the algorithm in Proposition \ref{vms}.

In physics, $\cM(\cQ)$ is called the {\red {\em vacuum moduli space}} (VMS) of the gauge theory \footnote{To be more precise, $\cM$ here defined is the mesonic, Higgs branch of the moduli space because the gauge invariants are built from taking traces. We could contract with other invariant tensors, for instance, contracting with Levi-Civita symbols to obtain the GIOs would give the baryonic moduli space.
The Jacobian variety $\left< \partial_{X_{ij}} W \right>$ is itself also interesting and the equations $\partial_{X_{ij}} W = 0$, which are collectively the F-terms, gives the so-called {\em master space} \cite{ot,Forcella:2008bb}
} because solving the F-terms and D-terms amounts to finding the space of solutions which the vacuum expectation values of the scalars parametrize the supersymmetric vacuum.

The rest of the this chapter is concerned with
\begin{question}\label{q:MQ}
When is the representation variety $\cM(\cQ)$ an affine variety that is (local) Calabi-Yau $M$? In particular, what is the relation between $\cQ$ and $M$ when $\dim_{\IC}(M) = 3$?
\end{question}
The skeptical reader may wonder why we have chosen to present varieties in this seeming convoluted way through representation of quivers.
Other than the fact that it is the central way to understand it in physics, we are assured by the universality of quiver moduli space
\cite{king,quivermoduli1,quivermoduli2,quivermoduli3,quivermoduli4,quivermoduli5}, in that {\it any} variety can be obtained by a suitably chosen quiver with superpotential \footnote{
The formal way to state this is, 
\begin{theorem}
Given any scheme $\cM$, there exists a quiver $\cQ$ and an ideal $\cJ$ in the path-algebra $k\cQ$ over the ground field $k$ whose moduli space of all indecomposable $k\cQ/\cJ$-modules (and even with dimension vector $(1,1,\ldots,1)$), is isomorphic to $\cM$.
\end{theorem}
Again, we relegated this important theorem to a footnote since the technicalities therein will intimate the neophyte.}.
In other words, we can always find some $\cQ$ and $W$ so that the VMS $\cM(\cQ)$ is a required variety.

\subsection{The Archetypal Example}
We have certainly bombarded the reader with volley of definitions and abstractions.
It is illustrative to take a look at a simple example, which will help to anchor us.
Let us take the ``clover'' quiver with a specific cubic superpotential:
\begin{equation}\label{c3}
\begin{picture}(150,80)
  \put(1,1){\includegraphics[trim=0mm 0mm 0mm 0mm, clip, width=1in]{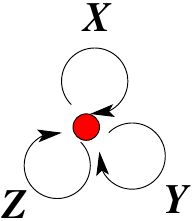}}
  \put(20,20){$N$}
  \put(100,20){$  W = \tr(XYZ - XZY) \ , $}
\end{picture}  
\end{equation}
In the above, there is a single node labelled $N$. Thus, our 3 arrows $X,Y,Z$ are $N \times N$ matrices.
The F-terms are obtained from the partials of $W$ with respect to the variables $X,Y,Z$: one note about taking a derivative by a matrix variable  
is that within the trace, we can cyclically permute the variables by property of matrix trace, so that we can by convention move the variable to be differentiated to the last position (this matrix derivative is sometimes called the cyclic derivative).
Performing this we obtain $\partial_X W = YZ-ZY$, $\partial_Y W = ZX-XY$ and $\partial_Z W = XY-YX$.
On the other hand $\tr(X), \tr(Y), \tr(Z)$ are clearly the 3 minimal generators for the GIOs.
Hence,
\begin{equation}
\cM(\cQ) \simeq \mbox{Spec} \left( \IC\left[\tr(X), \tr(Y), \tr(Z)\right] / \left< [X,Y], [Y,Z], [X,Z] \right>\right) \ .
\end{equation}

While the space may look rather complicated, the case of $N=1$ should be immediately clear to us.
Here, $X,Y,Z$ are just complex numbers and the F-terms provide no extra relations.
Hence, $\cM(\cQ)$ is parametrized freely by 3 complex variables, i.e.,
\begin{equation}
\cM(\cQ_{N=1}) = \mbox{Spec} \IC[X,Y,Z] = \IC^3 \ ,
\end{equation}
and we have retrieved our simplest non-compact Calabi-Yau 3-fold.
In general, one can check (though the defining equations become quite involved) that $\cM(\cQ_{N\geq 1}) \simeq (\IC^3)^N / \mathfrak{S}_N$, the $N$-th symmetric product of $\IC^3$, i.e., the direct product of $N$ copies of $\IC^3$, quotiented by the action of the symmetric group by permutations.

The strategy used above to compute $\cM(\cQ)$ explicitly is generally applicable \cite{vms} (formalized in \cite{LT} and algorithmized in \cite{Mehta:2012wk}) to obtaining $\cM(\cQ)$:
\begin{proposition}\label{vms}
  The VMS $\cM(\cQ)$ is realized as an affine algebraic variety by the following algorithm
  \begin{enumerate}
  \item Let $GIO_{min}$ be the set of minimal (Eulerian) cycles in $\cQ$, i.e., $\Phi_{r=1,\ldots,k}$, each of which is a polynomial in $X_{ij}$ and $k$, the total number of such cycles;
  \item Consider the ring map \footnote{
    As coordinate rings the map should go in the reverse direction so that as varieties the map goes as indicated, but we beg the readers' indulgence so that it is clear that the {\it image} is an affine variety in $\IC^k$.
    } from the quotient ring by the Jacobian ideal by $\Phi_r$ as 
    $$\IC[X_{ij}] / \left< \partial W \right> \stackrel{\theta}{\longrightarrow} \IC[\phi_r] \ ; $$
  \item $\cM$ is the image of this map, as an affine variety in $\IC^k$.
  \end{enumerate}
\end{proposition}
In the above example for $N=1$, we have $\Phi : \IC[X,Y,Z] \to \IC^3$. Later we will study more involved examples.

Though the above exercise may seem trivial, there is highly non-trivial mathematics and physics even for this example.
The quiver with superpotential given in \eqref{c3} is called {\bf $\cN=4$ super-Yang-Mills} theory in $(3+1)$-dimensions.
It is the unique conformal, supersymmetric, quantum field theory with maximal supersymmetry in 4-dimensions, and on it countless articles and books have been written.
The insight of \cite{Maldacena:1997re} was that it is precisely the world-volume theory on the D3-brane and there is a ``holographic'' duality between this gauge theory and the gravity in the bulk Calabi-Yau.
More precisely, the D3-branes furnish an asymptotic metric of $AdS_5 \times S^5$ the information of the full string theory on which is holographically projected onto the world-volume gauge theory.
The anti-de Sitter space (AdS) and the world-volume conformal field theory (CFT) is what engendered the name ``AdS/CFT''.
This $S^5$ factor is none other than the $S^5$ in \eqref{C3S5}.
We will return to this point later in the chapter.

Maldacena's duality is a set of precise statements on relating correlation functions and partition functions, in a particular limit of large $N$.
Our algebraic geometer's AdS/CFT, that $\cM(\cQ) \simeq \IC^3 = \mbox{Cone}(S^5)$, already remarkable, is only a tip of the iceberg  of correspondences.
But since this is a book on Calabi-Yau varieties and not on QFTs, we will content ourselves with playing on this oasis of a tip.

The point is that because the position of the brane is specified by a point in the transverse non-compact Calabi-Yau space $M$, the physical moduli, given by the vacuum expectation values (VEVs) of the scalar fields of the world-volume gauge theory, parametrize $M$ by furnishing its coordinates.
Now, the gauge theory on the brane is encoded by a quiver with representation variety $\cM(Q)=$VMS.
Therefore, tautologically, by construction,
\begin{equation}
M \simeq\cM(\cQ) = VMS \ .
\end{equation}
Therefore, string theory provides a natural answer to Question \ref{q:MQ}.

\section{Orbifolds and Quotient Singularities}\label{s:orb}
So far, our non-compact landscape consists of a single point, $\IC^3$.
What is the next natural candidate?
Since we are working locally, we are at liberty to allow certain singularities.
There is a class of generalization of manifolds which was called V-manifolds \cite{satake} but now better known as {\red orbifolds}.
These are perfectly adapted to our present needs.

Briefly, one considers the action of a discrete, finite group $G \acts \IC^n$ with a fixed locus, usually taken to be just a point, the origin, and form the quotient: the very fact there is a fixed locus means that the quotient is not smooth, quite contrary to what one is used to.
Indeed, in our discussion of projective spaces and toric varieties in the previous chapter, we remove the origin (Stanley-Reisner ideal) before the quotient so as to avoid singularities.

In order to preserve Calabi-Yau-ness, we must at least work within $SU(n)$ holonomy, i.e., we should only consider orbifolds of the form
\begin{equation}\label{Cnorb}
\cM \simeq \IC^n / G \ , \quad G \mbox{ discrete, finite, subgroup of }  SU(n) \ .
\end{equation}
It turns out that the condition is not sufficient for $\cM$ to be local Calabi-Yau, but is a necessary starting point.

In dimension \footnote{
In {\it real} dimension 1, there is only $\IR/ (\IZ/2\IZ)$, which is just a half-line with the point at the origin a singular point, coming from the real axis folded over.
} $n=1$, we essentially have only $\IC$ quotiented by a cyclic group so that the result is a pizza wedge but with the origin singular.

In dimension $n=2$, we are considering $\IC^2$ quotiented by discrete, finite, subgroups of $SU(2)$, whose classification goes back (at least) to Klein \cite{kleinIko}
\begin{equation}\label{ADE}
\begin{array}{|c|c|c|}
\hline
\mbox{Group} & \mbox{Name} & \mbox{Size} \\ \hline
A_n \simeq \IZ / (n+1)\IZ & \mbox{Cyclic} & n+1 \\ \hline
D_n & \mbox{Binary Dihedral} & 4n \\ \hline
E_6 & \mbox{Binary Tetrahedral} & 24 \\ \hline
E_7 & \mbox{Binary Octahedral (Cube)} & 48 \\ \hline
E_8 & \mbox{Binary Icosahedral (Dodecadedron)} & 120 \\ \hline
\end{array}
\end{equation}
For reference, we give all the explicit $2 \times 2$ generators for the groups in \eqref{HSADE} in the Appendix.

To see this, we recall that $SU(2)$ is the lift of $SO(3)$ by its centre $\IZ/2\IZ$, so we only need the regular symmetries in $SO(3)$, which are the two infinite families of the symmetries of the regular $n$-gon: the cyclic group of size $n$ and dihedral groups of size $2n$; 
as well as the symmetries of the 5 Platonic solids: the tetrahedron $T$, the cube $C$, the octahedron $O$, 
the dodecahedron $D$ and the icosahedron $I$.
However, $C$-$O$ and $D$-$I$ are graph dual to each other and share the same symmetry so we only have the 3 exceptional symmetry groups, viz., 
the alternating group $\mathfrak{A}_4$ of size 12, the permutation group $\mathfrak{S}_4$ of size 24, as well as $\mathfrak{A}_5$ of size 60.
Now all these groups need to be lift to $SU(2)$, and become their binary binary counterparts, except the cyclic case which is Abelian whereby affording no lift. 
The fact we have named the groups {\red ADE} is a whole story by itself to which we will devote the next subsection.

Geometrically, $\IC^2/(G \subset SU(2))$ should correspond to an interesting set of surface singularities.
To write their affine equations as hypersurfaces in $\IC^3$ is simple enough: one only needs to write down the list of generators of invariants under the group action (by a classical theorem of Hilbert-Noether, the ring of invariants is finitely generated) and find the defining relation amongst them.
This is a traditional problem in {\em polynomial invariant theory} and the textbook \cite{sturmfelsInv} offers a great account from a computational/Gr\"obner basis perspective (q.v.~Appendix \ref{ap:HS}).

\paragraph{The Ordinary Double Point: }
Let us illustrate the above discussions with the simple case of $A_1 = \IZ / 2\IZ$.
Let the coordinates of $\IC^2$ be $u,v$ and let $A_1 \acts \IC^2$ by $(u,v) \to (-u,-v)$.
Note that as a matrix group, the generator is {\scriptsize $\left(
\begin{array}{cc}
 -1 & 0 \\
 0 & -1 \\
\end{array}
\right)$} and we have to make sure that the generators, and hence all group elements, be unitary and have determinant 1, so as to guarantee the group is 
properly a subgroup of $SU(2)$.

Clearly, the basic invariants are $x = u^2$, $y = v^2$ and $z = uv$; in the sense that any other invariant of $A_1$ is a polynomial in these 3.
There is a single relation amongst the invariants, viz., $xy = z^2$, so we can write the equation defining $\IC^2 / A_1$, embedding into $\IC^3$ with coordinates $(x,y,z)$ as 
\begin{equation}\label{C2A1}
\IC^2 / A_1 \quad : \quad \quad \{ xy = z^2 \} \subset \IC[x,y,z] \ .
\end{equation}
This is perhaps the most well-known algebraic singularity and is known as the {\it ordinary double point}.

Doing the same for all the groups in \eqref{ADE} gives the list of well-known du Val singularities \cite{duVal}, and as with $A_1$, one can readily check that the affine equations of these as hypersurfaces $F(x,y,z)$ in $\IC^3$ are:
\begin{equation}\label{duValF}
\begin{array}{ll}
A_n: &xy+z^n=0\\
D_n: &x^2+y^2z+z^{n-1}=0\\
E_6: &x^2+y^3+z^4=0\\
E_7: &x^2+y^3+yz^3=0\\
E_8: &x^2+y^3+z^5=0 \ .
\end{array}
\end{equation}
One checks, by solving simultaneously for $F = \partial_x F = \partial_y F = \partial_z F = 0$, that the origin $(x,y,z) = (0,0,0)$ is a solution, meaning that it is a singular point.

\paragraph{Desingularization: }
As with all singularities, the standard procedure is to smoothen \footnote{
We do not have the space in this book to enter into the extensive subject of singularity resolution in mathematics, nor the physical interpretation in terms D-branes.
We mention here that there are 2 standard ways of ``resolving'' a singularity: (1) desingularization by deformation; and (2) blowup.
For K\"ahler spaces, as is the case with everything in this book, (1) is a complex structure deformation and (2) is a K\"ahler structure/metric deformation.
Deformation is easy to see. For example, $xy+z^n=0$ is singular at $(0,0,0)$ as discussed. Thus, we deform the origin by considering the variety $xy+z^n=\epsilon$.
Blow-up, on the other hand, is introduced in Appendix \S\ref{ap:dPF}, where we replace the origin by a $\IC\IP^1$.
The desingularization discussed here is done by blow-up and the extra divisor $D$, for $xy+z^n=0$, is the class of the $\IC\IP^1$.
} (or variously called desingularize, or resolve) it.
In performing resolution of a singularity $M$ to a smooth variety $\widehat{M}$, one compares the canonical bundle $K_{\widehat{M}}$ of 
$\widehat{M}$ with the canonical sheaf (we have to say sheaf here since $M$ is singular) $K_{M}$
\begin{equation}
K_{\widehat{M}} = f^*(K_M) + D 
\end{equation}
where $f : \widehat{M} \to M$ is the {\em resolution map} and the ``extra'' divisor $D$ is called the {\em discrepancy}.
Now $M$, being local Calabi-Yau, has $K_M = \cO_M$, and if the discrepancy is 0 and $K_{\widehat{M}} = \cO_{\widehat{M}}$ would mean that $\widehat{M}$ is Calabi-Yau.
Such a resolution has a cute name:
\begin{definition}
When the discrepancy divisor $D = 0$ and the canonical sheaf naturally pulls back as $K_{\widehat{M}} = f^*(K_M)$, the resolution $f : \widehat{M} \to M$ is called {\em crepant}. 
\end{definition}
That is, we have dropped the ``dis'' in ``discrepant'', as a play on the English.
It is these resolutions that we need for our Calabi-Yau purposes.

The list \eqref{duValF} is exhaustive in that we have (cf.~\cite{ADEsurface,K3db})
\begin{theorem}
The list \eqref{duValF} admit crepant resolutions to smooth K3 surfaces and vice versa, they are all (up to analytic isomorphism) the local models for K3 surfaces.
\end{theorem}
The situation is even better, explicit metrics are known for many of these \cite{kronheimer}, for the $A_n$ series it is the celebrated Eguchi-Hanson metric \cite{EHmetric}. In fact, for $A_1$ which we discussed above, the double point, the quotient $\IC^2/A_1$ is the total space of the cotangent bundle over $S^2$, and the Ricci flat metric for it was one of the very first to be constructed \cite{EHmetric}.

\subsection{McKay Correspondence}\label{s:mckay}

\begin{table}[h!!!]
\[
\begin{array}{c}
\widehat{A_{n\geq 1}} \quad
\begin{tikzpicture}
\draw[fill=black] 
(0,0)                         
	circle [radius=.1] node [above] {1} --
(1,0) 
	circle [radius=.1] node [above] {1} --
(2,0)
	circle [radius=.1] node [above] {1} 
(3,0)
	circle [radius=.1] node [above] {1} --
(4,0)
	circle [radius=.1] node [above] {1}    
;
\draw[dotted]
(2,0) -- (3,0)
;
\draw[fill=black]
(2,1)
	circle [radius=.1]
;
\draw
(2,1)
	circle [radius=.2] node [above=1mm] {1}
;
\draw[solid]
(0,0) -- (2,1)
;
\draw[solid]
(4,0) -- (2,1)
;
\end{tikzpicture}

\qquad

\widehat{D_{n \geq 4}} \quad
\begin{tikzpicture}
\draw[fill=black] 
(0,0)                         
	circle [radius=.1] node [above] {1} --
(1,0) 
	circle [radius=.1] node [above left] {2} --
(2,0)
	circle [radius=.1] node [above] {2} 
(3,0)
	circle [radius=.1] node [above right] {2} --
(4,0)
	circle [radius=.1] node [above] {1}
(1,1)
	circle [radius=.1] node [left] {1}
(3,1)
	circle [radius=.1] node [left] {}    
;
\draw[solid]
(3,0) -- (3,1)
;
\draw[dotted]
(2,0) -- (3,0)
;
\draw[solid]
(1,0) -- (1,1)
;
\draw[solid]
(3,0) -- (3,1)
;
\draw
(3,1)
	circle [radius=.2] node [right=1mm] {1}
;
\end{tikzpicture}
\\
\\
\widehat{E_6} \quad
\begin{tikzpicture}
\draw[fill=black] 
(0,0)                         
	circle [radius=.1] node [above] {1} --
(1,0) 
	circle [radius=.1] node [above] {2} --
(2,0)
	circle [radius=.1] node [above right] {3} --
(3,0)
	circle [radius=.1] node [above] {2} --
(4,0)
	circle [radius=.1] node [above] {1}
(2,1)
	circle [radius=.1] node [right] {2}
(2,2)
	circle [radius=.1] node [] {}
;
\draw[solid]
(2,0) -- (2,1)
;
\draw[solid]
(2,1) -- (2,2)
;
\draw
(2,2)
	circle [radius=.2] node [right=1mm] {1}
;
\end{tikzpicture}

\qquad

\widehat{E_7} \quad
\begin{tikzpicture}
\draw[fill=black] 
(-1,0)
	circle [radius=.1] node [above] {} --
(0,0)                         
	circle [radius=.1] node [above] {2} --
(1,0) 
	circle [radius=.1] node [above] {3} --
(2,0)
	circle [radius=.1] node [above right] {4} --
(3,0)
	circle [radius=.1] node [above] {3} --
(4,0)
	circle [radius=.1] node [above] {2} --
(5,0)
	circle [radius=.1] node [above] {1}  
(2,1)
	circle [radius=.1] node [right] {2}
;
\draw[solid]
(2,0) -- (2,1)
;
\draw
(-1,0)
	circle [radius=.2] node [above=1mm] {1}
;
\end{tikzpicture}
\\
\\
\\
\widehat{E_8}
\begin{tikzpicture}
\draw[fill=black] 
(0,0)                         
	circle [radius=.1] node [above] {2} --
(1,0) 
	circle [radius=.1] node [above] {4} --
(2,0)
	circle [radius=.1] node [above right] {6} --
(3,0)
	circle [radius=.1] node [above] {5} --
(4,0)
	circle [radius=.1] node [above] {4} --
(5,0)
	circle [radius=.1] node [above] {3} --
(6,0)
	circle [radius=.1] node [above] {2} --
(7,0)
	circle [radius=.1] node [] {}   
(2,1)
	circle [radius=.1] node [right] {3}
;
\draw[solid]
(2,0) -- (2,1)
;
\draw
(7,0)
	circle [radius=.2] node [above=1mm] {1}
;
\end{tikzpicture}
\end{array}
\]
\caption{
{\sf The extended (affine) ADE diagrams, with integer labels being the Coxeter numbers. The affine nodes are circled explicitly; without this node, the diagram is the Dynkin diagram of the ordinary ADE Lie algebra.}
\label{f:ade}}
\end{table}

The specialness of the ADE list is remarkable from many approaches in mathematics and this meta-pattern \cite{gannonCFT} appears in so many mysterious and inter-connected ways that it has inspired an entire discipline of {\it ADE-ology} \cite{ADEbook,HeMcKay}, a highlight of which is the {\red McKay Correspondence} on whose brief exposition we cannot resist but to give.

Take our discrete finite subgroup $G \subset SU(2)$, it has a defining 2 complex-dimensional representation ${\bf 2}$, which for the non-Abelian cases is irreducible and for the Abelian $A_n$-series, is the direct sum of 2 (conjugate) one-dimensional representations.
John McKay \footnote{
John's super-human ability to notice patterns from unthinkably disparate branches of mathematics, whereby giving profound new insight, is legendary.
Yet of all his correspondences, he seems most proud of the ADE one.
} performed an experiment: tensor ${\bf 2}$ with all the irreducible representations (irreps) $r_i$ of $G$, decompose this into irreps as
\begin{equation}\label{McKay2}
{\bf 2}\otimes {\bf r}_i=\bigoplus\limits_{j}a_{ij}^{{\bf 2}} {\bf r}_j
\end{equation}
and note the multiplicities $a_{ij}^{{\bf 2}} \in \IZ_{\geq 0}$.
He subsequently noticed that \cite{mckayADE} if one were to interpret $a_{ij}$ as the {\bf adjacency matrix} of a finite graph, they are
precisely the Dynkin diagrams of the affine ADE Lie algebras, as shown in Figure \ref{f:ade}.

Just to briefly recall some rudiments to clarify this amazing fact, we have
\begin{definition}
For a finite directed graph with $n$ nodes, the {\bf adjacency matrix}  $a_{ij}$ is an $n \times n$ matrix of non-negative integers whose  $(i,j)$-th entry counts the number of arrows from node $i$ to node $j$.
\end{definition}
In the Dynkin diagram case,  each line is understood to represent a bi-directional pair of arrows $i \leftrightarrow j$.
Next, an affine Lie algebra is an infinite dimensional extension of the ordinary Lie algebra whose Dynkin diagram has one more ``affine'' node which we circle in the Figure. 
Finally, of the complete set of semi-simple Lie algebras, the ADE ones are the {\it simply-laced}, i.e., there are no double or triple bonds, so that all simple roots are at 60 or 90 degrees from each other.

\subsubsection{McKay Quiver for $\IZ/(2 \IZ)$}
It is expedient to illustrate how to obtain the McKay quiver from \eqref{McKay2} for a simple case.
Take, for example, our ordinary double point in \eqref{C2A1}, which geometrically is the orbifold $\IC^2/A_1$ where $A_1 = \IZ/(2 \IZ)$, the cyclic group of order 2 acting on the complex coordinates $(u,v)$ of $\IC^2$ by inversion:
$(u,v) \to (-u, -v)$.
The defining (fundamental) representation of this $\IZ/(2 \IZ)$ is thus the negative of the identity matrix $- \II_{2\times2}$.
The group itself has only 2 elements, which we can denotes as $\{1, -1\}$.
The irreducible representation of $A_1$ are, since it is Abelian, all 1-dimensional.
There are two of them: 
(1) the trivial representation ${\bf 1}$ where all group elements are mapped to the identity (every finite group has this representation); and
(2) the representation ${\bf 1'}$ where the 2 elements are switched.
Subsequently, we have 2 nodes of the McKay quiver, which can be labelled ${\bf 1}$ and ${\bf 1'}$.

The fundamental {\bf 2} in this case is a direct sum ${\bf 1'} \oplus {\bf 1'}$, since we have both coordinates being negated (a non-$SU(2)$ action $(u,v) \to (u, -v)$, for example, would correspond to ${\bf 1} \oplus {\bf 1'}$).
Of all the ADE groups, only A-type has this split because it is Abelian; for the others {\bf 2} is a faithful, indecomposable, 2-dimensional defining representation.
Thus, \eqref{McKay2} reads
\begin{align}
\nn
({\bf 1'} \oplus {\bf 1'}) \otimes {\bf 1} &= {\bf 1'} \oplus {\bf 1'} \\
({\bf 1'} \oplus {\bf 1'}) \otimes {\bf 1'} &= {\bf 1} \oplus {\bf 1} \ .
\end{align}
Thus, for the node {\bf 1},  there are two arrows to node ${\bf 1'}$;
likewise, for the node ${\bf 1'}$, there are two arrows to node ${\bf 1}$.
Letting the pairs of bi-directional arrows be represented by a single line, the quiver looks like the Oxygen molecule
$\bullet - \bullet$.
This, is the Dynkin diagram for affine $\widehat{A_1}$.
In general, for $A_n = \IZ/((n+1) \IZ)$, one get a necklace of $n+1$ beads, which is the Dynkin diagram for affine $\widehat{A_n}$.

The remarkable observation of McKay in the late 1970s is that if we did this for all the discrete subgroups of $SU(2)$, the corresponding quivers are precisely the Dynkin diagrams of the (extended) ADE Lie algebras.
These are shown in Figure \ref{f:ade}.
The computational way to obtain the quivers from the decomposition is to use the character inversion formula  in \eqref{aijchar} (q.v., \cite{Hanany:1998sd} for a detailed exposition).

As is typical of correspondences from John McKay, the above is remarkable in that it relates two seemingly unrelated fields, here that of Lie algebras and of finite groups.
To this we can add geometry.
It turns out that in the resolution of the singularities in \eqref{duValF}, one perfroms blow-up and the exceptional curves are $\IP^1$s whose intersection matrix is precisely $a_{ij}$ \cite{ADEsurface}.
Thus, we have returned to our initial discussion about moduli space of quivers
\begin{proposition}
The representation variety $\cM(\cQ)$ for the affine Dynkin diagrams considered as quivers (the superpotentials are fixed and can be found in e.g.\cite{Johnson:1996py}) are the affine hypersurfaces in \eqref{duValF}, which are local Calabi-Yau 2-folds, i.e., K3 surfaces.
\end{proposition}

This beautiful web of relations quickly found its place in string theory \cite{ADEphysics,Johnson:1996py,Douglas:1996sw,Lawrence:1998ja},
with the word ``quiver'' introduced into physics by \cite{Douglas:1996sw} and the general method of computation set out in \cite{Lawrence:1998ja} and algorithmized in \cite{Hanany:1998sd}.
Recall our example of $\IC^3$. It can be thought of as an orbifold of $\IC^3$ with the trivial group $G =  \II$ and this gave us $\cN=4$ super-Yang-Mills theory in 4 dimension.
The discrete subgroups of $SU(2)$, then, furnish orbifolds of the form $\IC \times (\IC^2 / G)$ and these will give McKay quivers (each node also with an added self-adjoining loop due to the $\IC$ factor) which correspond to a special class of $\cN=2$ supersymmetric QFTs in 4 dimensions.

\subsection{Beyond ADE}
As so we can generalize.
A discrete subgroup $G$ of $SU(3)$ would give an orbifold of the form $\IC^3 / (G \subset SU(3))$.
Luckily, crepant resolutions also exist for these are they are indeed locally Calabi-Yau 3-folds \cite{reso3,crepant3}.
However, unlike the $n=2$ case where the crepant solution to K3 surfaces is unique, the $n=3$ case exist but is not unique and the resolutions are related to each other by flop transitions.

The $SU(3)$ subgroups were first classified at the turn of the twentieth century when matrix groups were still known as colineation groups \cite{blichfeldt}.
Again, the classification scheme follows the dichotomy: (1) infinite-families, and (2) finite exceptional cases.
In addition to the obvious cases of the $SU(2)$ subgroups which embed non-transitively (with an block-matrix structure) as well as the Abelian case of $\IZ/p\IZ \times \IZ/q\IZ$, the extra $SU(3)$ groups are
\begin{equation}
\begin{array}{c|c}
\mbox{Infinite Series} & \Delta(3n^2), \Delta(6n^2) \\ \hline
\mbox{Exceptionals} & 
\Sigma_{36 \times 3},
\Sigma_{60 \times 3},
\Sigma_{168 \times 3},
\Sigma_{216 \times 3}, \Sigma_{360 \times 3} \\
\end{array}
\end{equation}
The two infinite families $\Delta$ of size $3n^2$ and $6n^2$ are certain non-Abelian extensions of $A_n \times A_n$ and the sizes of the 5 exceptionals (the analogue of the symmetries of the Platonic solids) are marked as subscripts.

Using {\sf GAP} \cite{gap}, the representations for all these groups were worked out in \cite{Hanany:1998sd} and using the orthonomality of finite group characters, one can invert the generalization of \eqref{McKay2} (with a chosen defining representation ${\bf R}$ of $G$) as
\begin{equation}\label{aijchar}
{\bf R}\otimes {\bf r}_i=\bigoplus\limits_{j}a_{ij}^{{\bf R}} {\bf r}_j \qquad \Rightarrow \qquad 
a_{ij}^{{\bf R}} = 
\frac{1}{|G|}\sum\limits_{\gamma=1}^{r} r_{\gamma }\chi_{\gamma }^{{\bf R}}\chi_{\gamma }^{(i)}\ \overline{\chi_{\gamma }^{(j)}}
\end{equation}
where $\chi_{\gamma }^{(i)}$ are the entries in the finite character table of $G$, with $(i)$ indexing the irreps/rows and $\gamma$, the conjugacy classes/columns; the number of irreps and number of conjugacy classes are both equal to a positive integer $r$ and $r_\gamma$ is the size of the $\gamma$-th conjugacy class.

Using this, the analogues of the McKay ADE quivers were drawn (cf.~Fig.~5 in cit.~ibid.). In the physics, they correspond to $\cN=1$ super-conformal field theories in 4-dimensions.
In the mathematics, they are quivers whose $\cM(\cQ)$ are non-compact Calabi-Yau 3-folds.

Our non-compact landscape has therefore grown from a single example of $\IC^3$ to an infinite number.
On the physics side, there is a long programme to try to seriously construct the standard model from branes at singularities \cite{Randall:1999vf,branemodel1,branemodel2,branemodel3,branemodel4,branemodel5,branemodel6,branemodel7,branemodel8,branemodel9,branemodel10,Gmeiner:2005vz}.

To close the discussion on orbifolds, for $n>3$ in \eqref{Cnorb}, the situation is much more complicated.
Very little is known about which groups admit crepant resolution to Calabi-Yau manifolds.
Nevertheless, the $n=4$ case has also been classified in \cite{blichfeldt} and the McKay quiver for these, studied in \cite{Hanany:1999sp}.

\section{Toric Calabi-Yau Varieties}
As with the compact case, the combinatorics of toric geometry gives us the most fruitful method of construction.
The geometer need not be alarmed by the title of this subsection, while there are no compact Calabi-Yau manifolds which are toric (in fact they do not even admit any continuous isometries), the fact that we are only concerned with non-compact Calabi-Yau manifolds in this chapter salvages the situation
\cite{schencktoric} (cf.~\cite{Bouchard:2007ik} for a more immediate treatment).
Since we are dealing with a local, affine variety, the {\red toric variety} is described by a single cone.
We again refer to Appendix \ref{ap:toric} for notation and a rapid discussion on such cones.

The vanishing of the first Chern class translates to the fact that the end-points of the generators of the cone are co-hyperplanar.
That is, an affine toric variety of complex dimension $n$ is defined by a rational polyhedral cone in $\IR^n$, but for an affine Calabi-Yau manifold, the end-points lie on a hyperplane and therefore a lattice $(n-1)$-polyhedron suffices to encode it.
In summary,
\begin{theorem}
An affine (local) toric Calabi-Yau 3-fold $M$ is represented by a toric diagram $\cD$ that is a convex lattice polygon, defined up to $GL(2; \IZ)$.
In particular, this gives an infinite number of toric Calabi-Yau 3-folds. 
\end{theorem}

Now, $\IC^n$ is the prototypical example of an affine toric variety, so let us illustrate the above theorem with our familiar $\IC^3$.
The cone for $\IC^3$ has 3 generators, viz., the standard basis of $\IR^3$, which are clearly coplanar.
Therefore, the toric diagram for $\IC^3$ can simply be taken, after $GL(2;\IZ)$, to be the lattice triangle: $\{(0,0); \ (0,1); \ (1,0)\}$.

Next, it should be noted that the Abelian orbifolds to which we alluded earlier are all toric varieties, though the non-Abelian ones are not.
Specifically, letting the coordinates of $\IC^3$ be $(x,y,z)$, the 2 generators of $\IZ/p\IZ \times \IZ/q\IZ$ with $p,q \in \IZ_{+}$ can be chosen as
\begin{equation}
(x,y,z) \to (x, \omega_p y, \omega_p^{-1} z) \ , \qquad
(x,y,z) \to (\omega_q x, \omega_q^{-1} y, z)
\end{equation}
to ensure that the group embeds into $SU(3)$, where $\omega_p$ and $\omega_q$ are the primitive $p$-th and $q$-th roots of unity respectively.
In particular, $\IC^3$ itself and $\IC^3 / (\IZ/r \IZ)$ for $r \in \IZ_{\geq 2}$ are all toric varieties (the $\IC^3 / (\IZ/2 \IZ)$ case is actually just $\IC \times \IC^2 / (\IZ/2 \IZ)$ since there will always be one coordinate fixed).
The toric diagram for this Abelian orbifold is the lattice right-angle triangle whose legs (cathethi) are of length $p$ and $q$.
Consequently, by choosing large enough $p$ and $q$, any toric diagram is a sub-diagram of that of the Abelian orbifold.

\subsection{Cone over $\IP^2$}\label{s:dP0}
\begin{figure}[t!!!]
\begin{center}
$\begin{array}{l}\includegraphics[width=5in,angle=0]{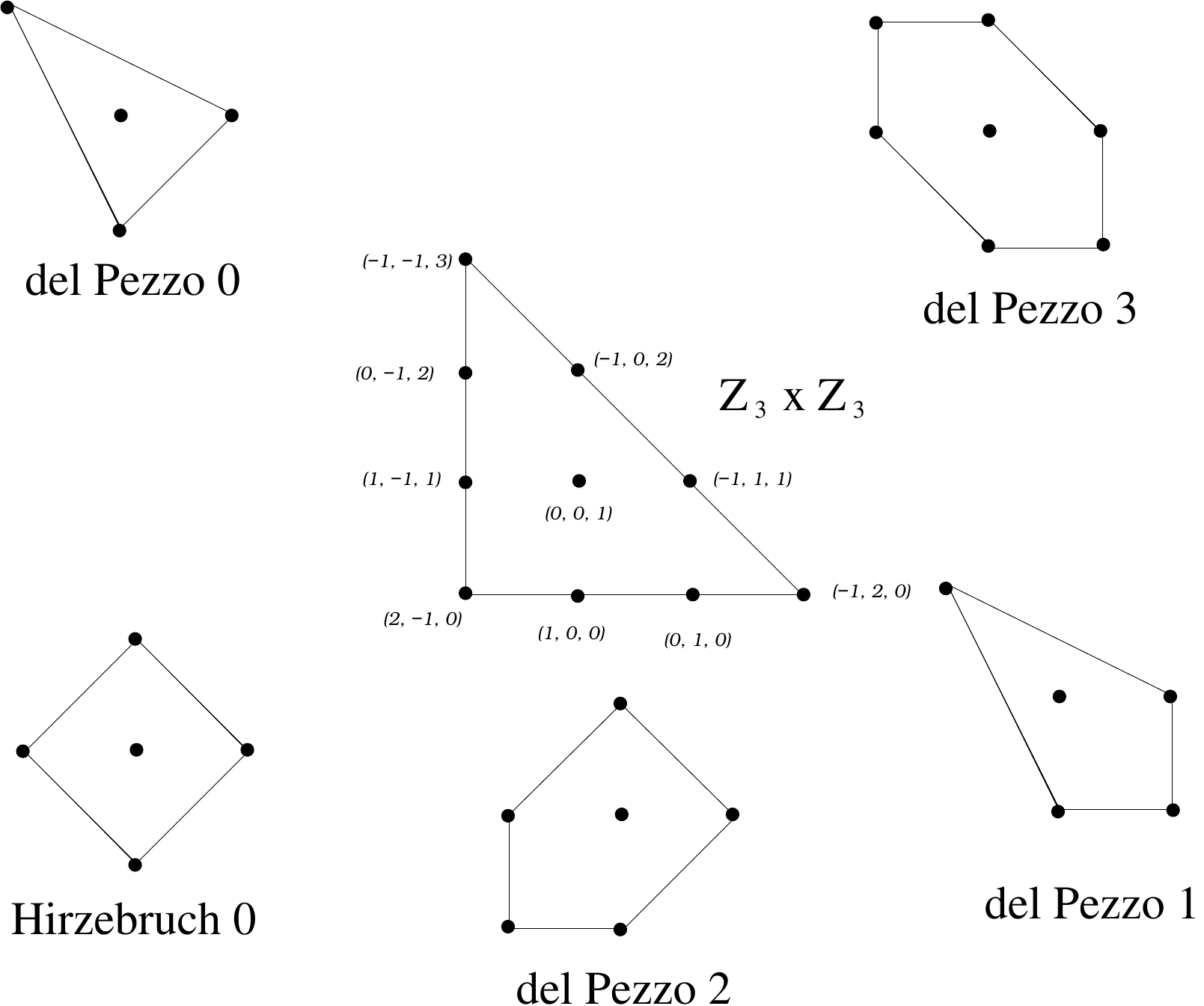}\end{array}$
\end{center}
\caption{
{\sf 
The toric diagram of $\IC^3 \slash (\IZ/3\IZ) \times (\IZ/3\IZ)$ as an affine cone; the coordinates of the vertices are just one choice, any $SL(3; \IZ)$ transformation is equivalent.
The subdiagrams are the cones over the  toric del Pezzo surfaces which are partial resolutions of $\IC^3 \slash (\IZ/3\IZ) \times (\IZ/3\IZ)$.
}
\label{f:z3z3reso}
}
\end{figure}

We return to our Question \ref{q:MQ} of relating the quiver and the Calabi-Yau 3-fold, which in the toric context translates to
\begin{question}\label{q:Mtoric}
What quiver with superpotential  $\cQ$  has its representation variety (VMS) $\cM(\cQ)$ that is an affine toric Calabi-Yau 3-fold with toric diagram $\cD$ (a lattice convex polygon)?
\end{question}
So far, we know the clover quiver \eqref{c3} corresponds to $\IC^3$, or 
\begin{equation}
\cD = \{(0,0); \ (0,1); \ (1,0)\} \ ,
\end{equation}
the (minimal) lattice triangle.
Also, we know that lattice triangle with legs $(p,q)$ corresponds to the McKay quiver for $\IZ/p\IZ \times \IZ/q\IZ$, which turns out to be the clover repeated $pq$ number of times, with all nodes tri-valent.
In physics, this is known as a {\it brane-box model} \cite{branebox}.
As an illustration, in Figure \ref{f:z3z3reso}, we draw the toric diagram for $\IC^3 \slash (\IZ/3\IZ) \times (\IZ/3\IZ)$, as well as how the cones over $\IP^2$ on which we now turn, as well as over its toric blow-ups (the toric del Pezzo surfaces).

It is expedient to illustrate all of this with a simple but important example.
Consider the action on $\IC^3[x,y,z]$ by a $\IZ/(3\IZ)$ action
\begin{equation}\label{z3action}
(x,y,z) \longrightarrow (\omega_3 x, \omega_3 y, \omega_3 z) \ , \quad \omega_3^3 = 1 \ .
\end{equation}
This is an $SU(3)$ action since the matrix Diag$(\omega_3,\omega_3,\omega_3)$ is special unitary.
There are 3 irreps for the Abelian group $\IZ/(3\IZ)$, all of dimension one, ${\bf 1}$, ${\bf 1}_{\omega_3}$, and ${\bf 1}_{\omega_3^2}$ and the action corresponds to a 3-dimensional reducible representation $R_3 = {\bf 1}_{\omega_3}^{\oplus 3}$.
The quiver can then be obtained from the analogue of \eqref{McKay2}, viz., 
\begin{equation}
R_3 \otimes {\bf r}_i=\bigoplus\limits_{j}a_{ij} {\bf r}_j \quad \Rightarrow \quad 
\mbox{Adjacency Matrix: } a_{ij} = \left(
\begin{array}{lll}
 0 & 3 & 0 \\
 0 & 0 & 3 \\
 3 & 0 & 0
\end{array}
\right) \ ,
\end{equation}
which says there are 9 arrows and 3 nodes with triplets going in a cycle.
We label the nodes from 1 to 3 and let $X_{ij}^{\alpha=1,2,3}$ denote the triplet of arrow from node $i$ to $j$.
The superpotential $W$ can also be found by projection on the clover using \cite{Lawrence:1998ja} and we have the quiver data
\begin{equation}
\begin{picture}(0,80)
  \put(-170,1){\includegraphics[trim=0mm 0mm 100mm 150mm, clip, width=1in]{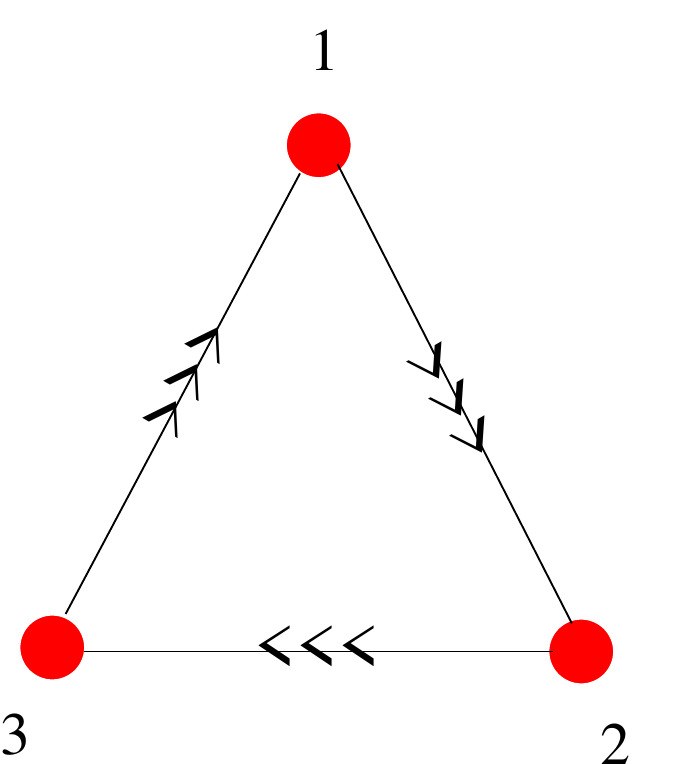}}
  \put(-180,40){$X_{31}$}\put(-140,-5){$X_{23}$}\put(-110,40){$X_{12}$}
  \put(-10,20){$  W=\sum\limits_{\alpha, \beta, \gamma = 1}^3 \epsilon_{\alpha\beta\gamma} X^{(\alpha)}_{12}
  X^{(\beta)}_{23} X^{(\gamma)}_{31}  \ , $}
\end{picture}  
\end{equation}
with arrows $X^{(\alpha)}_{12}, X^{(\beta)}_{23}, X^{(\gamma)}_{31}$ and the completely antisymmetric Levi-Civita symbol 
$\epsilon_{\alpha\beta\gamma}$.
Note that the dimension vector is $(1,1,1)$ and the $(1,2,3)$ are just index labels for the nodes.

Now, let us check that $\cM(\cQ)$ is as promised.
In order to do so, we apply Proposition \ref{vms} and 
to make everything crystal clear, we will again resort to Macaulay2 \cite{m2} for our illustration \footnote{
The map $\theta$ is actually a projection, and one can compactify the whole algorithm in one fell swoop using elimination, as is detailed in \S2.1 of \cite{Hauenstein:2012xs}. However, for illustrative purposes we will use the ring map presented in the text here.
}.
We begin by ordering the variables as
\begin{equation}
(X_{12}^1, X_{12}^2, X_{12}^3, X_{23}^1,X_{23}^2,X_{23}^3,X_{31}^1,X_{31}^2,X_{31}^3) \to (X_1, X_2, \ldots, X_9)
\end{equation}
and define the ring of the 9 arrow variables
\begin{verbatim}
R=ZZ/101[X_{1}..X_{9}];
\end{verbatim}

Next, there are clearly $3^3 = 27$ GIOs (minimal loops) $X_{12}^\alpha X_{12}^\beta X_{12}^\gamma$, which we index as
$y_1, y_2, \ldots, y_{27}$ (recall that 101 is just a prime of convenient choice):
\begin{verbatim}
S=ZZ/101[y_{1}..y_{27}];
gios = {
     X_{1}*X_{4}*X_{7}, X_{1}*X_{4}*X_{8}, X_{1}*X_{4}*X_{9}, 
     X_{1}*X_{5}*X_{7}, X_{1}*X_{5}*X_{8}, X_{1}*X_{5}*X_{9},
     X_{1}*X_{6}*X_{7}, X_{1}*X_{6}*X_{8}, X_{1}*X_{6}*X_{9}, 
     X_{2}*X_{4}*X_{7}, X_{2}*X_{4}*X_{8}, X_{2}*X_{4}*X_{9}, 
     X_{2}*X_{5}*X_{7}, X_{2}*X_{5}*X_{8}, X_{2}*X_{5}*X_{9}, 
     X_{2}*X_{6}*X_{7}, X_{2}*X_{6}*X_{8}, X_{2}*X_{6}*X_{9},
     X_{3}*X_{4}*X_{7}, X_{3}*X_{4}*X_{8}, X_{3}*X_{4}*X_{9}, 
     X_{3}*X_{5}*X_{7}, X_{3}*X_{5}*X_{8}, X_{3}*X_{5}*X_{9}, 
     X_{3}*X_{6}*X_{7}, X_{3}*X_{6}*X_{8}, X_{3}*X_{6}*X_{9}   };
\end{verbatim}

The superpotential expands to a six-term cubic
\begin{equation}
W = X_{23}^1 X_{31}^2 X_{12}^3-X_{23}^1 X_{31}^3 X_{12}^2+X_{23}^3 X_{31}^1 X_{12}^2+X_{23}^2 X_{31}^3 X_{12}^1-X_{23}^3 X_{31}^2 X_{12}^1-X_{12}^3 X_{23}^2 X_{31}^1 \ .
\end{equation}
Using our relabelling and taking the partial derivatives with respective to all 9 variables, we have the Jacobian ideal in $R$
\begin{verbatim}
jac = {
     -X_{6}*X_{8}+X_{5}*X_{9}, X_{6}*X_{7}-X_{4}*X_{9}, 
     -X_{5}*X_{7}+X_{4}*X_{8}, X_{3}*X_{8}-X_{2}*X_{9}, 
     -X_{3}*X_{7}+ X_{1}*X_{9}, X_{2}*X_{7}-X_{1}*X_{8}, 
     -X_{3}*X_{5}+X_{2}*X_{6}, X_{3}*X_{4}-X_{1}*X_{6}, 
     -X_{2}*X_{4}+X_{1}*X_{5}   };
\end{verbatim}

We are now ready to perform the map $\theta$:
\begin{verbatim}
M = ker(map(R/jac, S, gios));
\end{verbatim}
Note that we use the kernel rather than the image in Macaulay2 since, we mentioned in the footnote to Proposition \ref{vms}, the map between varieties is in the opposite direction from that between coordinate rings. 
As an ideal in $S$, $M$ is the affine variety $\cM(\cQ)$.
First, \verb|dim(M)| returns 3, which is good.
Next, we can see, using
\begin{verbatim}
minimalPresentation(M)
\end{verbatim}
what it actually is.  This minimal presentation strips off trivial linear relations and reduces the the number of variables in $S$.
In all, only 10 variables are left and $V$ is realized as the (non-complete) intersection of 27 quadrics in $\IC^{10}$,
which the astute algebraic geometer would recognize.
There is a standard Veronese embedding of $v : \IP^2 \hookrightarrow \IP^9$ of degree 3
\begin{align}
\label{Z3cubics}
v  : & \ \IP^2 \hookrightarrow \IP^9 \\
\nn
 ; & \ [z_0:z_1:z_2] \longrightarrow [z_0^3:z_0^2 z_1:z_0 z_1^2:z_1^3:z_0^2 z_2:z_0 z_1 z_2:z_1^2 z_2:z_0 z_2^2:z_1 z_2^2,z_2^3]
 \ ,
\end{align}
since there are exactly 10 degree 3 monomials in 3 variables.
In fact, these are precisely the 10 generators of the ring of invariants under the action of the $G = \IZ/3\IZ$ in \eqref{z3action} and there are 27 quadratic relations amongst them (the 27 quadrics returned from the minimal presentation).

Rephrasing all this in the affine language $M = \mbox{Spec} \IC[x,y,z]^G$, the spectrum of the $G$-invariant polynomial ring.
That is, $M \simeq \IC^3 / (\IZ/3\IZ)$, the orbifold as desired.
Furthermore, the cubics furnish the degree 3 sections of a line bundle over $\IP^2$, and thence, the orbifold is
\begin{equation}
\IC^3 / (\IZ/3\IZ) = \mbox{Cone}(\IP^2) \simeq Tot(\cO_{\IP^2}(-3)) \ ,
\end{equation}
 the total space of the anti-canonical bundle over $\IP^2$.
We can intuitively think of this as the negative curvature of the cone cancelling the positive curvature of $\IP^2$, giving a zero-curvature space which is Calabi-Yau.

Finally, as an Abelian orbifold, $M$ is toric, and the toric diagram can be retrieved readily. 
The 10 exponents of the invariant cubics from \eqref{Z3cubics} are 
\{
(3, 0, 0),\ (2, 1, 0),\ (1, 2, 0),\ (0, 3, 0),\ (2, 0, 1),\ (1, 1, 1),\ (0, 2, 1),\ (1, 0, 2),\ (0, 1, 2),\ (0, 0, 3)
\}
(one can check that as vectors they are co-planar).
The dual cone $\sigma^\vee$ to the cone $\sigma$ spanned by these 10 vectors can be readily found, again using {\sf SageMath} \cite{sage}, as was done in the digression in \S\ref{s:CYgradus}:
\begin{verbatim}
sigma = Cone([
     [3, 0, 0], [2, 1, 0], [1, 2, 0], [0, 3, 0], [2, 0, 1], 
     [1, 1, 1], [0, 2, 1], [1, 0, 2], [0, 1, 2], [0, 0, 3]   ]);
\end{verbatim}
and then
\verb|sigma.dual().rays()|
returns
$\{
(1,0,0), \ (0,1,0), \ (0,0,1)
\}$, which is the toric diagram: its endpoints are coplanar -- by appropriate $GL(2;\IZ)$ we can make the toric diagram as $\cD = \{(1,0), \ (0,1), \ (-1,-1)\}$.
Viewed from the plane on which the respective endpoints are coplanar, $\sigma$ and $\sigma^\vee$ are simply $\Delta_2$ and $\Delta^\circ_2$ in Figure \ref{f:DDdual} respectively, or, the top and bottom of Figure \ref{f:16delta2}.

\comment{
as the following algorithm, since the map $\theta$ is just a projection, which amounts to the elimination detailed below: 
\begin{itemize}
\item {\bf INPUT:}
\begin{enumerate}
\item Superpotential $W(\{X_i\})$, a polynomial in variables $X_{i=1, \ldots, n}$,
\item Generators of GIOs: $r_j(X_i)$, $j=1, \ldots, k$ polynomials in $X_i$;
\end{enumerate}

\item {\bf ALGORITHM:}
\begin{enumerate}
\item Define the polynomial ring $R = \IC[\phi_{i=1,\ldots,n}, y_{j=1,\ldots,k}]$,
\item Consider the ideal $I = \left< \frac{\partial {W}}{\partial X_i}; \ y_j - r_j(X_i) \right>$,
\item Eliminate variables $X_i$ from $I \subset R$ gives ideal $M$ in terms of $y_j$;
\end{enumerate}

\item {\bf OUTPUT:}
$M$ corresponds to the VMS as an affine variety in $\IC[y_1, \ldots, y_k]$.
\end{itemize}
}

\subsection{The Conifold}\label{s:coni}
If the toric diagram $\cD$ for $\IC^3$ is the minimal lattice triangle  $\{0,0), \ (1,0), \ (0,1) \}$, then one might wonder to what space - call it $\cC$ - the next simplest case of
$\cD = \{ (0,0), \ (1,0), \ (0,1), \ (1,1) \}$, the minimal square, might correspond.
The non-triangular shape already precludes the possibility of it being an orbifold of $\IC^3$ and we seem to have the beginning of another interesting family of toric Calabi-Yau 3-folds (simply enlargening the square indeed gives orbifolds of $\cC$).

The defining equation of $\cC$ can be readily found through toric methods, as in the previous subsection.
We first restore the third coordinate of the toric cone, say by adding height one for the $z$ direction, i.e.,
\begin{equation}
\cD = \{ (0,0,1), \ (1,0,1), \ (0,1,1), \ (1,1,1) \} \ ;
\end{equation} 
there are of course many $GL(3; \IZ)$-equivalent ways of doing this.
Next, we treat these vectors as exponents to 3 variables $(x,y,z)$, giving us the monomials 
$\{u,v,s,t \} := \{z, \ xz, \ yz, \ xyz \}$ which satisfy the single quadric relation
\begin{equation}\label{coni}
\{ u t = v s \} \subset \IC[u,v,s,t] \ ,
\end{equation}
In summary, this hypersurface in $\IC^4$, with $(0,0,0,0)$ as its singular point, is an affine, toric, Calabi-Yau 3-fold, called the {\red conifold} in the physics literature.
In the mathematics literature, it is just called a quadric hypersurface in $\IC^4$, the 3-dimensional version of the ordinary double point in \eqref{C2A1}.

Topologically, the conifold is the total space of the cotangent bundle over $S^2$ (as a local K3 surface), this is the total space of the cotangent bundle over $S^3$.
The metric (the 3-complex-dimensional analogue of Eguchi-Hanson \cite{EHmetric}), was found in \cite{Candelas:1989js}.
It is also a conical metric in the sense that the Calabi-Yau metric can be put in the form of \eqref{C3S5}, but now the base, instead of being $S^5$, is a 5-manifold known rather esoterically as $T^{1,1}$ \cite{Candelas:1989js}.
For a summary table of some of the most studied toric Calabi-Yau 3-folds and their toric diagrams, see p9 of \cite{He:2016fnb}.
Further pedagogical material on toric Calabi-Yau 3-folds, quivers and related physics and mathematics can also be found in cit.~ibid.

The conifold is one of the central objects in the mathematics and physics of mirror symmetry and geometric transitions and deserves a monograph by itself.
Sadly, due to spacetime constraints as well as the theme of this book, we shall only make two remarks.

\paragraph{Geometric Transitions: }
We can explicitly find the conifold -- and indeed any of the affine varieties in this chapter -- inside the compact ones from the previous chapter.
This justifies the name ``local'' to which we alluded in the beginning of the chapter.
For instance, $\cC$ can be a local singularity of our familiar quintic $Q$ from \S\ref{s:Q}.
Suppose a particular quintic (in the moduli space of complex structures, which we recall to be possible choices of monomials) looked like
\begin{equation}
\overline{Y} := \{ z_3 g(z_0, \ldots, z_4) + z_4 h(z_0, \ldots, z_4) = 0 \} \subset \IP^4_{[z_0: \ldots : z_4]} \ ,
\end{equation}
where $g$ and $h$ are homogeneous quartic polynomials in the projective coordinates.
Unlike the generic quintic, or the Fermat quintic, $\overline{Y}$ is singular, with its singular locus given as
\begin{equation}
\mbox{Sing}(\overline{Y})  =  \left\{ z_3 = z_4 = g(z_0, \ldots, z_4) = h(z_0, \ldots, z_4) = 0 \right\}
\end{equation}
that solves to $4^2 = 16$ points (nodes), being the intersection of two quartics in the 3 remaining homogeneous variables.
These nodes can be
\begin{enumerate}
\item Resolved to $Y$ by blow-up;
\item Smoothed to $\tilde{Y}$ by adding generic quintic monomials.
\end{enumerate}

From $Y$ to $\tilde{Y}$ is a geometric transition of the conifold type.
In relation to Conjecture \ref{yauconj}, it is believed by an optimism known as {{\bf \blue Reid's Fantasy}} (after Miles Reid), that all Calabi-Yau 3-folds can be related to each other by versions of such geometric transitions.

\paragraph{Sasaki-Einstein Manifolds: }
The other emblematic aspect of the conifold is its conical form of the metric, analogous to $\IC^3$ in \eqref{C3S5}.
This important property is one of a class of manifolds known as {\em Sasaki-Einstein}.
In general, suppose we have a K\"ahler $n$-fold $M$ with K\"ahler form $\omega$ which is a cone over a $(2n-1)$-dimensional real manifold $X$.
The metric takes the form
\begin{equation}
ds^2(M) = dr^2 + r^2 ds^2(X) \ , 
\end{equation}
with $r \in \IR_{\geq 0}$ the radial coordinate for the cone where $r=0$ is the tip.
The base $(2n-1)$-manifold is called Sasakian when $M$ is K\"ahler.
If in addition $M$ is Calabi-Yau, then $X$ is called Sasaki-Einstein.

The K\"ahler form $\omega$ can be written, for $\eta$ a global one-form on $X$, as
$\omega = - \frac{1}{2} d(r^2 \eta) = \frac{1}{2} i \partial \overline{\partial} r^2$.
Moreover, $X$ has a Killing vector field $R$ called the {\bf Reeb vector}, defined, for the complex structure $\mathcal{I}$ on $M$, as
$
R := \mathcal{I} \left( r \frac{\partial}{\partial r} \right)
$.

Now, when $M$ is furthermore toric Calabi-Yau, meaning that we have an integrable torus action $\mathbb{T}^n$ which leaves $\omega$ invariant, everything becomes even more explicit and the metric can be elegantly written down \cite{futaki}.
Taking $\partial /\partial \phi_i$ to be the generators of the torus action, with $\phi_i$ to be the angular coordinates,
allows for the introduction of symplectic coordinates $y_i$ defined as 
$y_i := -  \frac{1}{2} \langle r^2 \eta , \frac{\partial}{\partial \phi_i } \rangle$,
with $\left<~,~\right>$ the usual bilinear pairing between forms and vector fields.

Subsequently, the K\"ahler form and the metric become,  in these symplectic coordinates
\begin{equation}
\omega = d y_i \wedge d \phi_i \ , \qquad
d s^2 = G_{ij} d y_i d y_j + G^{ij}  d \phi_i  d \phi_j \ ,
\end{equation}
where $G^{ij}$ is the inverse of $G_{ij} := \partial_i \partial_j G$ for some symplectic potential $G$ determined from the complex structure as $\mathcal{I} = \left[
\begin{array}{cc}
0 & -G^{ij} \\
G_{ij} & 0 
\end{array}
\right]$. 
There is a long programme to understand the geometry of Sasaki-Einstein cones in relation to the world-sheet conformal field theory, generating such beautiful results as the equivalence of the minimization of the volume of the base $X$ and the maximization of certain central charge in the field theory
\cite{SEvol,Butti:2005vn,amax,Eager:2010yu,He:2017gam}.

Now, with non-compact manifolds, it is difficult to have a canonical notion of Hodge numbers and other such topological quantities.
Thus, while we have an infinite number of data points, it is hard to have a succinct analogue of the Hodge-pair plot of Figure \ref{f:KSplot}.
With a compact Sasaki-Einstein base, however, one {\it does} have an ordinary sense of Euler and Betti numbers.
The statistical studies of these, versus the normalized volumes for toric Calabi-Yau 3- and 4-folds coming from reflexive polytopes (cf.~the 16 polygons of Figure \ref{f:16delta2} and the analogous 4319 polyhedra) was initiated in \cite{He:2017gam}.

\paragraph{The Conifold Quiver: }
Returning to our question of quiver representations, what $(\cQ,W)$ has $\cC$ as its moduli space?
This problem was solved by \cite{Klebanov:1998hh} in one of the earliest examples of AdS/CFT beyond $\IC^3$.
The required data is a 2-noded quiver (with dimension vector $(N,N)$), with 4 arrows ($A_{1,2}$ and $B_{1,2}$, which are bi-fundamentals under $SU(N) \times SU(N)$) and 2-term quartic superpotential (upon expanding out the antisymmetric symbol $\epsilon$):
\begin{equation}
    \begin{array}{cc}
    \begin{array}{c}\includegraphics[trim=100mm 10mm 50mm 200mm, clip, width=1.5in]{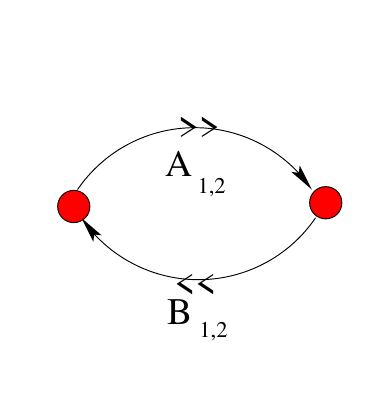}\end{array}
    &\qquad
         \begin{array}{l}
         \begin{array}{ccc}
         & SU(N) & SU(N) \\
         A_{i=1,2} & \fund & \antifund \\
         B_{j=1,2} & \antifund & \fund
         \end{array}\\
         ~\\
         \begin{array}{rcl}
         W &=& \tr(\sum\limits_{i,j,k,l=1,2} \epsilon_{il} \epsilon_{jk} A_i B_j A_l B_k) 
         \\
         &=& \tr(A_1 B_1 A_2 B_2 - A_1 B_2 A_2 B_1)
         \ .
         \end{array}
         \end{array}
       \end{array}
\end{equation}
One can check the VMS quite simply in this case.
Taking $N=1$, we again have $W=0$ so no further quotients by the Jacobian ideal is needed.
There are 4 minimal loops: $u = A_1B_1$, $t = A_2B_2$, $v = A_1B_2$ and $s = A_2B_1$, and they satisfy the one equation, the hypersurface in \eqref{coni}, as required for the defining quadric of $\cC$.
Taking $N>1$ would have given us the symmetric product $\cC^N / \mathfrak{S}_N$.

\subsection{Bipartite Graphs and Brane Tilings}
Having gained some confidence with examples, one might wonder whether there exists, on returning to the original Question \ref{q:MQ} of this chapter, an algorithm of translating between the quiver data and the Calabi-Yau data.
For orbifolds, we saw that we needed the character table of the finite group and then happily using McKay's method.
Surely, the rich combinatorics of toric varieties should facilitate this translation.
Indeed, the method was given in \cite{toricD3} and algorithmized in \cite{Feng:2000mi}, wherein the direction $\cQ \to \cD$, from the quiver (with superpotential) data to the toric diagram, was called, rather unimaginatively, the {\it forward algorithm} and the direction $\cD \to \cQ$, the {\it inverse algorithm}.

The forward algorithm is more or less  by direct computation, using a toric version of what was done in \S\ref{s:dP0}.
The inverse algorithm uses the facts that (1) every toric diagram is a sub-diagram of that of $\IC^3 / (\IZ/p\IZ  \times \IZ/q\IZ)$ for sufficiently large $p,q$ meaning that geometrically every affine toric Calabi-Yau 3-fold is a partial resolution of the said orbifold (corresponding to node deletion in its toric diagram);
(2) the quiver for the orbifold is known and it is simply the generalized McKay quiver with superpotential; and (3) the desired quiver can be obtained by deletion of the nodes which corresponds to removal of arrows (Higgsing in the field theory).

This inverse algorithm of ``geometrical engineering'' (a whimsical but very appropriate term first coined in \cite{Katz:1996fh}) the quiver/physics from the given Calabi-Yau geometry turns out to computationally expensive, with the bottle-neck being finding the dual cone for a given convex lattice polyhedral cone, which is exponential in complexity.
The break-through came in the mid-2000s when it was realized \cite{Hanany:2005ve,tilings} that a seemingly frivolous relation for the quiver $\cQ$ is actually of deep physical and mathematical  origin and consequence: whenever $\cM(\cQ)$ is toric, it was noted that
\begin{equation}\label{g=1}
N_0 - N_1 + N_2 = 0
\end{equation}
for $N_0$, the number of nodes, $N_1$, the number of arrows, and $N_2$, the number of monomial terms in the superpotential.
In our running examples, for $\IC^3$, this is $1-3+2=0$, for $\cC$, this is $2-4+2=0$, for Cone$(\IP^2)$, this is $3-9+6=0$.

Over the years, the {\it School of Hanany} 
\comment{
\footnote{
I am very honoured to be the primogenitus 
of this happy and industrious family whose siblings are unusually close and to whom Amihay Hanany, due to his calm demeanour, his insightfulness and his paterfamilias air almost biblical, is known affectionately as ``the Prophet''.
} 
}
pursued the mapping between toric Calabi-Yau manifolds and quiver representations relentlessly.
In brief, \eqref{g=1} is the Euler relation for a torus, and the quiver data (and superpotential) can be re-packaged into a bipartite tiling (called {\blue brane tiling} or {\blue dimer model}) of the doubly-periodic plane.
This bipartite-ness is profound and involves a plethora of subjects ranging from scattering amplitudes, to dessins d'enfants, to cluster mutation, etc. \cite{bipartite1,bipartite2,bipartite3,bipartite4,bipartite5,bipartite6,bipartite7}.

The bipartite tiling perspective on quiver representations and Calabi-Yau moduli spaces has grown into a vast field itself \cite{tilings+1,tilings+2,tilings+3,tilings+4,tilings+5,tilings+6,tilings+7,tilings+8}, on which there are excellent introductory monographs \cite{Kennaway:2007tq,Yamazaki:2008bt,Broom}, a short invitation by means of a conference report \cite{He:2016fnb}, and an upcoming book \cite{booksoon}. Again, with an apology we leave the curious reader to these references as well as two bird's-eye-view summary diagrams in Appendix \ref{ap:dimers}.

For now we will only summarize the method of translating between the quiver data and the bipartite graph data:
\begin{enumerate}
\item Consider a monomial term in $W$, if it comes with a plus (respectively minus) sign, draw a black (respectively white) node (the choice of colour is, of course, purely by convention), write all the variables in the monomial clockwise (respectively counter-clockwise) as edges around the node;
\item Connect all the black/white nodes by these edges.
  Because $W$ has the property (this comes from the fact that the VMS is a toric variety, corresponding to a so-called {\it binomial ideal}, q.v.,Appendix \ref{ap:toric}) that each variable appears exactly twice with opposite sign, we end up with a bipartite graph $\cB$;
\item The condition \eqref{g=1}  dictates that $\cB$ is a bipartite graph on $T^2$, or, equivalently, a tiling of the doubly periodic plane;
\item Each node in $\cB$ is a term in $W$ and thus a GIO, and edge is perpendicular to an arrow in $\cQ_1$ obeying orientation and each face in $\cQ$ corresponds to a node in $\cQ_0$; in other words, $\cB$ is a dual graph of $\cQ$;
\item  In particular, being the dual, $\cB$ has $N_2/2$ pairs of black/white nodes, $N_1$ edges and $N_0$ inequivalent (polygonal) faces. 
\end{enumerate}
Note that while the quiver has two pieces of information, the adjacency matrix and the superpotential, the tiling encodes both in one.
We illustrate the above procedure with our archetypal example of  $\IC^3$:
\begin{equation}\label{c3dimer}
\begin{array}{ccccc}
\begin{array}{c}
\begin{array}{c}\includegraphics[trim=0mm 0mm 0mm 0mm, clip, width=1in]{./PICTS/c3quiver}\end{array}
\\
W = \tr(XYZ - XZY)
\end{array}
& \longrightarrow
\begin{array}{c}\includegraphics[trim=0mm 0mm 0mm 0mm, clip, width=1in]{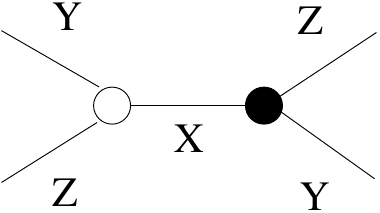}\end{array}
& \longrightarrow
& \begin{array}{c}\includegraphics[trim=0mm 0mm 0mm 0mm, clip, width=1.4in]{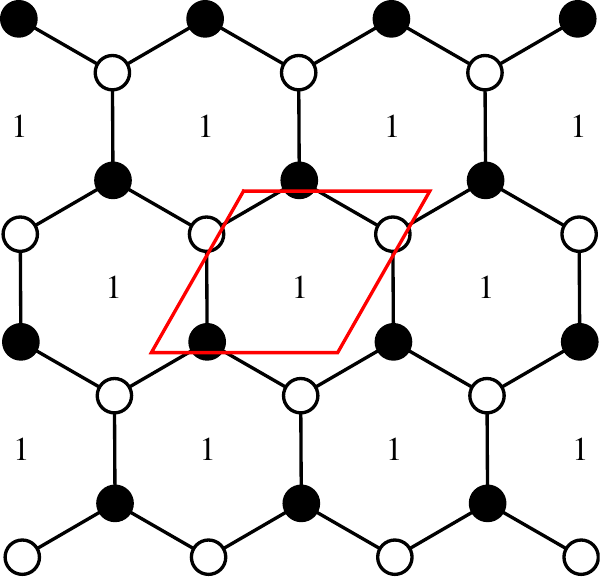}\end{array}
\end{array}
\end{equation}
We have marked the fundamental domain of $T^2$ with the red parallelogram; therein, there is 1 pair of black-white nodes, each of valency 3, corresponding respectively to the $+XYZ$ and $-XZY$ terms in $W$.
The edge together give the honeycomb tiling of the doubly periodic plane, with a single inequivalent face which is a hexagon marked ``1''.

The set of relations in \eqref{c3dimer} is only a corner of an intricate web of correspondence which we summarize in Figure \ref{f:c3summary} in Appendix \ref{ap:dimers}.
For further reference, the situation for our other favourite example, the conifold $\cC$, is shown in Figure \ref{f:conisummary} in the said appendix.

These explorations by physicists have generated sufficient interest in the mathematics community that teams of ``card-carrying'' algebraic geometers and representation theorists \cite{mathdimer1,mathdimer2,mathdimer3,mathdimer4,mathdimer5,mathdimer6,mathdimer7,mathdimer8,mathdimer9,mathdimer10,mathdimer11} have formalized the statement into
\begin{theorem}\label{toricThm}
Let $\cQ$ be a quiver with superpotential, which is graph dual to a bipartite graph drawn on $T^2$ according to the steps above, then the (coherent component of, i.e., the top dimensional irreducible piece of) the moduli space $\cM(\cQ)$ is an affine toric Calabi-Yau 3-fold.
\end{theorem}
A systematic probe of this toric landscape was initiated in \cite{Davey:2009bp} and updated using the latest algorithms and computer power in \cite{Franco:2017jeo}, marching upwards in incremental area of $\cD$.

We remark, therefore, that we have an infinite number of affine toric Calabi-Yau 3-folds, coming from (1) a lattice convex polygon $\cD$ as a toric diagram; or (2) a bipartite graph $\cB$ on $T^2$ as a quiver with superpotential.
Mapping between these two sets is intricate, it is generally believed that set (2) surjects onto set (1) and the orbit of quivers $\cQ$ or dimers $\cB$ mapping to the same $\cD$ are related to each other by cluster mutation (or known as Seiberg duality in the field theory).
As mentioned, we leave the full detail this toric-bipartite story to the wonderful reviews of \cite{Kennaway:2007tq,Yamazaki:2008bt}, the upcoming book \cite{booksoon}, or, for the impatient, the rapid report in \cite{He:2016fnb}.


\section{Cartography of the Affine Landscape}
We have taken our stroll in the landscape of non-compact, or affine, Calabi-Yau 3-folds, through the eyes of quiver representations, which {\it ab initio} may seem convoluted but turned out to luxuriate in an extraordinary wealth of mathematics and physics.
Indeed, the relaxation of compactness gave us not only explicit Ricci-flat metrics but also many infinite families of manifolds, exemplified by orbifolds and toric varieties.

Let us part with one last small but fascinating family, the del Pezzo surfaces, which we have encountered in many different circumstances (cf.~Appendix \ref{ap:dPF}).
These Fano surfaces are all of positive curvature, so a complex cone over them (think of these projective varieties simply affinized) with the tip at the origin, can be judiciously chosen to make the cone Calabi-Yau.
Computationally, we can ``affinize'' any compact variety projective $M$ to a non-compact one $\hat{M}$ rather easily by promoting the homogeneous coordinates of the projective space into which $M$ embeds to affine coordinates, i.e.,
\begin{equation}
M \subset \IP^n_{[z_0 : z_1 : \ldots : z_n]}  \longrightarrow \hat{M} \subset \IC^{n+1}_{(z_0 , z_1 , \ldots , z_n)} \ .
\end{equation}
In this sense, $\hat{M}$ is a complex cone over $M$ (not to be confused with the real cone which we discussed throughout the chapter for Sasaki-Einstein manifolds)  with the origin of $\IC^{n+1}$ as the tip.

In general, in this affinization $M$ and $\hat{M}$ cannot both be Calabi-Yau, one compact, and the other non-compact.
This is our case here, $M$ is a Fano surface and $\hat{M}$ is the affine Calabi-Yau 3-fold.
We already saw this in \S\ref{s:dP0} with the cone over $\IP^0$, which we recall is the first of the del Pezzo family (including Hirzebruch zero).
In this sense $Tot(\cO_{\IP^2}(-3))$ is a wonderful affine Calabi-Yau 3-fold, it is an orbifold, it is toric, and it is a del Pezzo cone.

Adhering to the notation of the Appendix, we can write $dP_0$ for $\IP^2$.
It turns out that $dP_{0,1,2,3}$ and $\IF_) = \IP^1 \times \IP^1$ are all toric, with their toric diagrams being number 1, 3, 6, 10 and 4 respectively in Figure \ref{f:16delta2}.
The higher ones with more blow-up points, $dP_{4,5,6,7,8}$ are not toric varieties, though their associated quivers were found in \cite{excepColl} using exceptional collections of sheafs.
As a historical note, the Calabi-Yau metrics for these del Pezzo cones - undoubtedly part of the tradition of the Italian School of algebraic geometry - were 
found by Calabi himself shortly after Yau's proof.

\subsection{Gorenstein Singularities}
Before closing, we make a brief remark about the singular nature of our manifolds.
In this chapter, except $\IC^3$, all the affine Calabi-Yau varieties are singular (at least) at the origin. 
We saw in \S\ref{s:orb} that orbifolds are by construction so.
Algebraic singularities in geometry is a complicated business and much effort has been devoted to their smoothing or resolutions.

The class of singularities of our concern are called {\bf Gorenstein}, which is a rather tamable situation.
The formal definition of Gorenstein-ness is intimidating and will take us too far into the inner sanctum of commutative algebra.
Roughly, a Gorenstein singularity is one outside of which there exists a global holomorphic form, or, from a sheaf-theoretical point of view, the affine scheme is Gorenstein if its canonical sheaf is a line bundle (of degree 0).
In other words, Gorenstein-ness is local Calabi-Yau-ness.

Luckily, there is an explicit computation that checks this crucial condition \cite{stanley}:
\begin{theorem}[R.~Stanley]
The numerator to the Hilbert series of a (graded Cohen-Macaulay domain) $R$ is palindromic iff $R$ is Gorenstein.
\end{theorem}
Recalling the basics of Hilbert series from Appendix \ref{ap:HS}, we can see this theorem in action.
We will not need the details of Cohen-Macaulay domains in this book; suffice it to say that coordinate rings of complex varieties have this property.
The Hilbert series for $\IC^3$ is just $1 / (1-t)^3$ from \eqref{HSCn}, the numerator is 1 and is trivially palindromic.
For the conifold $\cC$, it is $\frac{(1 - t^2)}{(1-t)^4}$ (coming for 4 generators obeying a single quadratic relation) and the numerator upon cancellation is $1+t$, which is palindromic.

The Hilbert series for $\IC^3/(\IZ/3\IZ)$ , using \eqref{molien}, is $\frac{1 + 7 t + t^2}{(1-t)^3}$ and the numerator is palindromic: the coefficients of the lowest and highest term are both 1.
In fact, for $dP_n$, the Hilbert series is $\frac{1 + (7-n) t + t^2}{(1-t)^3}$ \cite{pleth}.
We have included all the Hilbert series (Molien series) for the discrete finite ADE subgroups of $SU(2)$ in \eqref{HSADE} in the appendix.

In general, once we have some affine variety $M$, we can use the Gr\"obner basis methods detailed in the appendix to find the Hilbert series, and need only check the palindromy of the numerator to see whether $M$ is ``local Calabi-Yau''.

\subsection{The non-Compact Landscape}
\begin{figure}[t!]
	\centering
	\includegraphics[width=4in,angle=0]{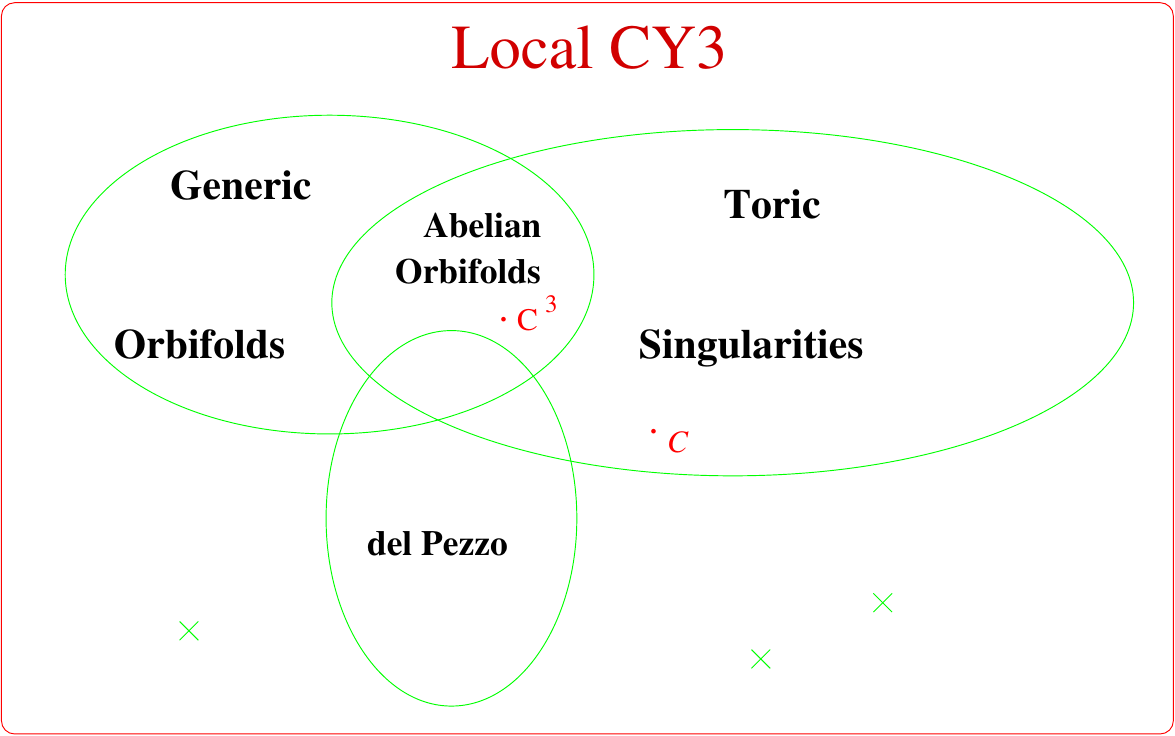}
	\caption{{\sf The landscape of non-compact, affine, local Calabi-Yau 3-folds.}}
	\label{f:landNonCompact}
\end{figure}

As we reach the end of our exposition, it is expedient to recall the vista of the non-compact Calabi-Yau 3-folds, in analogy to the Venn diagram in Figure \ref{f:landCompact} of \S\ref{s:compVenn}.
We show the landscape of non-compact Calabi-Yau 3-folds in Figure \ref{f:landNonCompact}.
As in the compact case, the Venn diagram is only topologically relevant and the size of the bubbles are not significant.
Several bubbles are meant to encapture an infinite number of spaces.

The analogue of the simplest starting point quintic $Q$ is here $\IC^3$, from which we have Abelian orbifolds of the form
 $\IC^3 / (\IZ/p\IZ \times \IZ/q\IZ)$ for any $p,q \in \IZ_{\geq 1}$, of which there is already an infinite number. Next, we have orbifolds of $\IC^3$ by discrete finite subgroups of $SU(3)$, which, in addition to the ADE subgroups of $SU(2)$, there are two infinite families, the delta-series, as well as a number of exceptionals.

The Abelian orbifold $\IC^3/(\IZ/ 3\IZ)$ is the cone (total space of the anticanonical bundle) over $dP_0 = \IP^2$, the first member of the very special family of del Pezzo surfaces. It is also a toric variety, as are all Abelian orbifolds. This toric family is again infinite in number: any convex lattice polygon is the toric diagram of an affine, toric Calabi-Yau 3-fold. The prototypical example in this toric class is $\cC$, the conifold, or the quadric hypersurface in $\IC^4$.

Of course, unlike local K3 surfaces, whose algebraic models can {\it only} be the ADE singularities, local Calabi-Yau 3-folds have no known complete characterization, the families chosen in the above Venn diagram are those which have been intensely studied, by mathematicians and physicists alike, and in particular, realize as the moduli space of quiver representations.
The crosses outside the bubbles are supposed to denote the plethora of other local Calabi-Yau models on which, though most certainly infinite in number, there have not been too intense an investigation.

\paragraph{Nunc Dimitis: }
With our tantalizing pair of plots of the Calabi-Yau landscape in Figures \ref{f:landCompact} and \ref{f:landNonCompact}  let us pause here.
We have taken a promenade in the land of CY$_3$, mindful of the intricate interplay between the mathematics and physics, emboldened by the plenitude of data and results, and inspired by the glimpses towards the yet inexplicable.

We have devoted a duo of chapters on cartography, first on the compact and second on the non-compact, gaining familiarity with the terrain of CICYs, KS hypersurfaces, orbifolds, del Pezzo cones, lattice polytopes, etc.
The landscape of Calabi-Yau manifolds will certainly continue to provoke further exploration, especially with the advance of ever new mathematics, physics and computing.

Assured by the abundance of data and the increasing computing power to calculate requisite quantities, we are perhaps simultaneously daunted by the complexities involved in the analysis, in the algorithmic sense of ``complexity''.
In the ensuing chapter, we shall speculate as to what might be done when confronted with the furrowed mountains of the Calabi-Yau landscape.

\chapter{Machine-Learning the Landscape}\label{s:learn}
\begin{chapquote}{David Hilbert}
``Die Physik ist f\"ur die Physiker eigentlich viel zu schwer.'' \footnote{
This is a famous quote from Hilbert at the turn of the 20th century, that ``physics has become much too difficult for physicists,'' meaning that physicists and mathematicians must work together to tackle the key problems of fundamental physics.
Indeed, the 20th century took this advice and the foundational questions in physics were inextricably linked to the deepest mathematics.
Perhaps, starting from the 21st century, data science and computer algorithms will become an integral part of mathematical physics and pure mathematics.
}
\end{chapquote}

And so we have partaken our excursion into the landscape of compact and non-compact Calabi-Yau manifolds, and like the keen geologist, soiled our hands with fascinating samples for scrutiny and experimentation, rather than the artist, whose primary concern is an impression of the magnificent vista.
Thus prepared, we enter the last chapter, where with the plenitude of data collected, we can speculate upon its treatment.
 We mentioned at the very outset of the preface that a Zeitgeist, cultural and intellectual, seems to be curiously present, which breathes its {\it spritus movens} onto scientific progress.
 Indeed, were we to examine our Scientific Age, it is doubtless that the omnipresence of data is what propels much of research, from exoplanets to the human genome to the outpouring particle-jets at CERN.
 
It is natural, therefore, to ask whether the most cutting-edge techniques from Data Science can be applied to the landscape of Calabi-Yau manifolds, to algebraic geometry, and indeed, to mathematical physics and pure mathematics in general.
This idea of machine-learning (ML) the Calabi-Yau landscape, and thence to mathematical structures \footnote{
For a quick summary and some speculations about the AI mathematician, q.v.~ \cite{myVideos,buzzard}.
} was initiated in \cite{He:2017aed,He:2017set}.
It was fortuitous that in 2017, the year that Sophia became the first non-human citizen, four independent collaborations \cite{He:2017aed,He:2017set,Krefl:2017yox,Ruehle:2017mzq,Carifio:2017bov} brought machine-learning to string theory.
It is interesting to note that the first annual ``Strings'' conference began in the 1980s, from which offshoots of ``String Phenomenology'' and ``String Mathematics'' emerged in the early 2000s and in 2017/8, began the new series of ``String Data''.
The reader is also referred to the ``String-Data Cooperation'' on Github, inaugurated by Sven Krippendorf et al.~in Munich \cite{GithubString}, which should grow to a marvelous repository of the data which we have discussed so far and much more beyond \footnote{
The reader is also referred to an exciting and relatively new field of topological data analysis \cite{TDA}, which has recently been applied to the landscape \cite{Cirafici:2015pky,Cole:2018emh}.
}.

Of course, in this respect, the mathematical physics community is rather behind the experimental, where the constant influx of data had long propelled the field into machine-learning techniques. 
For instance, the first AI-HENP (high energy and nuclear physics) seminar at CERN dates back to 1990 (interestingly around the same time as the establishment of the CICY data) and by 2010, the trigger system for particle detection already had machine-learning algorithms built in \cite{cernML}.
By now, we hope the reader is convinced that there is a vast landscape of {\bf mathematical data}, mostly compiled in the last decade or so, ranging from algebraic geometry, to representation theory, to number theory, etc.; the first two parts of this book had been using the Calabi-Yau data as a play-ground to illustrate various ideas in geometry and physics.

The initial investigations of 2017 have since been taken to various industrious and profitable ventures by many authors.
In the string context, these have included explorations in the cosmological landscape \cite{cosmo}, F-theory groups and matter \cite{Wang:2018rkk,Bies:2020gvf}, heterotic orbifold landscape \cite{Mutter:2018sra}, reinforcement learning of vacua from branes \cite{Halverson:2019tkf}, genetic algorithms on flux vacua \cite{Cole:2019enn}, standard model data mining and statistical predictions \cite{Parr:2019bta,Halverson:2020opj,Parr:2020oar,Deen:2020dlf,Larfors:2020ugo}, symmetries \cite{Krippendorf:2020gny}, and CFT properties \cite{Chen:2020dxg}, etc.
In the stringy-geometry direction in particular, these have ranged from machine-learning CICYs \cite{Bull:2018uow,Bull:2019cij,Erbin:2020srm,Erbin:2020tks,He:2020lbz}, triangulations of the Kreuzer-Skarke data-set \cite{Altman:2018zlc}, bundle cohomology \cite{Constantin:2018hvl,Klaewer:2018sfl,Brodie:2019dfx}, distinguishnig elliptic fibrations \cite{He:2019vsj}, classifying CYs \cite{Grimm:2019bey}, CY numerical metrics \cite{Ashmore:2019wzb}, or even to the linguistic structure of ArXiv mathematical physics sections \cite{He:2018dlv}, etc.

Remarkably, there has been daring proposals that quantum field theoretical structures, be it AdS/CFT
\cite{Hashimoto:2018ftp,Hashimoto:2019bih}, renormalization group flow \cite{Koch:2019fxy}, or QFT itself \cite{Halverson:2020trp}, are guises of neural networks.
In parallel, in the computational mathematical community, ML techniques have recently been applied to efficiently finding S-pairs in Gr\"obner bases \cite{mikeGB} and in computation in toric varieties \cite{lily}.

In the context of the structure of mathematics, there has been a programme on how different branches of mathematics respond to machine-learning \cite{myVideos}. One could imagine that number theory, especially unknown patterns in primes, would be the most resilient. This seems indeed the case, and we will discuss some of the experiments toward the end of the chapter.
Since 2017, preliminary explorations, in addition to the algebro-geometric ones above, have included machine-learning representation theory \cite{He:2019nzx} (surprisingly, simple finite groups seem to be distinguishable from non-simple ones by looking only at the Cayley table), knot invariants \cite{Jejjala:2019kio}, cluster mutation in quiver theory \cite{Bao:2020nbi}, combinatorics of graphs \cite{He:2020fdg} (where possession of Euler or Hamilon cycles, or graph Laplacian spectra, seem to be learnable from classifying the adjacency matrix), to number theory (the Birch-Swinnerton-Dyer Conjecture \cite{Alessandretti:2019jbs} and the Sato-Tate Conjecture \cite{He:2020kzg} in arithmetic geometry), etc.

In this final chapter, we shall let our imagination take the reins and be no longer like the meticulous naturalists, but rather the fearless frontiersmen, and roam freely in the landscape which we had charted in the previous chapters.
We will attempt to introduce some rudiments of machine-learning to the students in mathematics and theoretical physics communities who are also interested in explorations in data.
In keeping with the theme of this book, we will focus on Calabi-Yau data as a play-ground; we have used it as a kindergarten to learn some physics and some mathematics, now, we will use it to learn some modern data science.
Recently, there is an excellent monograph on introduction to data science to the theoretical physicist \cite{Ruehle:2020jrk} and the reader should refer to it for a self-contained and technical introduction.

\section{A Typical Problem}\label{s:prob}
First, let us motivate the reader with the problem at hand.
Reviewing the previous two chapters, we are encouraged by the ever-growing database of geometries \footnote{
In this book we have only focused on Calabi-Yau 3-folds for concreteness, there are many other data-sets which have been established, ranging from the closely related bundles over CY 3-folds, generalization of CICYs, etc., to diverse subjects as the ATLAS of finite groups, etc., as mentioned in the Preface.
}, mostly freely available online, as well as the computer software, especially the open-source {\sf SageMath} umbrella \cite{sage}, designed to calculate many of the requisite quantities.

Much of these data have been the brain-child of the marriage between
physicists and mathematicians, especially incarnated by applications of computational
algebraic geometry, numerical algebraic geometry and combinatorial geometry
to problems which arise from the classification in the physics and recast into a finite,
algorithmic problem in the mathematics. 
In principle, as far as addressing problems such as searching for the right bundle cohomology to give the Standard Model - which we recall was what inspired the field of Calabi-Yau data from the late 1980s -  is concerned, one could scan through configurations, find large clusters of computers \footnote{
I remember a wonderful quote from Dan Freedman's lectures when I was a PhD student at MIT, on addressing a difficult calculation, he, in his usual dry sense of humour, said, ``this is a problem perfectly adapted to large parallel clusters of graduate students.'' 
} and crunch away.

However, we had repeatedly alluded to the fact, even when demonstrating the powers of the likes of {\sf Macaulay2} \cite{m2} and {\sf SageMath} \cite{sage} in the foregoing discussions, that most of the core algorithms are exponential in running time and in memory usage.
This is the case for establishing Gr\"obner bases (due to getting all pairs of S-polynomials), for finding dual cones (due to finding all subsets of facets to check perpendicularity), for triangulation of polytopes (due to collecting all subsets of lattice points on the polytope), etc., which are the crux of the computational methods.
Even parallelizable computations in numerical algebraic geometry \cite{bertini,Mehta:2012wk} suffer from the need to find a large number of paths in homotopy continuation.

Confronted with this limitation, it would be helpful to step back and re-examine the desired result \cite{He:2017aed,He:2017set}.
One finds that regardless of the origin and the intent, the typical problem in computational algebraic geometry is one of the form
\begin{equation}\label{inout}
\fbox{\mbox{
{\LARGE
$\stackrel{INPUT}{\fbox{\mbox{integer tensor}}} \longrightarrow \stackrel{OUTPUT}{\fbox{\mbox{integer}}}$
}
}}
\end{equation}
We see this in all the cases presented hitherto (note that where the output is a list of integers, we can just focus on one at a time, conforming to the paradigm of \eqref{inout}), e.g.,
\begin{itemize}
\item Computing the Hodge number of a CICY. 
	\begin{description}
	\item[Input: ] an integer matrix whose row size ranges from 1 to 12, column size, from 1 to 15, and whose entries are from 0 to 5;
	\item[Output: ] a non-negative integer $h^{2,1}$ or a positive integer $h^{1,1}$.
	\end{description} 
\item Computing the cohomolgies of a line bundle $L$ over a CICY $M$.
	\begin{description}
	\item[Input: ] the CICY configuration matrix as above, plus a vector of the integer (negative also allowed) degrees of $L$; 
	\item[Output: ] a list of non-negative integers, the rank $h^*(M, L)$ of the cohomology of $L$.
	\end{description}
\item Computing the Hodge numbers and Chern classes of a KS Calabi-Yau 3-fold.
	\begin{description}
	\item[Input: ] an integer polytope, specified either by a list of integer 4-vectors or by the coefficients of the hyperplane inequalities defining the polytope; 
	\item[Output: ] integers $(h^{1,1}, h^{2,1})$, integer coefficients of $c_2$ (expanded into the basis of K\"ahler classes).	
	\end{description}
\item Computing the triple intersection numbers of a hypersurface in weighted projective $\IP^4$.
	\begin{description}
	\item[Input: ] a 5-vector of co-prime positive integers; 
	\item[Output: ] a list (totally symmetric 3-tensor) of non-negative integers.
	\end{description}
\item Computing the quiver from given affine toric Calabi-Yau 3-fold.
	\begin{description}
	\item[Input: ] the list of integer vertices of a convex lattice polygon, the toric diagram $\cD$; 
	\item[Output: ] the integer adjacency matrix of the quiver as well as the list of non-negative integer exponents of the monomial terms to include in the superpotential.
	\end{description}
\item Computing the number of cluster mutation/Seiberg dual phases of a quiver (whose moduli space of representations, the VMS, is toric Calabi-Yau 3-fold).
	\begin{description}
	\item[Input: ] the Kasteleyn adjacency matrix of the bipartite graph (brane-tiling) on the doubly periodic plane; 
	\item[Output: ] a positive integer.
	\end{description}
\end{itemize}
The list goes on and on. 
Such computations are also typical in standard-model building in string theory. This is because, due to the lack of analytic metrics of any kind on compact CYs, most calculations are, at the stage, interpretations of integer topological invariants like Hodge numbers or ranks of bundle cohomolgies, as particle numbers or charges.
We will, however, mention a little on numerical metrics towards the end of this chapter.

\subsection{WWJD}
As we enter the Age of Data Science and Artificial Intelligence (AI), one cannot resist how one might address problems of the type presented at the end of the previous section, without recourse to traditional methods, which are known to be computationally expensive.
Indeed, WWJD? {\bf {\red What Would JPython/Jupyter Do?}}

The joke is perhaps so esoteric that it might be lost on most.
First, {\sf JPython}, or its successor, {\sf Jython}, is the implementation of the {\sf Python} Programming language \cite{python} for the JAVA platform and {\sf Jupyter} \cite{jupyter} is an open-source project whose core language consists of {\sf Julia}, {\sf Python}, and {\sf R}, whose interface is where {\sf Python} and {\sf SageMath} \cite{sage} are usually run.

According to card-carrying computer scientists and data scientists, these are the platforms of their preference, and {\sf Python}, together with {\sf C++}, the programming languages of choice, one higher and one slightly lower level.
Second, in some circles, especially amongst Evangelical Christians, many wear T-shirts with the acronym WWJD, or ``What Would Jesus Do'', as a constant reminder on how to lead one's life 
\comment{
\footnote{
Though Catholics like myself may find this gauche, I do appreciate the nobility of the sentiment.
Nevertheless, I beg the reader for a moment's indulgence on my speculations on theology.
Perhaps a key difference between the Catholic and the Protestant is in the former's tormenting sense of worthless in failing to imitate Christ and the latter's over-optimism in being able to do so.
To quote Miguel de Unamuno who so eloquently puts it, ``\ldots  Protestantism, absorbed in this preoccupation with justification \ldots ends by neutralizing and almost obliterating eschatology; it abandons the Nicene symbol, falls into an anarchy of creeds, into pure religious individualism and a vague esthetic, ethical, or cultured religiosity. What we may call "other-worldliness" (Jenseitigkeit) was obliterated little by little by "this-worldliness" (Diesseitigkeit) \ldots''
}}.

So, what would one versed in Python do?
Let us make an analogy.
Suppose we are given the hand-written digits
\begin{equation}\label{digits}
\includegraphics[trim=0mm 0mm 0mm 0mm, clip, width=4in]{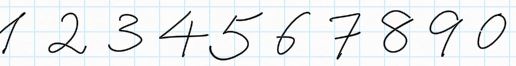}
\end{equation}
and we wish to let the computer recognize them.
The input is an image, which is nothing but a $m \times n$ matrix (indexing the pixels in a 2-dimensional grid) each entry of which is a 3-vector of a real value between 0 and 1, denoting the percentage of RGB values. Or, if we only wish to keep black-white (gray-scale) information as is sufficient for this case, each entry is then a real number between 0 and 1. In fact, it is even sufficient to only keep the information of whether each pixel is occupied so that the input is an $m \times n$ matrix of 0s and 1s.
The output is an integer from 0 to 9. In the computer science literature, this is called a {\it 10-channel output}.

As mathematicians or theoretical physicists, what might instinctively come to mind in order to solve this problem is to exploit geometry and find, say, a clever Morse function as we scan the input matrix row-wise and column-wise and detect the critical points; these points will be different because the topologies of the digits are all different.
Or, perhaps, one could compute the persistent homology, as it has become fashionable of late, of the pixels as a point-cloud of data.
The downside to all of these is that (1) the computation is very expensive, and (2) there is too much variation: how I write the digits will differ substantially (though not overwhelmingly) from how you would write them.
Upon reflection, this is rather like the situation of our concern: the computation, such as Gr\"obner bases, is too expensive and the input has some variation in configuration, e.g., CICY matrices are defined only up to permutations and splitting-equivalence, or a polytope is only defined up to $GL(n;\IZ)$.

How does your smartphone, or, indeed, Google, treat \eqref{digits}?
With today's access to data, one can readily proceed to repositories wherein there is a plenitude of writing samples.
For example, the NIST (National institute of Standards) database \cite{nist} has some $10^6$ actual samples, classified in  the form
\begin{equation}\label{sampledigits}
\includegraphics[trim=0mm 0mm 0mm 0mm, clip, width=5in]{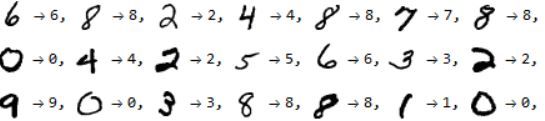} \ldots
\end{equation}
from which we see large (and in some sense also small) variation in hand-writing.

Returning to the earlier point of representation of pixelated images, one example of the number 3, for instance, would be 
\begin{equation}\label{digit3}
\fbox{$\begin{array}{c}\includegraphics[trim=15mm 0mm 0mm 0mm, clip, width=1in]{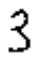}\end{array}$}
\to
{\tiny
\left(
\arraycolsep=1.2pt\def\arraystretch{0.2}
\begin{array}{cccccccccccccccccccccccccccc}
 1 & 1 & 1 & 1 & 1 & 1 & 1 & 1 & 1 & 1 & 1 & 1 & 1 & 1 & 1
   & 1 & 1 & 1 & 1 & 1 & 1 & 1 & 1 & 1 & 1 & 1 & 1 & 1 \\
 1 & 1 & 1 & 1 & 1 & 1 & 1 & 1 & 1 & 1 & 1 & 1 & 1 & 1 & 1
   & 1 & 1 & 1 & 1 & 1 & 1 & 1 & 1 & 1 & 1 & 1 & 1 & 1 \\
 1 & 1 & 1 & 1 & 1 & 1 & 1 & 1 & 1 & 1 & 1 & 1 & 1 & 1 & 1
   & 1 & 1 & 1 & 1 & 1 & 1 & 1 & 1 & 1 & 1 & 1 & 1 & 1 \\
 1 & 1 & 1 & 1 & 1 & 1 & 1 & 1 & 1 & 1 & 1 & 1 & 1 & 1 & 1
   & 1 & 1 & 1 & 1 & 1 & 1 & 1 & 1 & 1 & 1 & 1 & 1 & 1 \\
 1 & 1 & 1 & 1 & 1 & 1 & 1 & 1 & 1 & 1 & 1 & 1 & 1 & 1 & 0
   & 0 & 1 & 1 & 1 & 1 & 1 & 1 & 1 & 1 & 1 & 1 & 1 & 1 \\
 1 & 1 & 1 & 1 & 1 & 1 & 1 & 1 & 1 & 1 & 1 & 1 & 0 & 0 & 0
   & 0 & 0 & 1 & 1 & 1 & 1 & 1 & 1 & 1 & 1 & 1 & 1 & 1 \\
 1 & 1 & 1 & 1 & 1 & 1 & 1 & 1 & 1 & 1 & 1 & 0 & 0 & 0 & 1
   & 1 & 0 & 0 & 1 & 1 & 1 & 1 & 1 & 1 & 1 & 1 & 1 & 1 \\
 1 & 1 & 1 & 1 & 1 & 1 & 1 & 1 & 1 & 1 & 0 & 0 & 1 & 1 & 1
   & 1 & 0 & 0 & 1 & 1 & 1 & 1 & 1 & 1 & 1 & 1 & 1 & 1 \\
 1 & 1 & 1 & 1 & 1 & 1 & 1 & 1 & 1 & 1 & 1 & 1 & 1 & 1 & 1
   & 1 & 1 & 0 & 1 & 1 & 1 & 1 & 1 & 1 & 1 & 1 & 1 & 1 \\
 1 & 1 & 1 & 1 & 1 & 1 & 1 & 1 & 1 & 1 & 1 & 1 & 1 & 1 & 1
   & 1 & 0 & 0 & 1 & 1 & 1 & 1 & 1 & 1 & 1 & 1 & 1 & 1 \\
 1 & 1 & 1 & 1 & 1 & 1 & 1 & 1 & 1 & 1 & 1 & 1 & 1 & 1 & 1
   & 1 & 0 & 0 & 1 & 1 & 1 & 1 & 1 & 1 & 1 & 1 & 1 & 1 \\
 1 & 1 & 1 & 1 & 1 & 1 & 1 & 1 & 1 & 1 & 1 & 1 & 1 & 1 & 1
   & 0 & 0 & 1 & 1 & 1 & 1 & 1 & 1 & 1 & 1 & 1 & 1 & 1 \\
 1 & 1 & 1 & 1 & 1 & 1 & 1 & 1 & 1 & 1 & 1 & 1 & 0 & 0 & 0
   & 0 & 1 & 1 & 1 & 1 & 1 & 1 & 1 & 1 & 1 & 1 & 1 & 1 \\
 1 & 1 & 1 & 1 & 1 & 1 & 1 & 1 & 1 & 1 & 0 & 0 & 0 & 0 & 0
   & 1 & 1 & 1 & 1 & 1 & 1 & 1 & 1 & 1 & 1 & 1 & 1 & 1 \\
 1 & 1 & 1 & 1 & 1 & 1 & 1 & 1 & 1 & 1 & 1 & 1 & 1 & 0 & 0
   & 1 & 1 & 1 & 1 & 1 & 1 & 1 & 1 & 1 & 1 & 1 & 1 & 1 \\
 1 & 1 & 1 & 1 & 1 & 1 & 1 & 1 & 1 & 1 & 1 & 1 & 1 & 1 & 0
   & 0 & 1 & 1 & 1 & 1 & 1 & 1 & 1 & 1 & 1 & 1 & 1 & 1 \\
 1 & 1 & 1 & 1 & 1 & 1 & 1 & 1 & 1 & 1 & 1 & 1 & 1 & 1 & 1
   & 0 & 0 & 1 & 1 & 1 & 1 & 1 & 1 & 1 & 1 & 1 & 1 & 1 \\
 1 & 1 & 1 & 1 & 1 & 1 & 1 & 1 & 1 & 1 & 1 & 1 & 1 & 1 & 1
   & 1 & 0 & 0 & 1 & 1 & 1 & 1 & 1 & 1 & 1 & 1 & 1 & 1 \\
 1 & 1 & 1 & 1 & 1 & 1 & 1 & 1 & 1 & 1 & 1 & 1 & 1 & 1 & 1
   & 1 & 1 & 0 & 0 & 1 & 1 & 1 & 1 & 1 & 1 & 1 & 1 & 1 \\
 1 & 1 & 1 & 1 & 1 & 1 & 1 & 1 & 1 & 1 & 1 & 1 & 1 & 1 & 1
   & 1 & 1 & 0 & 0 & 1 & 1 & 1 & 1 & 1 & 1 & 1 & 1 & 1 \\
 1 & 1 & 1 & 1 & 1 & 1 & 1 & 1 & 1 & 1 & 1 & 1 & 1 & 1 & 1
   & 1 & 1 & 0 & 0 & 1 & 1 & 1 & 1 & 1 & 1 & 1 & 1 & 1 \\
 1 & 1 & 1 & 1 & 1 & 1 & 1 & 1 & 1 & 1 & 1 & 1 & 1 & 1 & 1
   & 1 & 1 & 0 & 0 & 1 & 1 & 1 & 1 & 1 & 1 & 1 & 1 & 1 \\
 1 & 1 & 1 & 1 & 1 & 1 & 1 & 1 & 1 & 0 & 0 & 1 & 1 & 0 & 0
   & 0 & 0 & 0 & 1 & 1 & 1 & 1 & 1 & 1 & 1 & 1 & 1 & 1 \\
 1 & 1 & 1 & 1 & 1 & 1 & 1 & 1 & 1 & 0 & 0 & 0 & 0 & 0 & 0
   & 0 & 1 & 1 & 1 & 1 & 1 & 1 & 1 & 1 & 1 & 1 & 1 & 1 \\
 1 & 1 & 1 & 1 & 1 & 1 & 1 & 1 & 1 & 1 & 1 & 1 & 1 & 1 & 1
   & 1 & 1 & 1 & 1 & 1 & 1 & 1 & 1 & 1 & 1 & 1 & 1 & 1 \\
 1 & 1 & 1 & 1 & 1 & 1 & 1 & 1 & 1 & 1 & 1 & 1 & 1 & 1 & 1
   & 1 & 1 & 1 & 1 & 1 & 1 & 1 & 1 & 1 & 1 & 1 & 1 & 1 \\
 1 & 1 & 1 & 1 & 1 & 1 & 1 & 1 & 1 & 1 & 1 & 1 & 1 & 1 & 1
   & 1 & 1 & 1 & 1 & 1 & 1 & 1 & 1 & 1 & 1 & 1 & 1 & 1 \\
 1 & 1 & 1 & 1 & 1 & 1 & 1 & 1 & 1 & 1 & 1 & 1 & 1 & 1 & 1
   & 1 & 1 & 1 & 1 & 1 & 1 & 1 & 1 & 1 & 1 & 1 & 1 & 1 \\
\end{array}
\right)}
\to
3
\end{equation}
A few technical remarks here.
In the NIST database, the digits are given in $28 \times 28$ pixel grayscales.
So each pixel, of the $28^2$ pixels, is represented by a real number from 0 to 1 (with 1 meaning white and 0, black) in order to designate the level of grey. This could be useful in that when people write, tapering off and making the stroke lighter differs for different digits; thus the level of greyscale is an extra feature in the input.
For simplicity, we have made the image purely black and white, whereby representing the digit by a binary $28\times28$ matrix (one can literally see this in the matrix of 1s and 0s above).
An ML algorithm would then have to be able to tell that this matrix corresponds to the number 3.

What the likes of smartphones performs the following four-step procedure
\begin{enumerate}
\item Acquire data: the collection of known cases (input $\to$ output), such as \eqref{sampledigits}, is commonly called {\blue training data};
\item Machine-Learn: this is the core algorithm which is slowly dominating all fields of human endeavour, from patient diagnoses, to LHC particle data analysis, to speech and hand--writing recognition, etc.
This chapter will be devoted to an invitation to this subject for mathematicians and theoretical physicists \footnote{
Indeed, for this audience which is versed in the elements of differential and algebraic geometry and/or quantum field theory and general relativity, the tools in the subject of machine-learning should present no conceptual challenge. This is part of the appeal of the field, it is remarkably simple but extraordinarily powerful and universally applicable, so powerful, in fact, that there is still much which seem almost magical and await statements of formal theorems and rigorous proofs.
};
\item Validate: once the machine/AI has ``learnt'' the training data, we can take a set of so-called {\blue validation data}, which, importantly, the machine has {\em not} seen before.  This is in the same format as the training data, with given input and ouput. Thus we can see how the machine performs on by checking the predicted output with the actual output;
\item Predict: If the validation is well-behaved, then the machine-learning is successful and we can use it to predict new output on input hithertofore unseen.
\end{enumerate}

We now move to a rapid introduction to machine-learning and the reader is referred to the canonical textbooks in \cite{ML}.
Furthermore, there have been several introductory monographs, tailored for the mathematician and the physicist \cite{MLmathphys}, which are also excellent resources.
Of course, the entire field of machine-learning has been around for decades and has been an indispensable tool in many areas of science.
The famous discovery of the Higgs boson, for instance, could not have been made, without machine-learning the patterns of particle jets.
Back in the early days, the human eye had to disentangle hundreds of trajectories in cloud chambers, the amount of data now clearly precludes this possibility.

The explosion in the last decade or so of machine-learning is due to an important advancement in hardware: the gaming industry has brought about the proliferation of graphic processing units (GPUs) in addition to the standard CPU, even to the personal computer.
Each GPU is a processor specialized in tensor transformations.  Essentially, every personal device now has become somewhat a super-computer with thousands of parallel cores.
This ready availability of the mini-super-computer has thus given a new incarnation to machine-learning whose algorithms have existed for decades but have been seriously hindered by the limitations of computing power.

Theoretical physics has been no exception in taking advantage of this explosion of technology \cite{physicslearn}.
The novelty of \cite{He:2017aed,He:2017set} is the proposal that the machine-learning paradigm can, too, be applied to fields of pure mathematics such as algebraic geometry.

\section{Rudiments of Machine-Learning}
Again, we refer the reader to \cite{Ruehle:2020jrk,MLmathphys} for marvelous introductions to ML and data science for the mathematician and the physicist.
There are canonical references in \cite{ML}, as well as a more recent and comprehensive introduction \cite{GBA}. 
In this book, we will largely focus on supervised ML because of the structure of the mathematical data and the type of problem we wish to study.
But first, let us comment on some generalities, especially some of the jargon necessary for the entrance into any subject.

\begin{definition}
A {\red point cloud} is a collection of data points, represented, typically, in $\IR^n$.
\end{definition}
The first step in data science is representation of the data, which itself is an art.
The point cloud is a standard way of collecting the features into vectors in Euclidean space (the $\IR^n$ is commonly called the ``feature space''), with flat metric \footnote{
Recently, there has been works by none other than Prof.~Yau himself in studying the ``manifold'' nature of data \cite{yauDL}.
It has been a particular honour to get him interested in the ML mathematical structures programme \cite{He:2020kzg}.
}.
For example, the matrix in \eqref{digit3} could be seen as a point in $\IR^{28^2}$ by flattening and concatenating the rows/columns.
Of course, and this is especially the case with image processing, one might wish to retain the matrix structure of the input, in which case we will be working with $\IR^{28} \otimes \IR^{28}$.
Much of ML is concerned with extracting information from the point cloud.
Typically, $n$ is very large (many features), and it is not immediately clear how to proceed; this has been called the ``curse of dimensionality''.

The data points can come with a label or not. The label could be a categorical value where some subsets of the data are assigned one of the discrete categories in a classification. It could also be a continuous value, in which case we have a map from $\IR^n \to \IR$.
When we have unlabeled data and we wish to find patterns, this goes under the general rubric of {\red unsupervised machine-learning}.
There are sophisticated methods such as clustering analysis, or autoencoder, etc.
In this book, due to the type of problems thus far discussed, we will actually only focus on {\bf supervised ML}, where we have {\it labeled data}, and we shall ``train'' and ``validate''.
The labels, as we will see, come from the results of often-intensive computation, such as topological invariants of varieties, or properties of groups, or arithmetic of elliptic curves, etc.
In this context, the representation of the data point is the ``input'' while the label is the ``output''.

\subsection{Regression \& A Single Neuron}
The steps outlined above in addressing \eqref{sampledigits} should remind the reader of first-year undergraduate statistics, it is what one does in {\bf regression}.
As a refresher, we leave a concrete example of logistic regression to Appendix \ref{ap:regression}.
In fact, supervised ML is but a massive generalization of regression in some sense.
In this spirit, we begin with
\begin{definition}\label{neuron}
A {\red neuron} or {\red perceptron} is a function $f(\sum_I w_I x_I + b)$ which is preferably but not necessarily analytic, whose argument is $x_I$, a tensor for some multi-index $i$ and whose range is typically in $[0,1] \subset \IR$.
The parameters $w_I$ are called weights and $b$, the bias or off-set.
\end{definition}
The multi-index is so that if the input is a vector, we have a single index, if it is a matrix, then $I$ is a pair of indices, etc.
We will loosely call $w_I$ a weight vector because it has a single multi-index.
The range is usually in $[0,1]$ in order to imitate the ``activation'' of an animal neuron, which ``fires'' or not according to stimuli.
Schematically, the perceptron looks like (for vector input $x_I$, and sigmoid function, for example)
\begin{equation}
\includegraphics[width=0.9\textwidth]{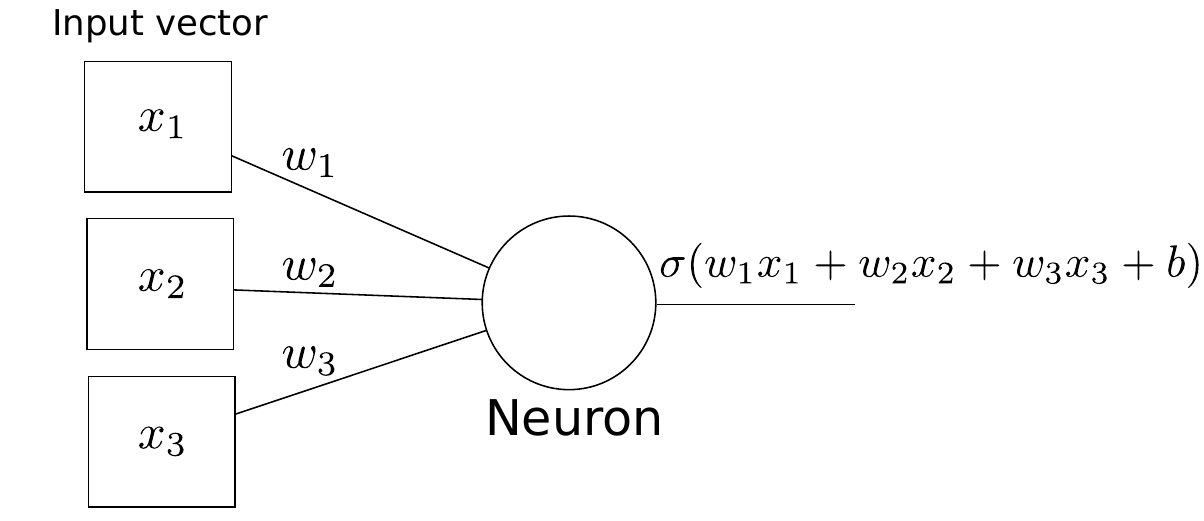}
\end{equation}
The bias $b$ is included to offset the resulting weighted sum so as to stay in the \textit{active region}. To be more explicit, consider the sigmoid
activation function. If we have a large input vector, without a bias, applying a sigmoid
activation function will tend to map the output to $1$ due to the large number
of entries in the sum, which may not be the correct response we expect from the
neuron. We could just decrease the weights, but training the neuron can
stagnate without a bias due to the vanishing gradient near $f(x)=1$ and $0$. 

Thus, standard choices of $f$ are the following
\begin{description}
\item[ Logistic Sigmoid]   $\left( 1 +  e^{-x} \right)^{-1}$

\item[ Hyperbolic tangent ] $\tanh(x) = \frac{e^{x} + e^{-x}}{e^{x} - e^{-x}}$

\item[ Softplus ] $\log \left( 1 + e^x\right)$, a ``softened'' version of {\bf ReLu} (Rectified Linear Unit):
	$\max(0,x)$

\item[ Parametric ReLu ] $R(x) = \left\{
\begin{array}{lll}
x \ ,  && x \geq 0 \\
\alpha x \ , && x < 0
\end{array}
\right.$

\item[ Softmax ] $x_i \rightarrow \frac{e^{x_i}}{\sum_i e^{x_i}}$

\item[ Maxout ]  $x_i \to \max_i x_i$

\item[ Identity ]  $x_i \to x_i$
\end{description}
Note that to all these shapes of $f$, weights and biases are added, so the last case of the ``identity'' activation really means an arbitrary affine transformation $x_i \to w_{ij} x_j + b_i$.

What is astounding that this idea of imitating the neuron in order to facilitate computation dates as far back as 1957, at the very dawn of computers \cite{CdS}:
Cadmium Sulfide photo-voltaic cells the size of a wall were linked up and stimulated in order to learn/produce pixelated images.
The nomenclature ``perceptron'', with its charming ``-tron'' ending probably gives away its 1950-60s origin.

With the neuron, the ``training'' proceeds as follows.
\begin{itemize}
\item The {\blue training data}: $D = \{ (x_I^{(j)}, d^{(j)} \}$ with input $x_I$ and {\it known} output $d^{(j)}$ where $j$ indexes over the number of data points;
\item Find an appropriate {\blue loss-function} (sometimes called the cost-function) whose minimization optimizes the the parameters $w_i$ and $b$.
We can define $\widehat{d^{(j)}} := f(\sum_I w_I x_I^{(j)} + b)$, as a ``fitted value''.
Then, typically, if $d^{(j)}$ is a continuous variable, we use the sum-squared-error (the standard deviation):
\begin{equation}\label{SD}
SD := \sum\limits_j \left(\widehat{d^{(j)}}   - d^{(j)} \right)^2 \ ,
\end{equation}
if $d^{(j)}$ is a discrete (catgorical) variable, especially in a binary classification where $d^{(j)} = 0,1$, we use the cross entropy:
\begin{equation}\label{CE}
CE := - \frac{1}{|D|} \sum\limits_j  d^{(j)} \log \widehat{d^{(j)}} + (1 - d^{(j)}) \log (1 -  \widehat{d^{(j)}} ) \ .
\end{equation}
Minimizing these with respect to the weights and biases fixes them;
\item the neuron is now ``trained'' and we can validate against {\it unseen} data.
\end{itemize}
Of course, this is precisely (non-linear) regression for model function $f$.
We remark that this, and most of ensuing discussions, is {\bf supervised ML}, in the sense that there are distinct pairs of inputs and outputs.
The machine is ``supervised'', as it were, to associate specific inputs with outputs. 
Again, the reader is referred to Appendix \ref{ap:regression} for details of a calculation.

\subsection{MLP: Forward Feeding Neural Networks}\label{s:MLP}
Next, we can generalize the neuron to
\begin{definition}
A collection of neurons is known as a {\bf layer} where the weight vector $w_I$ is promoted to a
weight matrix.
We denote the output of the $i$-th neuron in this layer as $f_i$ such that
\[
	f_i := f(\sum_I W_{iI} x_I + b_i)\,.
\]
\end{definition}

We can certainly string a sequence of layers in the next generalization by extending to several layers:
\begin{definition}
A {\bf multi-layer perceptron} (MLP) is a sequence of layers where the output of the previous layers is the input to the next layer, applying a different weight matrix and bias vector as we propagate through. 
All internal layers between the input and output layers are referred to as {\em hidden layers}. 
Denoting the output of the $i$-th neuron in the $n$-th layer as $f^n_i$, with $f^0_I = x_I$ as the input vector,
\[
	f_i^n = f(W_{ij}^n f_j^{n-1} + b_i^n)\,.
\]
\end{definition}

\begin{figure}[h!!!]
\begin{center}
$
\mbox{ Width }
\left\{
\underbrace{
\begin{array}{c}\includegraphics[width=0.6\textwidth]{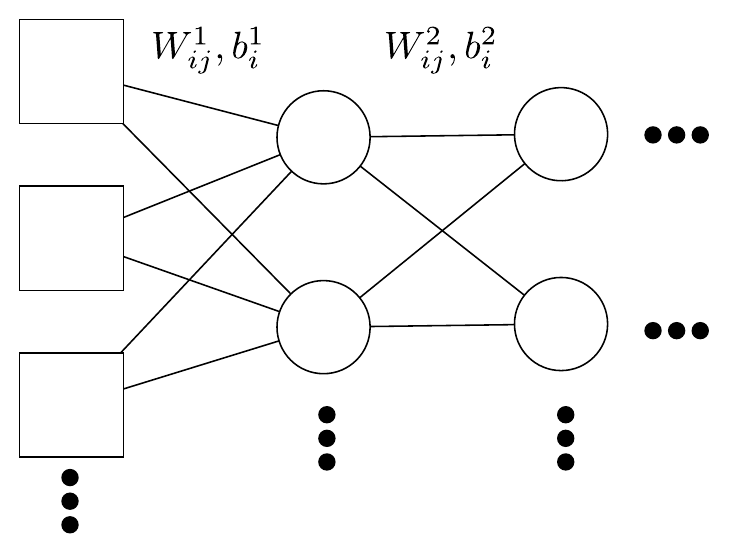}\end{array}}_{\mbox{Depth}}
\right.
$
\end{center}
\caption{
{\sf 
Schematic of a forward-feeding neural network (multi-layer perceptron).
Each node is a neuron, with appropriate activation function with its own weights $W$ and biases $b$.
}
}
\label{f:mlp}
\end{figure}

Schematically, the MLP looks like Figure \ref{f:mlp}, where the left-most layer is the {\it input layer}, the right-most, the {\it output layer} and every thing in between,{\it  hidden layers}.
The MLP is the simplest type of a {\red {\bf neural network}}, consisting of only forward-propagating layers, as one goes from left to right sequentially in the layers as indicated in Figure \ref{f:mlp}.
In general, a neural network is any finite directed graph of neurons and could consist of backward as well as forward propagation, even cycles in the graph are perfectly allowed.
What we therefore have is a {\red network} of neurons, called a neural network (NN).

If there are  many hidden layers, the neural network is usually called {\em deep} and the subsequent machine learning is referred to as {\red {\bf Deep Learning}}.
The number of neurons/nodes for a layer is called the {\bf width}.
We should point out that the words ``perceptron'' and ``MLP'' have since fallen out of fashion, today, we simply refer to them, respectively, as ``neuron''/``node'', and (multi-layer) ``forward feeding NN''.

The ``magic'' of the NN stems from the historical notion of {\it connectivism} \cite{GBA}, where computational power is expected to emerge from complexity and inter-connectivity of graphs.
To this effect we do have a powerful theorem, which basically shows that one can approximate any input-output (think of the Laurent series being able to approximate any analytic function).
There are various ways to phrase this universality, we take the following pair of statements:
\begin{theorem}[Universal Approximation Theorem]\label{thm:UAT}
We have the following approximations by feed-forward NNs:
\begin{description}
\item[Arbitrary Width] For every continuous function $f: \IR^d \to \IR^D$, every compact subset $K \subset \IR^d$, and every $\epsilon >0$,  there exists \cite{CybenkoHornik}
a continuous function $f_\epsilon : \IR^d \to \IR^D$ such that
$f_\epsilon = W_2 (\sigma ( W_1))$, where $\sigma$ is a fixed continuous function, $W_{1,2}$ affine transformations and composition appropriately defined, so that $\sup\limits_{x \in K} |f(x) - f_\epsilon(x) | < \epsilon$.
\item [Arbitrary Depth] Consider a feed-forward NN with $n$ input neurons, $m$ output neuron and an arbitrary number of hidden layers each with $n+m+2$ neurons, such that every hidden neuron has activation function $\varphi$ and every output neuron has activation function the identity
\footnote{Here $\varphi : \IR \to \IR$ is any non-affine continuous function which is continuously differentiable at least at one point and with non-zero derivative at that point.}
Then \cite{KidgerLyons}, given any vector-valued function $f$ from a compact subset $K \subset \IR^m$, and any $\epsilon > 0$, 
one can find an $F$, a NN of the above type, so that
$|F(x) - f(x)| < \epsilon$ for all $x \in K$.
\end{description}
\end{theorem}

\subsection{Convolutional Neural Networks}\label{s:CNN}
The MLP is one commonly used NN.
For image/tensor processing, another widely used one is the {\em Convolutional neural network} (CNN) which is an alternative type of network that thrives when inputs contain translational or rotational invariance, whereby making them particularly useful for image recognition.
We point out that for CNNs, it is important to retain the tensor structure of the input.
For example, an $n \times n$ matrix input such as \eqref{digit3} is {\it not} to be flattened into a data-point in $\IR^{n^2}$.  The neighbouring pixels to a pixel play an important role.

Like a fully connected, feed-forward layer, convolution layers use a set of neurons which pass a weighted sum through an activation function. 
However, neurons in convolution layers do not receive a weighted sum from all the neurons in the previous layer. Instead, a {\em kernel} (not to be confused with the kernel of matrix; this is more of a kernel in the sense of a discrete transform) restricts the contributing neurons.

To be more explicit, consider a two dimensional input (matrix). 
A {\em kernel} will be a grid sized $n \times n$ which convolves across the input matrix, taking the smaller matrix the grid is
covering as the input for a neuron in the convolution layer, as exemplified by the following which is a size $2 \times 2$ kernel
\begin{equation}	
\includegraphics[width=0.6\textwidth]{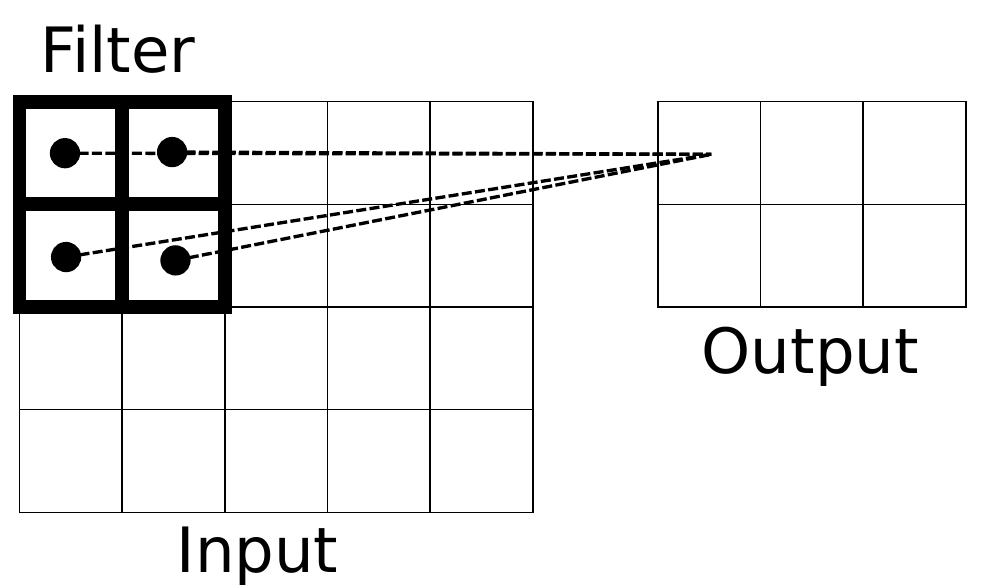}
\end{equation}
The output generated by convolving the kernel across the input matrix and
feeding the weighted sums through activations is called a {feature map}.
Importantly, the weights connecting the two layers must be the same, regardless
of where the kernel is located. Thus it is as though the kernel window is
scanning the input matrix for smaller features which are translationally
invariant.  For instance, when the kernel window moves one unit to the
right, the connected neuron in the output layer will be the centre square in the
top row.

In the running examples of the handwritten digit recognition problem, the network may learn to
associate rounded edges with a zero. What these features are in reality relies
on the weights learned during training. A single convolution layer will usually
use several feature maps to generate the input for the next layer. 
\comment{
Therefore when using convolution layers in a network, we introduce two new
hyperparameters --- the kernel size and the number of feature maps.
}

Of the multitude of NNs out there, in this book we will only make use of MLPs and CNNs, and with fairly small depth and width. Suffice it to mention in passing that there are recurrence NNs  (RNNs) which excel at series prediction, Boltzman machines which specialize in stochastic decision making,
autoencoders which find good lower dimensional representation of the input data, etc.

\subsection{Some Terminology}\label{s:MLterms}
It is expedient to summarize, at this point, some of the standard terms used in the ML community.
The vastness of the literature and the stylistic differences in phrasing from mathematics/physics texts may often intimidate the amateur \footnote{
Having myself suffered as an amateur to ML, at least let my pains be not in vain.
}; here, let us gather, at one glance, a small descriptive lexicon (not in an alphabetical order, but in an attempted narrative).

\begin{description}

\item[Architecture: ] The choice of layers and activation functions which determine a NN is called its architecture.
	The choice of this, of course, can significantly change the performance.

\item[Backward Propagation (Chain Rule): ]
In the actual optimization of the loss function, be it \eqref{SD} or \eqref{CE}, or otherwise, the standard numerical method (going back at least to Cauchy) is {\bf gradient descent} (or often called the method of {\it steepest descent}).
We recall this as follows
\begin{proposition} To find the (local) minimum $\vec{x}^*$ of a differentiable function $f(\vec{x}) : \IR^n \to \IR$, the iteration ($n = 1, 2, \ldots$, $\eta > 0$)
\[
\vec{x}_{n+1} := \vec{x}_n - \eta \nabla f(\vec{x}_n) 
\]
ensures that $f(\vec{x}_{n+1}) \leq \vec{x}_{n}$.
\end{proposition}
The step-size parameter $\eta$ is commonly called the {\bf {\blue Learning Rate}}.

In a multi-layer NN one might be overwhelmed by how many gradient descent to perform, since each node in each layer could come with its own optimization, constituting hundreds to thousands of activation functions and even more parameters.
Luckily, we are saved by the chain rule.
This gives us the flexibility to adjust the weights and biases of a neural network when training, in a consistent manner.
It is called  {\it back propagation} in the ML literature and, though a simple corollary of the chain rule, achieves parameter adjustments by optimiznig the loss function. 
It is so named as adjustments are first made to the last layer and then successive layers moving backwards through the hidden layers.

To illustrate, consider a neural network with $M$ layers and a mean squared error loss function, as in the standard deviation in \eqref{SD},
\begin{equation}
E := \frac{1}{N} \sum_{train}^N \left( \vec{ \sigma}^M-\vec{t} \right)^2 \ ,
\end{equation}
where $N$ is the number of training entries to which we sum and $\mathbf{t}$, the expected output for a given entry.
Taking derivatives, shifts in the last weight matrix become:
\begin{equation}
	\frac{\partial E}{\partial W_{ij}^M} =\frac{2}{N} \sum_{train} (\sigma_i^M - t_i) \ \sigma_i^{'M} \sigma_j^{M-1}.
\end{equation}
Likewise, working backwards, shifts in the second to last weight matrix:
\begin{equation}
	\frac{\partial E}{\partial W_{ij}^{M-1}} 
	=\frac{2}{N} \sum_{train}\sum_u\  (\sigma_u^M - t_u)\  \sigma_u^{'M} W_{ui}^{M} \sigma_i^{' M-1} \sigma_j^{M-2} .
\end{equation}
Defining
$\Delta_i^M := (\sigma_i^M - t_i) \sigma_i^{'M}$
and
$\Delta_i^m := \sum_u \Delta_u^{m+1} W_{ui}^{m+1} \sigma_i^{'m}$,
we can write by induction, for an arbitrary layer $m$,
\begin{equation}
	\frac{\partial E}{\partial W_{ij}^m} = \frac{2}{N} \sum_{train}
	\Delta_i^m \sigma_j^{m-1}  , \quad\quad
	\frac{\partial E}{\partial b_{i}^m} = \frac{2}{N} \sum_{train} \Delta_i^m.
\end{equation}

By utilizing our neural network's final output
and the expected output, we can calculate the $\Delta$s successively, starting from
the last layer and working backwards. We shift the weight values in the
direction the gradient is descending to minimise the error function. Thus shifts
are given by
\begin{equation}
	\Delta W_{i,j}^m = - \eta \frac{\partial E}{\partial W_{ij}^m}\,,\quad \quad\Delta
	b_i^m = - \eta \frac{\partial E}{\partial b_i^m} \ ,
\end{equation}
where $\eta$ is the learning rate. Care must be taken when
choosing the learning rate. A rate too small leads to slow convergence and the
possibility of becoming trapped in a local minimum. A rate too large leads to fluctuations
in errors and poor convergence as the steps taken in parameter space are too
large, effectively jumping over minima.

Note that parameter shifts are dependent on the gradient of the activation
function. For activation functions such as sigmoid or $\tanh$ this then drives the
output of a neuron to its minimal or maximal value, as parameter shifts become
increasingly small due to the vanishing gradient. This is advantageous in an
output layer where we may want to use binary classification. However, if neurons
in hidden layers are driven to their min/max too early in training, it can
effectively make them useless as their weights will not shift with any further
training. This is known as the {\em flat spot problem} and is why the ReLU activation
function has become increasingly popular.
Moreover, there is a version of the universality theorem \cite{Hanin} which essentially that ReLUs are already good enough:
\begin{theorem}[ReLU Universal Approximation]\label{thm:UATReLU}
For any Lebesgue-integral function $f: \IR^n \to \IR$ and any $\epsilon > 0$, there exists a fully connected ReLU NN $F$ with width of all layers less than $n+4$ such that $\int_{\IR^n} |f(x) - F(x)| dx < \epsilon$.
\end{theorem}

\item[Over-fitting: ]
In any statistical model, it is important that one does not over-fit, i.e., to allow for so many parameters in the model so as to render any data-set to be able to be fitted to any desired model; this must be avoided as the model might do extremely well to a training set and be completely useless for prediction.
In regression, as shown in the example in Appendix \ref{ap:regression}, a chosen function may not have too many parameters, but a NN could have tens of thousands of parameters, so it is important to check, especially in light of the Universal Approximation Theorems, that the NN could actually predict something meaningful.

For the neural network, {\em over-fitting} occurs during training when accuracy against the training data-set continues to grow but accuracy against unseen data stops improving.
The network is not learning general features of the data any more and  the complexity of the net architecture has more computing potential than
required. The opposite problem is {\em under-fitting}, using too small a network which is incapable of learning data to high accuracy.

An obvious solution to over-fitting is {\it early stopping}, cutting the training short
once accuracy against unseen data ceases to improve. However, we also wish to
delay over-fitting such that this accuracy is as large as possible after
stopping.

A common solution is called {\em Dropout}, which is a
technique where neurons in a given layer have a probability of being switched off
during one training round. This forces neurons to learn more general features
about the data-set and can decrease over-fitting \cite{Dropout}.
In practice, one adds a {\it dropout layer} to an NN which removes, at random, nodes from the previous layer.

\item[Cross-Validation: ]
Another important way to avoid over-fitting, as we had already discussed, is that the ML algorithm be trained and validated on complementary data-sets: a labeled data-set $\cD$ is split disjointly into $\cT \sqcup \cV$, the training and validation sets.
In order to obtain statistical error, one often performs {\bf $n$-fold} cross-validation, where $\cD$ is split into $n$ disjoint subsets. Training is then performed on $n-1$ of these chosen at random, and validated on the remaining 1.
This is repeated $n$ times to gather standard deviation on the performance.

\item[Goodness of Fit: ]
The performance of a validation can be measured by standard statistical methods.
Roughly, supervised ML go under 2 headings: {\bf regressors}, which treat continuous output/label, and {\bf classifiers}, which treat discrete/categorical output/label.
For continuous output, the {\it R-squared} is used. This is introduced for a regression in Appendix \ref{ap:regression} and we will see it again in \S\ref{s:goodness}.
For discrete output, one first uses {\it na\"{\i}ve precision}, which is simply the fraction of cases where the predicted category agrees with the actual category.
In order to have an idea of false negatives and false positives, one further creates a {\it confusion matrix}, from which one often computes the {\it F1-score} and {\it Matthews' coefficient} as a measure of how close the matrix is to being diagonal. All these quantities deserve a full treatment and we will define them properly in \S\ref{s:goodness}.
In all cases, these measures of goodness of fit, be they R-squared, F1-score, precision, Matthews' coefficient, etc., are customarily normalized so that they are valued in $[0,1] \subset \IR$, so that 1 means a perfect prediction/validation and 0 means complete uselessness.


\item[Hyper-parameters: ] The numerical quantities associated with the choice of architecture of NN, such as the number of nodes in a layer, the number of layers, the learning rate, the dropout rates, kernel sizes of CNNs, etc., are called the hyper-parameters. These are chosen before training and are distinguished from the parameters, such as the weights and biases, which are to be optimized during training. Tuning hyper-parameters also can significantly alter performance.

Several methods exist to optimize these hyper-parameters. For the
case of a few hyper-parameters, one could search by hand, varying parameters
explicitly and training repeatedly until an optimal accuracy is achieved. 
For the learning rate, for instance, a commonly used method is the so-called ADAM (adaptive moment estimation) optimizer \cite{adam}.

A grid search could also be used, where each parameter is scanned through a range
of values. However, for a large number of hyper-parameters permitting a large
number of values this quickly becomes an extremely time consuming task. Random
searches can often speed up this process, where parameter values are drawn from
a random distribution across a sensible range. 
Finally, random search coupled with a {\it genetic algorithm} is very much favoured,
which effectively begins as a random search but then makes an
informed decision of how to \textit{mutate} parameters to increase accuracy.


\item[Regularization : ]
Sometimes the parameters (weights and biases) can become too large in the loss function during optimization.
In order to control this, one defines an appropriate norm $| W |$ for a weight tensor $W$ (e.g., sum of squares over all entries), and adds a Lagrange multiplier term $\lambda | W | $ to the loss function.

\item[Data Enhancement : ]
When there is a paucity of data, one can often increase the size by {\it enhancement}, which means that we add equivalent representations of the same data point multiple times.
An example is where the input is a matrix and any row/column permutation thereof represents the same object.
We will encounter many of such cases later.

Data enhancement is particularly relevant for {\it imbalanced data} where the number of cases in a classification problem differs significantly across categories.
For instance, in a binary classification, suppose we have $10^4$ cases of ``1'' and $10^3$ cases of ``0'', an ML can obtain great accuracy in predicting all cases to be ``1'', which is very misleading.
In this case, we throw in equivalent representations for the input for cases corresponding to ``0'' so that each category has around $10^4$ cases. 
Alternatively, we can {\it under-sample} the ``1'' cases by randomly selecting around $10^3$ cases for both.
Such {\it balanced} data-sets can then be fed to a supervised ML algorithm.

\item[Training Rounds: ]
The training data $\cT$ is typically passed over to an ML multiple times; each such a time is called a round or {\bf epoch}.
Typically $|\cT|$ is large, so it is divided into {\bf batches}  (sometimes also called mini-batches) so that with an epoch, we have a number of iterations of passing $\cT$.
In particular, therefore, $|\cT| = (\mbox{ Batch size }) \times (\mbox{\# Iterations})$.

During training, one evaluates the loss-function on the validation set as well as the training set (or at least a sample of the validation set) to monitor progress as we go through the batches and the epochs; this loss is then plotted against training rounds/batches. This resulting pair of curves (loss for training and loss for validation) is called the {\bf learning curve} (sometimes also called the training curve), which helps with the visualization of how well the ML algorithm is behaving.
It is hoped that they are decreasing functions (up to stochastic fluctuations).
If the training curve is decreasing steadily for the training set but struggles to decrease for the validation set, this is often an indication of possible over-fitting.

There are also other possible measures of the learning curve, such as plotting the various measures of goodness of fit against increasing size of $\cT$ in a cross-validation, as we will later see. In this case, it is hoped that the learning curve is an increasing function (up to fluctuations and standard deviation) from 0 to 1.

\end{description}

\subsection{Other Common Supervised ML Algorithms}\label{s:nonNN}

Having familiarized ourselves with some basic terms, it is helpful to that the world of ML goes far beyond NNs, which we exemplified with MLPs and CNNs above.
In this section, we will introduce a few more which we will use later.

\subsubsection{Support Vector Machines}
One of the most widely used methods to address high dimensional labeled data is support vector machines (SVMs).
In contrast to NNs, SVMs take a purely geometric approach and has the advantage of {\bf interpretability}: one can tell the equation (hypersurface) which separates the data into regions.
SVM can act as both classifiers and regressors which we shall introduce sequentially.

\paragraph{SVM Classifiers: }
\begin{figure}[t!]
	\centering
	\includegraphics[width=0.46\textwidth]{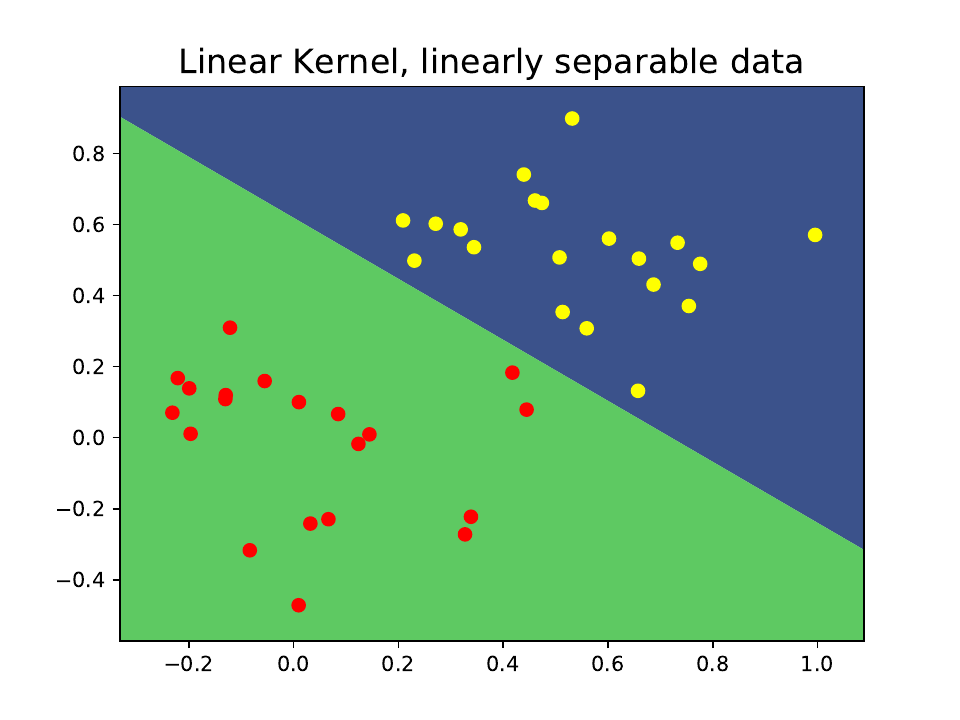}
	\includegraphics[width=0.46\textwidth]{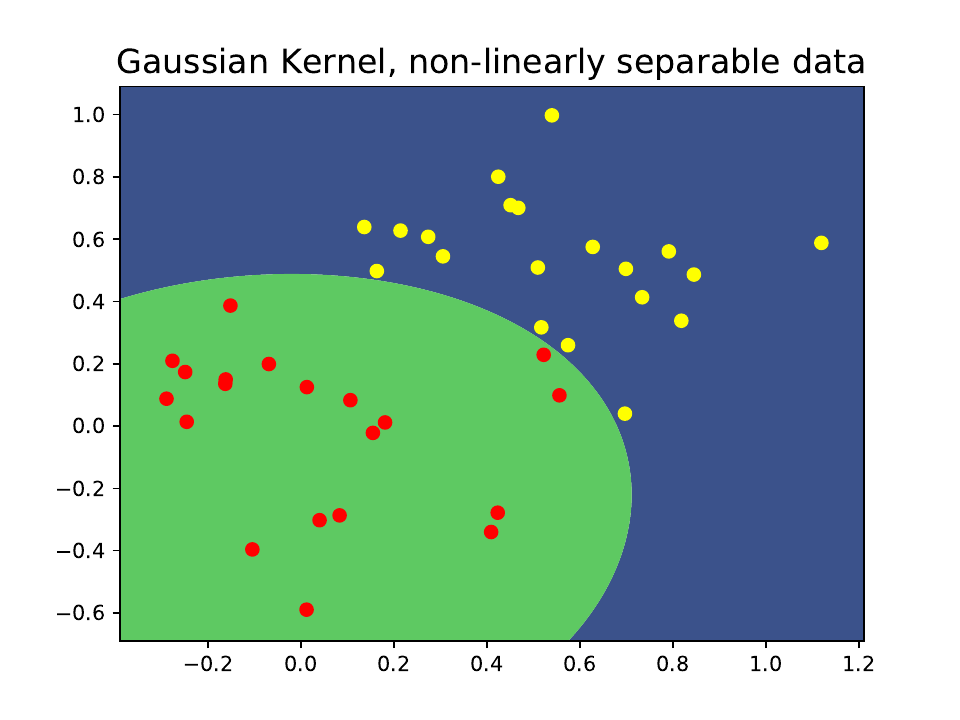}
	\caption{{\sf Example SVM separation boundary using different kernels.}}
	\label{fig:ex_svm}
\end{figure}
While a neural network classifier essentially fits a large number of parameters (weights and
biases) to obtain a desired function $f(\mathbf{v_{in}}) \in [0,1]$, a SVM tries to establish an optimal hyperplane separating clusters of points in the
{\em feature space}, the $n$-dimensional Euclidean space to which the
$n$-dimensional input vector belongs.
Points lying on one side of the plane are
identified with one class, and vice versa for the other.  

Thus, a vanilla SVM is only capable of acting as a binary classifier for linearly separable data. This is somewhat restrictive, but the approach can be generalised to
non-linearly separable data via the so called \textit{kernel trick} and likewise can be
extended to deal with multiple classes \cite{svm}.
We illustrate this in Figure \ref{fig:ex_svm}.
We wish to separate points $\mathbf{x_i}$ with a hyperplane based on a classification
of true/false, which we represent with the labelling $y_i=\pm1$; this is so-called binary feature.
 
First define a hyperplane
\begin{equation}
	\left\{ x \in \mathbb{R}^n | f(\mathbf{x}) = \mathbf{w \cdot x} + b = 0
	\right\},
	\label{eq:hplane_def}
\end{equation}
where $\mathbf{w}$ is the normal vector to the hyperplane.
Then we have that
\begin{definition}
{\bf Support vectors} are the points in the feature space lying closest to
the hyperplane on either side which we denote as $\mathbf{x^{\pm}_i}$.
\\
The {\bf margin} is the distance between these two vectors projected along
the normal $\mathbf{w}$, \textit{i.e.},
\begin{equation}
\mbox{Margin} := \mathbf{w \cdot (x^{+}_i-x^{-}_i})/|\mathbf{w}| \ .
\end{equation}
\end{definition}
There is typically not a unique hyper-plane we could choose to separate labeled
points in the feature space, but the most optimal hyper-plane is one which
maximizes the margin. This is because it is more desirable to have points
lie as far from the separating plane as possible, as points close to the
boundary could be easily mis-classified. Note that condition defining a
hyper-plane (\ref{eq:hplane_def}) is not unique as a re-scaling $\alpha(\mathbf{w
\cdot x}+b)=0$ describes the same hyper-plane. Thus we can re-scale the normal
vector such that $f(\mathbf{x^{\pm}_i})=\pm 1$ and the margin reduces to
\begin{align}
	\text{Margin} = \frac{2}{|\mathbf{w}|}.
\end{align}
Moreover, with such a re-scaling, the SVM acts as a classifier on an input
$\mathbf{x}$ through the function $\text{sgn}(f(\mathbf{x}))=\pm 1$.
In particular, for each point $(x_i, y_i)$,
$y_i = 1$ if $\mathbf{w \cdot x_i} + b \geq 1$ and 
$y_i = -1$ if $\mathbf{w \cdot x_i} + b \leq -1$, so that the product 
$y_i \times (\mathbf{w \cdot x_i} + b)$ is always $\geq 1$ for each $i$.
Maximizing the margin thus corresponds to minimizing $|\mathbf{w}|$, with the
constraint that each point is correctly classified. This wholly defines the
problem which can be stated as
\begin{align}\label{SVMhyper}
	&\text{Min    } 
	\frac{1}{2} |\mathbf{w}|^2  
	\text{    subject to    }\  
	y_i \times (\mathbf{w} \cdot \mathbf{x_i}+b) \geq 1.
\end{align}
This is a quadratic programming problem with well known algorithms to 
solve it. 
Reformulating this problem with Lagrange multipliers:
\begin{align}
\nonumber
	L &= \frac{1}{2} |\mathbf{w}|^2 - \sum_i 
	\alpha_i (y_i(\mathbf{w} \cdot \mathbf{x_i} + b) - 1)  \Rightarrow \\
	\frac{\partial L}{\partial \mathbf{w}} &= \mathbf{w} -
	\sum_i \alpha_i y_i \mathbf{x_i} = 0 , \quad
	\frac{\partial L}{\partial b} =
	-\sum_i \alpha_i y_i = 0, 
\end{align}
leads to the {\em dual problem} upon substitution:
\begin{question} Solve the quadratic programming:
\begin{align}
\nn
	&\text{Min} \quad
	\frac{1}{2} \sum_{i,j} 
	\alpha_i \alpha_j y_i y_j \mathbf{x_i} \cdot \mathbf{x_j}
	 - \sum_j \alpha_j \ , \\
	 &
	\quad\text{subject to} \quad 
	\alpha_j \geq 0 \text{,} \quad \sum_j \alpha_j y_j =0 \ .
\end{align}
With our classifying function now being $\text{sgn}(f(\mathbf{x})) = \text{sgn}
\left( \sum_i \left( \alpha_i y_i \mathbf{x_i \cdot x} \right) + b \right)$.
\end{question}
The {\sf Python} package {\sf Cvxopt} implements a
quadratic computing algorithm to solve such problems.

The dual approach is much more illuminating as it turns out the only $\alpha_i$
which are non-zero correspond to the support vectors \cite{svm} (hence the name
support vector machine). This makes SVMs rather efficient as unlike a neural
network which requires a vast amount of parameters to be tuned, a SVM is fully
specified by its support vectors and is ignorant of the rest of the data.
Moreover the quadratic programming optimization implemented via {\sf Cvxopt} ensures
the minimum found is a global one.

The dual approach also enables us to generalize to non-linearly separable data
rather trivially. 
In theory, this is achieved by mapping points in the feature
space into a higher dimensional feature space where the points are
linearly-separable, finding the optimal hyperplane and then mapping back into
the original feature space. 
However, in the dual approach, only the dot product
between vectors in the feature space is used. 

Therefore, in practice we can avoid the
mapping procedure as we only need to generalize the dot product in the higher
dimensional space, known as a {\bf kernel} (it is unfortunate that the kernel of an SVM, that of a CNN, and that of a matrix, all mean completely different things).
Hence, by replacing $\mathbf{x_i \cdot x}$ with $\text{Ker}(\mathbf{x_i,x})$ we can deal with non-linearly
separable data at almost no extra computational cost. This is known as the
\textit{kernel trick} which just replaces the usual dot product to common kernels such as
\begin{align} \label{SVMpar}
\nonumber
	\text {Gaussian:    }  \text{Ker}(\mathbf{x_i,x}) &= 	\exp \left(\frac{-|\mathbf{x_i-x}|^2}{2 \sigma} \right) , \\
	\text {Polynomial:    }  \text{Ker}(\mathbf{x_i,x}) &= \left(1+\mathbf{x_i \cdot x}\right)^n \ .
\end{align}
Indeed, the choice of kernel is another hyper-parameter.

We remark that SVMs reduce to our familiar linear regressor by finding a function $f(\mathbf{x}) =
\mathbf{w \cdot x} + b$ to fit to the data. Analogous to the above discussion,
one can frame this as an optimization problem by choosing the flattest line
which fits the data within an allowed residue $\epsilon$. Likewise one can make use of
Lagrange multipliers and the kernel trick to act as a non linear regressor as well.

Note in the above discussion we have avoided the concept of {slack}. In
order to avoid over-fitting to the training data, one can allow a few points in
the training data to be mis-classified in order to not constrain the hyper-surface
too much, allowing for better generalization to unseen data. In practice this
becomes quantified by replacing the condition $\alpha_i \geq 0$ with 
\begin{equation}\label{SVMslack}
0 \leq \alpha_i \leq C \ , 
\end{equation}
where $C$ is the {cost} variable \cite{svm}. 

\comment{
With regard to parameters, like a neural network, a SVM has a few hyper-parameters which need to be set
by hand outside of training, namely the cost variable and kernel parameters,
such as the standard deviation of the Gaussian kernel $\sigma$ or the degree of
the polynomial $n$ for the polynomial kernel.
}

\paragraph{SVM Regressors: }
Having discussed SVM as a classifier, its role as a regressor is similar.
The optimisation problem for a linear SVM regressor follows from finding the flattest
possible function $f(\mathbf{x}) =\mathbf{w \cdot x} + b$ which fits the data within a
residue $\epsilon$. As $|\nabla f |^2 = |\mathbf{w}|^2$, this flatness condition
reduces to the problem:
\begin{align}
	&\text{Min    } \label{SVMeps}
	\frac{1}{2} |\mathbf{w}|^2  
	\text{    subject to    } 
	-\epsilon \leq y_i - (\mathbf{w \cdot x_i} + b) \leq \epsilon \ .
\end{align}
Again, introducing Lagrange multipliers
\begin{align}
 \nn
	L &= \frac{1}{2} |\mathbf{w}|^2 - \sum_i 
	\alpha_i (y_i-(\mathbf{w} \cdot \mathbf{x_i} + b) + \epsilon)+\sum_i \alpha_i^\star (y_i-(\mathbf{w} \cdot \mathbf{x_i} + b) - \epsilon) 
	 \Rightarrow  \\
	\frac{\partial L}{\partial \mathbf{w}} &= \mathbf{w} -
	\sum_i (\alpha_i - \alpha_i^*) \mathbf{x_i} = 0 , \quad
	\frac{\partial L}{\partial b} =
	\sum_i (\alpha_i-\alpha_i^*) = 0 \ ,
\end{align}
leads to the dual problem:
\begin{question} Solve the optimization problem
\begin{align}
\nn
\text{ Min    } 
	&\frac{1}{2} \sum_{i,j} 
	(\alpha_i-\alpha_i^*) (\alpha_j-\alpha_j^*) y_i y_j \ \mathbf{x_i} \cdot
\mathbf{x_j}  +\epsilon \sum_i \left(\alpha_i+\alpha_i^* \right) + \sum_i  y_i(\alpha_i^*-\alpha_i) 
\\
& \mbox{subject to }  \quad \alpha_i,\alpha_i^* \geq 0\ , \quad  \sum_i (\alpha_i-\alpha_i^*) =0.
\end{align}
\end{question}
Thus, as with the classifier, this optimization problem can be
implemented with {\sf Cvxopt}. As the dual problem again only contains a dot product
between two entries in the feature space, we can use the kernel trick to
generalize this approach to fit non-linear functions. 

\subsubsection{Decision Trees}
Another widely used classifier is the decision tree.
Suppose we have a target variable $y$ and a set of input variables $x_i$. Typically, $y$ is \emph{discrete}.
However, we can treat a continuous variable as discrete by splitting it into appropriate intervals and taking the average of the interval to be its discretized value. The input, on the other hand, can be arbitrary.

A tree is then built from the input variables as a sequence of if/then/else statements.
Starting with $x_1$, we create a node and then travel down a ``branch'' depending on the value of $x_1$: if $x_1 \in [a_1, b_1]$ then we proceed to the first branch, 
if $x_1 \in [b_1, c_1]$, we proceed to a different branch, etc. At the end of each branch we create new node, say for $x_2$, from which new branches are created. As before, we then travel down one of these new branches depending on the value of $x_2$. Repeating this for all of the input variables gives a set of nodes partially connected by branches, giving a tree structure. The outermost nodes (the leaves) correspond to different predictions for the value of the output variable $y$. Once the tree structure is established, we can present it with a new input $x_i$ and then follow the tree structure to find the predicted output (the leaf) that it leads to. 

As in regression, the optimal parameters $(a_j, b_j, c_j \ldots)$ are determined by optimizing some goodness-of-fit score, for example the sum of squares of the error between the actual and predicted target variable. To prevent over-fitting, one often sets a maximum depth for the tree structure or a maximum number of leaves.

In a way, a decision tree can be understood as a highly non-analytic analogue of regression: whereas regression ultimately fits the target into some differentiable function $y = f(x_i)$ by minimizing sum squared errors, a decision tree writes down $y$ from $x_i$ via a sequence of discrete choices. Thus, when the output data $y$ is highly fluctuating with respect to $x_i$, regression is not so useful since an analytic function $f$ is difficult to find; slotting into a sequence of decisions is much more appropriate.

In order to improve the performance of predictions via decision trees, one can set up an ensemble (or forest) of trees, each with different decision criteria. The overall score of the prediction can be taken to be the sum over all trees.
Now, optimizing all parameters in all trees at once could become computationally intractable. Instead, one can take an additive strategy so that the predicted target $y$ is obtained by adding one tree at a time:
$y^{(j)} = y^{(j-1)} + g_j(x_i)$ where $g_j$ heuristically represents the function which captures the information about the tree at stage $j$.
The ensemble of trees is thus ``boosted'' iteratively and the tree to add at each stage is simply the one which optimizes the overall fitness score.

\subsubsection{Na\"{\i}ve Bayes Classifier}
Essentially, this classifier is a conditional probability problem.
Na\"{\i}ve Bayes is particularly simple but useful for categorical classification, and we now illustrate with a binary case.
Suppose our training set is of the form $\cT = \{t_i \to v_i\}$ where each $t_i = t_i^{(a)}$ is a list indexed by $a$ and $v_i \in \{0,1\}$.
First, we recall Bayes' theorem:
\begin{equation}
P(v|T) =\frac{P(T|v) P(v)}{P(T)} \ ,
\end{equation}
that the probability $P(v|T)$ of $v_i$ occurring contingent on some condition $T$, can be expressed in terms of the reverse $P(T|v|)$, and the (marginal) probabilities $P(v)$ and $P(T)$.

Now, $v$ is binary and suppose we are given a new input $T = T^{(a)}$ from the validation set, we wish to compare 
$P(0 | T^{(a)})$ and $P(1 | T^{(a)})$, whichever is larger will have the output being 0 or 1 correspondingly.
Since the denominators are the same, being some unknown quantity $P(T)$, we luckily do not need to consider it, and simply compare $P(T|0) P(0)$ versus $P(T|1) P(1)$.

Now comes the ``na\"{\i}ve'' part, which is remarkably powerful.
We assume that all entries of $t_i$ are independent, so that
\begin{equation}
P(T|v) = \prod\limits_a P(T^{(a)} | v) \ .
\end{equation}
Therefore, we only need to count in the training set:
$P(0)$ is the total percentage of cases with $0$ output, and likewise for $P(1)$;
$P(T|0) = \prod\limits_a P(T^{(a)} | 0)$ is the product over all percentages of components $T^{(a)}$ whose output is 0, and likewise for $P(T|1)$.

The only subtlety occurs is when a component $T^{(a)}$ has not appeared in the training set, i.e., if $P(T^{(a)} | v) = 0$ for some $T^{(a)}$ and $v$.
This is resolved by so-called {\it Laplace smoothing}: we add 1 to numerator to any counts $c$ (regardless of whether it is 0 or not), and - to balance it so that the percentage never exceeds 1 - add the total number of possible values $N_t$ of $t_i^{(a)}$ to any numerator $n$.
Thus, when we compute any of the percentages above, we always calculate
$(c+1) / (n + N_t)$, instead of $c/n$.
These smoothed percentages are what go into $P(T|0) P(0)$ and $P(T|1) P(1)$ before comparison.


\subsubsection{Nearest Neighbours}
A very efficient classifier that is quick to implement is the {\bf $k$-nearest-neighbour} algorithm.
Take a hyper-parameter $k$ and consider a random sample of $k$ labeled data-points from the training set, and suppose these are of the form $\{v_i \to f_i\}$ where $v_i \in \IR^n$ and $f_i$ is some output/label.
Fix an appropriate distance $d$.
Suppose the input is a vector of reals, then $d$ could simply be the Euclidean distance in $\IR^n$.
If the input were a binary vector, then the Hamming distance (number of positions where the 2 vectors differ) is a good choice.
Given a new vector $w$ from the validation set, we simply compute $d$ between $w$ and all $k$ vectors $v_i$ (neighbours) and the label/output for $w$ is that of the nearest by distance.

~\\

There are myriad more classifiers and regressors which the readers are encouraged to explore.
Hopefully, in this subsection and the previous, we have familiarize them them some of the most commonly used supervised-ML methods, NN and non-NN.

\subsection{Goodness of Fit}\label{s:goodness}
In this final section to introducing ML, as promised above, we define some standard measures of accuracies in a prediction, generally called {\blue goodness of fit}.
Common measures of the goodness of fit include (q.v.~\cite{encML})
\begin{definition}\label{measures}
Let $y_{i=1, \ldots, N}$ be the actual output for input $x_i$ in the validation set, and let $y^{pred}_{i}$ be the values predicted by the machine-learning model on $x_i$, then
\begin{description}
\item[(naive) precision] is the percentage agreement of $y_i$ with $y^{pred}_i$
	\[ 
	p := \frac{1}{N}|\{y_i =  y^{pred}_i\}| \ \in [0,1]; 
	\]
\item[root mean square error (RMS)]  $\left({ \frac{1}{N} \sum\limits_{i=1}^N (y^{pred}_i-y_i)^2 })\right)^{1/2}$ \ ;
\item[coefficient of determination] $R^2  := 1 - \frac{\sum\limits_{i=1}^N (y_i-y^{pred}_i)^2}{\sum\limits_{i=1}^N (y_i-\bar{y})^2} \in [0,1]$ \\
	where $\bar{y}$ is the mean over $y_i$ \ ;
\item[cosine distance] considering  $y_i$ and $y^{pred}_i$ as vectors $\overrightarrow{y}$ and $\overrightarrow{y^{pred}}$ respectively, compute the cosine of the angle between them so that 1 is complete agreement, $-1$, worst fit and 0, random correlation:
	\[
	d_C := \frac{\overrightarrow{y} \cdot \overrightarrow{y^{pred}}}{|\overrightarrow{y}| |\overrightarrow{y^{pred}}|} \in [-1,1] \ ;
	\]
\end{description}
\end{definition}

In the cases of categorical data where $y_i$ belong to finite, discrete categories, such as the binary case at hand. there are a few more measures.
First, we define
\begin{definition}
Let $\{x_i \to y_i\}$ be categorical data, where $y_i \in \{1,2,3,\ldots,k\}$ take values in $k$ categories.
Then, the $k \times k$ {\bf confusion matrix} has 1 added to its $(a,b)$-th entry, whenever there is an actual value of $y = a$ but is predicted to $y^{pred}=b$.
\end{definition} 
Ideally, of course, we want the confusion matrix to be diagonal.
In the case of binary $y_i$ as in \eqref{wp4data}, the $2 \times 2$ confusion matrix takes the rather familiar form and nomenclature
\begin{equation}\label{CM2x2}
\mbox{
\begin{tabular}{cc|c|c|}
	\cline{3-4} 
	&  & \multicolumn{2}{c|}{{Actual}} \\ \cline{3-4}
	& &  True  (1) & False (0)\\ \hline
	\multicolumn{1}{|c|}{{Predicted}} & True (1) & True Positive
	($tp$) & False Positive ($fp$) \\  \cline{2-4}
	\multicolumn{1}{|c|}{{Classification}} & False (0) & False Negative
	($fn$)& True Negative ($tn$)\\ \hline
\end{tabular}
}
\end{equation}

As mentioned above, in categorical classification, we have to be careful about {\bf imbalanced} data.
For instance, consider the case where only 0.1\% of the data is classified as true. In minimizing its loss function on training, the machine-learning algorithm could naively train a model which just predicts false for {\it any} input. Such a model would still achieve a 99.9\% accuracy as defined by $p$ in Definition \ref{measures}, but it is useless in finding the special few cases in which we are interested. 
Thus, naive $p$ is meaningless and we need more discriminatory measures.
These measure how diagonal the confusion matrix is.

\begin{definition}\label{scores}
For binary data, two commonly used measures of goodness of fit, in furtherance to naive accuracy are
\begin{description}
\item[F-Score: ] $F_1:= \frac{2}{ \frac{1}{TPR} + \frac{1}{Precision}}$ where \cite{F1}, using the confusion matrix \eqref{CM2x2},
	\[
	\begin{array}{cc}
	\text{TPR} := \frac{tp}{tp+fn} \ , \quad & \text{FPR} := \frac{fp}{fp+tn} \,,\\
	\text{Accuracy } p := \frac{tp+tn}{tp+tn+fp+fn} \ ,\quad &\text{Precision} := \frac{tp}{tp+fp}\,.
	\end{array}
	\]
	where TPR (FPR) stands for true (false) positive rate.
\item[Matthews' Correlation Coefficient: ] Using the confusion matrix  \eqref{CM2x2}, the Matthew's coefficient is the square root of the normalized chi-squared:
	\[
	\phi := \sqrt{\frac{\chi^2}{N}} = \frac{tp \cdot tn - fp \cdot fn}{\sqrt{(tp+fp)(tp+fn)(tn+fp)(tn+fn)}} \ .
	\]
	The definition is also generalizable to $k \times k$ confusion matrices \cite{phi}.
\end{description}
\end{definition}
We must guarantee that both $F_1$ and $\phi$ are close to 1, in addition to $p$ being close to 1, in order to conclude that the machine-learning model has predicted to satisfaction.

In addition to these measures of accuracy, we have notion of a confidence interval for the fit.
\begin{definition}
The Wilson confidence interval is $[\omega_, \omega_+]$ where
\[
\omega_\pm:=	\frac{p+\frac{z^2}{2n}}{1+\frac{z^2}{n}}
	\pm \frac{z}{1+ \frac{z^2}{n}} \left({
	\frac{p(1-p)}{n} + \frac{z^2}{4 n^2} }\right)^{1/2}\, .
\]
Here $n$ is the sample size, $p$ is the accuracy or probability of correct prediction within the sample, $z = \sqrt{2} Erf^{-1}(2P-1)$ is the probit for a normal distribution.
\end{definition}
This is a confidence interval in the sense that with probability $P$, the Wilson interval contains the mean of the accuracy from the population.

Finally, let us clearly define learning/training curves.
To avoid confusion, let us, in this book, call the plot of the loss-function during training as the training curve and the plot of appropriate goodness of fit as we increase training size, a learning curve.
In other words, we have the following.
Let $\{x_i \to y_i\}_{i=1,\ldots,N}$ be $N$ data-points.
We choose cross-validation by taking a percentage $\gamma N$ of the data randomly for a chosen $\gamma \in (0,1]$ as training data $\cT$, the complement $(1-\gamma)N$ data-points will be the validation data $\cV$.
Then, we have
\begin{definition}\label{def:tc}
For a fixed $\gamma$ and hence $\cT$, evaluate the loss function evaluated on $\cT$ and $\cV$, against increasing batch size. This produces a pair of curves called the {\bf training curve};
\end{definition}
as well as 
\begin{definition}\label{def:lc}
Compute a chosen goodness of fit from the above, it is a function $L(\gamma)$ of $\gamma$ as we increasing training size.
The {\bf learning curve} is the plot of $L(\gamma)$ against $\gamma$.
\end{definition}
In practice, $\gamma$ will be chosen at discrete intervals, for instance in increments of 10\% until the full data-set is attained for training.
Furthermore, we repeat each random sample $\gamma N$ a number of times, so the learning curve has error bars associated with each of its points.


\section{Machine-Learning Algebraic Geometry}
Armed with an arsenal of technique, we return to the theme of this book, viz., Calabi-Yau data.
We have seen in \S\ref{s:prob} a list of typical problems in computational algebraic geometry, inspired by physical need, all of which can {\it in principle} be addressed by brute-force methods ranging from Gr\"obner bases to polytope triangulations, all of which are exponential in complexity.

It is therefore natural to pose the adventurous query
\begin{question}
Can AI / ML help with algebraic geometry ?
\end{question}
Indeed, can machines help with problems in pure mathematics and theoretical physics, {\it without} explicit knowledge of the actual methods involved.
We know a plethora of cases which {\em has} been computed by traditional methods and wish to extrapolate to where computing power hinders us, by essentially bypassing - in a mysterious way, as we shall see - the frontal attacks offered by the likes of Gr\"obner bases.
This is precisely in analogy of the handwritten digit problem: to recognize a new, esoteric written digit is hard but furnished with tens of thousands of samples already classified by experience, to predict this new digit to good accuracy does not seem impossible. 
After all, there is inherent pattern, however big the variance, in written digits (or, for that matter, any manuscript).

There are inherent patterns in problems in mathematics, should some of these problems, too, not be amenable to this philosophy?

Of course, this philosophy is not new. It is simply the process of forming a conjecture.
There is a long tradition of experimental mathematics from which countless conjectures have been posited, many leading to the most profound consequences.
Consider the greatest mind of the 18th century, that of K.~F.~Gau\ss\, which can be considered as the best neural network of the time.
By looking at the distribution of primes, supposedly in his teens, Gau\ss\ was able to predict that the prime-counting function $\pi(x) \sim x / \log(x)$, a fact which had to wait over a century to be proven by de la Vall\'ee-Poussin and Hadamard, using complex analysis, a method entirely unknown to the 18th century.

The minds of the caliber of Gau\ss\ are rare, but neural networks of increasing sophistication now abound.
Could the combined efforts of AI at least stochastically achieve some correct predictions?
In this section, we will see that many, if not most of the problems discussed in this book, could indeed be machine-learned to great accuracy.

\subsection{Warm-up: Hypersurface in Weighted $\IP^4$}
Let us warm up with the simplest of the input configurations, viz., that of hypersurfaces in $W\IP^4$, discussed in \S\ref{s:WP4}, which is specified by a single 5-vector of co-prime positive integers.
To make things even simpler, suppose we wish for a {\em binary query} of whether $h^{2,1}$ is larger than a certain value, say 50.
We know that $h^{2,1} \in [1,491]$ with the histogram shown in Figure \ref{f:wp4dis}; 50 is a reasonable division point.
In terms of geometry we are looking for manifolds which have a relatively large/small number of complex structure deformations, a question of physical interest as well.

Our data, therefore, is of the form $\{ x_i \to y_i \}$:
\begin{equation}\label{wp4data}
\{1,1,1,1,1\} \to 1 \ , \ \
\{1,1,1,1,2\} \to 1 \ , \ \
\ldots \ ,
\{2,2,3,3,5\} \to 0 \ ,
\ldots
\end{equation}
totalling 7555.
The first entry, is the quintic $Q$ in regular $\IP^4$ and its $h^{2,1} = 101$ gives a positive hit (denoted as 1) on $h^{2,1} > 50$.
Further down the list, the degree 15 hypersurface in $W\IP^4[2,2,3,3,5]$ has $h^{2,1} = 43$, giving us the negative hit (denoted as 0).
As mentioned in  \S\ref{s:WP4}, the $h^{2,1}$ values were calculated meticulously by \cite{wp4,Candelas:1994bu} using Landau-Ginzberg methods (even on {\sf SageMath} (linked to {\sf Macaulay2} or {\sf Singular}), the computation using sequence-chasing is not straight-forward) and the results, though known, quite escapes the human eye in any discernible pattern.
Again, we note that while the Euler number $\chi$ can be expressed in terms of the weights, the individual Hodge/Betti numbers cannot be. This is a generic situation in algebraic geometry and is the story with index-type theorems: individual terms in (co-)homology are difficult to compute, but the index of an exact sequence typically reduces to an integral.

The classification problem in \eqref{wp4data} constitutes a labeled data-set and lends itself perfectly to supervised ML as a binary classification problem.
Of course, this is a somewhat contrived problem, but we will use it, if anything, to present the reader with some {\sf Python TensorFlow} and {\sf Wolfram Mathematica} syntax.

\subsubsection{ML $W\IP^4$-Hypersurfaces: Python} 
Let us take 25\% random samples as training data $\cT$, establish a neural network which is a simple MLP with 3 hidden layers: a linear layer of size $5 \times 50$ (i.e., a $5 \times 50$ matrix transformation of the 5-vector input), followed by an element-wise sigmoid transformation, and then a linear layer of size $50 \times 10$, before summing up to an integer.
This choice of hyper-parameter 50 is purely ad hoc and merely to illustrate.
A more serious thing to do would be, as mentioned above, to optimize the network structure and the hyper-parameters such as the sizes of the matrices in the linear layers.
To be completely explicit, we will explain the {\sf Python}  \cite{python} code in detail, as we have always done with software throughout the book.

As mentioned, the great thing about {\sf SageMath} \cite{sage} is that its core language is Python-based and we even execute Python from within {\sf SageMath} (by, for instance, calling \verb|sage -ipython| as with our early {\sf Macaulay2} \cite{m2} example in the previous chapters).
One could also run a {\sf Jupyter} notebook from which one can run the ensuing code, after all, WWJD is our guiding question.

We begin with the usual preamble of calling the necessary {\sf Python} packages:
`numpy' for numerical recipes, 'scipy' for scientific computing, 'matplotlib.pyplot' for plotting, `pandas' for data analysis and `random' for random number generation:
\begin{verbatim}
import numpy as np
import scipy as sp
import matplotlib.pyplot as plt
import pandas as pd
import random
\end{verbatim}

Next, we load the key packages, {\sf Tensorflow}, the premium {\sf Python} package for neural networks as well as {\sf keras}, which is the high-level interface (wrapper) for Tensorflow.
From {\sf keras}, a few standard ones of its layers (most of which we will not use for this example but keep here for illustration) and network structures
\begin{verbatim}
import tensorflow as tf
import keras
from keras.models import Sequential,load_mod
from keras.layers import Dense,Activation,Dropout
from keras.layers import Conv1D,Conv2D,MaxPooling1D,MaxPooling2D,Flatten
\end{verbatim}

Suppose now we have organized the data in the form (importing from \url{http://hep.itp.tuwien.ac.at/~kreuzer/CY/} on the `CY/4d' tab, for instance):
\begin{verbatim}
dataX = [[1, 1, 1, 1, 1], [1, 1, 1, 1, 2], [1, 1, 1, 1, 3], 
         [1, 1, 1, 2, 2], [1, 1, 1, 1, 4], ... ];
dataY = [1, 1, 1, 1, 1, ...];
\end{verbatim}
{\sf dataX} is the {\em ordered} list of 5-vector weights of $W\IP^4$ and {\sf dataY}, whether the corresponding CY3 has $h^{2,1} > 50$.
Note that there are 2867 cases of 1 and 4688 of 0, so this is a fairly balanced problem; it so happens in the ordering of the configurations, the first quite a few (the first being the quintic) all have $h^{2,1}>50$.

From this we split (achieved by the set difference) the data into training $\cT$ and validation $\cV$ sets as a 25-75\% split, and then convert $\cT$ and $\cV$ into arrays for pre-processing:
\begin{verbatim}
trainingindex = random.sample([k for k in range(len(dataX))],  
     round(len(dataX)/4)  );
validateindex = list(set([k for k in range(len(dataX))]) - 
     set(trainingindex));

trainingX = np.array([dataX[a] for a in trainingindex]);
trainingY = np.array([dataY[a] for a in trainingindex]);

validateX = np.array([dataX[a] for a in validateindex]);
validateY = np.array([dataY[a] for a in validateindex]);
\end{verbatim}

Finally, we are ready to establish our simple neural network:
\begin{verbatim}
network = Sequential()
network.add(Dense(50,activation=None,input_dim=5))
network.add(Activation('sigmoid'))
network.add(Dense(10,activation=None))
network.add(Dense(1))
\end{verbatim}
A few points of explanation.
The {\sf Sequential()} specification indicates that the neural network is a MLP.
The first (input) layer, {\sf Dense()}, is a fully-connected layer meaning that the input 5-vector will be mapped to 50 nodes in a fully-connected way, i.e., the weight-matrix will be dotted with the vector, and then, added with a bias, will return a 50-vector.
The specification of {\sf activation = None} means it is defaulted to the linear function $f(x) = x$.

The output of the input layer is fed to the first hidden layer, which is an element-wise sigmoid function $\sigma(x)$ defined under Definition \ref{neuron}, which is then sent to the second hidden layer, again a linear, fully-connected layer, this time outputting a 10-vector, before finally sent to the output layer which sums everything into a single number by the {\sf Dense(1)}.
Note that only the input layer requires a specification of its input-shape (of dimension 5) and the network will then automatically deal with the subsequent dimensions.
We will show a schematic of this shortly.

Having established the network structure, we can train simply by compiling and then fitting
\begin{verbatim}
network.compile(loss='mean_squared_error', optimizer = 'adam', 
                   metrics=['accuracy'])
network.fit(trainingX,trainingY,batch_size=32,epochs=100,verbose=1,
                   validation_data=(validateX,validateY))
\end{verbatim}
In the above, the {\sf batch\_size} specifies that the training data is passed in groups of 32 at a time and the {\it epochs} means that the data will be passed through the network 100 times to optimize the parameters.
The optimizer for the gradient descent is the widely used ADAM, as discussed in \S\ref{s:MLterms}.

Finally, we can use the fitted network to check the validation data by `predicting':
\begin{verbatim}
predictY = network.predict(validateX,verbose = 1)[:,0]
\end{verbatim}
where the \verb|[:,0]| is merely an artefact of the the output format of \verb|network.predict( )|, which is an array whose every entry is of the form such as
\verb|array([1.07], dtype=float32)|, so that we extract only the 0-th entry, viz., the number itself.

This last point is actually rather important: the predicted result is a floating number (real number up to, here, 32 significant digits).
Nowhere did we specify in the network structure that the result should be an integer 0 or 1.
The optimization performed only finds the best fit over the reals.
Therefore, the final answer needs to be rounded.
For instance, the error can be obtained by doing
\begin{verbatim}
error=0;
for a in range(len(train_predict)):
    if(np.round(train_predict[a] - validatey[a]) != 0):
        error+=1;
\end{verbatim}
The error turns out to be 375 here.

Let us pause a moment to reflect upon the above calculation.
The neural network randomly saw 25\% of the 7555 cases of whether $h^{2,1} > 0$, and when optimized, it predicted on the remaining {\it unseen} 75\% of the cases with only 375 errors, i.e., to an accuracy of $1 - 375/(7555\cdot 75\%) \simeq 93.3\%$, which is rather remarkable, considering the total computation time was less than a minute on an ordinary laptop \footnote{Of course, for this baby example, the actual calculation
of over 5,600 Hodge numbers for a $W\IP^4$-hypersurface (estimating whether $h^{2,1}>50$ has no short-cut) would not have been long in any event. Later, we will see how vast reductions in calculation time can be achieved.
\label{fn:wp4}
}

\subsubsection{ML $W\IP^4$-Hypersurfaces: Mathematica}
Before we go on to do more pedantic statistics, it is a comforting fact for the {\sf Mathematica} aficionado that as of version 11.2+ (timely released in 2017), machine-learning has been built into the core operating system and being such a high-level language, allows the above to be condensed into a few lines of code.
Note that, for clarity, we are not using any {\sf Mathematica} shorthand which would even shorten any of the list manipulation commands further.

The reader is highly encouraged to vary all the code below to adapt to his/her own needs, in architecture, hyper-parameters, as well as to explore the multitude of possibilities of which our naive example has touched only the surface.
Luckily, by now, {\sf Mathematica} version 12+ has detailed pedagogical documentation on ML, which, back in 2017, was very much lacking.

We have  data-set of the form
\begin{verbatim}
datah21 = {
  {1, 1, 1, 1, 1} -> 1, {1, 1, 1, 1, 2} -> 1, {1, 1, 1, 1, 3} -> 1, 
  {1, 1, 1, 2, 2} -> 1, {1, 1, 1, 1, 4} -> 1, ... 
  };
\end{verbatim}
from which we obtain $\cT$ and $\cV$ as
\begin{verbatim}
training = RandomSample[datah21, 2000];
validation = Complement[datah21, training];
\end{verbatim}
Then, the same neural network as in the previous section, can be specified by
\begin{verbatim}
net = NetChain[{LinearLayer[50], ElementwiseLayer[LogisticSigmoid], 
    LinearLayer[10], SummationLayer[]}, "Input" -> {5}];
\end{verbatim}

\comment{
net = NetChain[{LinearLayer[50], ElementwiseLayer[LogisticSigmoid], 
    LinearLayer[10],  LinearLayer[ ]}, "Input" -> {5}];
netall = NetTrain[net, training, All, ValidationSet -> validation, 
  Method -> {"ADAM", "LearningRate" -> 1/100}];
netall
}

This can be visualized by the \verb| Information[net, "SummaryGraphic"] | command:
\begin{equation}\label{NNwp4}
\includegraphics[trim=15mm 0mm 0mm 0mm, clip,width=1\textwidth]{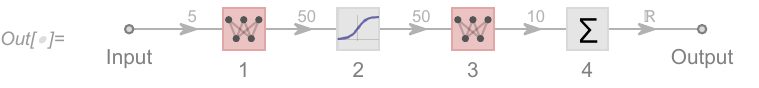}
\end{equation}
One can see the input, output, as well as the 3 hidden layers:
\begin{enumerate} 
\item Input: a vector in $\IR^5$ (for the NN, that the input is in $\IZ^5$ is, in this case, not exploited);
\item a fully-connected linear layer that takes the input in $\IR^5$ to $\IR^{50}$ (i.e., the most general $5 \times 50$ affine transformation).
\item an element-wise logistic sigmoid layer, i.e., $\IR^{50} \stackrel{\sigma}{\longrightarrow} \IR^{50}$;
\item another fully-connected linear layer, an affine transformation $\IR^{50} \to \IR^{10}$;
\item Output: a linear layer with no argument is a linear map from $\IR^{10}$ to $\IR$; the output should be a number.
\end{enumerate}
Note that the final output is in $\IR$, so we will need to take integer round-up.

Strictly, what we have done here is a {\bf neural regressor} since we are evaluating the actual value.
Of course, it is better use a {\bf classifier} and in particular to use cross-entropy as the loss function.
But, this example is purely to illustrate {\sf Python} and {\sf Mathematica} implementations, so we chose relatively simple commands.

\begin{figure}[t!!!]
	\includegraphics[trim=12mm 0mm 0mm 0mm, clip, width=5.5in]{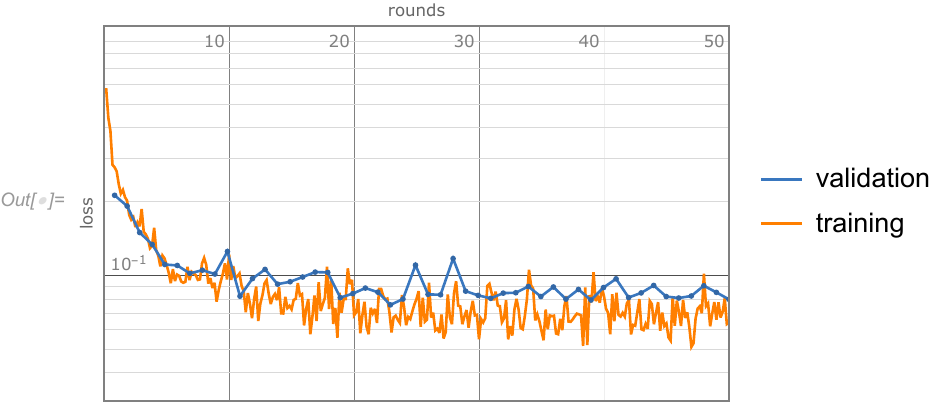}
	\caption{{\sf Training curve for NN \eqref{NNwp4} machine-learning whether $h^{2,1}>50$ for a hypersurface Calabi-Yau 3-fold in $W\IP^4$, for both training and validation sets as we feed batches into the training.}
	\label{f:wp4training}}
\end{figure}

The training done by
\begin{verbatim}
trainednet = NetTrain[net, training, ValidationSet -> validation, 
		Method -> {"ADAM", "LearningRate" -> 1/100}, MaxRounds -> 100];
\end{verbatim}
Here, we have used the ADAM optimizer, with learning rate $\eta = 0.01$, as well as 100 iterations of feeding the training set into batches.
We have specified to check validation during the training and can see the training curves during the training for the squared-error loss-function; this is shown in Figure \ref{f:wp4training} 
for both the training set and the validation set.
Indeed, they both converge toward 0.

We can check the predictions against the validation set by (note the integer round):
\begin{verbatim}
actual = Table[validation[[i, 2]], {i, 1, Length[validation]}];
predict = Table[Round[net[validation[[i, 1]]]],
          {i, 1, Length[validation]}];
\end{verbatim}
This gives us a confusion matrix 
{\scriptsize $\left(
\begin{array}{cc}
 3669 & 361 \\
 83 & 1931 \\
\end{array}
\right)$}, giving us na\"{\i}ve precision $0.926$, F-score $0.896$ and Matthews' $\phi=0.844$.

\begin{figure}[t!]
	\includegraphics[trim=0mm 0mm 0mm 0mm, clip, width=5.5in]{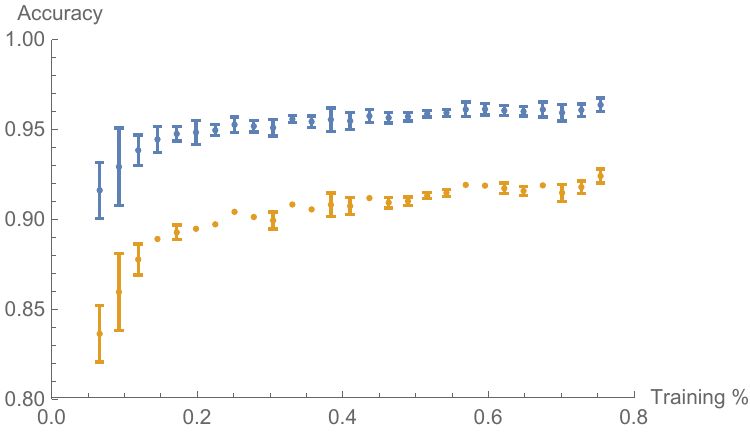}
	\caption{{\sf Learning curve for NN \eqref{NNwp4} machine-learning whether $h^{2,1}>50$ for a hypersurface Calabi-Yau 3-fold in $W\IP^4$. The blue (top curve) is the  na\"{\i}ve precision and the orange (bottom curve) is Matthews' $\phi$.}
	\label{f:wp4learning}}
\end{figure}

In the above, we set the training-validation split at around 20\%,
as a extra check, we can how the various goodness of fit behaves as we increase the training size.
The learning curve for this problem is presented in Figure \ref{f:wp4learning} (recall that we made, for clarity, a distinction between the learning curve and training curve in Definitions \ref{def:tc} and \ref{def:lc}).

We have repeated, at each incremental interval of 5\%, the training/prediction 10 times, which gives us the error bars.
There is a large error when the seen training data is less than 10\%, as expected -- the network simply has not seen enough data to make a valid prediction.
However, starting at 20\%, as discussed above, we are already looking at 96\% precision with Matthew's $\phi$-coefficient around $0.9$.
The curve is slowly but steadily climbing to around 97\% precision and $\phi \sim 0.92$ at around 80\% seen data.
All of this is indeed reassuring.
The neural network can estimate a solution to our complex structure problem to very high precision and confidence at orders of magnitude less computation time.

\subsubsection{non-NN ML methods}
As we mentioned in footnote \ref{fn:wp4}, to treat the present example with ML, especially NNs, is quite an overkill.
We have chosen this warm up not only to illustrate {\sf Python} and {\sf Mathematica} syntax, but also because of the simplicity of the representation of the input: a 5-vector of integers.
We close this subsection with some non-NN methods.

First, we can think the data as a labeled point-cloud in $\IR^5$ and apply SVM to see whether there is separation.
To this effect, the simple {\sf Mathematica} command (retaining the variables ``training'' and ``validation'' from the above)
\begin{verbatim}
cc = Classify[training, Method -> {"SupportVectorMachine",
              "KernelType" -> "RadialBasisFunction"}];

actual = Table[validation[[i, 2]], {i, 1, Length[validation]}];
predict = Table[cc[ validation[[i, 1]] ], {i, 1, Length[validation]}];
\end{verbatim}
In the above, we have specified the SVM kernel \eqref{SVMpar} to be Gaussian (here called ``RadialBasisFunction''; one could also choose ``Linear'' or ``Polynomial'').
The result is comparable to the NN in \eqref{NNwp4}, for example, we obtain na\"{\i}ve precision of around $0.936$.
Moreover, one could find the parameters in the Gaussian kernel by \verb|Information[cc, "GammaScalingParameter"]|,
which gives $\gamma := \frac{1}{2\sigma}$, and  \verb|Information[cc, "BiasParameter"]|, which gives $b=1$.
The biase is 1 here because we have labeled our categories as 0 and 1, while in the search for the support hyper-plane described in \eqref{SVMhyper}, they are $\pm 1$.

Of course, the problem at hand is simple enough that one could even try a straight-forward regression, similar to Appendix \ref{ap:regression}:
\begin{equation}
f(w_i) = \sigma \left( \sum\limits_{i=0}^4 w_i x_i + b\right) \ ,
\end{equation}
where $x_i$ are the 5 components of the input and $(w_i, b)$ are coefficients (weights and bias if you will) to be optimized.
Note that the sigmoid function $\sigma$ already has range $[0,1]$ so it is perfect for binary classifications.
Moreover, we are not doing cross-validation here since there are only 6 parameters and there is no danger of over-fitting some 8000 data points.

The {\sf Mathematica} command is 
\footnote{
The reason we had to restructure the data upfront is esoteric: regression in {\sf Mathematica} is not considered an ML algorithm so it does not like the arrow notation for ``input $\to$ output'', but rather prefers ``(input, output)''.
}
\begin{verbatim}
datah21regress = Table[ Flatten[ {datah21[[i,1]],  datah21[[i,2]]}], 
            {i,1,Length[datah21]} ];
vars = Table[x[i], {i, 1, 5}];
coefs = Flatten[ {Table[w[i], {i, 1, 5}], b} ];
model = NonlinearModelFit[datah21regress, 
  	  LogisticSigmoid[ Sum[w[i] x[i], {i, 1, 5}] + b ], coefs, vars ];
\end{verbatim}
From this, we can extract the R-squared as \verb|model["RSquared"]|, giving us $R^2 \sim 0.843$.
Again, we can take integer round to obtain the predicted \footnote{
We use {\sf model@@} rather than {\sf model[~]} because the input is a list and model is a function of 5 variables, so we need to ``thread'' the function over the input list.
} values:
\begin{verbatim}
predict = Table[ Round[ model@@(datah21[[i,1]]) ], {i,1,Length[datah21]}];
actual = Table[ datah21[[i, 2]], {i, 1, Length[datah21]} ];
\end{verbatim}
This gives us  na\"{\i}ve precision $0.919$, F-score $0.892$ and Matthews' $\phi=0.827$, which is comparable to the NN and the SVM.
As with the SVM, we have interpretability, the coefficients in the model can be explicitly found, using 
\verb|Normal[model]|, here, we find
\begin{equation}
b \sim 6.901, \
w_i \sim (2.529, -0.273, -0.060, 0.0079, 0.046) \ .
\end{equation}

\subsection{Learning CICYs}
Emboldened by the success in the warm-up, we can proceed to more sophisticated inputs.
A CICY, we recall, is an integer matrix with number of rows ranging from 1 to 12 and number of columns, from 1 to 15, and whose entries are 0 to 5.
As far as the computer is concerned, this is a $12 \time 15$ array of arrays (by down-right padding with zeros) with 6 discrete entries. In other words, it is a $12 \times 15$ pixelated image with 6 shades of grey (or 6 different colours).
For instance, this very colourful representation \cite{He:2017aed} for the CICY which is the complete intersection of 8 equations in $(\IP^1)^5 \times (\IP^2)^3$ would be
\begin{equation}\label{cicyPic}
\left(
\begin{array}{c|c}
{\scriptsize
\begin{array}{cccccccc}
 1 & 1 & 0 & 0 & 0 & 0 & 0 & 0 \\
 1 & 0 & 1 & 0 & 0 & 0 & 0 & 0 \\
 0 & 0 & 0 & 1 & 0 & 1 & 0 & 0 \\
 0 & 0 & 0 & 0 & 1 & 0 & 1 & 0 \\
 0 & 0 & 0 & 0 & 0 & 0 & 2 & 0 \\
 0 & 1 & 1 & 0 & 0 & 0 & 0 & 1 \\
 1 & 0 & 0 & 0 & 0 & 1 & 1 & 0 \\
 0 & 0 & 0 & 1 & 1 & 0 & 0 & 1 \\
\end{array}
}
&
0  \\ \hline
0 & 0
\end{array}
\right)_{12 \times 15}
\leadsto
\begin{array}{c}
\includegraphics[trim=0mm 0mm 0mm 0mm, clip, width=1.5in]{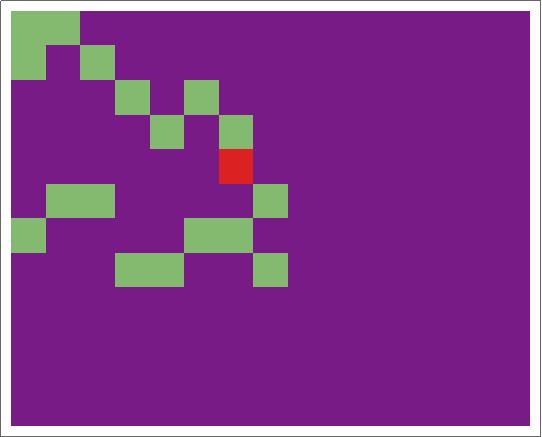}
\end{array}
\end{equation}
where purple is 0, green, 1 and red, 2.
This is a CY3 with Hodge pair $(h^{1,1},h^{2,1}) = (8,22)$ (and thus $\chi = -28$); that $h^{1,1}$ is equal to the number of product projective spaces means it is favourable.
For the reader's amusement, the {\em average} CICY as an image was drawn in {\it cit.~ibid.}

Now, we have shown in \S\ref{s:Q} how to compute individual $h^{1,1}$ or $h^{2,1}$ from the configuration matrix.
Even for the simplest case of the quintic, some substantial machinery such like long exact sequences in cohomology is needed. And, even if one were to do this on a computer using {\sf Macaulay2}, the Gro\"bner basis will render all but the simplest configurations intractable.
Yet, at the end of the day, we have an output as succinct as the number 22 (say, if we were interested in $h^{2,1}$).
How different indeed, is this association ``colourful matrix'' in \eqref{cicyPic} $\to 22$ any different from \eqref{digit3} in recognizing the digit 3?
Indeed, it was the similarity that prompted the investigations in \cite{He:2017aed}.

With such an image, one could use a CNN to process it, much like hand-writing recognition.
While we will not do so here, the point of presenting the manifold in this way is to emphasize the fact that the computer knows absolutely nothing about the algebraic geometry, or even that the image should correspond to a geometry.
We can associate the image with an integer (such as the topological numbers) to the image much as one associates a value to a hand-written scribble. 
Yet with enough samples of correct values we can teach the machine how to ``guess'' the right values.
This is doable because there is inherent pattern in algebraic geometry, regular enough to be ``learnt''.

\begin{figure}[h!!!!]
	\centering
		\includegraphics[width=0.7\textwidth]{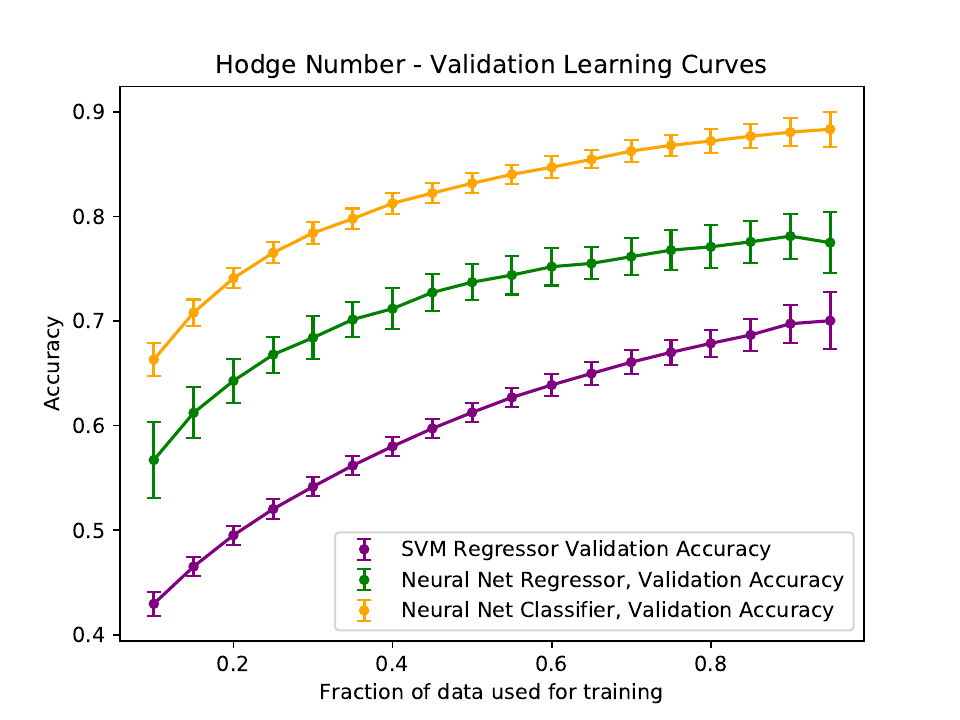}
	\vspace{5pt}
	\caption{{\sf Hodge learning curves generated by averaging over $100$ different
	random cross validation splits. The accuracy quoted for the
	$20$ channel (since $h^{1,1} \in [0,19]$) neural network
	classifier is for complete agreement across all $20$ channels.}}
	\label{f:hodgeCICY}
\end{figure}

In \cite{He:2017aed,He:2017set}, the analogous NN approach to testing CICY Hodge numbers as the one presented in detail above for $W\IP^4$ was performed.
A more insightful and comprehensive analysis  \footnote{
Though I am very much an amateur to data science, I must use this opportunity to thank my former Masters student Kieran Bull from Oxford, who was the torch-bearer for \cite{Bull:2018uow}. With admiration I joke with him that he eats 2000 lines of Python for breakfast, and, in retrospect, he probably did.
} specifically for CICYs was then done in \cite{Bull:2018uow,Bull:2019cij}.
The input data is a CICY configuration (image) and one can try the following structures to study not just whether one of the Hodge numbers exceeds a value, but the precise value of, say, $h^{1,1}$.
This is obviously of great importance as we are actually going to let the machine calculate the exact value of a topological invariant simply by experience:
\begin{description}

\item[NN regressor: ] the output is continuous (i.e., some real number which we will then round up);
Optimal neural network hyperparameters were found with a genetic algorithm, leading to an
overall architecture of five hidden layers with 876, 461, 437, 929, and 404 neurons, respectively.
The algorithm also found that a ReLU activation layer and dropout layer
of dropout probability 0.2072 between each neuron layer give optimal results.

\item[SVM: ] the output is one of the possible values for $h^{1,1}$, viz., an integer \footnote{
An artefact of the CICY data-set is that $h^{1,1} = 0$ is a marker for a trivial CY3, such as the direct product $(T^2)^3$.
} between 0 and 19, inclusive; here the hyper-parameters were found to be a Gaussian kernel with $\sigma=2.74$, as well as slack $C=10$ and margin $\epsilon=0.01$ (recall that these quantities were defined in \S\ref{s:nonNN}, in Eqs.~\eqref{SVMpar,SVMslack,SVMeps}).

\item[NN classifier: ] the output is a 20-channel classifier (like hand-written digits being a 10-channel classifier) with each neuron mapping to 0 or 1.
The optimal architecture can be found to be be four convolution layers with 57, 56, 55, and 43 feature
maps, respectively, all with a kernel size of $3 \times 3$. These layers were followed by two hidden fully-connected layers and the output layer, the hidden layers containing 169 and 491 neurons. ReLU
activations and a dropout of 0.5 were included between every layer, with the last layer using
a sigmoid activation.
\end{description}
Importantly, training with a laptop took on the order of 10 minutes and execution on the validation set, seconds.
The actual Hodge computation of the CICYs, on the other hand, was a stupendous task.

\begin{figure}[h!!!!]
	\includegraphics[width=0.31\textwidth]{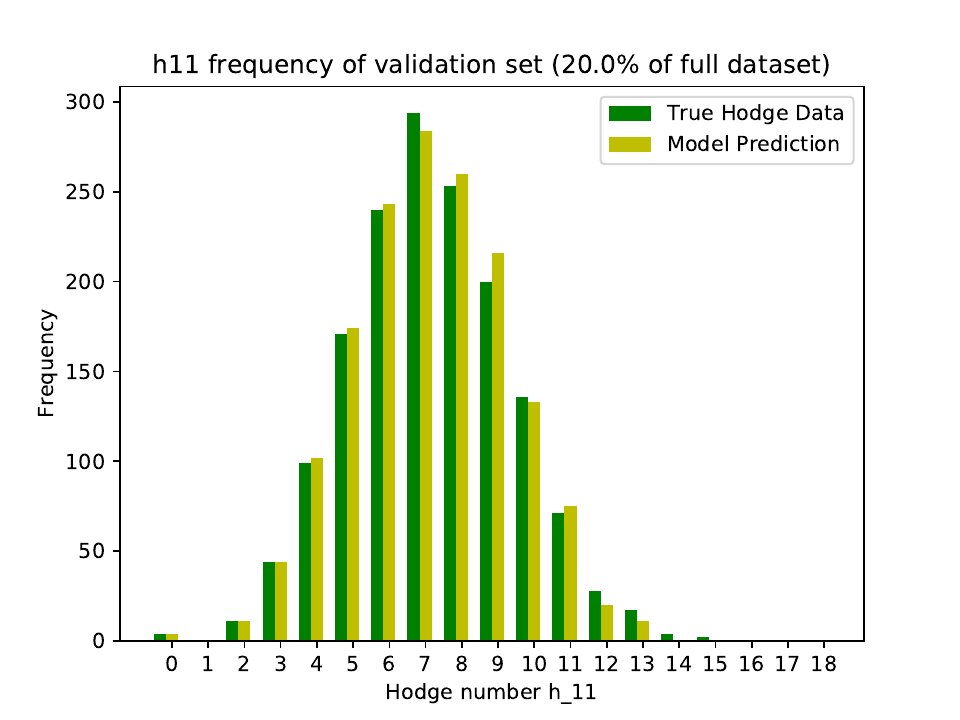}
	\includegraphics[width=0.31\textwidth]{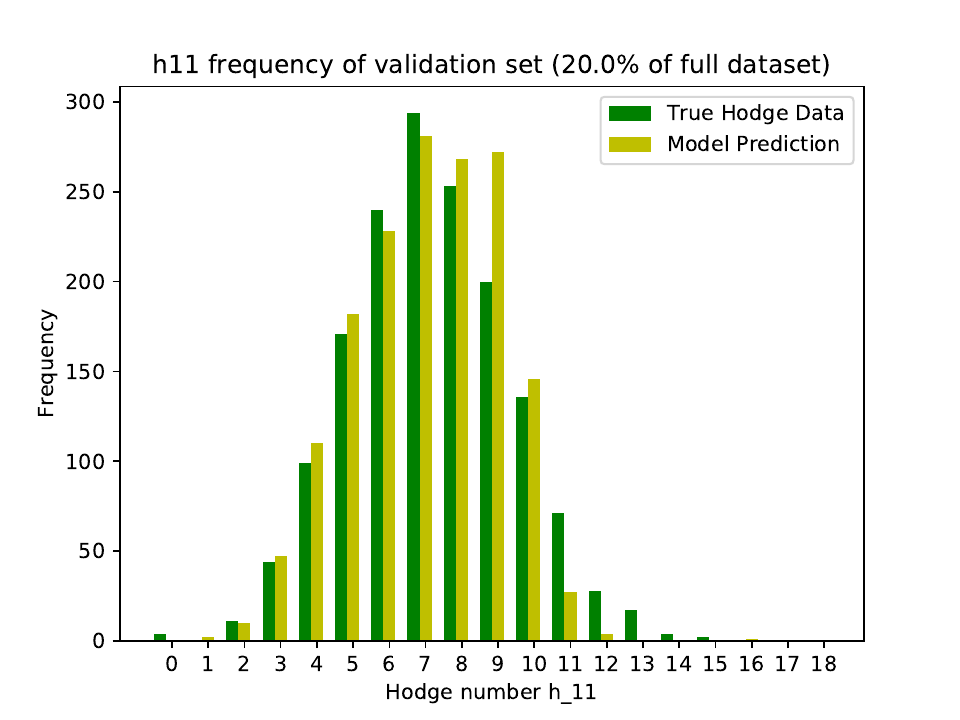}
	\includegraphics[width=0.31\textwidth]{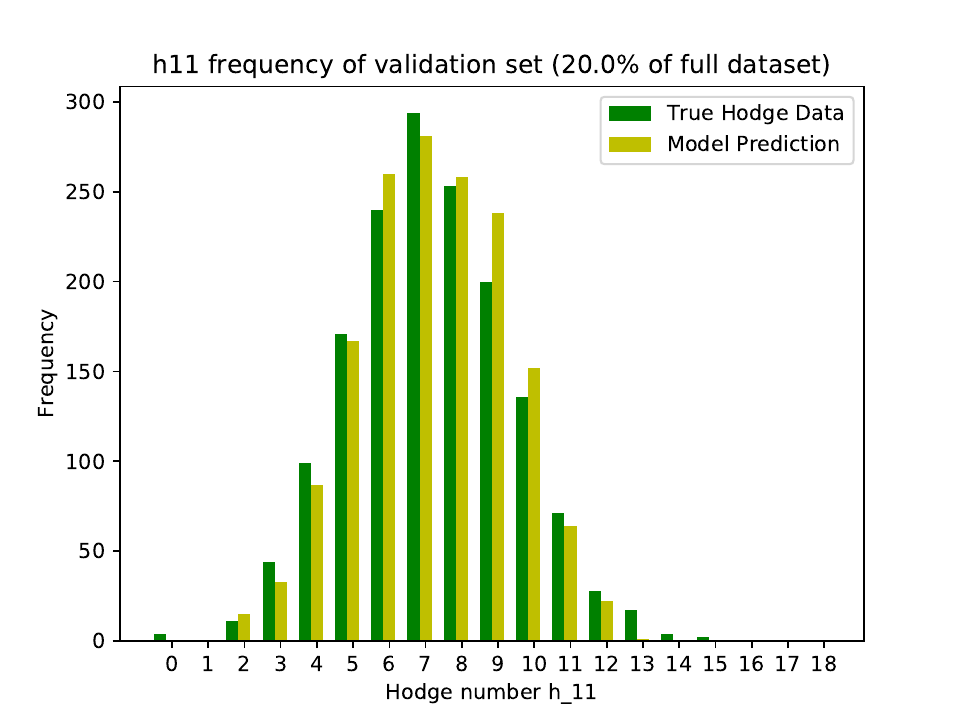}
	\caption{{\sf 
	The histogram of predicted versus actual CICY $h^{1,1}$ at 20\%-80\% training-validation split, 
	for (1) the neural network classifier, (2) regressor, and the SVM regressor.
	}}
	\label{f:CICYchannel}
\end{figure}

The performance is again quite satisfactory, for a 25\%-75\% split of training versus validation (the machine has only seen 1/4 of the data!) and we have  the various accuracy measures for the 3 above methods as
\begin{equation}
\mbox{
\begin{tabular}{c|ccccc}
	& Accuracy & RMS & $R^2$ & $\omega_-$ &$ \omega_+$  \\ \hline 
	NN Regressor & 0.78 $\pm$ 0.02 & 0.46 $\pm$ 0.05 & 0.72 $\pm$ 0.06 & 0.742 & 0.791 \\
	SVM Reg & 0.70 $\pm$ 0.02 & \textbf{0.53}$\pm$ \textbf{0.06} & \textbf{0.78} $\pm$ \textbf{0.08} & 0.642 & 0.697  \\
	NN Classifier & \textbf{0.88} $\pm$ \textbf{0.02} & - & - & \textbf{0.847} & \textbf{0.886}\\
\end{tabular}
}
\end{equation}
In the above, the error bars come from repetition over 100 random samples, the dashes for the RMS and $R^2$ are because these are only defined for continuous variables, and $\omega_{\pm}$ give the Wilson confidence interval.
Moreover, we have put in bold face the best performance.
Recently, a systematic study using the so-called Google Inception CNN was carried out \cite{Erbin:2020srm,Erbin:2020srm} on the CICYs and accuracies in the vicinity of 99\% has been achieved.

Reassured by the prediction based only on 1/4 of seen data,
the learning curves for the above 3 different methods are presented in Figure \ref{f:hodgeCICY}.
We see that interestingly the SVM seems to be the best performer, quite quickly reaching 90\% accuracy.
For reference, we also demonstrate, in Figure \ref{f:CICYchannel}, the comparative performance amongst the 3 methods at 20\% seen training data, for each of the 20 values of $h^{1,1}$, checked against the 80\% unseen validation data. We recall from Figure \ref{f:cicydis} that  the distribution for $h^{1,1}$ is approximately Gaussian, peaked at 7, for CICYs.

\subsection{More Success Stories in Geometry}\label{s:MLAG}
Over the last few years, there have been various success stories along the lines of the above in applying ML to problems in algebraic geometry, especially in the context of the CY landscape.
We will list some of the main directions below, in line with what we have covered in this book.

\paragraph{The KS Data-set: }
An enormous challenge with the KS data-set is that the half-billion polytopes,
each of which gives a certain Hodge pair according to \eqref{hKS}, but could lead to many topologically inequivalent CY3s with the same $(h^{1,1}, h^{2,1})$ . To compute the Chern numbers and the intersection form (q.v.~\eqref{CYdata}) requires full triangulation of the polytope (thus, the KS data-set actually produces several orders of magnitude more than 473,800,776 CY3s!).
While this has been done systematically for the low-lying cases \cite{rossaltman,Altman:2014bfa,Demirtas:2018akl}, to do so for the full list is many orders of magnitude beyond current computer abilities.
Could one train on these known cases in a supervised machine learning and predict what $c_2$ and $d_{rst}$ would most likely be and thus complete the topological data for the KS set \footnote{
If one has the $c_2$ and $d_{rst}$, one could pursue many serious phenomenological questions.
For example, certainly properties of the $d_{rst}$ (so-called ``Swiss-cheese'' \cite{Conlon:2005ki}) single out manifolds which can be used in cosmology.
However, the computation to decide whether a manifold is Swiss-cheese is very computationally intensive \cite{Gray:2012jy}.
Machine-learning the data-set in distinguishing such manifold while not be impeded by the expensive steps would be much desired.
}?
On a simpler level, could one predict how many triangulations - and hence different candidate CY$_3$ manifolds - there should be?
Recently, an ML estimate using equation learners, was nicely performed for the full KS set in \cite{Altman:2018zlc} and an astounding $10^{10^5}$ number of total triangulations was proposed.

\paragraph{The CICY Data-set: }
Other than ML of the CICY 3-fold Hodge numbers discussed in detail above, it is natural to also consider that of the CY4 version, a data-set of CICY4s which was created in \cite{Gray:2013mja}.
The Hodge numbers for these also responded well to MLPs and to a branched forward-feed NN \cite{He:2020lbz}.
In parallel, elliptic fibrations were classified within the CICY3 set by \cite{Anderson:2017aux}; again, with a simple MLP or an SVM, 
these can be distinguished \cite{He:2019vsj} to high accuracy.

\paragraph{Bundle Cohomology: }
There is, of course, nothing special about configuration matrices; these, as we saw, are degrees of the normal bundle in the embedding into the ambient projective space.
Line bundles, or sequences of line bundles, can be likewise represented.
Bundle cohomology is subtle and depends on regions in moduli space, these have nevertheless been computed over the years.
The novel approach of genetic algorithms were employed to compute bundle cohomologies in  \cite{Ruehle:2017mzq} and applied to the string landscape. 
Likewise, cohomology of bundles of CICY and KS manifolds  \cite{Klaewer:2018sfl,Larfors:2020ugo} respond well to ML algorithms.
Of particular note is \cite{Constantin:2018hvl,Brodie:2019dfx}, where, having obtained 100\% accuracy, the authors were able to conjecture new analytic expressions for bundle cohomology, which were then proven using classical algebraic geometry.

\paragraph{Numerical Metrics \& Volumes: }
Numerical volume of Sasaki-Einstein bases of affine CYs is an important quantity \cite{SEvol,Butti:2005vn,Eager:2010yu,He:2017gam}.
ML has been applied to these with success \cite{Krefl:2017yox}.
Likewise, volumes of hyperbolic knots and related Jones polynomials respond well to MLPs \cite{Jejjala:2019kio}.
Furthermore, we discussed in \S\ref{s:metric} the Donaldson algorithm for numerically computing CY metrics, and mentioned its complexity as more and more monomials are required.
With a decision tree classifier, one could speed this up to almost two orders of magnitude \cite{Ashmore:2019wzb}.

\paragraph{Outlook:}
However apparent the power of ML, much of it does remain mysterious: of all the intricate connections amongst the neurons or optimization of kernels and hyperplanes, what exactly is doing what? As Yau critiques, ``there are no theorems in machine-learning''.
Indeed, for a pure mathematician of his calibre, getting a topological invariant correct 99\% of the time may not be precise enough.
Nevertheless, getting such a quick estimate {\it entirely bypassing the expensive steps such as polytope triangulation or finding a Gr\"obner basis or finding the ranks of all the matrices in an exact sequence} is salient enough for practical purposes that the endeavour is well worth pursuing.
One concrete line of attack might be
\begin{quote}
Let P be a problem in algebraic geometry with a simple answer (such as an integer), reduce its computation to its constituent parts (typically finding kernels and cokenels of matrix maps \footnote{
Even finding the Gr\"obner basis, in a chosen basis of monomials in the variables, reduces to tensor manipulation.
In the case of toric varieties, the problem can be entirely phrased in combinatorics of polytopes whose vertices are extracted from the exponents of monomials \cite{toricAlg}.
}), test which of these is being out-performed by the likes of a neural network or SVM, and how.
\end{quote}

We will address the problem of ML to pure mathematics shortly, but, from an applicability point of view for the physics, the potential is already enormous.
As mentioned in the end of the first Chapter, Calabi-Yau compactifications with the naturally endowed tangent bundle constitute only a small portion of the heterotic landscape.
More sophisticated methods now use stable holomorphic vector bundles on manifolds with non-trivial fundamental group in obtaining the {\it exact} standard model. This is a much richer and phenomenologically interesting landscape of geometries which involve the much more intractable task of computing bundle-valued cohomology groups.
Initial attempts to finding exact standard-models from CY constructions using ML are already under way \cite{Ruehle:2017mzq,Carifio:2017bov,Mutter:2018sra,Otsuka:2020nsk,Deen:2020dlf,Bies:2020gvf}.
Indeed, once we have more efficient methods of numerical metrics \cite{Ashmore:2019wzb}, getting particle masses in realistic string standard models is finally within reach.

What about beyond Calabi-Yau geometry?  We know the string landscape extends far beyond Calabi-Yau manifolds and touch deeply upon $G_2$-manifolds, generalized K\"ahler manifolds, and more.
The paradigm of machine-learning algebraic geometry is clearly of interest to physicists and mathematicians alike, and applicable to varieties other than Calabi-Yau manifolds.
What can they say about computing topological or rational invariants in general?
We leave all these tantalizing challenges to the dedicated reader.

\section{Beyond Algebraic Geometry}
Encouraged by our string of successes, one might be led astray by optimism.
Whilst there is, by definition, patterns in all fields of mathematics, machine-learning 
 is not some omnipotent magical device, nor an oracle capable of predicting any pattern.
Otherwise, we might as well use it to guess the position of the Riemann zeros or similar such questions of profound consequences.
Indeed, one would expect problems in number theory to not be particularly amenable to our paradigm.

We therefore finish with a sanity check that machine-learning is not a magic black-box predicting all things. 
A reprobate \footnote{
In Calvinist heresy, a reprobate is a soul doomed for condemnation. I am grateful to my wife Dr.~Elizabeth Hunter He, an expert on Reformation history, for teaching me this.
} example which should be doomed to failure is the primes. 
Indeed, if a neural network could guess some unexpected
pattern in the primes, this would be a rather frightening prospect for mathematics.
Suppose we have a sequence of labeled data
\begin{equation}
\begin{array}{l}
\{2\} \to 3 \ ; \\
\{2, 3\} \to 5 \ ; \\
\{2, 3, 5\} \to 7 \ ; \\
\{2, 3, 5, 7\} \to 11 \ ; \ldots
\end{array}
\end{equation}
One can easily check that even with millions of training data, any neural network or SVM would be useless in prediction and that we are better off with a simple regression against the $n\log(n)$ curve in light of the Prime Number Theorem.

Perhaps the reader is worried about the huge variation in the output, that only some form of highly specialized regressor would do well (this had indeed been the experience from the various cases in \S\ref{s:MLAG}).
To re-assure the reader, let us make it as simple as possible.
We set up the data as follows:
\begin{enumerate}
\item Let $\delta(n) = 0$ or $1$ be the prime characteristic function so that it is 1 when an odd number $n$ is prime and 1 otherwise (there is no need to include even numbers);
\item Consider a ``sliding window'' of size, say, 100, and consider the list of vectors $\{\delta(2i+1), \ldots, \delta(2(i+100)+1) \}_{i = 1, \ldots, 50000}$ in $\IR^{100}$. This is our input;
\item For output, consider the prime characteristic of some distinct number from each windows, say, 
$\delta(2(i+100+k)+1)$ for $k=10000$.
\end{enumerate}
We thus have a binary classification problem of binary vectors, of the form ( we have ordered the set with a subscript for reference)
\begin{equation}
\begin{array}{l}
\{1, 1, 1, 0, 1, 1, \ldots, 1, 0, 1, 1, 0, 0 \}_1 \to 0 \ ; \\
\{1, 1, 0, 1, 1, 0,\ldots, 0, 1, 1, 0, 0, 0 \}_2 \to 0 \ ; \\
\ldots\\
\{1, 0, 0, 0, 0, 0,\ldots, 0, 0, 0, 0, 1, 0\}_{600} \to 1 \ ; \ldots
\end{array}
\end{equation}
Now, the primes are distributed fairly rarely ($\pi(x) \sim x/\log(x)$ amongst the integers), so we down-sample the 0-hits to around  9000 each of 0 and 1.
Applying various classifiers it was found that the k-nearest neighbour (using $k=50$ and Hamming distance) worked best, at na\"i{\i}ve precision around 0.77 and Matthews $\phi \sim0.60$.
Whilst this is perhaps somewhat surprisingly good \footnote{
This might have something to do with the fact that recognizing whether an integer is prime or not has (only recently!) been unconditionally determined by the AKS algorithm \cite{AKS} to be $\log(n)^{7.5}$.
We remark that the same binary classification experiment, using more erratic functions such as the Liouville lambda-function (which counts even/odd number of prime factors) instead of the prime-characteristic $\delta$, gave na\"i{\i}ve precision around $0.50$ and Matthews $\phi \sim0.001$; which means that the ML algorithm is doing as well as random guess.
}, and deserves further investigation, the performance is visibly not as good as the algebro-geometric experiments in this chapter, with precision in the $0.99$ range.

This principle of algebraic geometry being more susceptible to ML than number theory could be
approximately understood. At the most basic level, every computation in algebraic
geometry, be it a spectral sequence or a Gr\"obner basis, reduces to finding kernels
and cokernels of sets of matrices (over $\IZ$ or even over $\IC$), albeit of quickly forbidding
dimensions.
Matrix/tensor manipulation is exactly the heart of any machine-learning -- after all, this why the {\sf Python} package is called {\sf TensorFLow} -- and what ML is good at.
Number theory, on the other hand, ultimately involves patterns of prime numbers which, as is well
known, remain elusive.
Nevertheless, a natural question comes to mind as to 
\begin{quote}
How well do different branches of mathematics respond to ML?
\end{quote}
Is the intuitive hierarchy of difficulty, from numerical analysis to Diophantine problems, wherein the former has an abundance of methods, while the latter is provably undecidable by Matiyasevich (Hilbert 10-th), somewhat in line with the performance of ML algorithms?
This question is one of the author's current programmes and we shall report some preliminary results here.

\paragraph{Algebra \& Representation Theory: }
Trying a sample from the GAP database of finite groups and rings \cite{gap}, questions such (1) whether one could distinguish a Cayley multiplication table from a random collection of Latin squares (Sudokus), or (2) whether one could recognize a finite simple group by looking at its Cayley table (admittedly a difficult problem), were fed into the supervised ML paradigm \cite{He:2019nzx}.
With SVM classifiers, precision on the order of 0.90 and 0.99 can be achieved at 20-80 cross-validation for the 2 problems respectively. For continuous groups, simple MLPs were applied \cite{LieML} to some of the most important calculations for Lie groups, viz., tensor product decompositions and branching rules, these also reached accuracies in the 0.90 range.

In representation theory, particularly in relation to quivers, an important concept is {\it cluster mutation} (in physics, this is Seiberg duality and we alluded to this in Chapter 3), and there had been much work in establishing data-sets of equivalence classes under mutation, both in the physics and the mathematics communities.
A Na\"{\i}ve Bayes classifier \cite{Bao:2020nbi} was found to distinguish between various different mutation classes up to the 0.90 - 1.00 precision range.
Closely related is the ML of the plethystic programme \cite{PlethML} of Hilbert series (q.v.~Appendix \ref{ap:HS}), where $0.90-0.99$ accuracies can be achieved in recognition of information such as dimension, degree, and complete intersection.

\paragraph{Combinatorics: }
Using the Wolfram database of finite simple graphs \cite{wolfram}, experiments were carried out on how many properties can be predicted by ML from the adjacency matrix alone \cite{He:2020fdg}.
While the focus was mainly on Ricci-flatness of graphs, a discrete analogue of the CY property, common properties were also tested, primarily with an SVM.
It was found that cyclicity and Ricci-flatness can be learned with accuracies in the $0.92 \sim 0.95$ range, while classical problems such as existence of Euler and Hamiltonian cycles were in the $0.73 \sim 0.78$ range.
It is quite interesting how the Euler problem (which has linear complexity in number of edges by the Hierholzer algorithm) and the Hamiltonian problem (which is extremely hard and in fact NP-complete) have similar responses to an SVM classifier.

\paragraph{Number Theory: }
As mentioned from the outset, problems and patterns in number theory are subtle.
A frontal attack on the primes is ineffectual.
Likewise, it was found that ML predictions to relevant quantities in the Birch-Swinnerton-Dyer conjecture had very limited accuracies \cite{Alessandretti:2019jbs}.
However, some problems in arithmetic geometry, residing in between algebraic geometry and number theory, have met with surprising success.
The classification of degree of Galois extension from examining the permutation structure of {\it dessins d'enfants} have $0.90 \sim 0.93$ range of accuracies by simple ReLU MLPs \cite{He:2020eva}.
Classification of properties such as complex-multiplication and Sato-Tate groups for hyper-elliptic curves over the rationals, which ordinarily require highly refined techniques, can be classified, using Na\"{\i}ve Bayes, to $0.99 \sim 1.00$ accuracies \cite{He:2020kzg}. 
Similarly, finding the degree and rank of extensions of number fields over $\IQ$ from the zeta coefficients is also in the $>0.9$ range of accuracies \cite{1831055}.

\section{Epilogue}
We have followed the colourless skein of history throughout the book, keeping track of the discovery of the manifolds and the compilation of the data-sets in tune with the contemporaneous developments in mathematics and physics.
Though the area of machine-learning geometry and applying the latest results from data science to treat the landscape is still in its infancy, and much of its inner workings is still perplexing, its power is apparent and its potential, vast.
This new paradigm should prove useful as both a {\it classifer} and {\it predictor} of geometrical quantities.
The ability to rapidly obtain results stochastically, before embarking on to intensively compute from first principles is at once curious and enticing.
What we have sacrificed in complete confidence and analytical result, we have gained by improving the computation time by many orders of magnitude. And, for pure mathematics, when 100\% accuracies are reached, through unconventional ways, new conjectures and directions can be proposed.

To quote Boris Zilber, Professor of Mathematical Logic at Merton College, Oxford, ``you have managed syntax without semantics.''
This is an astutely appropriate analogy, not only for our present context but also for AI research ofr mathematics in general.
The prowess of machine-learning is precisely its not taking a reductionist framework where all intermediate quantities are computed analytically from first principles - much like our earlier comment of trying to recognize hand-writing by setting up Morse functions - but by gathering large quantities of data to probabilistically find latent patterns.

Nevertheless, the ``AI mathematician'' and the ``AI physicist'' are already being deviced, not only from the automated- theorem-proving point of view \cite{voevodskyy,ATP}, but from the ML perspective over the last year or so.
In physics, the NN {\it SciNet} has been employed to discover Newtonian mechanics from motion data \cite{scinet}, while AI Feymann \cite{Udrescu:2019mnk} has led to symbolic regression analysis of data leading to fundamental laws.
In mathematics, ML of symbolic algebra \cite{DLsymb} and the so-called {\it Ramanujan machine} in generating new (continued-fraction) identities \cite{ramanujanML} have already met with success.
Indeed, the linguistic structure of ArXiv, from its titles alone, upon using the NN {\it Word2Vec},  already tells us a lot about the field of study \cite{He:2018dlv}.

It is interesting to see that in the first decades of the 20th century, theoretical physicists had to learn algebraic geometry, and in its last decades, many mathematicians turn to learn some theoretical physics;
now, in the dawn of the 21st, it is AI which can be put to ``learn'' some tricks of the trade.
We are reminded of many species of primates who have learned to effectively use herbal medicine by trial and error without any idea of chemical composition. Perhaps confronting the vast landscape of mathematics and theoretical physics, the human mind is still learning by tentation.

\chapter*{Postscriptum}
Thus concludes our excursion of the {\it terra sancta} of Calabi-Yau manifolds.
From the compact landscape of CICYs and KS hypersurfaces, to the non-compact vista of quiver representations and Sasaki-Einstein cones, from the computational algebraic geometry of topological invariants, to the combinatorics of convex polytopes and rational cones, from the plethora of data compiled into readily accessible webpages to the {\it terra incognita} of machine-learning algebraic geometry, I hope the reader has beheld the richness of the grounds, fertile with mathematics, physics and data science.

Though from these alluring terrains we now pause to return to our mundanities, our orisons shall continue to drift to this lofty pursuit which transcends disciplines,
and so it is only fitting to close with an excerpt \footnote{
The Chinese poem is written in the antique style (1 millennium BC), rather than more familiar classical  style (the height of which is the 7th-8th centuries AD); I will attempt a translation here.
``The mysteries of the Cosmos, how grand thou art; 
the font of Truth, how beauteous thou art.
To quantize time and space, the Wise knows not the ways; 
the world though the dioptra, endless study fills the days. 
''.
} from the master himself, the ``Ode to Time and Space'' by Prof.~S.-T.Yau:

\comment{
\begin{quote}
\begin{CJK}{UTF8}{songti}
\ldots \\
大哉大哉，宇宙之謎。美哉美哉，真理之源。\\
時空量化，智者無何。管測大塊，學也洋洋。 \\
~\\
\qquad \qquad 丘成桐先生： 時空頌
~\\
\end{CJK}
\end{quote}
}

\begin{figure}[h!!!]
$
\begin{array}{c}\includegraphics[width=0.6\textwidth]{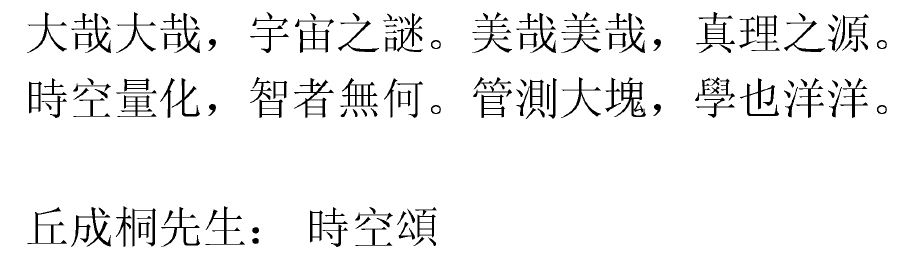}\end{array}
$
\end{figure}

\begin{appendix}
\renewcommand\chaptername{Appendix}
\chapter[Complex Geometry: Rudiments]{Some Rudiments of Complex Geometry}\label{ap:geo}
In this appendix, we give a lightning recapitulation of the key concepts from complex geometry which we use throughout the text.
Other than the canonical texts \cite{GH,hartshorne}, the reader is also referred to brief, self-contained expositions in \cite{Candelas:1987is,Bouchard:2007ik}, both of which do a wonderful job in rapid initiation.

\section{(Co-)Homology}
First, let us recall some standard facts on homology and cohomology of real manifolds.
For a compact, smooth, differential manifold $M$ of dimension $n$, we can define differential $p$-forms $\eta^{(p)}$ as elements of $\Omega^p := \bigwedge^p T^\vee_M$, the $p$-th antisymmetric power of the cotangent bundle $T^\vee_M$, as well as a differential operator $d : \Omega^p \rightarrow \Omega^{p+1}$ satisfying {\red $d^2 = 0$}.
Whence, we have the (de Rham) {\it cohomology} group $H^p_{dR}(M; K) = \ker(d)/ \im(d)$ for $p = 0, \ldots, n$, where $K$ is the set, typically $\IZ$ or $\IR$, wherein the coefficients for the linear combination of the elements take value.

We can take the dual to the above to define an appropriate {\it homology}, for instance, simplicial homology, where one approximates the cycles $C_q$ of $M$ of each dimension $q = 0, \ldots, n$ by the Abelian group freely generated  by the (standard) simplices at dimension $q$. Here we have a boundary operator $\partial : C_q \rightarrow C_{p-1}$, satisfying {\red $\partial^2 = 0$}, giving us the {\it homology} group $H_q(M; K) = \ker(\partial)/ \im(\partial)$ for $q = 0, \ldots, n$.
The duality is furnished by the inner product $H^p \times H_p \rightarrow K$, integrating a $p$-form over a $p$-cycle.
The non-degenerate, positive-definiteness of this inner product renders $H^p$ isomorphic to $H_p$.
The rank of $H^p$ (and thus also of $H_p$) is the $p$-th Betti number $b_p$ from which we obtain the Euler number of $M$ as
$\chi = \sum\limits_{p=0}^n (-1)^p b_p$.
Though we mention in passing, in the Prologue, bundle-valued cohomology groups, throughout this book we will not actually make use of them in any substantial way. Hence, we will not introduce \v{C}ech cohomology here.

Our running example of $T^2$, a manifold of (real) dimension 2, will serve to illustrate the above points.
We have 3 homology groups $H_{q=0,1,2}$, where $H_0$ is the class of the point (the torus is connected), and $H_2$, that of the entire $T^2$.
Thus $b_0 = b_2 = 1$.
Moreover, $H_1$ has two inequivalent generators: the 2 circles since $S^1 \times S^1 = T^2$. Thus, $b_1 = 2$.
The dual is to think of forms: let $x,y$ by the coordinates on $T^2$ (or of the plane of which $T^2$ is the quotient by a lattice), $H^0$ is just the constant function; $H^2$ is generated by $dx \wedge dy$; and $H^1$ is generated by $dx$ and $dy$. Thus again, $b^0 = b^2 = 1$ and $b^1 = 2$.
Importantly, $b_0 - b_1 + b_2 = 0$, which is the Euler number of $T^2$, consistent, via Gau\ss-Bonnet, with the fact the $T^2$ is Ricci-flat.

\section{From Real to Complex to K\"ahler}
Next, we recall the following sequence of specializations by imposing incremental structures to manifolds:
\begin{description}
\item[Riemannian] $M$ is endowed with a positive-definite symmetric metric $g$; 	
	With this metric, we can define the Hodge star operator $\star : \Omega^p \rightarrow \Omega^{n-p}$ as
	$\star(dx^{\mu_1} \wedge \ldots \wedge dx^{\mu_p}) :=
	\frac{\epsilon^{\mu_1\ldots\mu_n}}{\sqrt{|g|}} g_{\mu_{p+1}\nu_{p+1}} \ldots 
	g_{\mu_{n}\nu_{n}} dx^{\nu_{p+1}} \wedge \ldots \wedge dx^{\nu_n}$ so that a Laplacian $\Delta$ can be defined:
	\[
	\Delta_p =  dd^{\dagger} + d^{\dagger}d = (d + d^{\dagger})^2, \qquad
	d^{\dagger} := (-1)^{np+n+1} \star d \star \ .
	\]
	Importantly, Harmonic $p$-forms can be put into 1-1 correspondence with elements of cohomology, that is, 
	$\Delta_p \eta^{p} = 0$ $\stackrel{1:1}{\longleftrightarrow}$ $\eta \in H_{dR}^p(M)$.
	This allows us to solve for the zero-modes of the Laplacian to find representative in (co-)homology, rather than to compute the often more involved quotient ker/im. 
\item[Complex] $M$ is further endowed with a complex structure $J : T_M \rightarrow T_M$, inducing the split 
	$\Omega^p = \bigoplus\limits_{r+s=p} \Omega^{r,s}$ so that we have $(p,q)$-forms with $p$-holomorphic and $q$-antiholomorphic
	indices. Clearly, $n$ needs to be even at this point.
	Subsequently, the operator $d$ also splits to  $d = \partial + \bar{\partial}$, together with
	$\partial^2={\bar{\partial}}^2=\{\partial,\bar{\partial}\}=0$.
\item[Hermitian] The Riemannian metric on $M$ further obeys $g(J\_,J\_) = g(\_,\_)$, so that in local coordinates 
	$g_{\mu\nu} = g_{\bar{\mu}\bar{\nu}}=0$ and we only have the mixed components $g_{\mu\bar{\nu}}$.
	This is the Hermitian metric.
	At this point we can define {\blue Dolbeault Cohomology} $H^{p,q}_{\bar{\partial}}(X)$ as the cohomology of $\bar{\partial}$ 
	(we could equivalently use $\partial$) with Laplacian 
	$\Delta_{\partial} := \partial \partial^{\dagger} + \partial^{\dagger} \partial$ and
	similarly $\Delta_{\bar{\partial}}$.
\item[K\"ahler] With the Hermitian metric one can define a K\"ahler form $\omega := i g_{\mu\bar{\nu}} dz^{\mu} \wedge 
	dz^{\bar{\nu}}$ such that $d\omega=0$.
	This in particular implies that $\partial_{\mu} g_{\nu \bar{\lambda}} = \partial_{\nu} g_{\mu \bar{\lambda}}$ so that 
	$g_{\mu\bar{\nu}} = \partial \bar{\partial} K(z,\bar{z})$ for some scalar function (the K\"ahler potential) $K(z,\bar{z})$.
	Now, the Laplacian becomes $\Delta = 2\Delta_{\partial} = 2\Delta_{\bar{\partial}}$, and we have {\red Hodge decomposition}
	\begin{equation}\label{hodgedecomp}
		H^i(M; \IC) \simeq \bigoplus_{p+q = i}H^{p,q}(M) = \bigoplus_{p+q = i}H^{q}(M, \bigwedge^p T_M^\vee) \ , 
		\quad
		i = 0, 1, \ldots, n \ .
	\end{equation}
	In the above, we have written ordinary complex-valued cohomology in terms of Dolbeault cohomology in the first equality and in terms of bundle-valued cohomology in the second, where $T_M^\vee$ is the cotangent bundle of $M$, dual to the tangent bundle $T_M$, and $\wedge^p$ gives its $p$-th wedge (anti-symmetric tensor) power. 
\end{description}

\section{Bundles and Sequences}\label{s:bundles-seq}
The only bundles we use in the book are the tangent, the cotangent and line bundles over a projective space $A$, their restrictions to an algebraic variety $M$ living in $A$, as well as the normal bundle of $M$.
Thus, instead of boring the reader with standard definitions, let us simply illustrate with the case of projective space as well as a variety of degree $d$ therein.

Let us begin with $\IC\IP^1$. This is just $S^2$ as a real manifold.
To see this, we notice that $\IC\IP^1$ has 2 local patches by definition. The homogeneous coordinates are 
$[z_0:z_1]$, so patch $U_0$ is where $z_0 \neq 0$ and the local affine coordinates are $(1, z_1/z_0)$. Similarly, patch $U_1$ is where $z_1 \neq 0$ and the local coordinates are $(1, z_0/z_1)$. Clearly, each patch is homeomorphic to $\IC$, parametrized by a single complex coordinate, respectively $w_0 :=  z_1/z_0$ and $w_1 := z_0/z_1$. The transition function between the two patches is thus $w_1 = 1 / w_0$ at the intersection $U_0 \cap U_1 = \IC^* = \IC\IP^1 \backslash \{0, \infty \}$.

On the other hand, $S^2$ has 2 famous local patches, the north and south.  Embed $S^2$, with unit radius and center at the origin, into $\IR^3$, with Cartesian coordinates $(x,y,z)$. The two patches are the upper and lower hemispheres.
Projection from the north pole at $(0,0,1)$ puts the upper hemisphere in bijection with the plane $z=0$ through the equator, giving stereographic coordinates $(X_N, Y_N) := (\frac{x}{1-z}, \frac{y}{1-z})$.
Similarly, projection from the south pole at $(0,0,-1)$ puts the lower hemisphere in bijection with the $z=0$ plane with stereographic coordinates $(X_S, Y_S) := (\frac{x}{1+z}, -\frac{y}{1+z})$.
The complexification $Z := X_N + i Y_N$ and $W :=  X_S + i Y_S$ then furnishes coordinates $Z$ and $W$ respectively for the 2 patches, with transition function $W = \frac{x - i y}{1+z} = \frac{1-z}{1+z} (X_N - i Y_N) 
= \frac{X_N -i Y_N}{X_N^2 + Y_N^2}= 1/Z$. 

Thus, we see the same two patches, with the same transition functions between  $\IC\IP^1$ and $S^2$, whereby identifying them as (real) manifolds.
Unfortunately, the above algebra does not work for any of the higher even-dimensional spheres, and $S^2$ is the only sphere which is complex (and in fact also K\"ahler: being complex dimension 1, the differential of the K\"ahler 2-form trivially vanishes).

Let us proceed to {\it bundles} on $\IC\IP^1$.
Of course, there is the tangent bundle $T_{\IC\IP^1}$, which is  locally a 1 complex dimensional vector space, with coordinate $\frac{\partial}{\partial w_i}$ for $i=0,1$ for each of the 2 patches. Dual thereto is the cotangent bundle $T^\vee_{\IC\IP^1}$, a dual 1 complex dimensional vector space (locally), with coordinate $dw_i$.

The final important example of a vector bundle we will use is the bundle $\cO_{\IC\IP^1}(k)$ called the {\it line bundle of degree $k$}, which, when without ambiguity, is often just written as $\cO(k)$.
The construction, adhering to the 2 aforementioned patches with coordinates $w_0$ and $w_1$, is as follows.
Consider $U_{i=0,1} \times \IC$ with coordinates $(w_i, \zeta_i)$ so that locally each is $\IC \times \IC$.
Glue them by the transition function
$w_0 = 1/w_1$ and $\zeta_0 = \zeta_1 / (w_1)^k$ for some $k \in \IZ$ at the intersection.
This is the vector bundle of rank 1 (meaning that the local vector space is of complex dimension 1) and is denoted as
the line bundle $\cO_{\IC\IP^1}(k)$.

All these above discussions are readily generalized to $\IC\IP^n$, with $n+1$ local patches.
Some relevant properties of line bundles over projective space are
\begin{itemize}
\item The case of $k=0$, i.e., $\cO_{\IC\IP^n}$, is the trivial bundle over $\IC\IP^n$;
\item The global sections (i.e., globally defined continuous maps from the base $\IC\IP^n$ to the bundle) for $k>0$ are the homogeneous degree $k$ polynomials in the projective coordinates $w_0, w_1, \ldots w_n$, i.e., with monomials $w_0^{k_0} \ldots w_n^{k_n}$ as the basis; the trivial bundle $k=0$ has only zero section and $\cO_{\IC\IP^n}(k)$ has no sections for $k<0$;
The space of global sections is written as the bundle-valued cohomology group $H^0(\IC\IP^n, \cO_{\IC\IP^n}(k))$;
\item The {\it dual} bundle to $\cO_{\IC\IP^n}(k)$ is $\cO_{\IC\IP^n}(-k)$;
\item The canonical bundle $K := \wedge^n T^\vee_{\IC\IP^n}$ is simply the line bundle 
$\cO_{\IC\IP^n}(-n-1)$.
We see that $\IC\IP^n$ is Fano in that the anticanonical bundle $K^\vee$ is positive degree.
\item As degree $k$ polynomials are obtained by multiplication of the variables, we have that
	$\cO(k) = \cO(1)^{\otimes k}$, the tensor product over the degree 1 line-bundle;
\item Line bundles are rank 1, we can also form rank $m$ bundles by direct sum (Whitney sum) $\cO(k_1) \oplus \cO(k_2) \oplus \ldots \oplus \cO(k_m)$.
\end{itemize}

Everything we have said so far is on $\IC\IP^n$.
The entire field of algebraic geometry is about realizing manifolds as polynomials in appropriate ambient space.
In this book, we are only concerned with K\"ahler manifolds $M$ embedded into projective space (i.e., realized as projective varieties) or into toric varieties \footnote{Note that though infinite numbers of toric varieties are projective varieties, not all are.}, we can obtain bundles on $M$ by restriction.
For example, the CY$_1$ as an elliptic curve $\cE$ in $\IC\IP^2$ is a cubic hypersurface.
We can restrict the bundle $\cO_{\IC\IP^2}(k)$ on $\cE$. Care must be taken in general to ensure that this restriction remains a bundle (rather than a sheaf).
The most important fact we will need here is that the {\it normal bundle} which defines $\cE$ is the bundle prescribed by the degree of the embedding hypersurface, in other words, $N_{\cE} = \cO_{\IC\IP^2}(3)$.

We mentioned above that one can obtain higher rank bundles from the Whitney sum, which locally is just the direct sum of the vector spaces. Another non-trivial way is to build them from {\em exact sequences}.
Instead of launching into the vast subject of homological algebra, it suffices to review some generalities.
Suppose we have objects  (be they vector spaces, bundles, groups, etc.) $A_i$ as well as maps $f_i : A_i \to A_{i+1}$ so that a sequence can be written as
\begin{equation}\label{sequenceAi}
\ldots \stackrel{f_{i-1}}{\longrightarrow} A_i \stackrel{f_{i}}{\longrightarrow} A_{i+1} \stackrel{f_{i+1}}{\longrightarrow} \ldots \ .
\end{equation}
If $\ker(f_{i+1}) = \im(f_{i})$ for each $i$, then the sequence is called {\red exact}.
The most important case is when we have 3 objects $A,B,C$, maps $f: A \to B$, $g : B \to C$, and
\begin{equation}
0 \hookrightarrow A \stackrel{f}{\longrightarrow} B \stackrel{g}{\longrightarrow} C \to 0 
\qquad
\ker(g) = \im(f)\ .
\end{equation}
Here the first map from 0 is the trivial inclusion and the last map to 0 is the trivial map.
Such a case\footnote{
The prototypical example is 
$0 \stackrel{i}{\hookrightarrow} \IZ \stackrel{f}{\longrightarrow} \IZ \stackrel{g}{\longrightarrow} \IZ/(2\IZ) \stackrel{p}{\to} 0$ where $f$ is multiplication by $2$ and $g$ is integer modulo 2.
We see that (a) $\im(i) = \{0\}$ and $\ker(f) = \{0\}$; 
(b) $\im(f) = 2 \IZ$ since $f$ is multiplication by 2 and $\ker(g) = 2 \IZ$ since only even numbers reduce to 0 modulo 2; and (c) $\im(g) =  \{0,1\} = \IZ/(2\IZ)$ and $\ker(p) = \IZ/(2\IZ)$ since $p$ is the zero map so its kernel is the full pre-image.
Hence, the sequence is exact at every stage and we have a short exact sequence.
}
is called a {\it short exact sequence}.

The most important sequence of bundles we will use in this book is the Euler sequence for $\IC\IP^n$:
\begin{equation}
0 \to \cO_{\IC\IP^n} \to \cO_{\IC\IP^n}(1)^{\oplus (n+1)} \to T_{\IC\IP^n} \to 0 \ ,
\end{equation}
from which we have the useful corollary that for a projective variety $M \subset \IC\IP^n$,
\begin{equation}
0 \to T_M \to \left. T_{\IP^n}\right|_M \to N_{M / \IP^n} \to 0 \ .
\end{equation}
This states that the tangent bundle of the ambient space $\IC\IP^n$, when restricted to $M$, is composed of the tangent bundle and the normal bundle of $M$ inside the ambient space.
Note that locally, the middle term is just the direct sum \footnote{think of embedding $S^2$ into $\IR^3$, the direct sum of tangent space at a point and the normal vector together compose $\IR^3$}, but the structure of the bundle dictates that  globally this ``splitting'' does not hold and we need the short exact sequence.

Of course, not all sequences are exact.
When it is not so but nevertheless $\im(f_{i}) \subset \ker(f_{i+1})$ in \eqref{sequenceAi}, we can do something interesting.
Note that this subset condition, written in a more familiar way, is simply $f^2 = 0$.
Hence, we can define the homology of a sequence as $H_i := \ker(f_{i+1}) / \im(f_{i})$.
With this we are familiar.
When $A_i$ are the (vector) spaces of $i$-forms and $f_i$ are the differentiation operators $d_i$, $H_i$ is (de Rham) cohomology.
When $A_i$ are the freely generated Abelian groups of $i$-simplices and $f_i$ are the boundary operators $\partial_i$ ,
$H_i$ is (simplicial) homoogy.

The final fact on bundles and sequences we need in this book is the following.
Given a short exact sequence of bundles $0 \to V_1 \to V_2 \to V_3 \to 0$ on a manifold $M$, this induces a long exact sequence in bundle-valued cohomology $H(M, V)$:
\begin{equation}
\begin{array}{cccccccc}
0  & \to & H^0(M, V_1) & \to & H^0(M, V_2) & \to & H^0(M, V_3) & \to \\
    & \to & H^1(M, V_1) & \to & H^1(M, V_2) & \to & H^1(M, V_3) & \to \\
\end{array}
\end{equation}
The fancier way of saying this is that the cohomological functor $H^*(M, \bullet)$ is covariant on short exact sequences.

\section{Chern Classes}
One of the most distinguishing properties of vector bundles on manifolds is their characteristic classes, especially the Chern class, which we use in this book.
In this section, we will begin with the differential-geometric definition, before proceeding to the algro-geometric.

The Ricci curvature 2-form also assumes a particularly simple form for $M$ K\"ahler, which is local coordinates, is
\begin{equation}
R = -i \partial_{\mu} \bar{\partial}_{\bar{\nu}} \log \det (g) dz^\mu \wedge d\bar{z}^{\bar{\nu}} \ .
\end{equation}
This is a $(1,1)$-form which as a real 2-form is closed.
Thence, we can define the {\red Chern classes} $c_k(M) = c_k(T_M) \in H^{2k}(M)$ as
\begin{equation}
\det\left( \II_{n \times n} + \frac{it}{2\pi} R \right) = \sum\limits_{k=0}^{n/2} c_{k}(M) t^k \ .
\end{equation} 

As we will always take a more algebraic rather than differential approach, we can think of the Chern classes more axiomatically \cite{hartshorne}. 
In the above we defined the Chern classes of $M$, understood as those of the tangent bundle $T_M$; they can be defined for arbitrary bundles.
\begin{definition}\label{chern}
Let $E$ be a complex vector bundle of rank $r$ over our manifold $M$ (i.e., locally, $E \simeq M \times \IC^r$), then the Chern classes 
$c_k(E) \in H^{2k}(M; \IZ)$, together with their associated formal sum (the total Chern class)
\[
c(E) := c_0(E) + c_1(E) + \ldots + c_r(E)
\ ,
\] 
obey the axiomata
\begin{itemize}
\item Leading term: $c_0(E) = 1$ for any $E$;
\item Naturality: if $f: N \to M$ is a continuous morphism from manifold $N$ to $M$ and $f^*(E) \to N$ is the pull-back vector bundle, then
	\[
	c_k(f^*E) = f^* c_k(E) \ ;
	\]
\item Whitney sum: If $F \to X$ is another complex bundle on $M$, then
	\[
	c(E \oplus F) = c(E) \wedge c(F) \qquad \leadsto \qquad c_k(E \oplus F) = \sum\limits_{i=0}^k c_i(E) \wedge c_{k-i}(F) \ ;
	\]
\item Normalization: For complex projective space $\IC\IP^n$, let $H$ be the Poincar\'e dual to the hyperplane class $\IC\IP^{n-1} \subset \IC \IP^n$, then for the degree 1 bundle $\cO_{\IP^n}(1)$ whose transition functions are linear (the dual $\cO_{\IP^n}(-1)$ is the tautological bundle),
	\[
	c(\cO_{\IP^n}(1)) =  1 + H \ .
	\]
\end{itemize}
\end{definition}
We remark that the Whittney sum generalized to short exact sequence of bundles (a {\it splitting principle} due to Grothendieck) gives
\begin{equation}\label{ssChern}
0 \to E \to G \to F \to 0 \quad \Rightarrow \quad c(G) = c(E) \wedge c(F) \ .
\end{equation}
Furthermore, for the {\em dual bundle} $E^\vee$, we have that
\begin{equation}
c_i(E^\vee) = (-1)^i c_i(E) \ .
\end{equation}

We can reorganize the Chern class into a {\em character} from $(\oplus, \otimes) \to (+,\wedge)$ called the {\it Chern character} $\ch(~)$.
\begin{equation}
\ch(E \oplus F) = \ch(E) + \ch(F) \ , \qquad \ch(E \otimes F) = \ch(E) \wedge \ch(F) \ ,
\end{equation}
The total Chern class does not enjoy this nice property, 
For a line bundle (i.e., rank 1) $L$, the Chern character is
$\ch(L) := \exp\left( c_1(L) \right)$, and more generally for a rank $r$ bundle $E$, using the splitting principle into line bundles, $E = \bigoplus\limits_{i=1}^r L_i$, we have that
\begin{align}
\nn
\ch(E) & := \sum\limits_{i=1}^r \exp\left( c_1(L_i) \right) = \sum\limits_{m=0}^\infty \frac{1}{m!} \sum\limits_{i=1}^r c_1(L_i)^m 
 =  \ch_(E) + \ch_1(E) + \ldots \quad \\
 \nn
 \mbox{ where } 
 \\
 \nn
& \ch_0(E) = \mbox{rk}(E), \quad \\
\nn
& \ch_1(E) = c_1(E), \quad   \label{Ch} \\
\nn
& \ch_2(E) = \frac12(c_1(E)^2-2c_2(E)), \\  & \ch_3(E) = \frac16(c_1(E)^3 - 3 c_1(E)c_2(E) + 3 c_3(E)) \ ,  \ldots
\end{align}
The recasting of $\ch_i$ in terms of the $c_i$ is evidently an exercise in Newton symmetric polynomials.
The Chern character of more sophisticated combinations such as antisymmetric products, crucial in physics, can be similarly obtained (cf.~Appendix A of \cite{Donagi:2004ia}).

Finally, we will also be making use of the Todd class.
It is yet another combination of the Chern classes, which is multiplicative over $\otimes$, i.e.,
\begin{equation}
\td(E \oplus F) = \td(E) \wedge \td(F) \ ,
\end{equation}
Much like the Chern class, we have the expansion:
\begin{align}
\nn
\td(E) &= 1 + \td_1(E) + \td_2(E) + \td_3(E) + \ldots \ , \mbox{ where } \\
\nn
& \td_1(E) = \frac12 c_1(E), \quad \\
\nn
& \td_2(E) = \frac{1}{12} (c_2(E)+c_1(E)^2), \quad \\
& \td_3(E) = \frac{1}{24} (c_1(E)c_2(E)) \ , \ldots
\label{Td}
\end{align}
In general, the above are obtained from the series expansion of $x(1-e^{-x})^{-1}$.
One can invert the above to obtain, for instance, the top Chern class, from an identity of Borel-Serre (cf.~3.2.5 of \cite{fultonIT})
\begin{equation}\label{BS}
\sum\limits_{i=0}^r (-1)^i \ch(\bigwedge^i E^\vee) = c_r(E) \td(E)^{-1} \ . 
\end{equation}

\section{Covariantly Constant Spinor}\label{ap:geoSpinor}
We discussed in the Prologue that physicists first came across Ricci-flat, K\"ahler manifolds by considering heterotic string compactifications.
While we refer the reader to the excellent pedagogical material in Vol. 2 of \cite{gsw}, especially the chapters on algebraic geometry, it is helpful to summarize the key arguments here.

In brief, the authors of \cite{Candelas:1985en} take the effective 10-dimensional action 
\begin{equation}
S \sim \int d^{10} x \sqrt{g} e^{-2\Phi} \left[R + 4 \partial_\mu \Phi \partial^\mu \Phi - \frac12 |H_3'|^2) - \frac{1}{g_s^2} \tr|F_2|^2 \right]+SUSY
\end{equation}
of the heterotic string (where we write only the bosonic part and $+SUSY$ is understood to be the supersymmetric completion) and considers \footnote{
One way to think of supersymmetry, regardless of experimental results and from a philosophical point of view, is to ask ``are complex numbers part of reality?''
We know that the complex numbers complete the reals in a canonical way, and has become indispensable even to engineers.
Supersymmetry completes QTFs in a canonical way, and should likewise be thought of as the natural realm in which to study elementary physics.
} the supersymmetric variation $\delta$ on the various fields:
\begin{equation}
	\begin{array}{lcl}
	\mbox{gravitino} && \delta_{\epsilon} \Psi_{m'=1,\ldots,10} = 
		\nabla_{m'} \epsilon - \frac14 H_{m'}^{(3)} \epsilon \\
	\mbox{dilatino} && \delta_{\epsilon} \lambda = -\frac12 \Gamma^{m'} \partial_{m'} \Phi \ \epsilon 
		+ 	\frac14 H_{m'}^{(3)} \epsilon \\
	\mbox{adjoint Yang-Mills} && \delta_{\epsilon} \chi = -\frac12 F^{(2)} \epsilon\\
	\mbox{Bianchi} && dH^{(3)} = \frac{\alpha'}{4} [ \tr(R\wedge R) - \tr(F\wedge F)]	
	\end{array}
\end{equation}
and demand that these all vanish so that supersymmetry is preserved.
Assume that the 3-form field $H^{(3)} = 0$ (this is a strong assumption and can be generalized, for which there has been much recent work, cf.~e.g.~\cite{delaOssa:2014lma}), we have that upon dimensional reduction $\IR^{1,9} \simeq \IR^{1,3} \times M$ for some compact (internal) 6-manifold $M$, 
\begin{equation}
0 = \delta_{\epsilon} \Psi_{m'=1,\ldots,10} = \nabla_{m'} \epsilon =  \nabla_{m'} \xi(x^{\mu=1,\ldots,4}) \eta(y^{m=1,\ldots,6}) \ , 
\end{equation}
giving us the Killing spinor equation on $M$.

Now, the external space, being Minkowski, is flat so that $[\nabla_{\mu}, \nabla_{\nu}] \xi(x) = 0$.
Thus, the internal space $M$ is also Ricci-flat and admitting not only a spinor (and is thus a spin manifold), but in fact a covariant constant spinor
\begin{equation}
\nabla_m \eta = 0 \ .
\end{equation}
Defining $\eta_+^* = \eta_-$ to split chirality to the original 10-dimensional spinor as
$\epsilon(x^{1,\ldots,4},y^{1,\ldots,6}) = \xi_+ \otimes \eta_+(y) + \xi_- \otimes \eta_-(y)$, we have the following set of definitions which sequentially specialize $M$:
\begin{itemize}
\item Define $J_m^n = i \eta_+^{\dagger} \gamma_m^n \eta_+ = 
	-i \eta_-^{\dagger} \gamma_m^n \eta_-$, one can check that $J_m^n J_n^p = -\delta_m^n$ for Clifford Gamma matrix $\gamma_m^n$.
	Therefore, $(M, J)$ is {\blue almost-complex};
\item However,  $\eta$ covariantly constant $\leadsto \nabla_m J_n^p = 0 \leadsto \nabla N_{mn}^p = 0$ so that the
	Nijenhuis tensor $N_{mn}^p := J_m^q \partial_{[q} J^p_{n]} - (m \leftrightarrow n)$.
	Therefore, $(M, J)$ is {\blue complex}
	(in particular, one can choose coordinates $(z,\bar{z})$ so that $J_m^n = i\delta_m^n, \ 
	J_{\bar{m}}^{\bar{n}} = i\delta_{\bar{m}}^{\bar{n}}, \
	J_{\bar{m}}^{n} = J_{m}^{\bar{n}} = 0$, rendering the transition functions holomorphic);
\item Define $J = \frac12 J_{mn} dx^m \wedge dx^n$ with $J_{mn} := J_m^k g_{kn}$.
	One can check that $dJ = (\partial + \bar{\partial})J = 0$, so that $(M,J)$ is {\blue K\"ahler}.
\end{itemize}

Now, for a spin 6-manifold, the holonomy group is generically $SO(6) \simeq SU(4)$.
The existence of the covariantly constantly spinor implies that the holonomy is reduced: in the decomposition
${\bf 4} \to {\bf 3} \oplus {\bf 1}$ of $SU(4) \rightarrow SU(3)$, this covariantly constant spinor corresponds to the {\bf 1}.
Thus, in summary, our internal manifold $M$ is K\"ahler, dim$_{\IC}=3$, and with $SU(3)$ holonomy.

\section{A Lightning Refresher on Toric Varieties}\label{ap:toric}
This section, as mentioned in the main text, is not meant to be a review on toric geometry, of which there are may excellent sources in the literature \cite{fultontoric,schencktoric,Bouchard:2007ik,Closset:2009sv,toricAlg}.
We will use the quadric hypersurface, the ``conifold'',
\begin{equation}\label{coniabcd}
\{ab = cd \} \in \IC^4_{(a,b,c,d)} \ ,
\end{equation}
to illustrate the 3 equivalent definitions of an affine toric variety.
As in the standard case for manifolds, these affine patches, with the appropriate transition functions, can be glued to a compact toric variety.

\paragraph{(i)~Combinatorics of Lattice Cones: }
The most standard definition of an affine toric variety is via
\begin{definition}
A convex polyhedral cone is the set $\sigma := \{r_1 v_1 + \ldots + r_s v_s \in V : r_i \in \IR_{\geq 0} \}$ for $v_i \in V$.
Here $V = N_{\IR} = N \otimes_{\IZ} \IR$ is a vector space obtained from a lattice $N \simeq \IZ^n$. 
\end{definition}
In other words, $\sigma$ is the cone with tip at the origin and whose edges are rays determined by lattice vectors.
The lattice $N$ has its dual lattice $M = \hom(N, \IZ)$ by the inner product $\left<M,N\right>$, as well as its own associated vector space $M_{\IR} = M \otimes_{\IZ} \IR$.

The {\bf dual cone} $\sigma^\vee$ is the set of vectors in $M_{\IR}$ whose inner product with all vectors in $\sigma$ are non-negative.
The affine toric variety $U_{\sigma}$ for $\sigma$ is then
\begin{equation}\label{affinetoric}
U_\sigma := \mbox{Spec} \left( \IC[\sigma^\vee \cap M] \right) \ , 
	\qquad \sigma^\vee \cap M := \{u \in M : \left< u, v \right> \geq 0 ~ \forall v \in \sigma \} \ .
\end{equation}
In the physics literature, the set of end points of the integer vector generators of  $\sigma^\vee \cap M$ is called the {\bf toric diagram}.

\begin{figure}
(a) $\begin{array}{c}\includegraphics[width=3in]{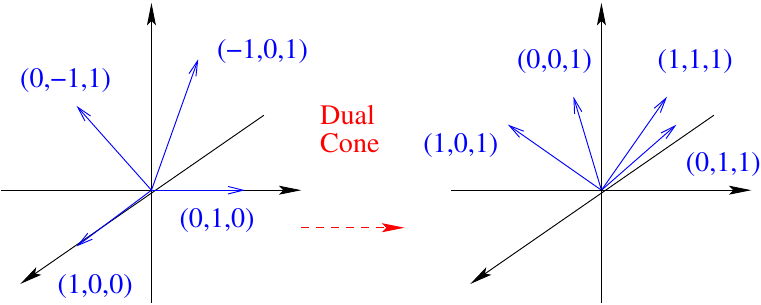}\end{array}$
\qquad
(b) $\begin{array}{c}\includegraphics[width=1in]{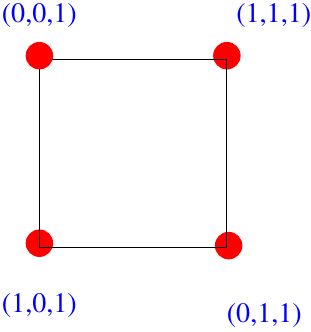}\end{array}$
\caption{{\sf
(a) the cone and dual cone; (b) the toric diagram, for the conifold.
}
\label{f:toriceg}}
\end{figure}

\subsection{Digressions on Spec and all that}\label{ap:spec}
Since we have made repeated use of the symbol ``Spec'' in this book so it might be worth to explain it a little and here is as good a place as any other.
We first recall that in a commutative (every ring in this book is commutative, in fact, we will almost exclusively consider polynomials rings) ring $(R,+,\times)$, an ideal $I$ is a subgroup under $+$ closed under $\times$. That is, $(I,+)$ is a subgroup of $(R,+)$ such that for every $r \in R$ and every $x \in I$, we have that $r x \in I$.
A maximal ideal is, as the name suggests, a proper ideal such that no other proper ideal contains it.

For instance, in the polynomial ring $\IC[x]$, (i.e., set of polynomials in a single variable $x$ with complex coefficients that constitute a ring by ordinary $+$ and $\times$), the set of polynomials which are of the form $(x-1) f(x)$ (i.e., polynomials with $(x-1)$ as a factor) is an ideal which is maximal.
This ideal is denoted as $\left< x-1 \right>$. 
The set of polynomials of the form $(x-1)(x-2) f(x)$ is also an ideal, denoted as $\left< (x-1)(x-2) \right>$.
But it is not maximal because $\left< (x-1)(x-2) \right> \subset \left< (x-1) \right>$.

Now, we get to the heart of the matter by defining \footnote{
Sometimes this is written explicitly as $\spec_M(R)$ to show that we are dealing with the set of maximal ideals.
Another commonly used concept is $\spec_P(R)$, the set of {\it prime ideals} in $R$. Since we are not going into prime ideals here, our $\spec$ will merely refer to $\spec_M$.
}
\begin{definition}
$\spec(R) := \{ \mbox{maximal ideals in } R \} \ ,$ 
called the (maximal) spectrum of the ring $R$.
\end{definition}
Whilst this may seem like an artificially constructed set, the geometric meaning is profound.
Returning to our example of $R = \IC[x]$, it is clear that every maximal ideal is of the form $\left< (x-a) \right>$ for an arbitrary $a \in \IC$. In other words, because we have complete factorization of polynomials over $\IC$, any polynomial can be uniquely factorized into linear factors -- the fundamental theorem of algebra helps us here and this is, again, the reason we work over $\IC$, rather than over $\IR$, say.
Hence, $\spec(\IC[x]) \simeq \IC$.
In general, we have that, for the polynomial ring in $n$ variables,
\begin{equation}\label{specCn}
\spec(\IC[x_1, x_2, \ldots, x_n]) \simeq \IC^n \ .
\end{equation}
The key point is that we now have a purely algebraic way of describing the geometrical familiar object: the affine space $\IC^n$.

As this book is about polynomial systems in complex variables defining manifolds as varieties, it is natural to ask how to extend \eqref{specCn} to an algebraic variety $X \subset \IC^n$.
First, as mentioned in the main text, a polynomial system $S$ itself constitutes an ideal: its vanishing locus which defines the manifold remains so by any multiplication by an element of the polynomial ring. 
It is standard to denote this ideal as $I(X) \in \IC[x_1, x_2, \ldots, x_n]$.

As groups can quotient normal subgroups, so too can rings quotient ideals.
Consider, therefore, the quotient ring $R_X = \IC[x_1, x_2, \ldots, x_n] / I(X)$.
In this ring, polynomial $+$ and $-$ proceed as normal, but modulo the defining polynomial system $S$.
For example, suppose we have a hypersurface in $\IC^2$ defined by $y= x^2$, then whenever we see an $x^2$, we need to substitute it by $y$.
This quotient ring $R_X$ thus captures the property of the variety $X$ and is called the {\red coordinate ring} of $X$.
The set of maximal ideals on $R_X$ are likewise those on $ \IC[x_1, x_2, \ldots, x_n]$, but modulo the defining polynomials \footnote{
Note that in order to simplify discussions, we have glossed over technicalities of ideals defined, for example, by powers of the polynomial system, which obviously give the same geometry.
In other words, we have bypassed the discussions on radicals of ideals and the Nullstellensatz as well as distinctions between varieties and schemes.
}.
Thus, in a similar fashion, we have a correspondence between points on the variety $X$ and $\spec(R_X)$.

Everything said above was for affine (non-compact) varieties, there is an graded analogue for $\spec$, which is ``Proj'', which should be applied to projective (compact) varieties. For example, the projective space itself, is
\begin{equation}
{\rm Proj}(\IC[x_0 :  x_1 : \ldots : x_n])  \simeq \IC\IP^n \ .
\end{equation}
This is the starting point for playing with varieties in software such as {\sf Macaulay2} \cite{m2}.

\subsection{Example: The Conifold}
The diagrams in Figure \ref{f:toriceg} will reify the above concepts.
In part (a), the cone is spanned by the four vectors 
\begin{equation}
\{(0,-1,1),\ (-1,0,1), \ (0,1,0),\ (1,0,0) \}
\end{equation}
in $\IZ^3$ (so the resulting variety is complex dimension 3).
One can readily check that the dual cone is spanned by the four vectors 
\begin{equation}\label{dualconeConi}
\{(0,0,1),\ (1,1,1), \ (0,1,1),\ (1,0,1) \} \ ,
\end{equation}
so that all pair-wise inner products of the generators of $\sigma^\vee$ and $\sigma$ are non-negative.
In part (b), the end points  of the generators of $\sigma$ and $\sigma^\vee$ are co-planar, as is in congruence with the fact that our hypersurface is Calabi-Yau. The diagram is a square of unit area and the last coordinate is 1 (of course, any $GL(3;\IZ)$ transformation on the diagram gives the same variety).
In fact, dropping the last coordinate gives the convex polygon, here the square, which is the diagram used in the main text.

We treat the coordinates of these generators of the dual cone as exponents of 3 complex variables $(x,y,z)$, giving us 
\begin{equation}
S_\sigma = \langle a = z,\ c = yz,\ b = xyz, \ d = xz \rangle \ .
\end{equation}
Finally, computing the maximal spectrum of the polynomial ring $\IC[S_\sigma]$ means that we 
need to compute $\spec \IC[a = z,\ c = yz,\ b = xyz, \ d = xz]$.
In other words, as introduced in \S\ref{ap:spec}, we need to find any basic relations among $(a,b,c,d)$.
Clearly, there is one quadratic relation: $(z)(xyz) = (yz)(xz)$, this gives the hypersurface
\eqref{coniabcd}, the conifold as desired.

\paragraph{(ii)~K\"ahler Quotients: }
The second definition is a direct generalization of projective space $\IP^n$, which we recall to be  $\IC^{n+1}$ (with coordinates $x_{i=0,\ldots,n}$ quotiented by a $\lambda \in \IC^*$ action $x_i \mapsto \lambda x_i$.
Now, consider $\IC^q$ with homogeneous (Cox) coordinates $x_{i=0,\ldots,q}$ and a $(\IC^*)^{q-d}$ action
\begin{equation}
x_i \mapsto \sum\limits_a \lambda_a^{Q_{i=1...q}^{a=1...q-d}} x_i \ , \qquad \lambda_a \in \IC^* \ , Q_i^a \in \IZ \ ,
\end{equation}
where $Q$ is called a charge matrix, the integer kernel of which is given by
\begin{equation}
\sum\limits_{i=1}^d Q_i^a v_i = 0 \ ,
\end{equation}
and the generators $v_i$ is the toric diagram.

For our running example, take $Q = [-1,-1,1,1]$, as a $\IC^*$ action on $\IC^4$, so that
{\scriptsize $
\ker(Q) = 
\left(
\begin{array}{llll}
 1 & 0 & 0 & 1 \\
 1 & 0 & 1 & 0 \\
 1 & 1 & 1 & 1
\end{array}
\right)
$}
and the 4 columns are precisely \eqref{dualconeConi}.

\paragraph{(iii)~Computational Definition: } 
This is the simplest definition, the defining equation $ab - cd$ for the conifold is known as a {\it binomial ideal}, i.e., every defining equation is of the form of ``monomial = monomial''. It turns out that \cite{toricAlg} every such polynomial ideal defines a toric variety.
Interestingly, this condition is crucial to the bipartite interpretation of quiver representation variety for toric CY3, as in Theorem \ref{toricThm}.

\subsection{From Cones to Fans}
Everything said so far about toric varieties has been about {\it affine} ones, where a single (convex, polyhedral, lattice) cone gives a monomial ideal from \eqref{affinetoric}.
From the point of view of local Calabi-Yau space, this is sufficient.
Indeed, as mentioned in the main text, so long as the end-points of the generators of the cone are co-hyperplanar, the result is affine CY.
However, this condition, as we will see shortly, immediately rules out compact toric CY:
\begin{theorem}
There are no compact {toric} Calabi-Yau manifolds.
\end{theorem}
Note that it is certainly fine for the {\it ambient} variety to be toric, such as $\IC\IP^4$ for the quintic as discussed in detail in \S\ref{s:Q}.

As one forms a (compact) manifold out of affine patches by gluing, so too can one construct a compact toric variety by gluing affine ones.
Let us take $\IC\IP^2$ as a simple example.
We recall from \S\ref{s:bundles-seq} that $\IC\IP^2$ has 3 patches: using the homogeneous coordinates $[z_0: z_1: z_2]$, they are $U_{i = 0,1,2} \simeq \IC^2$ with coordinates $w_j^{(i)} = z_j / z_i$ for $i \neq 0$.
The transition function from $U_{i_1}$ to $U_{i_2}$ at their intersection where $z_{i_1}, z_{i_2} \neq 0$ is $w_j^{(i_1)} = w_j^{(i_2)} z_{i_2} / z_{i_1}$.

All of this information can be phrased torically.
Let us look back to Figure \ref{f:toricP2}.
There are 3 cones $\sigma_{0,1,2}$.
The cone $\sigma_0$ has generators $(0,1)$ and $(-1,-1)$; its dual cone has generators $(-1,0)$ and $(-1,1)$, so the coordinate ring is $\IC[1/x, y/x]$.
The cone $\sigma_1$ has generators $(1,0)$ and $(-1,-1)$; its dual cone has generators $(0,-1)$ and $(1,-1)$, so the coordinate ring is $\IC[1/y, x/y]$.
The cone $\sigma_2$ has generators $(0,1)$ and $(1,0)$; its dual cone has generators $(1,0)$ and $(0,1)$, so the coordinate ring is $\IC[x,y]$.
Thus, taking the maximal spectrum, each is just $\IC^2$.
The 3 cones together form a {\bf fan}.
Thus, if we defined $x := z_1/z_0$ and $y := z_2/z_0$, the cones $\sigma_{0,1,2}$ can be identified with the above ordinary patches $U_{1,2,0}$ respectively.

We have therefore given a toric description of $\IC\IP^2$ as a compact variety.
The fan can be re-packaged into a convex lattice polytope by the face-fan construction discussed near Figure \ref{f:f:toricP2}.
Some important results (q.v.~\cite{fultontoric,schencktoric}) are the following (note that since our expositions here are informal, we have not used the standard but technical terms such as the $M$ and $N$ lattices).
\begin{theorem}
The $n$ complex dimensional toric variety is compact if the fan is complete, in the sense that the collection of cones covers all of $\IR^n$.
\end{theorem}
For example, we can see the 3 cones of $\IC\IP^2$ covering the plane.
Thus, the co-hyperplanar condition already renders a compact toric CY impossible. 
\begin{theorem}
The toric variety is smooth if each cone is regular, in the sense that the determinant of the matrix, formed by generators as lattice vectors, is $\pm1$.
\end{theorem}

\section{Dramatis Personae}\label{ap:dPF}
Recall the trichotomy of $g=0$, $g=1$ and $g>1$ from Figure \ref{f:RiemannSurface}.
As Riemann surfaces/complex manifolds, these are respectively $S^2 \simeq \IP^1$ (positive curvature), $CY_1 \simeq T^2$ (zero curvature) and hyperbolic (negative curvature).
This trichotomy \cite{hartshorne} in complex dimension 1 persists to general K\"ahler manifolds and are referred to as (1) {\red Fano}; (2) {\red Calabi-Yau} and (3) {\red general type}, the first two falling into some classifiable families whilst the third usually proliferates untamably.
More strictly, for a (complete) complex manifold $M$ and consider its anti-canonical sheaf $K_M^\vee := (\wedge^n T^\vee_M)^{\vee}$, if it is trivial, $M$ is Calabi-Yau and if it is ample, then $M$ is Fano.

In complex dimension two, the Calabi-Yau case consists of $T^4$ and K3, while the Fano family also enjoys a wealth of structure.
Clearly, $\IP^2$ and $\IP^1 \times \IP^1$ are Fano, the first since $K_{\IP^n}^{\vee} = \cO_{\IP^n}(n+1)$, and the second, essentially by K\"unneth.
Interestingly, there are 8 more. To introduce these, we need the concept of a {\red blow-up}.

In brief, a blow-up of a point in an $n$-fold is to replace the point with a $\IP^{n-1}$.
The general definition is rather abstruse (cf.~\S7, circa Prop 7.11 of \cite{hartshorne} but locally the definition is straight-forward.
Consider a codimension $k$ subvariety $Z \subset \IC^n$ with affine coordinates $x_{i=1, \ldots, n}$, defined, for simplicity, by $x_1 = x_2 = \ldots = x_k = 0$.
Let $y_i$ be the homogeneous coordinates of $\IP^{k-1}$, then the blow-up $\widetilde{\IC^n}$ of $\IC^n$ along $Z$ is defined by the equations 
\begin{equation}
\{x_i y_j = x_j y_i \}_{i,j=1,\ldots,k} \subset \IC^n \times \IP^{k-1} \ .
\end{equation}
Thus defined, away from $Z$, $\widetilde{\IC^n}$ is just $\IC^n$ as the non-identical-vanishing of $x_i$ fixes $y_i$ to be a single point on $\IP^{k-1}$ (i.e., we project back to $\IC^n$), but along $Z$, where $x_i$ vanish, we have arbitrary $y_i$, parametrizing $\IP^{k-1}$.
Thus, we have ``replaced'' $Z$ by $\IP^{k-1}$.

\paragraph{del Pezzo Surfaces: }
We apply the aforementioned to $\IP^2_{[X_0:X_1:X_2]}$.
Take the point $P = [0:0:1]$ without loss of generality, and work on the affine patch $X_2 \neq 0$ so that $(x,y) = (X_0/X_2, X_1/X_2) \in \IC^2$.
We blow the point $P$ up by ``replacing'' it with a $\IP^1_{[z:w]}$.
Then the blowup of $\IP^2$ at $P$ is given by 
\begin{equation}
\{ \big(  (x,y); \  [z:w] \big) : x z + y w = 0 \} \subset \IC^2 \times \IP^1 \ .
\end{equation}
At points away from $P$ where $(x,y)$ are not simultaneously 0, we can fix a value of $[z:w]$ but at $P$ where $(x,y)=(0,0)$, $[z:w]$ are arbitrary and parametrizes a full $\IP^1$.
The result is a smooth complex surface, denoted $dP_1$, which is $\IP^2$ blown up at a point.

Now, take $r$ generic points \footnote{
Generic means that no 3 points are colinear, no 4 points are coplanar, etc., so that the points are in general position, i.e., their projective coordinates can be taken as $n$ random triples $[x_i:y_i:z_i]$.
} on $\IP^2$, and perform successive blow-up.
We denote the result as $dP_r$, the $r$-th del Pezzo surface (with $r=0$ understood to be $\IP^2$ itself).
Letting $\ell$ be the hyperplane class of $\IP^1$ in $\IP^2$ and $E_{i=1,\ldots,r}$ be the {\blue exceptional divisors} corresponding to the various $\IP^1$ blow-ups, the intersection numbers are
\begin{equation}\label{intdPr}
\ell \cdot \ell=1,\quad \ell \cdot E_i=0,\quad E_i \cdot E_j=-\delta_{ij} \ ,
\end{equation}
and the Chern classes are given by
\begin{equation}
c_1(dP_r) = 3\ell - \sum_{i=1}^r E_i,\quad c_2 (dP_r) = 3+r.
\end{equation}
The canonical class $K_{dP_r}$ is $c_1(dP_r)$ and its self-intersection is the {\bf degree}:
using \eqref{intdPr}, we see that the degree is $c_1(dP_r)^2 = 9-r$.
For instance, the degree 3 surface in $\IP^2$, on which there are the famous 27 lines, corresponds to $r=6$ blowups.

From the degree, we see the $r \leq 8$, thus there are only $8+1$ del Pezzo surfaces ($r=0,\ldots,8$) which are Fano \footnote{
When $r=9$, we have $c_1(dP_r)^2 = 0$ so that $dP_9$ is not Fano. It has been jocundly called half-K3 because while the first Chern class is not zero, it ``squares'' to zero. In the mathematics literature, del Pezzo surfaces and Fano surfaces are synonymous in that Fano varieties in complex dimension 2 are called del Pezzo, but in the physics literature, $dP_9$ has been added to the list of del Pezzo surfaces.
}.
It is remarkable that they correspond to the exceptional simply-laced series in semi-simple Lie algebras, viz., 
$E_{r = 8,7,6,\ldots,1}$ with the convention that
\begin{equation}
E_{5,4,3,2,1} := (D_5, D_4, A_2 \times A_1, A_2, A_1) \ ,
\end{equation}
 in the precise sense that the matrix of intersection forms in \eqref{intdPr} is the adjacency matrix of the Dynkin diagram for affine $E_r$ (with the extra affine node corresponding to $\ell$).
Finally, we remark that $dP_{r=1,2,3}$ are toric varieties.

\paragraph{Hirzebruch Surfaces: }
The other of the two basic Fano surfaces, $\IP^1 \times \IP^1$, also belongs to a notable family.
Consider surfaces which are $\IP^1$-fibrations over $\IP^1$.
It turns out that such surfaces are classified by a single integer, viz, the number of self-intersections of the class of the base \cite{fultontoric}, a positive integer $r$, and the surface is called the $r$-th Hirzebruch surface, denoted $\IF_r$. 

The simplest case is when the fibration is trivial and the surface is $\IF_0 := \IP^1 \times \IP^1$.
In general, let  $f$ be the class of the fiber $\IP^1$ and $b$, the class of the base $\IP^1$, then the intersection numbers are
\begin{equation}
f \cdot f = 0 \ , \quad f \cdot b = 1 \ , \quad b \cdot b = - r \ .
\end{equation}
The Chern classes are
\begin{equation}
c_1(\IF_r)  = 2b + (r+2) f \ , \quad c_2(\IF_r) = 4 \ .
\end{equation}
All members of the Hirzebruch family are toric (q.v.~\S1.1 of \cite{fultontoric}).

Of the family $\IF_r$, it turns out that only $r=0$ is Fano.
Clearly, we could perform successive one-point blowups on $\IP^1 \times \IP^1$.
However, $\IF_0$ blown up at 1 point is just $dP_2$ (this can be seen from the toric diagrams).
Therefore, in all we have 10 Fano surfaces, organizing themselves in a curious fashion:
\begin{equation}\label{fanoS}
\begin{array}{cccccccccccc}
&& &&\IF_0 &&&&&&\\
&& &&\downarrow &&&&&&\\
dP_0 = \IP^2 & \to & dP_1  & \to & dP_2 & \to & \ldots & \to & dP_7 & \to & dP_8
\end{array}
\end{equation}
where ``$\to$'' means blow-up by one point.
Finally, we remark that of this sequence, $dP_{0,1,2,3}$ and $\IF_0$ are toric varieties but not the others.
The toric diagrams for these are giving in Figure \ref{f:z3z3reso} in the text.
Incidentally, it was noticed in \cite{Iqbal:2001ye} that this is exactly the structure of M-theory/string theory compactifications on successive circles, a mysterious duality indeed!

Thus, in analogy of the trichotomy of Riemann surfaces/complex curves of Figure \ref{f:RiemannSurface}, the situation in complex dimension 2 is\begin{equation}
\begin{array}{c|c|c}
\begin{array}{c}
\mbox{10 Fano surfaces in } \eqref{fanoS} \\
dP_{n=0, \ldots, 8}, \ \IP^1 \times \IP^1
\end{array}
& 
\begin{array}{c}
CY_2 = \\
T^4, \ K3 
\end{array}
& \mbox{Surfaces of general type} 
\\ \hline
+ \mbox{ curvature} & 0 \mbox{ curvature} & - \mbox{ curvature}
\end{array}
\end{equation}

The complex dimension 3 case is, on the other hand, much more complicated and this book is devoted to the middle column of CY3s, of which there already is a superabundance.

\section{The Kodaira Classification of Elliptic Fibrations} \label{ap:Kodaira}
We mentioned elliptically fibered CYs in the text and stated that this can be achieved by promoting the coefficients in the Weierstra\ss\ equation to functions.
In this appendix, let us illustrate this explicitly with elliptically fibered surfaces, especially as an invitation to Kodaira's beautiful classification \cite{kodaira}, which is yet another emergence of the ADE meta-pattern (q.v.~\S\ref{s:mckay}).

Let us consider a complex surface $S$ which is an elliptic fibration over a $\IP^1$ base whose projective coordinate is $w$.
This means that we can write the equation of $S$ in Weierstra\ss\ form \eqref{weierstrass} as (at least in the $z=1$ patch):
\begin{equation}
y^2 = 4x^3 - g_2(w) x  - g_3(w) \ ,
\end{equation}
where $g_2$ and $g_3$ are arbitrary polynomials in $w$. Globally, we really should think of the complex coordinates $(x,y,z)$ as being sections of the bundles $(L^{\oplus 2}, \ L^{\oplus 3}, L)$ where $L = \cO_{\IP^1}(2)$ is the anti-canonical bundle of $\IP^1$, but we will work in the $z=1$ affine patch for simplicity.

Now, every isomorphism class of an elliptic curve $y^2 = 4x^3 - g_2 x  - g_3$  is determined by the famous $j$-invariant (into which this book sadly does not have the space to go) $j := g_2^3 / (g_2^3 - 27 g_3^2)$.
Thus, for an elliptic surface, the j-invariant becomes an meromorphic function in $w$, as a map from $\IP^1$ to $\IP^1$:
\begin{equation}
j(w) := \frac{g_2(w)^3}{g_2(w)^3 - 27 g_3(w)^2} \ , \qquad
\Delta(w) := g_2(w)^3 - 27 g_3(w)^2 \ .
\end{equation}
It is a fascinating fact that this is a clean, trivalent Belyi map (q.v.~\S\ref{ap:dessin} and \cite{He:2020eva}).
When the denominator $\Delta$ vanishes the curve (elliptic fiber) becomes singular.
This can be easily checked: consider the RHS $f = 4x^3 - g_2 x  - g_3 = 0$ and $f' = 12 x^3 - g_2$. the elliptic curve becomes singular when $f=f'=0$, which upon eliminating $x$, gives the condition $\Delta = 0$.

Kodaira then proceeded to classify all possible singularity types by considering what happens to $g_2$ and $g_3$ on the singular locus $\Delta(w) = 0$.
To give a trivial example, suppose $y^2 = 4 x^3 - w x$ so that $g_2 = w$, $g_3 = 0$, and $\Delta = w^3$.
The discriminant $\Delta$ vanishes at 0, so the singular locus is the point $\{w=0\}$, where the order of vanishing of $g_2$ is 1, that of $g_3$ is infinity, and that of $\Delta$, 3.
A simple transformation renders the equation to be $(y/2)^2 = x(x^2 - w/4)$, which, upon redefining $y/2 =  s$, $u=x$ and $v = x^2 - w/4$ gives the equation $s^2 = uv$, which is the surface singularity $A_1$ (q.v.~\eqref{duValF}).

In general, massaging the Weierstrass form into surface singularity types is highly non-trivial and requires the so-called Tate-Nagell algorithm. The insight of Kodaira is the classification of the complete singularity types of the elliptic fibration.
Defining ${\rm Ord}(f(w))$ as the order of vanishing of a function $f(w)$ at $w_0$, i.e., the Taylor series of $f(w)$ starts with $\cO(w - w_0)^{{\rm Ord}(f(w))}$, we have
\begin{theorem}[Kodaira]
The elliptic fibration types are determined by the order of vanishing of $g_2, g_3, \Delta$ at the singular locus $\Delta = 0$, and are the following
\[
\begin{array}{|c|c|c|c|c|} \hline
   {\rm Ord}(g_2) 
 & {\rm Ord}(g_3)
 & {\rm Ord}(\Delta)
 & \mbox{Kodaira Notation}
&  \mbox{Singularity type}
\\ \hline
\ge 0&\ge 0&0& \mbox{smooth}& -  \cr
 0   &    0&n&I_n   & A_{n-1} \cr
\ge 1&    1&2&II    & \mbox{none} \cr
\ge 1&\ge 2&3&III  & A_1     \cr
\ge 2&    2&4&IV    & A_2    \cr
    2&\ge 3&n+6&I^*_n&D_{n+4} \cr
\ge 2&    3&n+6&I^*_n&D_{n+4} \cr
\ge 3&    4&8& IV^*   & E_6   \cr
    3&\ge 5&9&III^*   & E_7  \cr
\ge 4&    5&10&II^*    & E_8   \cr
\hline
\end{array}
\]
\end{theorem}
Beautifully, we see another emergence of the ADE meta-pattern (q.v.~\S\ref{s:mckay}).

In the above, we illustrated with an elliptic surface as a fibration over $\IP^1$. 
For 3-folds, the Weierstrass coefficients will depend on more base variables. 
In the text in \S\ref{s:ellipticCY3}, the base types are enumerated into 4 families of surfaces.
Suppose their coordinates are $(w_1, w_2)$, then $g_2, g_2, \Delta$ will be polynomials in these.

\chapter[Gr\"obner Bases]{Gr\"obner Bases: The Heart of Computational Algebraic Geometry}
\label{ap:gb}
As much as almost any problem in linear algebra begins with Gaussian elimination, when encountering a system of polynomials - to which almost any query in algebraic geometry be reduced - one begins with the Gr\"obner Basis.
This analogy is actually strict: for a system of polynomials all of degree 1 (i.e., back to a linear system), the Gr\"obner basis (in elimination ordering) is Gaussian elimination to triangular form.

As promised in the text, we cannot possibly not go into a brief digression into Gr\"obner bases, which lies at the root of all the algorithms used in this book, the reader is referred to \cite{CLO} for a marvellous account of the subject of computational geometry.
For diehards in {\sf Mathematica}, a package has been written \cite{Gray:2008zs} 
to link it with {\sf Singular} so that much of the algorithms can be performed with a more familiar front-end.

As always, we exclusively work over the polynomial ring $\IC[x_i]$, and with $x_i \in \IC$, which greatly simplifies things.
First, we need to define an ordering of the monomials, denoted as $\textbf{x}^\alpha \prec \textbf{x}^\beta$ where $\textbf{x}^\alpha$ is the standard vector exponent notation understood to be the monomial $x_1^{\alpha_1} x_2^{\alpha_2} \ldots x_n^{\alpha_n}$.
This is a {\em total order} in the sense that if $u \preceq v$ and $w$ is any other monomial,  then $uw \preceq vw$ (it is in fact furthermore well-ordered in that $1 \preceq u$ for any $u$).
There are many ways to define such an ordering and indeed the Gr\"obner basis is dependent on such a choice.
The final conclusions in the geometry must, of course, be independent of any ordering choice.
The most common choices of ordering are
(we will use the four monomials $\{x_1x_2^2x_3, x_3^2, x_1^3, x_1^2 x_2^2 \}$ to illustrate each case):
\begin{description}
\item[lex] or Lexicographic ordering, which, like a dictionary, compares the exponent of $x_1$, and if equal, proceeds to $x_2$,  etc.,
	so that the power of $x_{i<j}$ dominates over $x_j$; 
	e.g., $x_1^3 \succ x_1^2 x_3^2 \succ x_1x_2^2x_3  \succ x_3^2$;
\item[grlex] or Graded (or degree)-lexicographic ordering, which compares {\em total degree} first and in the case of ties, proceeds to lex;
	e.g., $x_1^2x_3^2 \succ x_1x_2^2x_3 \succ x_1^3 \succ x_3^2$ (tie on highest degree, 4, in the first two, and proceeded to lex so that $x_1$ dominated);
\item[grevlex] or Graded (or degree) reversed lexicographic ordering, which compares total degree first and then uses the reversed order of lex for ties;
	e.g., $x_1x_2^2x_3 \succ x_1^2x_3^2 \succ  x_1^3 \succ x_3^2$;
\item[lexdeg] or Elimination (or block) ordering, which divides the variables into ordered blocks, each of which has a chosen ordering, usually grevlex;
	e.g., suppose we wish to eliminate $x_1$, we create two blocks:  $\{x_1\}$ and $\{x_2, x_3\}$ so that lex ordering is considered for $\{x_1\}$ before $\{x_2, x_3\}$;
	e.g., $x_1^3 \succ x_1^2 x_3^2 \succ x_1x_2^2x_3  \succ x_3^2$.
\end{description}

With a monomial ordering fixed, we have a definite {\em leading term} of any polynomial $f$, denoted $LT(f)$, which is the monomial, together with its coefficients, maximal according to $\succ$.
Suppose we are given $F = \{f_i\}$, a system of multivariate polynomials.
The Gr\"obner basis $G$ of $F$ is obtained by the {\red {\bf Buchberger Algorithm}}  as follows.
\begin{enumerate}
\item Set $G := F$ to initialize;
\item For every pair $f_i, \ f_j \in G$, find the leading terms $LT(f_i)$ and $LT(f_j)$, as well as their least common multiple \footnote{
While finding the LCM of two polynomials involves a generalization of the Euclidean division algorithm, since we are only finding the LCM of two monomials, this is easily done by comparing exponents.
}  $a_{ij} := LCM\left(LT(f_i),\ LT(f_j) \right)$;
\item Compute the S-polynomial $S_{ij} := \frac{a_{ij}}{LT(f_i)} f_i - \frac{a_{ij}}{LT(f_j)} f_j$ so that the leading terms cancel by construction;
\item Divide $S_{ij}$ by every element of $G$ to see if there is a remainder, i.e., write $S_{ij} = \sum\limits_{g \in G} p_g g + r$. This is called reduction of $S_{ij}$ over the set $G$. The remainder $r$ is so that none of its terms divides any of the leading terms from $p_g$, in analogy to arithmetic division.
\item If there is non-trivial remainder: $r \neq 0$, then add $r$ to $G$;
\item Repeat until all pairs are considered in the augmented $G$ (i.e., including the generated remainders included into $G$ as we go along) and until no new remainders are generated.
\end{enumerate}
The final list of polynomials $G$ (which in general will have more elements than $F$) is the Gr\"obner basis.

There are many implementation of the Buchberger Algorithm, exemplified by the {\sf GroebnerBasis[ ]} command in {\sf Mathematica}, the {\sf gb( )} command in {\sf Macaulay2}, the {\sf groebner( )} comand in {\sf Singular}.
Let us illustrate with a very simple example, step by step:
\begin{enumerate}
\item INPUT: $G = F = \{-12 xy  + 3 y^2, \ 4x - 6y\}$, with lex ordering $x \succ y$, the leading terms are 
$LT(F_1) = -12xy$ and $LT(F_2) = 4x$;
	
\item There is one S-polynomial between the 2 elements of $I$. First, $a_{12} = LCM(xy, x) = xy$ (where the polynomial LCM is done without the coefficient, but this a mere convention and including the coefficient will just give an overall numerical factor), so
$S_{12} = \frac{a_{12}}{LT(F_1)} F_1 - \frac{a_{12}}{LT(F_2)} F_2 = \frac{5y^2}{4}$.
Note that here, by design, the leading terms cancelled;
Reducing $S_{12}$ by the 2 elements of $G$ gives non-trivial remainder for both, so $S_{12}$ should be kept.

\item At this point $G$ is augmented to $G = \{-12 xy  + 3 y^2, \ 4x - 6y, \ \frac{5y^2}{4}\}$, with leading terms
$LT(G_{1,2}) = LT(F_{1,2}), LT(G_3) = G_3$.
We now have 2 new S-polynomials to compute:
$a_{13} = xy^2 \leadsto G_{13} = -\frac{y^3}{4}$
and 
$a_{23} = xy^2 \leadsto G_{23} = -\frac{3 y^3}{2}$.
Both $G_{13}$ and $G_{23}$, when reduced on $G$ have zero remainder because $a_{13} = -\frac15 y G_3$ and $a_{23} = -\frac65 G_3$.
Thus nothing new can be added and $G$ is the final Gr\"obner basis.
\end{enumerate}

\section{An Elimination Problem}
Of the many wonders of the Gr\"obner basis, let us only illustrate one, viz., elimination, on our favourite quintic from
\eqref{specialQ} (note that we have for convenience included a factor of 5 in front of the complex structure parameter $\psi$)
\begin{equation}
Q := \sum\limits_{i=0}^4 z_i^5 - 5 \psi \prod\limits_{i=0}^4 z_i \ , \quad \psi \in \IC \ .
\end{equation}
Suppose we wish to check the conditions on $\psi$ for which $Q$ is non-singular.
To do this, we compute the Jacobian of $Q$ and form an ideal $\cI$ together with $Q$:
\begin{equation}
\cI = \left\{
\sum\limits_{i=0}^4 z_i^5 - 5 \psi \prod\limits_{i=0}^4 z_i \ , \
\left( 5 z_j^4 - 5 \psi \prod\limits_{k=0, k\neq j}^4 z_k \right)_{j=0,\ldots, 4}
\right\} \ .
\end{equation}
We now consider lexdeg ordering by having two blocks: variables $z_{1, \ldots, 4}$ and $\{z_0\}$, leaving $\psi$ as a free complex parameter, and establish the Gr\"obner basis $G(\cI)$ for the ideal $\cI$, which we then intersect with $\IC[z_0]$.
That is,  we eliminate the variables $z_{0, \ldots, 4}$.
We find that:
\begin{equation}\label{smoothQ}
G(\cI) \cap \IC[z_0] = (\psi^5 - 1) z_0^{16} \ .
\end{equation}
Suppose we are at a generic point $z_0 \neq 0$, this means that if $\psi$ is not a 5-th root of unity, then the above is not zero, meaning that there are no simultaneous solutions to $Q$ and its Jacobian vanishing.
In summary then, $Q$ is smooth so long as $\psi^5 \neq 1$.

\section{Hilbert Series}\label{ap:HS}
The Hilbert Series is an important quantity that characterises an algebraic variety.  
It is not a topological invariant in that it depends on the embedding under consideration \cite{CLO,m2book}.
For a complex variety $M$ in $\IC[x_1,...,x_k]$, the Hilbert series is the generating function for the dimensions of the graded pieces:
\begin{equation}
H(t; M) = \sum\limits_{i=0}^{\infty} (\dim_{\IC} M_i) t^i ~,
\end{equation}
where $M_i$, the $i$-th graded piece of $M$ can be thought of as the number of independent degree $i$ polynomials on the variety $M$.
The coefficients $\dim_{\IC} M_i$ (as a function of $i$) is known as the {\blue Hilbert function}.
We remark that ordinarily, the Hilbert series is defined for {\it projective} varieties, especially for varieties defined in weighted projective space. 
Since much of the calculations performed in the physics literature is for {\it affine} varieties, we ``decompactify'' by considering the affine cone over the projective variety for which we compute the Hilbert series, but retain the grading (weights).

An important computational fact about the Hilbert series is a classic result due to Macaulay (q.v.~\cite{HScompute}):
\begin{theorem}
The Hilbert series of the initial ideal $in(I)$, the ideal generated by leading terms $\left<LT(f_i) \right>$, is the same as that of the ideal $I = \left<f_i\right>$ itself.
\end{theorem}
This provides a Gr\"obner basis and practical method to calculate the Hilbert series for any algebraic variety.

Let us turn to some examples.
Consider the simplest case of $M = \IC$ (as a cone over a point if you wish), given as the maximal spectrum $\IC = \mbox{Spec}(\IC[x])$, i.e., $\IC$ being parametrized by a single complex number $x$.
Clearly, at degree $i$, there is only a single monomial $x^i$.
Thus, $\dim_{\IC} M_i = 1$ for all $i \in \IZ_{\geq 0}$ so that the Hilbert series becomes $H(t; \IC) = (1-t)^{-1}$.
In general, we have that
\begin{equation}\label{HSCn}
H(t; \IC^n) = (1-t)^{-n} \ .
\end{equation}
For multi-graded rings with pieces $X_{\vec{i}}$ and grading $\vec{i}=(i_1,\dots,i_k)$, the Hilbert series takes the form $H(t_1,\dots,t_k; M) = \sum\limits_{\vec{i}=0}^{\infty} \dim_{\mathbb{C}}(X_{\vec{i}} ) t_1^{i_1}\dots t_k^{i_k}$ and becomes multi-variate.
This is sometimes called refinement \cite{pleth} and one could ``unrefine'' it to a univariate Hilbert series by, for instance, setting all $t_k = t$.

A useful property of $H(t)$ is that for algebraic varieties it is always rational function in $t$
and can be written in two ways: 
\begin{equation}\label{hs12} 
H(t; M) = \left\{
  \begin{array}{ll}
  \frac{Q(t)}{(1-t)^k} \ , & \mbox{ Hilbert series of the first kind} ~;\\
  \frac{P(t)}{(1-t)^{\dim({\cal M})}} \ , & \mbox{ Hilbert series of the second kind} ~.  
  \end{array} \right.  
  \end{equation} 
Importantly, both $P(t)$ and
$Q(t)$ are polynomials with {\em integer} coefficients.  The powers of
the denominators are such that the leading pole captures the dimension
of the embedding space and the manifold, respectively.

For a Hilbert series in second form, 
\begin{equation}\label{hilb2prop} 
H(t; M) =
\frac{P(1)}{(1-t)^{\dim(M)}} + \ldots ~, \qquad P(1) = {\rm  degree}(M) ~.  
\end{equation} 
  In particular, $P(1)$ always equals the degree of the variety.
We recall that when an ideal is
  described by a single polynomial, the degree of the variety is
  simply the degree of the polynomial.  In the case of multiple
  polynomials, the degree is a generalisation of this notion: it is the number of points at which a generic line intersects the
  variety.

In the context of \S\ref{s:coni} in the text, one of the important expansions of the Hilbert series is a Laurent
expansion about $1$.
When $M$ is a non-compact Calabi-Yau 3-fold, as a cone over a Sasaki-Einstein base $Y$.
The coefficient of the leading pole in the Laurent expansion can be
interpreted as the volume of $Y$.
This volume is in turn related to the central charges of supersymmetric gauge theory
({\em cf.}~ \cite{SEvol, Forcella:2008bb,He:2017gam}).

\paragraph{Molien Series: }
When $M$ is an orbifold (not necessarily Calabi-Yau), of the form $\IC^n/G$ for some finite group $G$ acting on the $n$ coordinates, the Hilbert series is just the Molien \cite{sturmfelsInv} series for $G$:
\begin{equation}\label{molien}
H(t; \IC^n/G) = \frac{1}{|G|}\sum_{g \in G} \frac{1}{\det(\II - t g)} \ ,
\end{equation}
where all group elements $g$ are $n \times n$ matrices denoting the action on the coordinates.
We can immediately check that for $G = \II$, the trivial group, we have $|G| = 1$, and that 
$\det(\II - t g) = \det((1-t) \II_{n\times n}) = (1-t)^n$ and the Molien series agrees with the Hilbert series for $\IC^n$ in \eqref{HSCn}.

For reference, we tabulate the Hilbert series for the {\blue ADE} subgroups of $SU(2)$, with
$\omega_n := e^{\frac{2\pi i}{n}}$ and
\begin{equation}
\begin{array}{l}
S := \frac12\mat{-1+i & -1+i \\ 1+i & -1-i}, \quad
T := \mat{i & 0 \\0& -i}, \\
U := \frac{1}{\sqrt{2}}\mat{1+i & 0 \\ 0 & 1-i}, \quad
V := \mat{\frac{i}{2} & \frac{1-\sqrt{5}}{4}- i \frac{1+\sqrt{5}}{4} \\
  -\frac{1-\sqrt{5}}{4}-i \frac{1+\sqrt{5}}{4} & -\frac{i}{2}} \ ,
\end{array}
\end{equation}
as follows:
\begin{equation}\label{HSADE}
\begin{array}{|c|c|c|c|c|} \hline
G & |G| & \mbox{Generators} & \mbox{Equation}
& \mbox{Molien/Hilbert } HS(t; G) \\ \hline
\hat{A}_{n-1} & n & \gen{\mat{\omega_n & 0 \\ 0 & \omega_{n}^{-1}}} &
uv = w^n
&
{\Large \mbox{$\frac{(1+t^n)}{(1-t^2)(1-t^n)}$}}
\\ \hline
\hat{D}_{n+2} & 4n & \gen{\mat{\omega_{2n} & 0 \\ 0 & \omega_{2n}^{-1}},
  \mat{0 & i \\ i & 0}} &
u^2 + v^2w = w^{n+1}
&
{\Large \mbox{$\frac{(1+t^{2n+2})}{(1-t^4)(1-t^{2n})}$}}
\\ \hline
\hat{E}_6 & 24 & \gen{S,T} & u^2+v^3+w^4=0
&
{\Large \mbox{$\frac{1 - t^4 + t^8}{1 - t^4 - t^6 +  t^{10}}$}}
\\ \hline
\hat{E}_7 & 48 & \gen{S,U} & u^2+v^3+vw^3=0 &
\begin{array}{c}\\
{\Large \mbox{$\frac{1 - t^6 + t^{12}}{1 - t^6 - t^8 + t^{14}}$}}
\end{array}
\\ \hline
\hat{E}_8 & 120 & \gen{S,T,V} & u^2+v^3+w^5=0 &
\begin{array}{c}\\
{\Large \mbox{$\frac{1 + t^2 - t^6 - t^8 - t^{10} + t^{14} + t^{16}}
{1 + t^2 - t^6 - t^8 - t^{10} - t^{12} + t^{16} + t^{18}}$}}
\\
\end{array}
\\ \hline
\end{array}
~\\
\end{equation}

\paragraph{Toric Hilbert Series: }
When $M$ is toric Calabi-Yau $n$-fold as prescribed by a toric diagram which is a convex lattice polyhedron $\Delta_{n-1}$, its Hilbert series is also easy to compute \cite{SEvol}.
The fully refined (multi-graded) version is succinctly obtained from the triangulation of $\Delta_{n-1}$ as follows 
\begin{equation}\label{toricHS}
  H(t_1,\dots,t_n ; M(\Delta_{n-1})) =
  \sum\limits_{i=1}^{r} \prod\limits_{j=1}^{n} (1-\vec{t}^{~\vec{u}_{i,j}})^{-1} \ ,
\end{equation}
Here, the index $i=1,\dots,r$ runs over the $n-1$-dimensional simplices in the (fine, regular, stellar) triangulation and $j=1,\dots,n$ runs over the faces of each such simplex. Each $\vec{u}_{i,j}$ is an integer $n$-vector, being the outer normal to the $j$-th face of the fan associated to $i$-th simplex. 
$\vec{t}^{~\vec{u}_{i,j}} := \prod\limits_{a=1}^{n} t_a^{u_{i,j}(a)}$ multiplied over the $a$-th component of $\vec{u}$.

\paragraph{General Case: }
For a general variety, defined by a polynomial ideal, one first finds the Gr\"obner basis.
Then, the Hilbert function $\dim_{\IC} M_i$ is the number of monomials of degree $i$ that are not a multiple of any leading monomial in the Gr\"obner basis.

\paragraph{The Plethystic Programme: }
As discussed in the very beginning of Chapter 2, there is an intimate relation between the geometry of representation variety of the quiver and the gauge invariants of the quantum field theory.
A so-called plethystic programme \cite{pleth} was established to harness the Hilbert series of $\cM(\cQ)$ which gave
some intriguing properties of the Hilbert series of algebraic varieties in general.
We will present some key points as observations since a rigourous treatment is yet to be fully administered.
We begin with a few definitions:
\begin{definition}
  Given a smooth function $f(t)$, its plethystic exponential \cite{fulton} is the formal series
  \[
  PE[f(t)] = \exp\left( \sum_{n=1}^\infty \frac{f(t^n) -
  f(0)}{n} \right) \ .
  \]
\end{definition}
One can readily check from this definition by direct manipulation of the series expansions (assuming reasonable regions of convergence), that
\begin{proposition}
  If $f(t)$ has Taylor series $f(t) = \sum\limits_{n=0}^\infty a_n t^n$, then
  \begin{itemize}
  \item There is an Euler-type product formula\\
    $g(t) = PE[f(t)] = \prod\limits_{n=1}^\infty (1-t^n)^{-a_n}$;
  \item There is explicit inverse (called plethystic logarithm) such that\\
    $f(t) = PE^{-1}(g(t)) = \sum\limits_{k=1}^\infty \frac{\mu(k)}{k} \log (g(t^k))$
  \end{itemize}
\end{proposition}
In the above, $\mu(k)$ is the standard M\"obius function for $k \in \IZ_+$ which is 0 if $k$ has repeated prime factors and is $(-1)^n$ if $k$ factorizes into $n$ distinct primes (also with the convention that $\mu(1)=1)$.
From the physical perspective, the plethystic exponential relates single-trace to multi-trace operators in the gauge theory, as mentioned in the dictionary between quiver representation and supersymmetric gauge theory in \S\ref{s:geoAdS}.

Now, let us apply this formalism to a Hilbert series  $H(t; M)$ (whose Taylor coefficients are by definition non-negative integers):
\begin{observation}\label{plog}
  Given Hilbert series $H(t; M)$ of an algebraic variety $M$, the plethystic logarithm is of the form
  \[
  PE^{-1}[H(t; M)] = b_1 t + b_2 t^2 + b_3 t^3 + \ldots
  \]
  where all $b_n \in \IZ$ and a positive $b_n$ corresponds to a generator in coordinate ring of $M$ and a negative $b_n$, a relation.
  In particular, if $M$ is complete intersection, then $PE^{-1}[H(t; M)]$ is a finite polynomial.
\end{observation}

For example, our quadric hypersurface $\cC$ in $\IC^4$ from \eqref{coni} has Hilbert series
\begin{equation}
H(t; \cC) = (1 - t^2) / (1 - t)^4 = \sum\limits_{n=0}^\infty (n+1)^2 t^n \ ,
\end{equation}
which can be obtained from \eqref{toricHS} because $\cC$ is a toric variety.
One checks that
\begin{equation}
PE^{-1}[H(t; \cC)] = 4t - t^2 \ ,
\end{equation}
 meaning that there are 4 generators in degree 1 (corresponding to the 4 complex coordinately of $\IC^4$) with 1 relation at degree 2 (meaning that $\cC$ is a single quadratic hypersurface, and hence complete intersection, in $\IC^4$).
Thus, one could have retrieve this Hilbert series by simply writing down $4t - t^2$ and taking the plethystic exponential.

On the other hand, for Cone($dP_0$)$= \IC^3 / (\IZ/3\IZ)$, the Hilbert series is the Molien series
\begin{equation}
H(t; \IC^3 / (\IZ/3\IZ)) = \frac{1+7t+t^2}{(1-t)^3} = \sum\limits_{n=0}^\infty \frac12 (2 + 9 n + 9 n^2) t^n \ ,
\end{equation}
and its plethystic logarithm is
\begin{equation}
PE^{-1}[H(t; \IC^3 / (\IZ/3\IZ))] = 1 + 10 t + 28 t^2 + 55 t^3 + 91 t^4 + 136 t^5 + 190 t^6 + \ldots
\end{equation}
which is non-terminating, in congruence with the fact that the variety, as discussed in \S\ref{s:dP0}, is not complete intersection.
While for these examples, it seems $PE^{-1}$ only terminates for complete intersections, but how one might disentangle positive and negative contributions to each $b_n$ is, as far as we are aware, not in general known.

\chapter{Brane Tilings}\label{ap:dimers}
\begin{figure}[h!!!]
\begin{center}
$
\begin{array}{c}\includegraphics[trim=0mm 0mm 0mm 0mm, clip, width=5.5in]{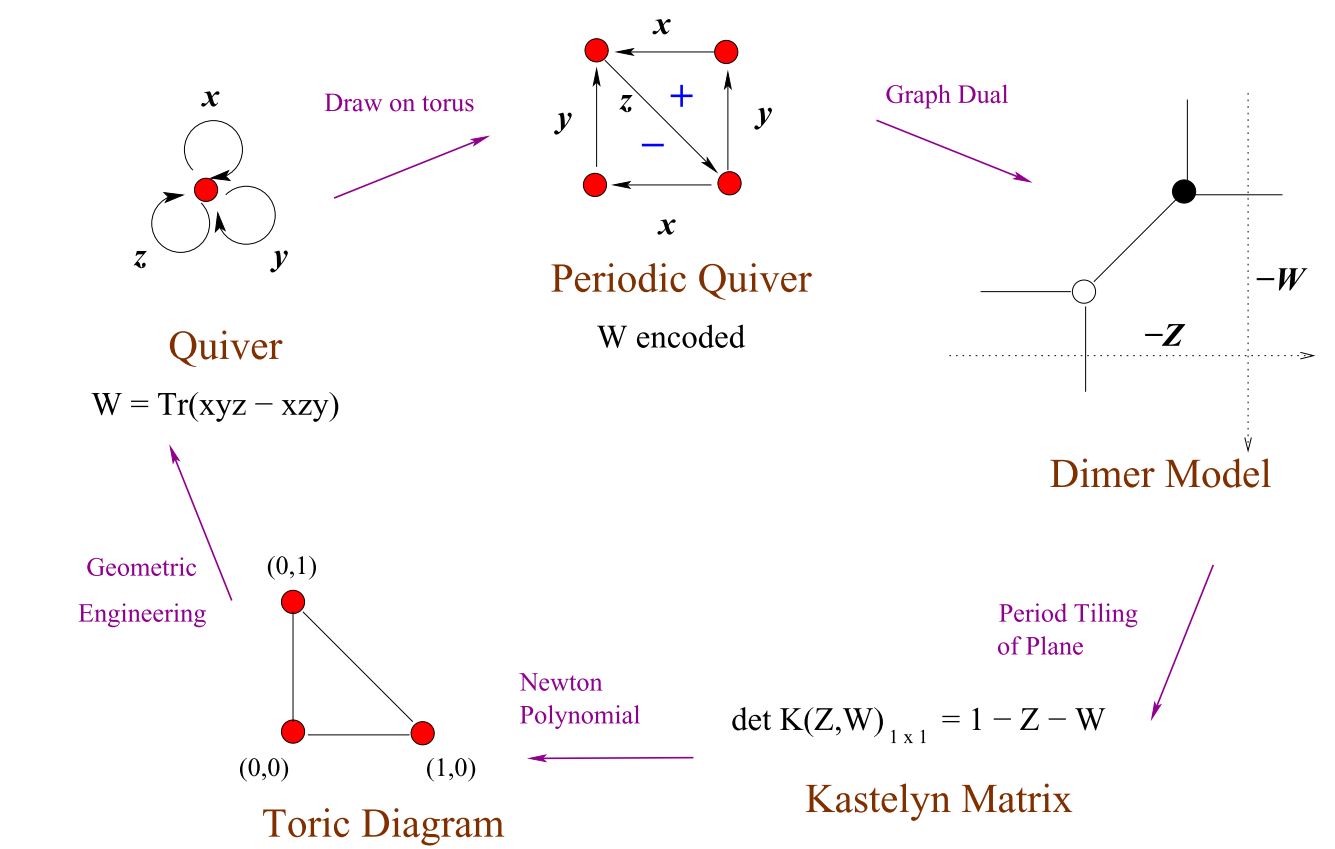}\end{array}
$
\end{center}
\caption{
{\sf 
The web of correspondences of the quiver, bipartite graph and toric Calabi-Yau moduli space for $\IC^3$.
}
}
\label{f:c3summary}
\end{figure}

In this brief appendix, we give a summary diagram of some of the inter-connections in the quiver-bipartite tiling-toric Calabi-Yau correspondence.
In Figure \ref{f:c3summary}, we present the quiver with superpotential for $\cN=4$ super-Yang-Mills theory, whose moduli space of representations trivially the non-compact toric Calabi-Yau 3-fold $\IC^3$.
This can be encoded into a single periodic quiver with the plus/minus term of the superpotential captured by the anti-clockwise/clockwise 3-cycles and identifying the 4 nodes into 1.
Subsequently, this can be graph-dualized to a trivalent bipartite graph on $T^2$ as the dimer model/brane-tiling.
The perfect matchings of a dimer model can be found by the Kastyleyn matrix whose determinant, as a bivariate polynomial, is exactly the Newton polynomial for the toric diagram for $\IC^3$.

\begin{figure}[h!!!]
\begin{center}
$
\begin{array}{c}\includegraphics[trim=0mm 0mm 0mm 0mm, clip, width=5.5in]{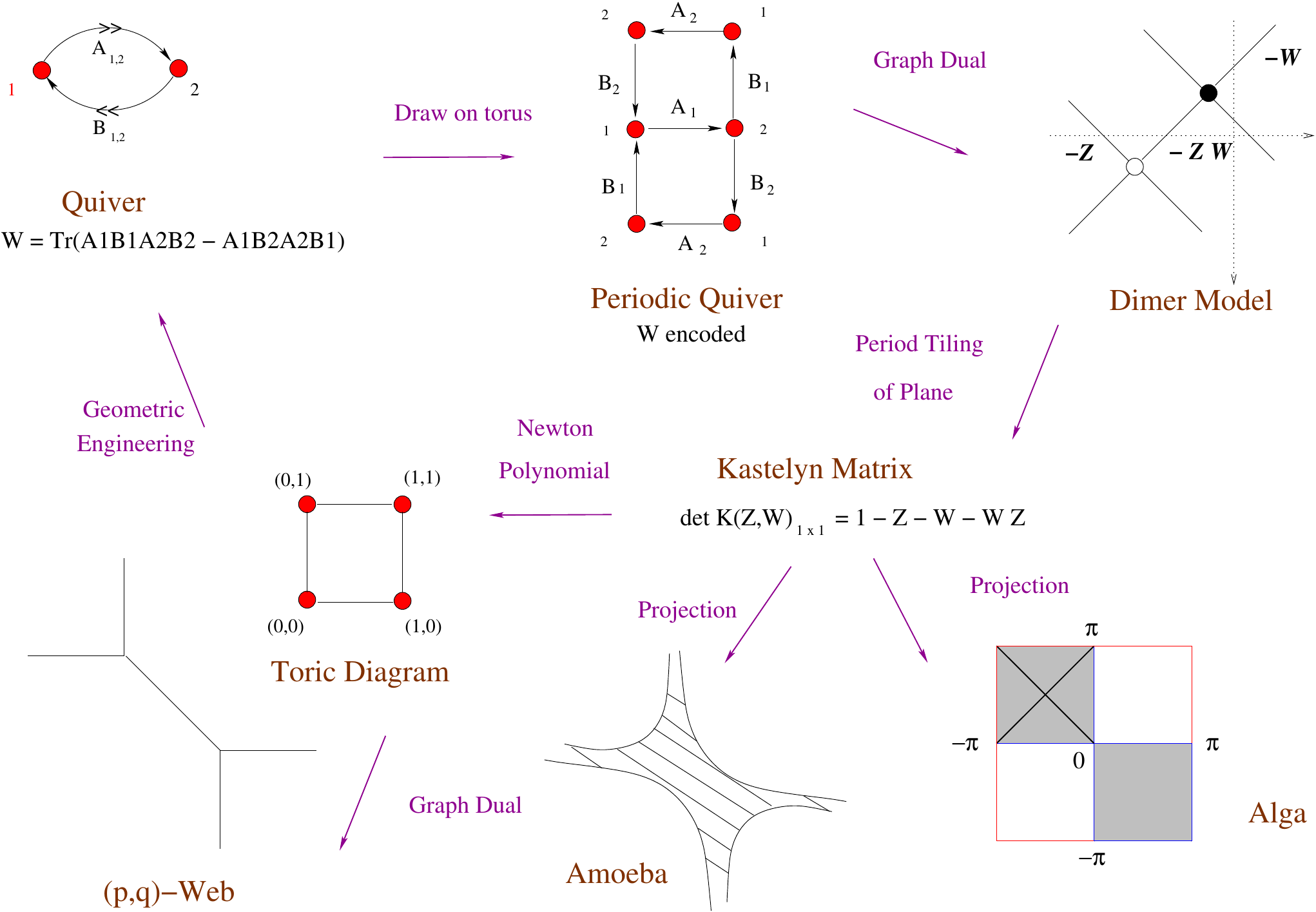}\end{array}
$
\end{center}
\caption{
{\sf 
The web of correspondences of the quiver, bipartite graph and toric Calabi-Yau moduli space for $\cC$.
}
}
\label{f:conisummary}
\end{figure}
Likewise in Figure \ref{f:conisummary}, we preesnt the same for the conifold, the quadric hypersurface in $\IC^4$, which is also a local toric Calabi-Yau 3-fold.
We have also added the two projections of the Newton polynomial, the so-called amoeba and alga (co-amoeba) projections.
The reader is referred to the brief lecture \cite{He:2016fnb} as well as the excellent monographs \cite{Kennaway:2007tq,Yamazaki:2008bt} for further details.

\section{Dessins d'Enfants}\label{ap:dessin}

It is impossible to resist, when speaking of bipartite graphs on Riemann surfaces, to at least introduce Grothendieck's {\bf dessin d'enfant}, one of the most profound pieces of modern mathematics \cite{leila}, residing in the intersections between number theory, geometry and combinatorics.
Indeed, in his own immortal words,
\begin{quote}
This discovery, which is technically so simple, made a very strong impression on me, and it represents a decisive turning point in the course of my reflections, a shift in particular of my centre of interest in mathematics, which suddenly found itself strongly focused. {\em I do not believe that a mathematical fact has ever struck me quite so strongly as this one, nor had a comparable psychological impact.}

This is surely because of the very familiar, non-technical nature of the objects considered, of which any child's drawing scrawled on a bit of paper (at least if the drawing is made without lifting the pencil) gives a perfectly explicit example. To such a dessin we find associated subtle arithmetic invariants, which are completely turned topsy-turvy as soon as we add one more stroke.
\flushright{A.~Grothendieck}
\end{quote}

FIrst, there is the extraordinary theorem of Belyi \cite{belyi}:
\begin{theorem}
Let $\Sigma$ be a smooth compact Riemann surface, then $\Sigma$ has an algebraic model over $\overline{\IQ}$ IFF there is surjective map $\beta: \Sigma \to \IP^1$ ramified at exactly 3 points.
\end{theorem}
There are several points of note about this theorem.
First, the IFF translates the analytic property of ramification (i.e., in local coordinates the Taylor series of $\beta$ starts from order 2) to the number-theoretical property of the algebraic closure of $\IQ$.
Second, the proof, which uses classical methods essentially known to Riemann, arose as late as 1980.

Using $SL(2;\IC)$ on $\IP^1$, we can take the three branch points to be $(0,1,\infty)$.
Then the theorem states that $\Sigma$, defined as a polynomial in $\IC[x.y]$ (say, as a hyper-elliptic curve $y^2 = p(x)$), has coefficients as algebraic numbers IFF one can find a rational function $f(x,y) : \Sigma \to \IP^1$ whose Taylor series has no linear term at $0,1,\infty$.

Grothendieck's insight was to realize that the Belyi map gives an embedded graph on $\Sigma$ as follows:
\begin{definition}
  Consider $\beta^{-1}(0)$, which is a set of points on $\Sigma$ that can be coloured as black, and likewise $\beta^{-1}(1)$, white.
  The pre-image of any simple curve with endpoints 0 and 1 on $\IP^1$ is a bipartite graph embedded in $\Sigma$ whose valency at a point is given by the ramification index (i.e., order of vanishing of Taylor series) on $\beta$.
  This is the {\bf dessin d'enfant}.
\end{definition}
Restricted by Riemann-Hurwitz, $\beta^{-1}(\infty)$ is not an independent degree of freedom, but is rather taken 1-to-1 to the faces in the bipartite graph, with the number of sides of the polygonal face being twice the ramification index.
One can succinctly record the valencies as
\[
\left[
  r_0(1), r_0(2), \ldots, r_0(B) \ \big| \
  r_1(1), r_1(2), \ldots, r_1(W) \ \big| \
  r_\infty(1), r_\infty(2), \ldots, r_\infty(I)
  \right]
\]
where $r_0(j)$ is the valency (ramification index) of the $j$-th black node, likewise $r_1(j)$, that for the $j$-th white node, and $r_\infty(j)$, half the number of sides to the $j$-th face.
This nomenclature is called the {\em passport} of the dessin; it does not uniquely determine it since one further needs the connectivity between the white/black nodes, but it is nevertheless an important quantity.
The {\em degree} of the Belyi map is the row-sum $d = \sum\limits_j r_0(j) = \sum\limits_j r_1(j) = \sum\limits_j r_\infty(j)$ and is the degree of $\beta$ as a rational function.
Finally, Riemann-Hurwitz demands that
\begin{equation}
2g-2 = d - (B+W+I)
\end{equation}
where $g$ is the genus of $\Sigma$ and $B,W,I$ are respectively the number of pre-images of $0,1,\infty$.
Of course, our focus will be on $g=1$, balanced (i.e., $B=W$) dessins, for which the total number of pre-images of $0,1,\infty$ is equal to the degree.
We emphasize that the dessin is {\em completely rigid}, fixing the coefficients in the defining polynomial of $\Sigma$ to be specific - and oftentimes horrendous - algebraic numbers with no room for complex parameters.
That is, given a bipartite graph on a Riemann surface, there is, up to coordinate change (birational transformation), a unique algebraic model and associated Belyi map which gives the graph as a dessin; even the slight change in the numerical coefficients in either will result in an utterly different graph.

In passing we mention that there is a purely combinatorial way to encode dessins - though, of course, in bypassing the analytics of Belyi maps, the algebro-number-theoretic information is somewhat lost.
Nevertheless, the ramification structure is captured completely, unlike the rather coarse encoding by passports.
One simply labels all edges, say $1,2, \ldots, d$, of the graph on $\Sigma$, and work within the symmetric group $\mathfrak{S}_n$ using the standard cycle notation for the elements.
Choose a {\em single} orientation, say clockwise, and for each black node, write the cycle $(e_1 e_2 \ldots e_i)$ going clockwise where $e_j$ is the integer label for the said edge.
Then form an element $\sigma_B \in \mathfrak{S}_d$ which is the product over all cycles for all black nodes.
Similarly, do this for the white nodes (also going clockwise for the edges), forming $\sigma_W \in \mathfrak{S}_d$.
Finally, define $\sigma_\infty \in \mathfrak{S}_d$ such that
\begin{equation}
\sigma_B \sigma_W \sigma_\infty = Id \ ;
\end{equation}
this so-called {\bf permutation triple} captures the structure (valency and connectivity) of the dessin. Indeed, when $\sigma_\infty$ is itself expressed in terms of product of cycles - as every element of a symmetric group ultimately must - each cycle is associated to a face of the dessin, with its length being half the number of sides of the polygon.

Let us re-consider the haxagon example of \eqref{c3dimer}.
Obverse the Belyi pair of elliptic curve $\Sigma$ and Belyi map $\beta$
\begin{equation}\label{BelyiC3}
y^2 = x^3 + 1 \ , \quad \beta = \frac12 ( 1 + y) \ .
\end{equation}
We see that the pre-image of 0 has $y = -1$, whence $(x,y) = \beta^{-1}(0) = (0,-1)$ on the elliptic curve $y^2 = x^3 + 1$. Choosing local coordinates $(x,y) = (0 + \epsilon, -1 + \delta)$ for infinitesimals $(\epsilon, \delta)$, we have that $-2 \delta = \epsilon^3$. Therefore locally the map is $\beta = \frac{\delta}{2} \sim -\frac{\epsilon^3}{4}$ and thus the ramification index for this single pre-image of 0 is 3.
Similarly, the pre-image of 1 has $y = +1$, i.e., $(x,y) = (0,1)$ is the single pre-image of 1, where local coordinates can be chosen as $(0 + \epsilon, 1 + \delta)$, so that $\beta \sim \frac{\epsilon^3}{4}$. Hence, the ramification index is also 3 for this single pre-image of 1.
Finally, $(\infty, \infty)$ is the pre-image of $\infty$ where the local coordinates $(\epsilon^{-1}, \delta^{-1})$ can be chosen so that $(x,y) \sim (\epsilon^{-2}, \epsilon^{-3})$. Hence, the ramification index at $\infty$ is also 3.
Thus, the passport is $[3 \big| 3 \big| 3]$ and the maps is degree 3.
In summary
\[
\begin{array}{|c|c|c|c|c|c|}
\hline
\mathbb{T}^2: y^2 = x^3 + 1 & \stackrel{\beta=\frac12(1+y)}{\longrightarrow} & \mathbb{P}^1 & \mbox{Local Coordinates on }\mathbb{T}^2 & \mbox{Ram.~Index}(\beta) \\ \hline\hline
(0,-1) & \sim -\frac14{\epsilon^3} & 0 & (x,y) \sim (\epsilon,-1-\frac12\epsilon^3) & 3 \\ \hline
(0,1) & \sim \frac14{\epsilon^3} & 1 & (x,y) \sim (\epsilon,1+\frac12\epsilon^3) & 3 \\ \hline
(\infty,\infty) & \sim \frac12 \epsilon^{-3} & \infty & (x,y) \sim (\epsilon^{-2},\epsilon^{-3}) & 3 \\ \hline
\end{array}
\]
Therefore, what we have is a bipartite graph on $T^2$ specified by the elliptic curve in \eqref{BelyiC3}, with a single pair of black/white nodes, each trivalent.

\chapter{Remembering Logistic Regression}\label{ap:regression}
It is expedient to recollect some elements of non-linear regression, especially with the logistic sigmoid function $\sigma(z) = (1 + \exp(-z))^{-1}$.
We will phrase this problem from an introductory statistics course in as much ML language as possible.

\begin{figure}[h!!!]
(a)
$
\begin{array}{c}\includegraphics[trim=10mm 0mm 0mm 0mm, clip, width=2.3in]{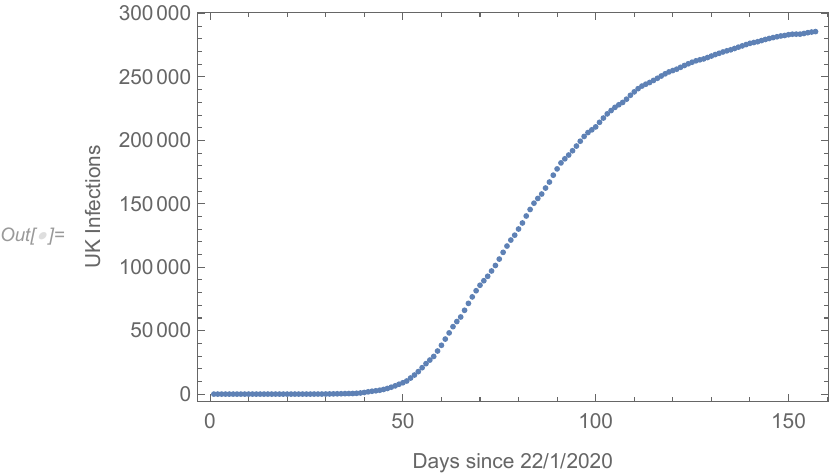}\end{array}
$
(b)
$
\begin{array}{c}\includegraphics[trim=10mm 0mm 0mm 0mm, clip, width=2.3in]{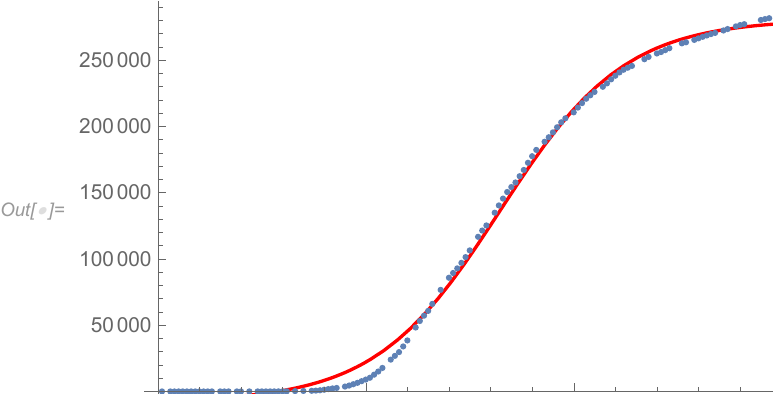}\end{array}
$
\caption{
{\sf 
(a) UK daily cumulative infection of COVID19 since Jan 22nd, 2020;
(b) A fit by a logistic sigmoid.
}
}
\label{f:covid}
\end{figure}

First, let us take some actual data. As it is rather topical, let us take some COVID19 data from the Johns Hopkins University COVID resource center \url{https://coronavirus.jhu.edu/}.
In Figure \ref{f:covid} (a), we plot the (cumulative)  number of infections $I$ in the UK since the beginning of records (Jan 22nd, 2020) by the day $t$. 
Thus, we a labeled dataset $\cD = \{ t \to I \}$, a total of 157 points.
Inspecting the curve, a logistic sigmoid function seems a good fit.
We introduce weights and biases
\begin{equation}
\sigma_{a,b,c,d}(t) = a (1 + b \exp(- c t) )^{-1} + d \ , \qquad (a,b,c,d) \in \IR \ .
\end{equation}
Ordinarily, we would try to fit the curve to the entire set of points; since there are only 4 parameters $(a,b,c,d)$, there is no danger of over-fitting here.
In a more complicated supervised ML situation, there might be thousands of parameters, and we thus adopt the strategy of cross-validation.
It is, of course, an overkill in the present example, but let us thus proceed for illustrative purposes.

We split $\cD$ into a training set $\cT$, a random sample of, say, $20\%$ data.
Then, we minimize a cost-function, which here can be taken to be the sum squared:
\begin{equation}
F(a,b,c,d) := \sum\limits_{ (t \to I) \in \cT } \left[ ( a (1 + b \exp(- c t) )^{-1} + d)  - I \right]^2 \ ,
\end{equation}
in order to find $(a,b,c,d)$.
For example, we find (on one random $20\%$ sample), 
$(a,b,c,d) = (288681.0, 241.259, 0.0666724,  -7933.96)$.
In part (b) of the Figure, we plot the fitted curve against the complement validation set $\cV = \cD - \cT$.
The missing points in the (blue) scatter plot are precisely the ones which we used for the training set.

While we see that the fit is fairly reasonable, we need to quantify a measure of ``goodness of fit''.
For continuous variables, such as $I$ here, a commonly used one is the {\bf coefficient of determinantion}, otherwise known as ``R-squared'', defined as
\begin{equation}
R^2 := 1 - \frac{\sum_i (y_i - f_i)^2}{\sum_i (y_i - \bar{y})^2 } \ ,
\end{equation}
where $y_i$ are the outputs (in the above, the list of $I$ values), $\bar{y}$ is the mean over $y_i$ and $f_i$ are the fitted values (in the above, the list of $\sigma(t)$).
The numerator $\sum (y_i - f_i)^2$ is thus, for the above example, our list of $F(a,b,c,d)$ at the optmized values of $(a,b,c,d)$.
A value of $R^2 =1$ would be a perfect fit, and a value of $R^2=0$ is the baseline where the model always predicts the mean.
In our example above, we find that $R^2 \sim 0.999$.
We remark that in ML, such measures are ordinarily only computed for the validation set.

What we have done in this example, was to phrase a simple non-linear regression in the terminologies of  ML. In fact, where there is a single neuron, with a sigmoid activation function, this problem is exactly what supervised ML is for a neural ``network'' with a single node.

\chapter[Homage to {\sf SageMath}]{A Computational Compendium: Homage to {\sf SageMath}}
As promised in the text, we demonstrate in this appendix some of the powers of {\sf SageMath}, an over-arching, open-source software which takes full advantage of the various software we have mentioned throughout the book.
As always, we illustrate with explicit examples, paying attention to how it links to the software.

The dedicated are encouraged to download the latest version of {\sf SageMath} onto their laptops from
\begin{center} \url{https://www.sagemath.org/} \end{center}
Those who wish for a more casual acquaintance and immediate calculation can go to the Cloud version at
\begin{center} \url{https://cocalc.com/} \end{center} 
One creates a free account, hits the ``Create Project'' tab, followed by the ``New'' tab, within which one then chooses the option to create a ``Sage worksheet''.
This brings us to the worksheet interface immediately.

\section{Algebraic Varieties}
We begin with some problems in standard algebraic geometry, and our go-to programme \cite{m2} is {\sf Macaulay2}, which we can easily call within {\sf SageMath}:
\begin{verbatim}
1. macaulay2('R = ZZ/101[x,y,z]');
2. macaulay2('testid = ideal(z*y^2 - x^3 - 4*x*z^2 - z^3)');
3. macaulay2('ell = Proj( R / testid )');
\end{verbatim}
In line 1, we first create a polynomial ring $R$ in 3 complex variables $(x,y,z)$, with coefficients in $\IZ/(101 \IZ)$.
In {\sf Macaulay2} it is customary to work with coefficients modulo some prime, here $101$, to avoid rapid growth of the size of coefficients. In other words, we use $\IZ/(101 \IZ)[x,y,z]$ to model $\IC[x,y,z]$. In case we hit bad primes, we could always redo the computation over a few different coefficient fields to make sure all geometric quantities stabilize.
That is, we could try \verb|ZZ/101[x,y,z]|, \verb|ZZ/541[x,y,z]|, etc. and repeat the calculation to make sure that we obtain the same answer.

In line 2, we create an ideal, called \verb|testid|, defined by a single (homogeneous) cubic in the 3 variables.
In line 3, the projective algebraic variety \verb|ell| is created from the coordinate ring \verb|R/testid| by taking \verb|Proj|.
If one wished to create an affine algebraic variety, one would have used \verb|macaulay2.Spec(R/testid)|.
To create an ideal defined by multiple polynomials, we simply list these polynomials separated by comma, e.g.,
\begin{verbatim}
4. macaulay2('testid2 = ideal(z*y^2 - x^3 - 4*x*z^2 - z^3, x+y+z)');
5. macaulay2('pt = Proj(R/testid2)');
\end{verbatim}

Indeed, \verb|ell| is now an elliptic curve as a hypersurface in $\IC\IP^2$ with homogeneous coordinates $[x:y:z]$.
If one wished to play with $\IC\IP^2$ itself, one would execute \verb|Proj(R)| directly without going to the quotient (coordinate) ring.
To quickly check that the dimensions are correct, we execute
\begin{verbatim}
6. macaulay2.dim(ell)
7. macaulay2.dim(pt)
\end{verbatim}
which return 1 and 0 respectively (these are understood to be the complex dimensions).

Let us check that the elliptic curve is smooth with
\begin{verbatim}
8. macaulay2('singularLocus(ell)');
\end{verbatim}
which returns \verb|Proj ( R / 1)|, meaning that the corresponding ideal is $\left<1\right> \in R$ which is clearly never vanishing. In other words, the singular locus is the null set.

We can further check that \verb|ell| is Calabi-Yau explicitly. First, we construct the canonical bundle
\begin{verbatim}
9. macaulay2('cb = cotangentSheaf(ell)');
10. macaulay2('can = exteriorPower(1, cb)');
11. macaulay2('rank(can)');
\end{verbatim}
where we check that the canonical bundle $\wedge^{1}T^*$, the top wedge product of the cotangent bundle $T^*$, is rank 1. Of course, here in dimension 1, this is an over-kill. In dimension $n$, we do need to compute $\wedge^{n}T^*$ using \verb|exteriorPower(n, _)|.

In particular, line 10 returns, for the canonical bundle $K_{ell}$ as $\cO_{ell}$, which is one of the equivalent definitions of Calabi-Yau-ness.
In any event, we can go to compute all the requisite cohomology groups
\begin{verbatim}
11. macaulay2('HH^0(can)');
12. macaulay2('HH^1(can)');
\end{verbatim}
giving, as expected, 1 and 1.

Finally, we can compute the Hilbert series of this variety, embedded in $\IP^2$:
\begin{verbatim}
13. macaulay2('hilbertSeries ideal(ell)');
\end{verbatim}
which returns $(1- T^3) / (1- T)^3$, being that of a single cubic in 3 (projective) variables with the same grading 1.

We make some parenthetical remarks here.
In the above we had all computations done {\it within} {\sf Macaulay2}.
One can actually pass variables freely {\it between} {\sf SageMath} and an external software, in which case, we can redo the above with (note the {\sf Python} syntax in commands such as \verb|macaulay2.ideal()|):
\begin{verbatim}
R = macaulay2('ZZ/101[x,y,z]');
x = macaulay2('x'); y = macaulay2('y'); z = macaulay2('z');
testid = macaulay2.ideal(z*y^2 - x^3 - 4*x*z^2 - z^3);
ell = macaulay2.Proj(R/testid);
cot = macaulay2.cotangentSheaf(ell);  
can = macaulay2.exteriorPower(1, cot);
macaulay2.cohomology(0, cot); macaulay2.cohomology(1, cot)
\end{verbatim}
The advantage of this approach, of course, would be the flexibility to manipulate these objects within {\sf SageMath}.

Second, we could just as easily go to {\sf Singular} as the preferential computational geometry software \cite{singular}, in which case the syntax for creating the variety and testing its dimension would be something like:
\begin{verbatim}
r = singular.ring(101, '(x,y,z)', 'lp');
testid = singular.ideal(z*y^2 - x^3 - 4*x*z^2 - z^3, x+y+z);
singular.dim(testid)
\end{verbatim}
The reader is referred to \cite{Gray:2008zs} for linking {\sf Singular} to {\sf Mathematica}.

\section{Combinatorics \& Toric Varieties}
Luckily, manipulations of polytopes and related toric varieties have been built into {\sf SageMath} so there is no need to even call external software.
In fact, all reflexive polytopes for dimension $n=2,3$ are also built in (the half billion $n=4$ cases are, unfortunately, too large a set).
Let us try $\IP^2$ as a toric variety.
There are 3 equivalent ways to define it:
\begin{verbatim}
1. p2dual = ReflexivePolytope(2,0);
2. p2 = p2dual.polar();
3. V1 = ToricVariety( NormalFan(p2) );
4. V2 = ToricVariety( Fan( [
  	           Cone([(1,0), (0,1)]), 
  	           Cone([(1,0), (-1,-1)]), 
  	           Cone([(0,1), (-1,-1)])] ) );
5. V3 = toric_varieties.P2();
\end{verbatim}
In lines 1 and 2, we pull up the 2-dimensional reflexive polytope in {\sf SageMath}'s database, number 0 (there are 16 reflexive polygons, so the index goes from 0 to 15; for \verb|ReflexivePolytope(2,n)|, $n$ ranges from 0 to 4318).
This happens to be the polar dual polytope to that for $\IP^2$, thus we take the dual back in line 2.
Next, in line 3, we construct the normal fan to this polygon and construct the toric variety \verb|V1|.
Alternatively, we can be explicit and write down the list of cones, and the subsequent fan, to obtain \verb|V2| in line 4.
Finally, $\IP^2$ is standard enough that it is in the small built-in  database of \verb|toric_varieties|.
One can readily check, e.g., by \verb|V1.fan().rays()|, or \verb|V3.fan().cones()|, that the 3 definition do coincide.

We could further visualize the fan (line 6) and the polygon (line 7) by
\begin{verbatim}
6. V1.fan().plot()
7. LatticePolytope(p2).polyhedron().plot()
\end{verbatim}
Moreover, we can check the dimension of the variety with \verb|dimesion(V1)|, giving 2 as required.

Next, we can start experimenting in algebraic geometry. One extracts (co-)homology classes by
\begin{verbatim}
8. hh = V1.cohomology_ring()
9. hh.gens()
\end{verbatim}
Line 8 produces the cohomology ring $H^*(\IP^2; \IZ)$. We know this to be $\IZ[x] / \left< x^3 \right>$, with generator $x$ corresponding to the class $H^2(\IP^2; \IZ)$; indeed, $\IC\IP^2$ is a 4-manifold, with $H^p(\IP^2; \IZ) = 0$ for $p = 1,3$ and $\IZ$ for $p = 0,2,4$.
One sees that line 9 produces a single generator.

From the above, one can proceed to intersection theory, and hence integration:
\begin{verbatim}
10. D = V1.divisor(0);
11. V1.integrate( hh(D)^2 );
\end{verbatim}
We create a divisor (co-dimension 1 sub-manifold) $D$, which for the toric variety, corresponds to the 3 rays.
In line 10, we take the first of them and then take the corresponding element in the cohomology ring \verb|hh|, a real 2-form. The square (wedge product) gives a 4-form, which can then be integrated over the whole manifold, here returning the (normalized) value of 1.

\section{Representation Theory}
The go-to program in finite group theory, another subject extensively discussed in this book, is perhaps {\sf GAP} \cite{gap}. Again, {\sf SageMath} readily incorporates the functionalities thereof.
We begin by creating a finite group. To be concrete, let us take the $\widehat{E_8}$ example from \eqref{HSADE}:
\begin{verbatim}
1. GG = gap('Group( [
   1/2*[[-1+E(4),-1+E(4)],  [[1+E(4),-1-E(4)]], [[E(4),0],[0,-E(4)]], 
   [[E(4)/2,(1-ER(5))/4-E(4)*(1+ER(5))/4], 
        [-(1-ER(5))/4-E(4)*(1+ER(5))/4,-E(4)/2]]   ]); ');
2. gap.IdGroup(GG);
3. gap.StructureDescription(GG);
\end{verbatim}
In line 1, we have called {\sf GAP} from within {\sf SageMath} and the command to create a finite group is simply
\verb|Group( )|.  The 3 matrix generators are $\left<S,T,V\right>$ as given in \eqref{HSADE}.
Note that in {\sf GAP}, $E(n)$ means the primitive $n$-th root of unity (hence $E(4)$ is their notation for $i$) and $ER(n)$ means $\sqrt{n}$.
In line 2, we use the command \verb|IdGroup( )| which identifies {\sf GAP}'s internal database of small groups.

As mentioned in the text, this is a database of {\it all} finite groups up to order 2000, except that of order 1024, a staggering total of 423,164,062 groups (there are, in addition, few more families).
The command returns \verb|[120, 5]|, which is number 5 (in their internal library) of the groups of order 120; the order is certainly correct.
We further identify the group using line 3, which returns \verb|SL(2, 5)|, which is the finite group $SL(2; \IF_5)$, the
special linear group defined on the finite field of 5 elements. This is indeed an equivalent description for the binary icosahedral group.
Matrix groups is only one of the many ways to encode groups. {\sf SageMath} and {\sf GAP} can as easily handle explicit presentation from free groups, or permutations in terms of cycle notation.

We could obtain standard group theoretic properties, such as
\begin{verbatim}
4. tt = gap.CharacterTable(GG);
5. gap.Display(tt);
6. gap.IsSimple(GG);
\end{verbatim}
where lines 4 and 5 display the ordinary character table of the group and line 6 checks whether it is a simple finite group.

To obtain the Molien series (Hilbert series) from the action by a particular representation, we execute:
\begin{verbatim}
7. irr = gap.Irr( tt );
8. gap.MolienSeries( irr[2] );
\end{verbatim}
where line 7 gives the list of all irreducible representations and line 8 takes the second (the defining 2-dimensional irrep) and  computes the Molien series according to \eqref{molien}.

\section{Number Theory \& More}
Of course, of all the branches of pure mathematics and of the tens of thousands of commands, we have only given a glimpse of {\sf SageMath}'s power and versatility, but hopefully enough to get the non-experts started with explorations and sufficient elements to allow them to experiment with the concepts presented in this book.
With the growing importance of the interactions between number theory and physics, as well as the beginnings of machine-learning in number theory \cite{Alessandretti:2019jbs,He:2020kzg}, we end our homage to {\sf SageMath} with some of its abilities therein to whet the reader's appetite.

The important LMFdB  (L-functions and modular forms database ), to which we alluded several times in the book \cite{lmfdb}, has much of its contents also incorporated into  {\sf SageMath}.
For instance
\begin{verbatim}
1. E = EllipticCurve('37a')
\end{verbatim}
returns the elliptic curve number \verb|'37a'| which is the so-called Cremona label.
We can see explicitly that this is the elliptic curve $y^2 + y = x^3 - x$ in Tate form.
One can extract many of our favourite quantities such as
\begin{verbatim}
2. E.analytic_rank()
3. E.lseries()
4. E.j_invariant()
5. E.conductor()
6. E.short_weierstrass_model()
\end{verbatim}
which are, respectively, the analytic rank, L-series, j-invariant, condutor and Weierstra\ss\ model.

\end{appendix}



\end{document}